\documentclass[12pt]{article}

\usepackage{graphicx,color}			
\usepackage{subcaption}			
\usepackage{flafter}			
\usepackage{amsmath}						
\usepackage{array}							
\newcolumntype{?}[1]{!{\vrule width #1}}	
\usepackage{booktabs} 
\usepackage{amsfonts} 
\usepackage[noabbrev]{cleveref} 
\usepackage{bbm} 
\usepackage[a4paper, total={6.5in, 9in}]{geometry} 
\usepackage[T1]{fontenc} 
\usepackage{float}  
    
\def\rf#1{(\ref{eq:#1})}
\def\lab#1{\label{eq:#1}}

\def\br{\begin{eqnarray}}
\def\er{\end{eqnarray}}
\def\be{\begin{equation}}
\def\ee{\end{equation}}
\def\({\left(}
\def\){\right)}

\def\rlx{\relax\leavevmode}

\def\ve{\varepsilon}

\def\tG{{\widetilde \Gamma}}

\newcommand{\sbr}[2]{\left\lbrack\,{#1}\, ,\,{#2}\,\right\rbrack}

\def\IZ{\rlx\hbox{\sf Z\kern-.4em Z}}
\def\IR{\rlx\hbox{\rm I\kern-.18em R}}
\def\IC{\rlx\hbox{\,$\inbar\kern-.3em{\rm C}$}}
\def\one{\hbox{{1}\kern-.25em\hbox{l}}}

\begin{document}
	\begin{titlepage}
		\vspace*{-1cm}

		\vskip 3cm

		\vspace{.2in}
		\begin{center}
			{\large\bf Quasi-integrability of deformations of the KdV equation}
		\end{center}
		
		\vspace{.5cm}
		
		\begin{center}
			F. ter Braak$^{\dagger}$, L. A. Ferreira$^{\dagger \dagger}$ and W. J. Zakrzewski$^{\dagger}$

			\vspace{.5 in}
			\small
			
			\par \vskip .2in \noindent
			$^{(\dagger)}$~Department of Mathematical Sciences,\\
			University of Durham, Durham DH1 3LE, U.K.\\
			
			\par \vskip .2in \noindent
			$^{(\dagger \dagger)}$Instituto de F\'\i sica de S\~ao Carlos, IFSC/USP,\\
			Universidade de S\~ao Paulo,  \\ 
			Caixa Postal 369, CEP 13560-970, S\~ao Carlos-SP, Brazil \\

			\normalsize
		\end{center}
		
		\vspace{.5in}	
		
		\begin{abstract}
			We investigate the quasi-integrability properties of various deformations of the Korteweg-de Vries (KdV) equation, depending on two parameters $\varepsilon_1$ and $\varepsilon_2$, which include among them the regularized long-wave (RLW) and modified regularized long-wave (mRLW)  equations. We show, using analytical and numerical methods, that the charges, constructed from a deformation of the zero curvature equation for the KdV equation, are asymptotically conserved for various values of the deformation parameters. By that we mean that, despite the fact that the charges do vary in time during the scattering of solitons, they return after the scattering to the same values they had before it. That property was tested numerically for the scattering of two and three solitons, and analytically for the scattering of two solitons in the mRLW theory ($\varepsilon_2=\varepsilon_1=1$). We also show that the Hirota method leads to analytical one-soliton solutions of our deformed equation for $\varepsilon_1 = 1$,  and any value of $\varepsilon_2$. We also mention some properties of soliton-radiation interactions seen in some of our simulations.
		\end{abstract}
	\end{titlepage}
	
	\section{Introduction}
	\setcounter{equation}{0}

	The objective of the present paper is to study the quasi-integrability properties of deformations of the Korteweg-de Vries (KdV) equation \cite{kdv,dickey} that include as particular cases the so-called regularized long-wave equation (RLW), proposed by  Peregrine~\cite{Peregrine} and T. B. Benjamin, J. L. Bona and J. J. Mahoney~\cite{Benjamin}, and also the modified regularized long-wave equation (mRLW) introduced and studied by J. D. Gibbon, J. C. Eilbeck and R. K. Dodd~\cite{Gibbon}. Concretely, the model we consider involves a scalar field $u$ satisfying the equation
	\begin{equation}
			u_t + u_x + \left[ \frac{\alpha}{2}\, u^2 +{\varepsilon}_2\, \frac{\alpha}{4} \, w_x \,v_t  + u_{xx} - \varepsilon_1 \left( u_{xt}  + u_{xx} \right)\right]_x =0 \,,	\label{5.05}
		\end{equation}
In which~$\varepsilon_1$,~$\varepsilon_2$, and~$\alpha$ are real parameters, and where 
\begin{equation}
			u = w_t = v_x  \,. \label{soliton_fields}
		\end{equation}

Note that the integrable KdV equation~\cite{kdv,dickey}, corresponds to the case where the deformation parameters vanish, {\it i.e.} $\varepsilon_1=\varepsilon_2=0$, and it is given by\footnote{The KdV equation as presented here is not in its standard form, but one can perform a change of variables to get the standard form of the KdV equation (see, for instance, the notes by M. Dunajski for more information~\cite{Dunajski}). The same applies for the RLW and mRLW equations. In fact, to get the original notation used by the authors one should choose $\alpha=12$, in (\ref{5.02}), and $\alpha=8$ in (\ref{mrlw}).}  
		\begin{equation}
			u_t + u_x + \left[ \frac{\alpha}{2} u^2 + u_{xx} \right]_x = 0 \,. \label{5.01}
		\end{equation}
		As is well known, this equation describes non-linear waves in shallow waters traveling in the positive direction of the $x$-axis only. If one considers the linearization of the KdV equation one finds that its traveling wave solutions satisfy a dispersion relation of the form $\omega=k-k^3$, and so the phase velocity $\omega/k=1- k^2$, and group velocity $\frac{d\,\omega}{d\,k}=1-3\,k^2$, which become negative, and in fact unbounded,  for large enough $k$. 
		
		Motivated by this fact, Peregine~\cite{Peregrine} and  T. B. Benjamin, et.al.~\cite{Benjamin} proposed the so-called regularized long wave equation (RLW) 
		\begin{equation}
		u_t + u_x + (\frac{\alpha}{2} u^2 - u_{xt})_x = 0 \,, \label{5.02}
		\end{equation}
	which corresponds to (\ref{5.05}) for the case $\varepsilon_1=1$ and $\varepsilon_2=0$. The advantage of the RLW equation over the KdV equation is that the RLW equation yields a dispersion relation of the form $\omega=k/(1+k^2)$, and so a phase velocity that is bounded and tends to zero for short wavelengths. Its disadvantage is that the RLW equation is not integrable and that it possesses only one analytical solution, namely the one-soliton solution. The two and three-soliton  solutions for RLW are only known numerically and were constructed by Eilbeck and McGuire~\cite{Eilbeck, eilbeck2}.  

	The mRLW  equation, introduced by Gibbon,  Eilbeck and  Dodd~\cite{Gibbon}, can be written in terms of the $u$ field as  
\be
u_t +u_x + 
 \left[ \frac{\alpha}{2} \, u^2 + \frac{\alpha}{4}\,  w_x\, v_t    -  u_{xt}  \right]_x =0
 \label{mrlw}
\ee
and so it corresponds to (\ref{5.05}) for the case $\varepsilon_1=\varepsilon_2=1$. The linearized version of such an equation has the same dispersion relation as the RLW equation, and the same exact analytical one-soliton solution as RLW. The remarkable property of the mRLW equation, however, is that it also possesses analytic two-soliton solutions, even though it is not integrable in the sense of possessing an infinite number of conserved quantities. 

The analytical one-soliton solution for the RLW equation, and the analytical one- and two-soliton solutions for the mRLW  can be constructed using the Hirota direct method where the relation between the $\tau$-function and the $u$-field is of the form $u\sim -\(\ln \tau\)_{xt}$. Therefore, by changing the $\tau$-function as $\tau \rightarrow f\(x\)\,g\(t\)\,\tau$, does not change the solution for the $u$-field. So, integrating (\ref{soliton_fields}) one gets that $w\sim -\(\ln \tau\)_x + h\(x\)$, and $v\sim -\(\ln \tau\)_t + j\(t\)$. Consequently, as long as Hirota's solutions are concerned the integration ``constants''  $h\(x\)$ and $j\(t\)$, can be  reabsorbed in the redefinition of the $\tau$-function.  Therefore, when constructing the soliton solutions, either analytically or numerically, it is convenient to work with the field $q$ defined as 
\be
u\equiv - \frac{8}{\alpha}\,q_{xt}.
\label{fieldqdef}
\ee
Dropping the integration ``constants'' as explained above one can then write $w_x = -\frac{8}{\alpha}\,q_{xx}$ and $v_t=-\frac{8}{\alpha}\,q_{tt}$. Replacing the $u$-field by the $q$-field into (\ref{5.05}) one gets an equation for $q$ which can in fact be written as the $x$-derivative of the equation 
\be
q_{tt}+q_{xt}-4\,q_{xt}^2-2\,\varepsilon_2\,q_{xx}\,q_{tt}+q_{xxxt}-\varepsilon_1\(q_{xxtt}+q_{xxxt}\)=0.
\label{deformedqkdv}
\ee
Therefore, any solution of (\ref{deformedqkdv}) leads to a solution of (\ref{5.05}), when the integration ``constants'' in $w$ and $v$ are absorbed as explained above. So in this paper we base our discussion on the study of (\ref{deformedqkdv}).

First, we show in this paper that (\ref{deformedqkdv}) admits an analytical Hirota one-soliton solution for the case $\varepsilon_1=1$ given by
\be
q=\frac{3}{\(2+\varepsilon_2\)}\,\ln \tau 
\label{onegen}
\ee
with
\be
\tau= 1+e^{k\,x-\omega\,t+\delta}\;;\qquad \qquad\omega=\frac{k}{1-k^2}\;;\qquad \qquad\varepsilon_1=1.
\ee
In terms of the $u$-field this solution takes the form
\be
u=\frac{8}{\alpha}\, \frac{3}{\(2+\varepsilon_2\)}\,\frac{k}{4\(1-k^2\)}\, {\rm sech}^2\left[\frac{1}{2}\(k\,x-\omega\,t+\delta\)\right].
\label{uonesoleps2}
\ee

Note that (\ref{uonesoleps2}) constitutes a one-parameter family of one-soliton solutions, labeled by the deformation parameter $\varepsilon_2$. It interpolates between the one-soliton solution of the RLW model (for $\varepsilon_2=0$ and $\alpha=12$) constructed  in \cite{Peregrine,Benjamin}, and the one-soliton solution of the mRLW model (for $\varepsilon_2=1$ and $\alpha=8$) in \cite{Gibbon}.  Note also that, as pointed out in \cite{Gibbon},  the expressions for the one-soliton solutions of the RLW and mRLW, indeed look the same for the $u$-field, when the rescaling parameter $\alpha$ is chosen as above, {\it i.e.} $\alpha=12$ for RLW and $\alpha=8$ for mRLW.

The analytical Hirota two-soliton solution, however, exists only for the case $\varepsilon_1=\varepsilon_2=1$, and that is the solution constructed in \cite{Gibbon} given by
	\begin{equation}
		q = \ln \left( 1 + e^{\Gamma_1} + e^{\Gamma_2} + A_{12}\, e^{\Gamma_1 + \Gamma_2} \right)  
		 \;; \qquad\qquad\qquad \varepsilon_1=\varepsilon_2=1,
		\label{1.2}
	\end{equation}
	where
	\begin{equation}
		\Gamma_i = k_i x - \omega_i t + \delta_i \,,\qquad\qquad\qquad
		\omega_i = \frac{k_i}{1-k_i^2} \,,\qquad\quad  i = 1,2 \,, \label{1.3}
	\end{equation}
	and
	\begin{equation}
		A_{12} = - \frac{(\omega_1 - \omega_2)^2(k_1 - k_2)^2 + (\omega_1 - \omega_2)(k_1 - k_2) - (\omega_1 - \omega_2)^2}{(\omega_1 + \omega_2)^2(k_1 + k_2)^2 + (\omega_1 + \omega_2)(k_1 + k_2) - (\omega_1 + \omega_2)^2} \,.
		\label{a12intro}
	\end{equation}

	The previous paper by two of us (FtB and WJZ)~\cite{first_paper},  looked at the scattering properties of various soliton-like configurations  of the modified regularized long-wave (mRLW) equation, originally introduced and studied by J. D. Gibbon, J. C. Eilbeck and R. K. Dodd~\cite{Gibbon}, and given by equation (\ref{deformedqkdv}) for $\varepsilon_1=\varepsilon_2=1$. As stated in the aforementioned paper, the three soliton-solutions are not known and, in fact, they cannot be found using Hirota's method.

	The main results of the previous paper involved the observation that the scattering of two initially well-separated solitons, which, in fact, is described by an exact solution of the equations of motion (see equation~(\ref{1.2})), could be approximated very well by the numerical time-evolution of a linear superposition of two one-soliton solutions. In fact the approximation was so good that it was virtually impossible to see any difference between them. Then this paper also extended this idea to the simulation of various three-soliton interactions by numerically evolving linear superpositions of three (initially well-seperated) single-soliton solutions.  Such scatterings were also found to be very elastic (in the sense that the solitons preserved their shapes and velocities and there was no visible loss of radiation). Moreover, the phase shifts in the scatterings of three solitons were, again, very well approximated by the sums of the successive two two-soliton scatterings. This property holds for integrable models and the fact that it  holds also for the mRLW equation, which is not an Hirota integrable system, suggests that the mRWL equation may be `close' to being integrable.  

	To test this further we have decided to extend the investigations presented in the previous paper by considering this model as a perturbation of the integrable KdV equation since then we can look at the conserved quantities of the integrable model and see how they change when we perturb this model to become the mRLW model, or for that matter any other model `nearby'. We discuss all of this in more detail in the next section.
	
		The paper is organized as follows: In the next section we introduce our Lax-Zakharov-Shabat equation which we then use to 
		study the quasi-integrability of the model described by equation~(\ref{5.05}) for various values of $\varepsilon_1$ and $\varepsilon_2$. We perform the usual
		gauge transformations and finally define the charges which are truly conserved when  $\varepsilon_1=\varepsilon_2=0$ ({\it i.e.}, for the KdV model). In section~\ref{sec:parity} we introduce the parity argument for the travelling wave solutions of (\ref{5.05}) and show that if the
		multi-soliton field configurations of the model possess this symmetry, then the charges are quasi-conserved. The next section
		discusses the Hirota method of finding solutions of some nonlinear equations and appling it to our equation (\ref{5.05}). In it we show that
		this method gives a one soliton solution for any $\varepsilon_2$ with $\varepsilon_1=1$ being constant. The next two sections discuss
		the analytical quasi-integrability of the mRLW model and present arguments for the integrability based on the eveness parity properties 
		of the multi-soliton functions. Section~\ref{sec:kdvsolitons} discusses the soliton solutions of the KdV equation, obtained via the Hirota method, and the parity properties of two- and three-soliton solutions of this model. The lengthy section~\ref{sec:numerical} presents some results of our numerical simulations. These simulations were performed using a specially
		constructed numerical program based on implicit and explicit methods of solving (\ref{deformedqkdv}). The calculations involved studying the time evolution of various field configurations initially corresponding to two- or three-soliton systems and then 
		checking whether the observed results supported quasi-integrability of the model for various values of $\varepsilon_1$ and $\varepsilon_2$. We finish the paper with our conclusions and a few short appendices presenting more information about our numerical techniques and providing some details on the construction of conserved or quasi-conserved
		quantities discussed in section~\ref{sec:anomalouszc} of the paper.

	\section{The anomalous Lax-Zakharov-Shabat equation}
	\label{sec:anomalouszc}
	\setcounter{equation}{0}

		Our motivation is to study equation~(\ref{5.05}) in the context of quasi-integrability as a deformation of the integrable KdV equation expressed by equation~(\ref{5.01}). To this end, we start by constructing the quasi-zero curvature equation, and subsequently produce the quasi-conserved charges from it. To do this we introduce the Lax potentials $A_x$ and~$A_t$, as 
		\br
			A_x &=& -\left[b_1 - \frac{\alpha}{12} u \left(b_{-1}-F_{-1}\right)   \right] \label{5.8.1},\\
			A_t &=& -\left[-4 b_3 - b_1 + \frac{\alpha}{6} u_x F_0 - \frac{\alpha}{3} u F_1  + \frac{G}{2} \left(b_{-1}-F_{-1}\right)\right] \,,
			\label{5.8.1.b}
		\er
		where the generators $b_{2n+1}$ and $F_n$, are defined in the appendix \ref{app:loopalgebra}, and where
		\begin{equation}
			\begin{aligned}
				\begin{split}
					G \equiv & \frac{\alpha^2}{18} u^2 + \frac{\alpha^2}{24} \varepsilon_2 w_x v_t - \frac{\alpha}{6} \varepsilon_1 u_{xt}  + \frac{\alpha}{6} \left(1 - \varepsilon_1\right) u_{xx}  + \frac{\alpha}{6} u  \,.
				\end{split}
			\end{aligned}	
			\label{gdef}			
		\end{equation}
The curvature of the Lax connection is given by 
\begin{equation}
			F_{tx} = \partial_t A_x - \partial_x A_t + [A_t, A_x] = -X \,F_0 + \frac{\alpha}{12}Y \left(b_{-1} - F_{-1}\right) \,,
			\label{curvaturelax}
		\end{equation}
where $Y=0$ corresponds to the equation of motion of our deformed model (\ref{5.05}) 
\begin{equation}
			Y \equiv u_t + u_x + \left[ \frac{\alpha}{2} u^2  + \varepsilon_2 \frac{\alpha}{4} w_x v_t  + u_{xx}  - \varepsilon_1 \left( u_{xt}  + u_{xx} \right)\right]_x \,, \label{5.8.3}
		\end{equation}
		and where 
		\begin{equation}
			X \equiv \frac{\alpha}{6}\left[\frac{\alpha}{4} \varepsilon_2 w_x v_t - 
			\varepsilon_1 \left( u_{xt}  + u_{xx} \right) \right]. \label{5.8.2}
		\end{equation}
 Thus, if we now assume that~$u$ is a solution of the equations of motion  ({\it i.e.}, $Y=0$), then~$X$ represents the anomaly of the zero curvature equation. In the case of the KdV equation ({\it i.e.}, $\varepsilon_2 = \varepsilon_1 = 0$), the anomaly vanishes resulting in the well-known infinite number of truly conserved charges. 
 

	\subsection{Quasi-conserved charges}
		
		In order to construct the quasi-conserved charges we follow the procedure discussed in~\cite{us}, which is an adaptation to quasi-integrable theories, of the so-called abelianization procedure for exactly integrable theories \cite{drinfeld,oliveabelian,afgzcharges}. The key ingredient is that the generator $b_1$ appearing in $A_x$ in (\ref{5.8.1}),  is a semi-simple element of the loop algebra and so splits this algebra into its kernel and image under its adjoint action as explained in the appendix \ref{app:loopalgebra}.  Then, we perform a gauge transformation to rotate the Lax potentials into an infinite abelian subalgebra of the $\mathfrak{sl}(2)$ algebra, generated by $b_{2n+1}$. The gauge transformation transforms the potentials as 
		\begin{equation}
			A_x \to a_x = gA_x g^{-1} - \left(\partial_x g \right) g^{-1}  \,, \label{5.9.1}
		\end{equation}
		\begin{equation}
			A_t \to a_t = gA_t  g^{-1} - \left(\partial_t g \right) g^{-1} \,, \label{5.9.2}
		\end{equation}
		where the group element $g$ is chosen to be
		\begin{equation}
			g = \exp \left( \sum\limits_{n=1}^\infty \zeta_n F_{-n} \right) \equiv \exp\left( \sum\limits_{n=1}^\infty \mathcal{F}_{-n}\right) \,. 
			\label{groupgdef}
		\end{equation}
		Performing the expansion we find that $a_x$ can be expressed as
		\begin{equation}
			\begin{aligned}
				\begin{split}
					a_x = & A_x + \left[\sum\limits_{n=1}^\infty \mathcal{F}_{-n}, A_x \right] + \frac{1}{2!} \left[\sum\limits_{n=1}^\infty \mathcal{F}_{-n}, \left[\sum\limits_{m=1}^\infty \mathcal{F}_{-m}, A_x\right]\right] + \dots 
					\\& - \sum\limits_{m=1}^\infty \partial_x \mathcal{F}_{-m} - \frac{1}{2!} \left[\sum\limits_{n=1}^\infty \mathcal{F}_{-n}, \sum\limits_{m=1}^\infty \partial_x \mathcal{F}_{-m} \right]
					\\& - \frac{1}{3!} \left[\sum\limits_{k=1}^\infty \mathcal{F}_{-k}, \left[\sum\limits_{n=1}^\infty \mathcal{F}_{-n}, \sum\limits_{m=1}^\infty \partial_x \mathcal{F}_{-m} \right]\right] + \dots 
				\end{split}
			\end{aligned}
		\end{equation}
Using the algebra defined by \crefrange{5.3}{5.8}, we can  write $a_x$ as
		\begin{equation}
			\begin{aligned}
				\begin{split}
					a_x = & -b_1
					\\& + \zeta_1\,[b_1, {F}_{-1}]
					\\& + \zeta_2\,[b_1, {F}_{-2}] + \frac{\alpha}{12}u \left(b_{-1} - F_{-1}\right) - \frac{1}{2} \zeta_1^2\,\left[ {F}_{-1} , \left[{F}_{-1},b_1 \right] \right]
					-\partial_x \zeta_1\;{F}_{-1}
					\\& + \dots, \label{5.10}
				\end{split}
			\end{aligned}
		\end{equation}	
		where we have written down all the terms of the same grade on the same line and the grading is defined by the grading operator $d$ given in (\ref{gradingop}).
Note that the parameter $\zeta_{k}$,  in the expansion (\ref{5.10}), multiplying the commutator $[b_1, F_{-k}]$,  and so first appears among the terms of grade $-k+1$. Thus, one can choose the parameters $\zeta_k$, recursively, to cancel the image part of $a_x$ at the grade $-k+1$, {\it i.e.} its  component in the direction of ${F}_{-k+1}$. The expressions for the parameters $\zeta_{k}$, obtained this way, are given in the appendix \ref{app:parameters}, and they are polynomials in $u$ and its $x$-derivatives. Having chosen the parameters $\zeta_{k}$ this way, the expression for $a_x$ becomes
		\begin{equation}
			a_x = -b_1 + \sum\limits_{n=0}^\infty a_x^{(-2n-1)}b_{-2n-1} \,,
			\label{axabeliana}
		\end{equation}
		where the first values of $a_x^{(-2n-1)}$ are given by
		\begin{align}
			a_x^{(-1)} & = \frac{\alpha }{2^2\,3} \, u, 
			\\ a_x^{(-3)} & = \frac{\alpha^2 }{2^5\,3^2} \,u^2, 
			\\ a_x^{(-5)} & = \frac{\alpha^3 }{2^7\,3^3}\;u^3 + \frac{\alpha^2 }{2^7\,3^2} \,u u_{xx}, 
			\\ a_x^{(-7)} & = \frac{5 \alpha^4 }{2^{11}\,3^4}\,u^4 + \frac{\alpha^3 }{2^7\,3^3}\,u^2 u_{xx} + \frac{1 }{2^9\,3^2} \,\(\alpha^3\,u u_x^2 + \alpha^2 u u_{xxxx}\). 
		\end{align}

Note that in our gauge transformation we have not used the equations of motion $Y=0$, with $Y$ given in (\ref{5.8.3}). In the transformation  (\ref{5.9.2}) of the $A_t$ component of the Lax connection, the group element $g$ is already fixed, but we still can use the equations of motion $Y=0$ to simplify it, {\it i.e.} we are performing an on-shell gauge transformation. The `on-shell  result is then given by
\be
a_t= 4\, b_3 + b_1+\sum_{n=0}^{\infty} a_t^{(-2n-1)}\, b_{-2n-1}
+\sum_{n=-2}^{\infty} c_t^{(-n)}\, F_{-n}.
\label{abelat}
\ee
In the appendix \ref{app:parameters} we give explicit expressions for the first few  quantities $a_t^{(-2n-1)}$ and $c_t^{(-n)}$. Note that, due to the anomaly $X$, the quantities $c_t^{(-n)}$ do not vanish, and so the potential $a_t$ is not really rotated to the abelian sub-algebra generated by $b_{2n+1}$. That is a difference with respect to the case of exactly integrable theories, but it will not be a concern for us  as we explain below. 
		
The `on-shell  gauge-transformed curvature then becomes 
		\begin{equation}
			F_{tx} \to f_{tx} = \partial_t a_x - \partial_x a_t + [a_t, a_x] = g F_{tx} g^{-1} = -X g F_0 g^{-1}\,
			\label{transformedcurvature1}
		\end{equation}
		and so it takes the form
		\begin{equation}
			f_{tx} = -X \left(\sum\limits_{n=0}^\infty \gamma^{(-2n-1)}b_{-2n-1} + \sum\limits_{n=0}^\infty \beta^{(-n)}F_{-n} \right) \,,
			\label{transformedcurvature2}
		\end{equation}
		where we have assumed that the equations of motion are satisfied ({\it i.e.},~$Y=0$), and where 
\br
\gamma_{-1}&=& 0,
\nonumber\\
\gamma_{-3}&=& -\partial_x\left[\frac{ \alpha}{2^3\,3} u\right]
\label{gammaresult},\\
\gamma_{-5}&=&-\partial_x\left[\frac{\alpha^2}{2^6\,3}  u^2 +\frac{\alpha}{2^5\,3}  u_{xx}\right],
\nonumber\\
\gamma_{-7}&=& -\partial_x\left[\frac{5 \alpha^3 }{2^8\,3^3}\,u^3+\frac{5\,\alpha^2}{2^8\,3^2} 
   u_x^2+\frac{5 \alpha^2}{2^7\,3^2}\,u_{xx}+\frac{\alpha}{2^7\,3}  u_{xxxx}\right].
\nonumber
\er
In addition we have that  $\beta_0=1$, and $\beta_{-m}=0$ for $m=1,2,3,4$. The first nonvanishing $\beta_{i}$ is $\beta_{-5}$, and it is given by 
\be
\beta_{-5}= \partial_x\left[\frac{\alpha^2}{2^7\,3^2}\, u^2\right].
\ee

From the commutation relations \crefrange{5.3}{5.8} one can deduce that the commutators of a given element $b_{2n+1}$ with any element of the algebra never produce an element of the abelian sub-algebra generated by   $b_{2n+1}$. Therefore, the commutator $[a_x, a_t]$ does not contain any terms in the direction of the $b_{-2n-1}$ generators, since $a_x$ lies in this abelian sub-algebra. Thus, if we now consider only the terms in the direction of the $b_{-2n-1}$ generators of the gauge transformed curvature, we find that
		\begin{equation}
			 \partial_t a_x^{(-2n-1)} - \partial_x a_t^{(-2n-1)}=-X  \gamma^{(-2n-1)}  \,,\qquad \qquad\forall n \in \mathbb{Z}^+_0 \,,
		\end{equation}
		which can be rewritten as
		\begin{equation}
			\frac{dQ^{(-2n-1)}}{dt} = \left. a_t^{(-2n-1)} \right|^\infty_{x=-\infty} +\alpha^{(-2n-1)} \,, \label{5.12}
		\end{equation}
		where
		\begin{equation}
			Q^{(-2n-1)} \equiv  \int\limits^\infty_{-\infty} \! \mathrm{d} x \, a_x^{(-2n-1)}  \qquad \text{and} \qquad  \alpha^{(-2n-1)} \equiv - \int\limits^\infty_{-\infty} \! \mathrm{d} x \, X \gamma^{(-2n-1)} \,.
			\label{chargesdef}
		\end{equation}
		Since all the terms of the parameters~$\zeta_n$  depend on~$u$ and its $x$-derivatives, and~$u \to 0$ when~$x \to \pm \infty$, we see that~$g \to \one$ as~$x \to \pm \infty$. This implies that
		\begin{equation}
			\lim_{x \to \pm \infty} a_t = \lim_{x \to \pm \infty} A_t \,
		\end{equation}
		and so from (\ref{5.8.1.b}), we get that 
		\begin{equation}
			\left. a_t^{(-2n-1)} \right|^\infty_{x=-\infty} = 0.
		\end{equation}
		Hence, equation~(\ref{5.12}) becomes
		\begin{equation}
			\frac{dQ^{(-2n-1)}}{dt} = \alpha^{(-2n-1)} \,.
			\label{quasiintegralconservation}
		\end{equation}
		and so the quantities $Q^{(-2n-1)}$ are candidates for our quasi-conserved charges.

		It is important to point out that the Lax potential $A_x$, given in (\ref{5.8.1}),  does not depend upon the deformation parameters $\varepsilon_1$ and $\varepsilon_2$, and so it is the same as the Lax potential for the integrable KdV equation. Therefore, the charge densities $a_x^{(-2n-1)}$ and, consequently, the charges  $Q^{(-2n-1)}$, are the same as those for the KdV theory.  The dependence upon the deformation parameters in (\ref{quasiintegralconservation}), comes only from the anomaly $X$.  Furthermore, we note that the lowest of these charges is exactly conserved, {\it i.e.} 
		\begin{equation}
			Q^{(-1)} = \frac{\alpha}{12} \int\limits^\infty_{-\infty} \! \mathrm{d} x \, u \,,\qquad\qquad \qquad \frac{d\, Q^{(-1)}}{d\,t}=0.
		\end{equation}
		Indeed, from  (\ref{gammaresult}) one sees that  $\gamma^{(-1)} = 0$, and this implies that $Q^{(-1)}$ is conserved and so $u$ is a density of a conserved quantity for any value of~$\varepsilon_1$,~$\varepsilon_2$, and~$\alpha$. 
		
		The remaining charges are not properly conserved. However, as we show below, for some special soliton solutions such charges are quasi-conserved. By that we mean that in the process of a soliton scattering these charges do vary in time but, after the scattering, they all return to the same values they had prior to this scattering ({\it i.e.} the values before and after the scattering are the same). This property is not well understood yet, but we have found that it is accompanied by a space-time parity symmetry of the soliton solutions, and this is useful in trying to gain an understanding why the anomalies of the charges vanish when integrated over time during the whole scattering process. In section~\ref{sec:parity} we explain how this works for the case under consideration here.

		\section{The parity argument}
		\label{sec:parity}
		\setcounter{equation}{0}

Here we explain in detail how the quasi-integrability concept is related  to the existence of some parity symmetries  of the soliton solutions. Before we show how it works for some specific soliton solutions, let us give the general argument for the deformations of the KdV theory given by the equation (\ref{5.05}). 

\subsection{Parity argument for the traveling wave solutions} 
\label{sec:parityonesol}

Consider a class of traveling wave solutions $u\equiv u\(x-\frac{\omega}{k}\,t+\delta\)$ of the equation (\ref{5.05}) which, of course, includes in it the one-soliton solutions as its particular cases. For a fixed value of the time $t$, we define the shifted space coordinate as
\be
{\bar x}=x-\frac{\omega}{k}\,t+\delta\
\ee
and introduce the space parity transformation 
\be
P_{{\bar x}}:\qquad\qquad {\bar x}\rightarrow -{\bar x}.
\label{spaceparitydef}
\ee
The only hypothesis that we are making here is that the traveling wave solution is invariant under parity, {\it i.e.} that
\be
P_{{\bar x}}\(u\)=u.
\label{hypotesisuonesol}
\ee
Since $u=w_t=v_x$, it turns out that ($\partial_x=\partial_{{\bar x}}$)
\be
P_{{\bar x}}\(w\)=w,\qquad\quad P_{{\bar x}}\(v\)=-v\qquad\qquad \mbox{\rm and so}\qquad 
P_{{\bar x}}\(w_x\)=-w_x,\quad\qquad P_{{\bar x}}\(v_t\)=-v_t.
\ee
In consequence, we see that $u_{xt}=-\omega \, u_{{\bar x}{\bar x}}$ and so $P_{{\bar x}}\(u_{xt}\)=u_{xt}$.
Thus, the anomaly $X$ given in (\ref{5.8.2}),  for such type of  solutions satisfies
\be
P_{{\bar x}}\(X\)=X.
\ee
Note that the anomalies coefficients, $\gamma$'s given in (\ref{gammaresult}), always involve an odd number of $x$-derivatives of the field $u$, and so 
\be
P_{{\bar x}}\(\gamma_{-2n-1}\)=-\gamma_{-2n-1}.
\ee
Thus we see that
\be
\int_{-\infty}^{\infty}dx \, X\, \gamma_{-2n-1} =\int_{-\infty}^{\infty}d{\bar x} \, X\, \gamma_{-2n-1}
 =0
 \label{vanishanomalyonesol}
\ee 
and so the charges (\ref{chargesdef}) are exactly conserved for all   traveling wave solutions, and so also for the one-soliton solutions.

\subsection{Parity argument for multi-soliton fields}
\label{sec:paritymultisoliton}

Now we consider the space-time parity transformation around a given point $(x_{\Delta}\, , \, t_{\Delta})$, of space-time 
\be
P : \qquad \qquad \({\tilde x}\,,\,{\tilde t}\)\rightarrow \(-{\tilde x}\,,\,-{\tilde t}\)\;; \qquad\qquad
{\tilde x}=x-x_{\Delta}\; \qquad {\tilde t}=t -t_{\Delta}.
\label{paritydef}
\ee
Again we make the hypothesis  that the soliton solution we are considering  is even under such a parity transformation, {\it i.e.} 
\be
P\(u\)=u,
\label{parityu}
\ee
where $u$ in (\ref{parityu}) is evaluated on that particular soliton solution. 
In addition, since $w=\int dt\, u$, and $v=\int dx\,u$ (see (\ref{soliton_fields})), we see that
\be
P\(w\)=-w,\qquad\qquad\hbox{and}\qquad\qquad P\(v\)=-v.
\label{paritywv}
\ee
From (\ref{gdef}), (\ref{5.8.3}) and (\ref{5.8.2}) we have that
\be
P\(G\)=G,\qquad\qquad\qquad P\(Y\)=-Y,\qquad\qquad\qquad P\(X\)=X.
\label{gyxparitya}
\ee

Next we consider the following order 2 automorphism ({\it i.e.} $\sigma^2=1$) of the $\mathfrak{sl}(2)$ loop algebra
\be
\sigma\(T\)=e^{i\,\pi\,d}\, T\,e^{-i\,\pi\,d},
\ee
where $d$ is the grading operator defined in (\ref{gradingop}). Then,
\be
\sigma\(b_{2n+1}\)=-b_{2n+1},\qquad\qquad\quad \quad 
\sigma\(F_{2n+1}\)=-F_{2n+1},\qquad\qquad\quad \quad 
\sigma\(F_{2n}\)=F_{2n}.
\ee
We also consider the combination of these two operations, the parity and the automorphism:
\be
\Omega\equiv P\, \sigma.
\ee
One can check that the Lax operators (\ref{5.8.1}) and (\ref{5.8.1.b}), when evaluated on soliton solutions satisfying (\ref{parityu}) and (\ref{paritywv}), satisfy
\be
\Omega\(A_{\mu}\)=-A_{\mu},\qquad\qquad\qquad \mu=x\, , \, t,
\ee 
and so the curvature (\ref{curvaturelax}) satisfies
\be
\Omega\(F_{tx}\)=F_{tx}.
\ee
Let us rewrite (\ref{axabeliana}) as
\be
a_x= \sum_{n=-1}^{\infty} {\hat a}_x^{(-2n-1)}\; ;\qquad\hbox{where}\qquad 
{\hat a}_x^{(1)}=-b_1\;;\qquad\hbox{and}\qquad  {\hat a}_x^{(-2n-1)}=a_x^{(-2n-1)}\, b_{-2n-1}.
\lab{abeliana2}
\ee
Since $\Omega\(b_1\)=-b_1$, it follows that 
\be
\Omega\(\sbr{b_1}{{\cal F}_{-n}}\)= -\sbr{b_1}{\Omega\({\cal F}_{-n}\)},
\ee
where  ${\cal F}_{-n}$ is defined in (\ref{groupgdef}), 
and so 
\be
\(1+\Omega\)\(\sbr{b_1}{{\cal F}_{-n}}\)=\sbr{b_1}{\(1-\Omega\)\({\cal F}_{-n}\)}.
\ee
From (\ref{5.10}) we see that ${\hat a}_x^{(0)}=\sbr{b_1}{{\cal F}_{-1}}$, and so
\be
\(1+\Omega\){\hat a}_x^{(0)}=\sbr{b_1}{\(1-\Omega\)\({\cal F}_{-1}\)}
\lab{omega0}.
\ee
Since ${\hat a}_x^{(0)}$ lies in the kernel and $\Omega$ maps the kernel into the kernel, it follows that the l.h.s. of \rf{omega0} also lies in the kernel. But the r.h.s. of the same equation clearly lies in the Image. Therefore, we conclude that both sides vanish. Since  $\(1-\Omega\)\({\cal F}_{-1}\)$ cannot lie in the kernel we conclude that 
\be
\(1-\Omega\)\({\cal F}_{-1}\)=0\; ; \qquad\qquad\qquad {\rm or} \qquad\qquad  \Omega\({\cal F}_{-1}\)={\cal F}_{-1}
\label{omegaf1}
\ee
Using (\ref{omegaf1}),  we get from the third line of (\ref{5.10}) that
\be
\(1+\Omega\){\hat a}_x^{(-1)}=\sbr{b_1}{\(1-\Omega\)\({\cal F}_{-2}\)}.
\lab{omega1}
\ee
Since ${\hat a}_x^{(-1)}$ lies in the kernel, we get that the l.h.s. of \rf{omega1} lies in the kernel, and its r.h.s. lies in the image. Therefore both side have to vanish and so similarly to the previous case we conclude that
\be
\Omega\({\cal F}_{-2}\)={\cal F}_{-2}.
\ee
Continuing this process we conclude that $\Omega\({\cal F}_{-n}\)={\cal F}_{-n}$, for any positive integer $n$, and so the group element $g$ defined in (\ref{groupgdef}) satisfies
\be
\Omega\(g\)=g.
\lab{ginvariant}
\ee
Indeed from \rf{zetaresults} we observed that
\be
P\(\zeta_{n}\)=\(-1\)^n \, \zeta_{n}\qquad\qquad \qquad \mbox{\rm and so}\qquad\qquad\qquad
\Omega\(\zeta_{n}\, F_{-n}\)=\zeta_{n}\, F_{-n}.
\ee
Thus we see that
\be
\Omega\(g\, F_0\,g^{-1}\)=g\, F_0\,g^{-1}
\ee
and so from (\ref{transformedcurvature1}) and (\ref{transformedcurvature2})  we get that
\be
P\(\gamma_{-2n-1}\)=-\gamma_{-2n-1}. 
\ee
Thus we see that all the anomalies appearing in (\ref{quasiintegralconservation}) satisfy (see (\ref{gyxparitya})) 
\be
P\(X\, \gamma_{-2n-1}\)=-X\,\gamma_{-2n-1}.
\ee
In consequence,  integrating the r.h.s of this expression over a rectangle with centre at  the origin of the $({\tilde x}\,,\,{\tilde t})$ coordinates (see (\ref{paritydef})) we get
\be
\int_{-{\tilde t}}^{{\tilde t}} dt \, \int_{-{\tilde x}}^{{\tilde x}}dx\,X\, \gamma_{-2n-1}=0
\ee
Sending ${\tilde x}$ to infinity we find that the charges (\ref{chargesdef}) satisfy the mirror type symmetry (see (\ref{quasiintegralconservation})) 
\be
Q^{(-2n-1)}\({\tilde t}\)=Q^{(-2n-1)}\(-{\tilde t}\).
\label{mirrorproperty}
\ee
Thus, for any soliton solution satisfying the property (\ref{parityu}), all the charges $Q^{(-2n-1)}$ are quasi-conserved, and the sector defined by such soliton solutions constitutes a quasi-integrable sector of the theory.

\section{The analytical Hirota soliton solutions}	\label{The_analytical_Hirota_soliton_solutions}
\setcounter{equation}{0}

Next we investigate the existence of analytical soliton solutions for the deformed KdV equation (\ref{deformedqkdv}), or equivalently (\ref{5.05}).  To do this we introduce the Hirota $\tau$-function as 
\be
q=\beta\,\ln \tau
\label{taubetadef}
\ee
for some parameter $\beta$ to be appropriately chosen later.  Putting (\ref{taubetadef}) into (\ref{deformedqkdv}) one gets the following Hirota's equation
\br
&-&\tau ^2 \left[2 \beta {\ve}_2 \tau_{tt} \tau_{xx}+(4 \beta-2 \ve_1)
   \tau_{xt}^2+\tau_{t} \left(-2 \ve_1 \tau_{xxt}-(\ve_1-1) \tau_{xxx}+\tau_{x}\right)
  \right. \nonumber\\
   &-&2 \left.\ve_1
   \tau_{x} \tau_{xtt}-\ve_1 \tau_{tt} \tau_{xx}-3 (\ve_1-1) \tau_{xt} \tau_{xx}-3 \ve_1 
   \tau_{x} \tau_{xxt}+\tau_{t}^2+3 \tau_{x} \tau_{xxt}\right]
   \nonumber\\
   &+&2 \tau  \left[(\beta {\ve}_2-\ve_1) \tau_{xx}
   \tau_{t}^2+(\beta {\ve}_2-\ve_1) \tau_{tt} \tau_{x}^2
   \right. \nonumber\\
  & +& \left. \tau_{x} \tau_{t} \left(4 (\beta-\ve_1) \tau_{xt}-3 (\ve_1-1) \tau_{xx}\right)
   -3 (\ve_1-1) \tau_{x}^2 \tau_{xt}\right]
   \nonumber\\
   &-& 2 \tau_{t} \tau_{x}^2
   \left[(\beta ({\ve}_2+2)-3 \ve_1) \tau_{t}-3 (\ve_1-1) \tau_{x}\right]
   \nonumber\\
   &+& \tau ^3 \left(-\ve_1 \tau_{xxtt}-\ve_1 \tau_{xxxt}+\tau_{tt}+\tau_{xt}+\tau_{xxxt}\right)=0.
   \label{floristaueq}
   \er
Next we take the one-soliton ansatz 
\be
\tau=1+\eta\, e^{k\,x-\omega\,t +\delta}
\ee
and insert it into (\ref{floristaueq}). We consider the expansion of the equation in powers of $\eta$.  In the lowest order ($\eta^0$) the equation is automatically satisfied. In the next order ($\eta^1$) we find that the equation is satisfied if either $\omega =0$ or
\be
\omega=\frac{k+\(1-\ve_1\)\,k^3}{1-\ve_1\,k^2}. 
\ee
The order $\eta^2$ terms show that the equation is satisfied if
\be
\frac{2 k^4 \left[(\ve_1-1) k^2-1\right] \left[\beta ({\ve}_2+2) \left((\ve_1-1)
   k^2-1\right)+3\right]}{\left(\ve_1 k^2-1\right)^2}=0.
   \label{solutionforkonesol}
   \ee
If we do not want  $k$ to depend on $\ve$'s then we have to take
\be
\ve_1=1\;;\qquad\qquad\qquad\qquad\qquad \beta= \frac{3}{2+{\ve}_2}.
\ee
With that choice one can check that all the higher order terms, in powers of $\eta$, vanish automatically, showing that the truncation of the series leads to an exact solution. Thus, we see that we have found a family of one-soliton solutions, parameterized by ${\ve}_2$, and given by
\be
u=\frac{8}{\alpha}\,\frac{3}{\(2+{\ve}_2\)}\, \frac{k^2}{4\(1-k^2\)}\,\frac{1}{\left[\cosh\(\frac{\Gamma}{2}\)\right]^2}\;;\qquad\qquad {\rm for}\qquad \qquad \ve_1=1
\label{niceonesoliton}
\ee
with
\be
\Gamma=k\,x-\omega\,t+\delta\;;\qquad\qquad \qquad\omega=\frac{k}{1-k^2},
\ee
where we have absorbed the $\eta$ parameter into $\delta$, by writing $\eta=e^{{\tilde \delta}}$ and shifting $\delta+{\tilde \delta}\rightarrow \delta$. This  is the solution given in (\ref{uonesoleps2}). Note that this solution, written in terms of the $q$ field, takes the form: 
\be
q=\frac{3}{\(2+{\ve}_2\)}\,\ln\(1+e^{\Gamma}\)
=\frac{3}{\(2+{\ve}_2\)}\left[\ln 2+ \frac{\Gamma}{2}+\ln\cosh\(\frac{\Gamma}{2}\)\right]. \label{anal_hirota}
\ee

Note also that, as we have stated in the introduction,  for ${\ve}_2=0$ we get the RLW one-soliton solution, and for ${\ve}_2=1$, we get the mRLW one-soliton. In between we get a whole new family of one-soliton solutions. The one-soliton solution for $\ve_1\neq 1$, would need $k$ to depend on the deformation parameters $\ve_1$ and $\ve_2$ (see (\ref{solutionforkonesol})), and so it appears to be unphysical.

We have also checked that applying the above Hirota-type procedure for a two-soliton type ansatz, {\it i.e.} one that expands $\tau$ up to a second order in $\eta$, leads to a solution only for the case ${\ve}_2=\ve_1=1$ and $\beta= 1$. In such a case, the Hirota's equation (\ref{floristaueq}) to 
\br
\tau_{t}^2+\tau_{x} \tau_{t}-2 \tau_{xxt} \tau_{t}+2 \tau_{xt}^2-2 \tau_{x} \tau_{xtt}+
\tau_{tt} \tau_{xx}
  -\tau  \left(\tau_{tt}+\tau_{xt}-\tau_{xxtt}\right)=0.
   \er
Its two-soliton solution is
\be
\tau=1+ e^{\Gamma_1}+ e^{\Gamma_2}+A_{12}\,e^{\Gamma_1+\Gamma_2}\; ; 
\qquad\qquad{\rm for}\qquad \qquad {\ve}_2=\ve_1=\beta= 1
\label{twosoltaumrlw}
\ee
with
\be
\Gamma_i= k_i\,x-\omega_i\,t+\delta_i\qquad\qquad\qquad\qquad \omega_i=\frac{k_i}{1-k_i^2}\qquad\qquad i=1,2
\ee
and 
\br
A_{12}=-\frac{(k_1-k_2)^2 \left[k_1^3 k_2-k_1^2
   \left(k_2^2-1\right)+k_1 k_2
   \left(k_2^2-4\right)+k_2^2-3\right]}{(k_1+k_2)^2
   \left[k_1^3 k_2+k_1^2 \left(k_2^2-1\right)+k_1 k_2
   \left(k_2^2-4\right)-k_2^2+3\right]}.
   \label{a12section}
\er
This is precisely the two-soliton solution given in (\ref{1.2})-(\ref{a12intro}), and first found in \cite{Gibbon}. One can check that by replacing $\omega_i$ given in (\ref{1.3}) and putting it into (\ref{a12intro}), one finds that $A_{12}$ given in (\ref{a12section}) is the same as the one in (\ref{a12intro}).

\section{The analytical quasi-integrability of the mRLW theory}	
\label{sec:analyticalmrlw}
\setcounter{equation}{0}

In subsection \ref{sec:parityonesol} we have shown that if the field $u$ evaluated on a one-soliton is even under the space parity (\ref{spaceparitydef}) (see (\ref{hypotesisuonesol})), then the anomalies vanish (see (\ref{vanishanomalyonesol})), and so all the charges $Q^{(-2n-1)}$, introduced in (\ref{chargesdef}), are exactly conserved. Note that the family of one-soliton solutions (\ref{niceonesoliton}) are even under the parity (\ref{spaceparitydef}), and so this infinity of charges are exactly conserved for such one-soliton solutions. 

Let us now analyse the two-soliton solution (\ref{twosoltaumrlw})-(\ref{a12section}). Denoting 
\be
A_{12} = e^{\Delta}
\ee
we can write the two-soliton tau-function (\ref{twosoltaumrlw}) as 
\br
\tau=1+  e^{\Gamma_1}+ e^{\Gamma_2}+e^{\Gamma_1+\Gamma_2+\Delta}
=2\,e^{z_{+}}\left[\cosh z_{+}+  e^{-\Delta/2}\, \cosh z_{-}\right],
\label{twosoltaumrlw2}
\er
where we have defined
\be
z_{+}=\frac{1}{2}\(\Gamma_1+\Gamma_2+\Delta\)\qquad\qquad\hbox{and}\qquad\qquad
z_{-}=\frac{1}{2}\(\Gamma_1-\Gamma_2\).
\ee
Note that if $k_1=k_2$ we see that $A_{12}=0$ and so the two-soliton solution reduces to a one-soliton solution. Therefore, for truly two-soliton solutions, {\it i.e.} when $k_1\neq k_2$, $z_{+}$ and 
$z_{-}$ can be considered independent space-time variables, {\it i.e.} they are linearly  independent   combinations  of $x$ and $t$.  Then we have that
\be
\partial_x z_{\pm}= \frac{1}{2}\,\(k_1\pm k_2\)\equiv k_{\pm}\qquad\qquad\hbox{and}\qquad\qquad
\partial_t z_{\pm}= -\frac{1}{2}\,\(\omega_1\pm \omega_2\)\equiv -\omega_{\pm}.
\ee

From (\ref{fieldqdef}) and (\ref{taubetadef}), for $\beta=1$, we find that the two-soliton solution (\ref{twosoltaumrlw})-(\ref{a12section}) for the $u$-field can be rewritten as (see (\ref{twosoltaumrlw2}))
\br
u&=&-\frac{8}{\alpha}\;\frac{\(\tau\, \tau_{xt}-\tau_x\,\tau_t\)}{\tau^2}
\nonumber\\
&=&\frac{8}{\alpha}\;\frac{e^{-\Delta/2}}{\left[\cosh z_{+}
+  e^{-\Delta/2}\, \cosh z_{-}\right]^2}\left[
e^{\Delta/2}k_{+}\omega_{+}+e^{-\Delta/2}k_{-}\omega_{-}
\right. \nonumber\\
&+&\left. \(k_{+}\omega_{+}+k_{-}\omega_{-}\)\cosh z_{+}\,\cosh z_{-}-\(k_{-}\omega_{+}+k_{+}\omega_{-}\)\sinh z_{+}\,\sinh z_{-}\right].
\label{utwosol}
\er

To find the point of space-time around which we perform our parity transformation, we consider the linear relation among $z_{\pm}$ and $x$ and $t$, {\it i.e.}
\br
\(\begin{array}{c}
z_{+}\\
z_{-}
\end{array}\) =\(
\begin{array}{cc}
k_{+}& -\omega_{+}\\
k_{-}&-\omega_{-}
\end{array}\)\(
\begin{array}{c}
x\\
t
\end{array}\)
+\(
\begin{array}{c}
\frac{\Delta}{2}+\delta_{+}\\
\delta_{-}
\end{array}\),
\er
where we have denoted $\delta_{\pm}=\(\delta_1\pm\delta_2\)/2$. Then
\br
\(\begin{array}{c}
{\tilde x}\\
{\tilde t}
\end{array}\)\equiv
\(\begin{array}{c}
x-x_{\Delta}\\
t-t_{\Delta}
\end{array}\)= \(
\begin{array}{cc}
k_{+}& -\omega_{+}\\
k_{-}&-\omega_{-}
\end{array}\)^{-1}
\(\begin{array}{c}
z_{+}\\
z_{-}
\end{array}\)
\er
with
\br
x_{\Delta}=\frac{\omega_{+}\,\delta_{-}-\omega_{-}\(\frac{\Delta}{2}+\delta_{+}\)}{k_{+}\,\omega_{-}-k_{-}\,\omega_{+}}\;;\qquad\qquad\qquad
t_{\Delta}=\frac{k_{+}\,\delta_{-}-k_{-}\(\frac{\Delta}{2}+\delta_{+}\)}{k_{+}\,\omega_{-}-k_{-}\,\omega_{+}}.
\er
Thus, the space-time parity transformation
\be
P:\qquad\qquad\qquad \(z_{+}\,,\,z_{-}\)\rightarrow \(-z_{+}\,,\,-z_{-}\)
\ee
is of the same form as (\ref{paritydef}), and from (\ref{utwosol}) we see that the field $u$ evaluated on the two-soliton solution of the mRLW model is invariant under this parity, and so it satisfies the hypothesis made in (\ref{parityu}), {\it i.e.} that $P\(u\)=u$. So, as shown in subsection \ref{sec:paritymultisoliton}, all the charges $Q^{(-2n-1)}$, defined in (\ref{chargesdef}), satisfy the mirror symmetry described in (\ref{mirrorproperty}), and they are asymptotically conserved in the scattering of two solitons. 

We have thus presented an analytical proof of the quasi-integrability of the  mRLW theory. 
It is worth adding that this is the first analytical proof of the quasi-integrability of a (non-integrable) field theory in $1+1$ dimensions
(as in this case we have an analytical form of a two-soliton solution).  

\section{The parity versus dynamics argument}
\label{sec:paritydynamicsargument}
\setcounter{equation}{0}

We shall now check whether  the dynamics of the deformed model (\ref{5.05}) favours or not the correct parity property of the field $u$ (\ref{parityu}), so as to make the infinite set of charges $Q^{(-2n-1)}$, defined in (\ref{chargesdef}), quasi-conserved. The plan of our approach  is to write a given solution of (\ref{5.05}) as a perturbative expansion around an exact solution of the integrable KdV equation. In addition, if that exact solution does satisfy the desired property (\ref{parityu}) under the parity, we want to understand how the higher terms in the expansion behave under this parity. Our studies will show that the dynamics of the deformed model (\ref{5.05}) does indeed favour the parity property (\ref{parityu}) in a quite non-trivial and interesting way.

  To make the perturbative expansion simpler we parameterise our two deformation parameters as 
\be
\ve_1=\ve\,\sin \xi \; ; \qquad\qquad\qquad {\ve}_2=\ve\,\cos \xi.
\ee
Then (\ref{5.05}) becomes
 \be
u_t +u_x + 
 \left[ \frac{\alpha}{2}\, u^2   + u_{xx} +\ve\(
 \cos\xi\,  \frac{\alpha}{4}\,  w_x\, v_t    - 
   \sin\xi \( u_{xt}  + u_{xx} \)\)\right]_x =0.
   \lab{floris3}
   \ee

Next we expand the field $u$ as 
\be
u=u^{(0)}+\ve\, u^{(1)}+ \ve^2\,u^{(2)} + \ldots,
\ee
where each $u^{(i)}$, in general, depends upon $\xi$. Thus,  we find that 
\be
w=\int dt\,u= \int dt\, u^{(0)}+\ve\, \int dt\,u^{(1)}+ \ve^2\,\int dt\,u^{(2)} + \ldots
\equiv w^{(0)}+\ve\, w^{(1)}+ \ve^2\,w^{(2)} + \ldots
\ee
and
\be
v=\int dx\,u= \int dx\, u^{(0)}+\ve\, \int dx\,u^{(1)}+ \ve^2\,\int dx\,u^{(2)} + \ldots
\equiv v^{(0)}+\ve\, v^{(1)}+ \ve^2\,v^{(2)} + \ldots
\ee

Next we split all fields into their eigen-function parts under the parity (\ref{paritydef}). Thus 
\be
u^{(i,\pm)}= \frac{1}{2}\,\(1\pm P\)u^{(i)}
\ee
and the same for $w$ and $v$. Of course, the $\ve^0$ part of \rf{floris3} is the KdV equation, {\it i.e.}
 \be
u_t^{(0)} +u_x^{(0)} + 
 \left[ \frac{\alpha}{2}\, \(u^{(0)}\)^2   + u_{xx}^{(0)} \right]_x =0
   \lab{eq0}
   \ee

The next order - $\ve^1$ equation is then given by
 \be
u_t^{(1)} +u_x^{(1)} + 
 \left[ \alpha\, u^{(0)}\, u^{(1)} + u_{xx}^{(1)} +\cos\xi\,  \frac{\alpha}{4}\,  w^{(0)}_x\, v^{(0)}_t    - 
   \sin\xi \( u^{(0)}_{xt}  + u^{(0)}_{xx} \)\right]_x =0.
   \lab{eq1}
   \ee
   We

 Next we split this equation into its even and odd parts under the parity. The odd part is given by
  \br
&&u_t^{(1,+)} +u_x^{(1,+)} + 
 \left[ \alpha\,\( u^{(0,+)}\, u^{(1,+)}+u^{(0,-)}\, u^{(1,-)}\) + u_{xx}^{(1,+)} 
  \right.\nonumber\\
&& + \left. \cos\xi\,  \frac{\alpha}{4}\,  \(w^{(0,+)}_x\, v^{(0,+)}_t+w^{(0,-)}_x\, v^{(0,-)}_t\)
  -     \sin\xi \( u^{(0,+)}_{xt}  + u^{(0,+)}_{xx} \)\right]_x =0
   \lab{eq1odd}
   \er 
   and the even part takes the form
 \br
&&u_t^{(1,-)} +u_x^{(1,-)} + 
 \left[ \alpha\,\( u^{(0,+)}\, u^{(1,-)}+u^{(0,-)}\, u^{(1,+)}\) + u_{xx}^{(1,-)} 
  \right.\nonumber\\
 &&+\left. \cos\xi\,  \frac{\alpha}{4}\,  \(w^{(0,+)}_x\, v^{(0,-)}_t+w^{(0,-)}_x\, v^{(0,+)}_t\)
 -     \sin\xi \( u^{(0,-)}_{xt}  + u^{(0,-)}_{xx} \)\right]_x =0.
   \lab{eq1even}
   \er   

Let us now suppose that the zeroth order solution is even under the parity, {\it i.e.} that
\be
P\(u^{(0)}\)=u^{(0)}\;;\qquad\qquad\qquad \mbox{\rm and so}\qquad u^{(0,-)}=0.
\lab{parity1}
\ee
In this case we find that
\be
w^{(0,+)}=v^{(0,+)}=0
\lab{parity2}
\ee
In consequence, we see that \rf{eq1odd} becomes
 \br
u_t^{(1,+)} +u_x^{(1,+)} + 
 \left[ \alpha\, u^{(0,+)}\, u^{(1,+)} + u_{xx}^{(1,+)} \right]_x&=&-\left[
 \cos\xi\,  \frac{\alpha}{4}\,  w^{(0,-)}_x\, v^{(0,-)}_t
-  \sin\xi \( u^{(0,+)}_{xt}  + u^{(0,+)}_{xx} \)\right]_x 
\nonumber\\
u_t^{(1,-)} +u_x^{(1,-)} + 
 \left[ \alpha\, u^{(0,+)}\, u^{(1,-)} + u_{xx}^{(1,-)} \right]_x &=&0.
\lab{eq1split}
   \er 

Note that $u^{(1,+)}$ satisfies an inhomogeneous equation and $u^{(1,-)}$ a homogeneous one. Thus, we can have a solution for which
\be
u^{(1,-)}=0\qquad\qquad \qquad \mbox{\rm and so} \qquad\qquad w^{(1,+)}=v^{(1,+)}=0.
\lab{parity3}
\ee
The order $\ve^2$ part of \rf{floris3} takes the form  
\br
&& u^{(2)}_t +u^{(2)}_x + 
 \left[ \frac{\alpha}{2}\, \(\(u^{(1)}\)^2+ 2\, u^{(0)}\,u^{(2)}\)   + u^{(2)}_{xx} 
 \right.\nonumber\\
 &&+ \left.
 \cos\xi\,  \frac{\alpha}{4}\,  \(w^{(0)}_x\, v^{(1)}_t+ w^{(1)}_x\, v^{(0)}_t\)    - 
   \sin\xi \( u^{(1)}_{xt}  + u^{(1)}_{xx} \)\right]_x =0.
   \lab{eq2}
   \er
Splitting \rf{eq2} into its even and odd parts one gets that the odd part becomes
\br
&& u^{(2,+)}_t +u^{(2,+)}_x + 
 \left[ \frac{\alpha}{2}\, \(\(u^{(1,+)}\)^2+\(u^{(1,-)}\)^2+ 2\, \(u^{(0,+)}\,u^{(2,+)}+u^{(0,-)}\,u^{(2,-)}\)\) 
 \right.    \lab{eq2odd}\\
 &&+ \left. u^{(2,+)}_{xx} 
  + \cos\xi\,  \frac{\alpha}{4}\,  \(w^{(0,+)}_x\, v^{(1,+)}_t+ w^{(1,+)}_x\, v^{(0,+)}_t+
 w^{(0,-)}_x\, v^{(1,-)}_t+ w^{(1,-)}_x\, v^{(0,-)}_t\)   
 \right. \nonumber\\
 && - \left.  
   \sin\xi \( u^{(1,+)}_{xt}  + u^{(1,+)}_{xx} \)\right]_x =0
  \nonumber
   \er
   while the even part takes the form
\br
&& u^{(2,-)}_t +u^{(2,-)}_x + 
 \left[ \frac{\alpha}{2}\, \(2\, u^{(1,+)}\,u^{(1,-)}+ 2\, \(u^{(0,+)}\,u^{(2,-)}+u^{(0,-)}\,u^{(2,+)}\)\)   + u^{(2,-)}_{xx} 
 \right. 
 \lab{eq2even}\\
 &&+ \left.
 \cos\xi\,  \frac{\alpha}{4}\,  \(w^{(0,+)}_x\, v^{(1,-)}_t+ w^{(1,+)}_x\, v^{(0,-)}_t+
 w^{(0,-)}_x\, v^{(1,+)}_t+ w^{(1,-)}_x\, v^{(0,+)}_t\)  
 \right. \nonumber\\
 &&  - \left.  
   \sin\xi \( u^{(1,-)}_{xt}  + u^{(1,-)}_{xx} \)\right]_x =0.
  \nonumber
   \er

Note that if \rf{parity1}, \rf{parity2} and \rf{parity3} are all satisfied, then these equations become
 \br
 &&u^{(2,+)}_t +u^{(2,+)}_x + 
 \left[ \alpha\, u^{(0,+)}\,u^{(2,+)}   + u^{(2,+)}_{xx} \right]_x =
 \nonumber\\
 &&-\left[ \frac{\alpha}{2}\,\(u^{(1,+)}\)^2 
 +
 \cos\xi\,  \frac{\alpha}{4}\,  \(w^{(0,-)}_x\, v^{(1,-)}_t+ w^{(1,-)}_x\, v^{(0,-)}_t\)    - 
   \sin\xi \( u^{(1,+)}_{xt}  + u^{(1,+)}_{xx} \)\right]_x, 
  \nonumber\\
&&u^{(2,-)}_t +u^{(2,-)}_x + 
 \left[ \alpha\, u^{(0,+)}\,u^{(2,-)}   + u^{(2,-)}_{xx} \right]_x =0.  
   \er
So, again $u^{(2,+)}$ satisfies an inhomogeneous equation and $u^{(2,-)}$ a homogeneous one. So again, we can have a solution where
\be
u^{(2,-)}=0\qquad\qquad \qquad \mbox{\rm and so} \qquad\qquad w^{(2,+)}=v^{(2,+)}=0.
\lab{parity3}
\ee
Continuing that process it seems that we can always have a solution in which the $u$ field is even under our parity. Such a solution of the perturbed model fits into the scheme presented in  subsection \ref{sec:paritymultisoliton}  and so all the charges $Q^{(-2n-1)}$ 
(from their infinite set)are quasi-conserved, {\it i.e.} they satisfy the property (\ref{mirrorproperty}). In this sense, the dynamics of the perturbed model favours the even $u$ field, since as we saw there cannot exist pure odd $u$ field solution.  Of course, there can exist mixed solutions, {\it i.e.} $u$s  with an even plus a (perhaps small) odd parts. 
 
We do not understand yet how one could fine tune the perturbed solution to be purely even under the parity. Note that the condition of making the odd part of the solution vanish at each order of perturbation, does not work like an initial condition. So, we cannot prepare the solution at a given initial time $t_0$ and guarantee that it will evolve in time keeping its evenness. The conditions discussed above are imposed at all times. However, as our analysis has shown that a purely odd solution cannot exist, perhaps there might be a mechanism making the odd part of the solution small (maybe its appearance is not energetically favorable). We have not found such a mechanism yet. However, our numerical simulations, which we describe below, show that if we start with a seed configuration which is even under the parity, the numerical evolution of the configuration, under the perturbed dynamics, essentially keeps the configuration even. This is a very intriguing property of our quasi-integrable field theories, and we shall describe it, in more detail, in the next sections.

\section{The exact Hirota's soliton solutions for the KdV equation}
\label{sec:kdvsolitons}
\setcounter{equation}{0}

In section \ref{sec:paritydynamicsargument} we have shown how the dynamics of the perturbed equation (\ref{5.05}) favours the even property of the $u$ field under the parity.  Since our discussion involved a perturbative expansion around an exact solution of the KdV equation  (\ref{5.01}), in this section we discuss the properties of the exact multi-soliton solutions under the space-time parity transformation (\ref{paritydef}).  In order to do that, we construct these solutions using the Hirota's method.

We introduce the Hirota's $\tau$-function for the KdV equation (\ref{5.01}) as
\be
u=\frac{12}{\alpha}\, \partial_x^2\ln \tau
\label{tauurel}
\ee
Putting this expression into (\ref{5.01}) we get the Hirota's equation for KdV in the form
\br
&&2 \tau_{x}^3+8 \tau_{xxx} \,\tau_{x}^2
-\left[6\, \tau_{xx}^2+\tau  \left(2 \tau_{xt}+3 \tau_{xx}+5 \tau_{xxxx}\right)\right] \tau_{x}
  \nonumber\\
   &+&\tau_{t} \left(2 \tau_{x}^2-\tau  \tau_{xx}\right)
   +\tau  \left[2 \tau_{xx} \tau_{xxx}+\tau  \left(\tau_{xxt}+\tau_{xxx}+\tau_{xxxxx}\right)\right]=0.
   \label{kdvtaueq}
\er
The one-soliton solution of this equation is given by 
\be
\tau_{\rm one-sol}= 1+  e^{k\, x-\omega\,t+\delta} \qquad\qquad\qquad {\rm with}\qquad \omega=k\(1+k^2\)
\ee
and so
\be
u_{\rm one-sol}=\frac{3}{\alpha}\,\frac{k^2}{{\rm cosh}^2\left[\(k\, x-\omega\,t+\delta\)/2\right]}.
\ee
So we see that the KdV one-soliton is invariant under the space parity $P_{{\bar x}}$, defined in (\ref{spaceparitydef}).

The two-soliton solution of the KdV equation corresponds to 
\be
\tau_{\rm two-sol}= 1+  e^{\Gamma_1} +  e^{\Gamma_2} +e^{\Gamma_1+\Gamma_2+\Delta} 
\ee
where
\be
\Gamma_i=k_i\, x-\omega_i\,t+\delta_i\;; \qquad\qquad \omega_i=k_i\(1+k_i^2\)\;;\qquad \qquad
 e^{\Delta}=\frac{\(k_1-k_2\)^2}{\(k_1+k_2\)^2}. 
\ee
with $i=1,2$. To understand its parity properties we introduce
\be
z_{+}=\frac{1}{2}\(\Gamma_1+\Gamma_2+\Delta\)\qquad\qquad\hbox{and}\qquad \qquad
z_{-}=\frac{1}{2}\(\Gamma_1-\Gamma_2\)
\ee
and so we find that
\br
\tau_{\rm two-sol}= 2\, e^{z_{+}}\left[\cosh z_{+} +e^{-\Delta/2}\, \cosh z_{-}\right].
\er

From (\ref{tauurel}) we find that
\br
u_{\rm two-sol}&=&\frac{12}{\alpha}\,
\frac{1}{\left(e^{\Delta /2} \cosh (z_{+})+\cosh (z_{-})\right)^2}\left[
e^{\Delta /2} \left(k_{-}^2+k_{+}^2\right) \cosh (z_{-}) \cosh(z_{+})+k_{-}^2 \cosh ^2(z_{-})
\right.\nonumber\\
&-&\left. \left(k_{-} \sinh
   (z_{-})+k_{+} e^{\Delta /2} \sinh (z_{+})\right)^2+k_{+}^2 e^{\Delta }
   \cosh ^2(z_{+})\right],
\er
where $k_{\pm}\equiv \(k_1\pm k_2\)/2$. Thus, the KdV two-soliton solution is even under the parity transformation $P : \;\;  \(z_{+}\,,\,z_{-}\)\rightarrow \(-z_{+}\,,\,-z_{-}\)$, which is of the same form as (\ref{paritydef}).

Next we look at the three-soliton solution which corresponds to 
\br
\tau_{\rm three-sol}&=& 1+  e^{\Gamma_1} +  e^{\Gamma_2}+  e^{\Gamma_3} 
\nonumber\\
&+&\frac{\(k_1-k_2\)^2}{\(k_1+k_2\)^2}\,e^{\Gamma_1+\Gamma_2}
+\frac{\(k_1-k_3\)^2}{\(k_1+k_3\)^2}\,e^{\Gamma_1+\Gamma_3}
+\frac{\(k_2-k_3\)^2}{\(k_2+k_3\)^2}\,e^{\Gamma_2+\Gamma_3}
\nonumber\\
&+&\frac{\(k_1-k_2\)^2}{\(k_1+k_2\)^2}\,
\frac{\(k_1-k_3\)^2}{\(k_1+k_3\)^2}\,\frac{\(k_2-k_3\)^2}{\(k_2+k_3\)^2}\,
e^{\Gamma_1+\Gamma_2+\Gamma_3}
\er
with $\Gamma_i=k_i\, x-\omega_i\,t+\delta_i$,  $ \omega_i=k_i\(1+k_i^2\)$,  $i=1,2,3$. To simplify the expressions we
define $\Delta_{ij}$  by 
\be
\frac{\(k_i-k_j\)^2}{\(k_i+k_j\)^2} = e^{\Delta_{ij}}
\ee
and so
\br
\tau_{\rm three-sol}&=& 1+  e^{\Gamma_1} +  e^{\Gamma_2}+ e^{\Gamma_3} 
+e^{\Gamma_1+\Gamma_2+\Delta_{12}}
+e^{\Gamma_1+\Gamma_3+\Delta_{13}}
+e^{\Gamma_2+\Gamma_3+\Delta_{23}}
\nonumber\\
&+&e^{\Gamma_1+\Gamma_2+\Gamma_3+\Delta_{12}+\Delta_{13}+\Delta_{23}}
\nonumber\\
&=& e^{\(\Gamma_1+\Gamma_2+\Gamma_3+\Delta_{12}+\Delta_{13}+\Delta_{23}\)/2}\left[
e^{\(\Gamma_1+\Gamma_2+\Gamma_3+\Delta_{12}+\Delta_{13}+\Delta_{23}\)/2}+
e^{-\(\Gamma_1+\Gamma_2+\Gamma_3+\Delta_{12}+\Delta_{13}+\Delta_{23}\)/2}
\right.\nonumber\\
&+&\left. e^{\(\Gamma_1-\Gamma_2-\Gamma_3-\Delta_{12}-\Delta_{13}-\Delta_{23}\)/2}
+e^{\(-\Gamma_1+\Gamma_2+\Gamma_3-\Delta_{12}-\Delta_{13}+\Delta_{23}\)/2}
\right.\nonumber\\
&+&\left. e^{\(-\Gamma_1+\Gamma_2-\Gamma_3-\Delta_{12}-\Delta_{13}-\Delta_{23}\)/2}
+e^{\(\Gamma_1-\Gamma_2+\Gamma_3-\Delta_{12}+\Delta_{13}-\Delta_{23}\)/2}
\right.\nonumber\\
&+&\left. e^{\(-\Gamma_1-\Gamma_2+\Gamma_3-\Delta_{12}-\Delta_{13}-\Delta_{23}\)/2}
+e^{\(\Gamma_1+\Gamma_2-\Gamma_3+\Delta_{12}-\Delta_{13}-\Delta_{23}\)/2}
\right]
\label{analysetau}\\
&=&2\,e^{z_{+}}\left[ \cosh z_{+}+e^{-\(\Delta_{12}+\Delta_{13}\)/2}\,\cosh z_{-}^{(1)}
+e^{-\(\Delta_{12}+\Delta_{23}\)/2}\,\cosh z_{-}^{(2)}
\right. \nonumber\\
&+& \left. e^{-\(\Delta_{13}+\Delta_{23}\)/2}\,\cosh z_{-}^{(3)}
\right],
\nonumber
\er
where we have defined
\br
z_{+}&=& \(\Gamma_1+\Gamma_2+\Gamma_3+\Delta_{12}+\Delta_{13}+\Delta_{23}\)/2,
\nonumber\\
z_{-}^{(1)}&=& \(-\Gamma_1+\Gamma_2+\Gamma_3+\Delta_{23}\)/2,
\nonumber\\
z_{-}^{(2)}&=&\(\Gamma_1-\Gamma_2+\Gamma_3+\Delta_{13}\)/2,
\nonumber\\
z_{-}^{(3)}&=&\(\Gamma_1+\Gamma_2-\Gamma_3+\Delta_{12}\)/2
\er
and so $z_{+}=z_{-}^{(1)}+z_{-}^{(2)}+z_{-}^{(3)}$. We can put
\be
z_{-}^{(i)}= k^{(i)}\,x-\omega^{(i)}\,t+\Delta^{(i)}\qquad\qquad\qquad i=1,2,3
\label{zminusi}
\ee
where
\br
k^{(1)}&=& \frac{1}{2}\, \(-k_1+k_2+k_3\),\quad\quad 
\omega^{(1)}= \frac{1}{2}\, \(-\omega_1+\omega_2+\omega_3\),\quad\quad
\Delta^{(1)}=\frac{1}{2}\, \(-\delta_1+\delta_2+\delta_3+\Delta_{23}\),
\nonumber\\
k^{(2)}&=& \frac{1}{2}\, \(k_1-k_2+k_3\),\qquad\quad 
\omega^{(2)}= \frac{1}{2}\, \(\omega_1-\omega_2+\omega_3\),\qquad\quad
\Delta^{(2)}=\frac{1}{2}\, \(\delta_1-\delta_2+\delta_3+\Delta_{13}\),
\nonumber\\
k^{(3)}&=& \frac{1}{2}\, \(k_1+k_2-k_3\),\qquad\quad 
\omega^{(3)}= \frac{1}{2}\, \(\omega_1+\omega_2-\omega_3\),\qquad\quad
\Delta^{(3)}=\frac{1}{2}\, \(\delta_1+\delta_2-\delta_3+\Delta_{12}\).
\nonumber
\er
From (\ref{analysetau}) we find that
\be
\ln \tau_{\rm three-sol}= \ln 2 + z_{+} + \ln F
\ee
with 
\be
F=\cosh z_{+}+e^{-\(\Delta_{12}+\Delta_{13}\)/2}\,\cosh z_{-}^{(1)}
+e^{-\(\Delta_{12}+\Delta_{23}\)/2}\,\cosh z_{-}^{(2)}
+e^{-\(\Delta_{13}+\Delta_{23}\)/2}\,\cosh z_{-}^{(3)}.
\lab{fdefdelta}
\ee
Therefore, from (\ref{tauurel}), we get
\be
u_{\rm three-sol}=\frac{12}{\alpha}\, \partial_x^2\ln F=\frac{12}{\alpha}\left[\frac{\partial_x^2 F}{F}-\(\frac{\partial_xF}{F}\)^2
\right]
\lab{tauurel3sol}
\ee
with
\br
\partial_xF&=&\(k^{(1)}+k^{(2)}+k^{(3)}\)\,\sinh z_{+}+k^{(1)}\,e^{-\(\Delta_{12}+\Delta_{13}\)/2}\,\sinh z_{-}^{(1)}
\nonumber\\
&+&k^{(2)}\,e^{-\(\Delta_{12}+\Delta_{23}\)/2}\,\sinh z_{-}^{(2)}
+k^{(3)}\,e^{-\(\Delta_{13}+\Delta_{23}\)/2}\,\sinh z_{-}^{(3)}
\er
and
\br
\partial_x^2F&=&\(k^{(1)}+k^{(2)}+k^{(3)}\)^2\,\cosh z_{+}+\(k^{(1)}\)^2\,e^{-\(\Delta_{12}+\Delta_{13}\)/2}\,\cosh z_{-}^{(1)}
\nonumber\\
&+&\(k^{(2)}\)^2\,e^{-\(\Delta_{12}+\Delta_{23}\)/2}\,\cosh z_{-}^{(2)}
+\(k^{(3)}\)^2\,e^{-\(\Delta_{13}+\Delta_{23}\)/2}\,\cosh z_{-}^{(3)}.
\er

One way of implementing the parity argument for the KdV three-soliton solution is to have the inversion (change of signs) of all three $z_{-}^{(i)}$'s, {\it i.e.}  $z_{-}^{(i)}\rightarrow - z_{-}^{(i)}$, for $i=1,2,3$, which would also imply $z_{+}\rightarrow -z_{+}$. However,
we are performing our calculations in two dimensions and the three $z_{-}^{(i)}$'s cannot be linearly independent. Each $z_{-}^{(i)}$ defines a straight line in the $\(x\,,\,t\)$ plane, and if they are to be simultaneously inverted we need these three lines to cross at the same point, and so all three $z_{-}^{(i)}$'s should vanish at that point. So, there should exist a point $\(x_{\Delta}\,,\,t_{\Delta}\)$ such that
\br
\(\begin{array}{ccc}
k^{(1)}&-\omega^{(1)}& \Delta^{(1)}\\
k^{(2)}&-\omega^{(2)}& \Delta^{(2)}\\
k^{(3)}&-\omega^{(3)}& \Delta^{(3)}
\end{array}\)\,
\(\begin{array}{c}
x_{\Delta}\\
t_{\Delta}\\
1
\end{array}\) =0.
\er
For this to happen, the determinant of this $3\times 3$ matrix should vanish, and this implies 
 \br
\(\Delta^{(1)} +\Delta^{(2)}\) \left(k_1^3 k_2-k_1k_2^3\right)
+\(\Delta^{(1)} +\Delta^{(3)}\) \left(k_3^3 k_1-k_3 k_1^3\right)
+\(\Delta^{(2)} +\Delta^{(3)}\) \left(k_2^3 k_3-k_2 k_3^3\right)
  =0.
  \nonumber
   \er

 Note that if we choose any pair of $k_i$'s equal, we reduce the three-soliton solution to a two-soliton solution. So, we need $k_1\neq k_2\neq k_3$. One way of satisfying this is to choose $\delta_i$, $i=1,2,3$, in such a way that
\be
\Delta^{(1)}=\Delta^{(2)}=\Delta^{(3)}=0.
\lab{nicecondparity}
\ee
Note that in such a case the $z_{-}^{(i)}$'s, given in (\ref{zminusi}), (and so $z_{+}$) become homogeneous in $x$ and $t$, {\it i.e.} 
\be
z_{-}^{(i)}= k^{(i)}\,x-\omega^{(i)}\,t\qquad\qquad\qquad i=1,2,3
\lab{zminusi2}
\ee
Therefore, the parity transformation
\be
P:\qquad\qquad\qquad \(x\,,\,t\)\rightarrow \(-x\,,\,-t\)
\label{paritythreesoliton}
\ee 
is sufficient to have what we want, {\it i.e.} $ z_{-}^{(i)}\rightarrow - z_{-}^{(i)}$, and so $z_{+}\rightarrow -z_{+}$. 

The condition \rf{nicecondparity} leads to the linear system
\br
\(\begin{array}{ccc}
1&-1&-1\\
-1&1&-1\\
-1&-1&1
\end{array}\)\,
\(\begin{array}{c}
\delta_1\\
\delta_2\\
\delta_3
\end{array}\)
=\(\begin{array}{c}
\Delta_{23}\\
\Delta_{13}\\
\Delta_{12}
\end{array}\)
\er
and so
\br
\delta_1&=&-\frac{1}{2}\,\(\Delta_{13}+\Delta_{12}\)
= -\frac{1}{2}\,\ln\left[\frac{\(k_1-k_3\)^2}{\(k_1+k_3\)^2}\,\frac{\(k_1-k_2\)^2}{\(k_1+k_2\)^2}\right],
\nonumber\\
\delta_2&=&-\frac{1}{2}\,\(\Delta_{23}+\Delta_{12}\)
= -\frac{1}{2}\,\ln\left[\frac{\(k_2-k_3\)^2}{\(k_2+k_3\)^2}\,\frac{\(k_1-k_2\)^2}{\(k_1+k_2\)^2}\right],
\\
\delta_3&=&-\frac{1}{2}\,\(\Delta_{23}+\Delta_{13}\)
= -\frac{1}{2}\,\ln\left[\frac{\(k_2-k_3\)^2}{\(k_2+k_3\)^2}\,\frac{\(k_1-k_3\)^2}{\(k_1+k_3\)^2}\right].
\nonumber
\er
Denoting
\be
\Gamma_i= \tG_i+\delta_i \qquad\qquad\qquad{\rm with}\qquad\qquad \tG_i=k_i\, x-\omega_i\,t
\ee
we get  from (\ref{analysetau})  that
\br
\tau_{\rm three-sol}&=& 1+e^{\tG_1+\tG_2+\tG_3}+  e^{-\(\Delta_{13}+\Delta_{12}\)/2}\,\(e^{\tG_1}+e^{\tG_2+\tG_3}\) 
+  e^{-\(\Delta_{23}+\Delta_{12}\)/2}\,\(e^{\tG_2}+e^{\tG_1+\tG_3}\)
\nonumber\\
&+& e^{-\(\Delta_{23}+\Delta_{13}\)/2}\,\(e^{\tG_3} +e^{\tG_1+\tG_2}\).
\er
One can check that  all these cases correspond to the three soliton colliding simultaneously at the origin $x=0$ and $t=0$. The conclusion is therefore that for the KdV three-soliton solution in which the three solitons collide at the same point in space-time, the field $u$ is even under the parity  (\ref{paritythreesoliton}), when evaluated on such a  solution.


	\section{Numerical analysis of quasi-conserved charges}
	\label{sec:numerical}
\setcounter{equation}{0}
	
	In this section we discuss results of some of oyr numerical simulations of equation~(\ref{deformedqkdv}) using various values of $\varepsilon_1$ and $\varepsilon_2$. We particularly focus on investigating the corresponding first non-trivial quasi-conserved charge~$Q^{(-3)}$, defined in equation~(\ref{chargesdef}). The numerical scheme used to approximate equation~(\ref{deformedqkdv}) is very similar to the algorithm used to solve the mRLW equation~\cite{first_paper}. We have adjusted this scheme appropriately, which is discussed in appendix~\ref{Numerical_method}, for the simulations presented in this section.

	We have used several different initial conditions depending on the values of $\varepsilon_1$ 
and $\varepsilon_2$ used for each simulation, and we discuss these initial conditions in more detail in 
the subsections below. We have performed the numerical experiments for two-soliton simulations for a range of values for $\Gamma_1$ and $\Gamma_2$ but, for consistency, all the plots presented in this section regarding two-soliton simulations are obtained using the same values regardless of the initial conditions used, and similarly for all the three-soliton simulations ($\Gamma_1$, $\Gamma_2$ and $\Gamma_3$) presented in this section. For all the simulations presented in this section we have used a grid spacing $h=0.1$ and time level $\tau = 0.001$. More details on the variables used for the simulations discussed in this section can be found in appendix~\ref{summary_of_variables}.
		
	\subsection{Quasi-conserved charges of the mRLW equation}
	
	In this subsection we discuss the results of our investigations of the quasi-conserved charges of the mRLW equation. To perform the simulations of the mRLW equation, we have set $\varepsilon_1 = \varepsilon_2 = 1$ in equation~(\ref{deformedqkdv}). 

	\subsubsection{Two-soliton solutions of the mRLW equation}
		
\begin{figure}[t!]
		\centering
		\hspace*{-0.1cm}
		\begin{subfigure}{.34\textwidth}
			\centering
			\includegraphics[scale=0.34]{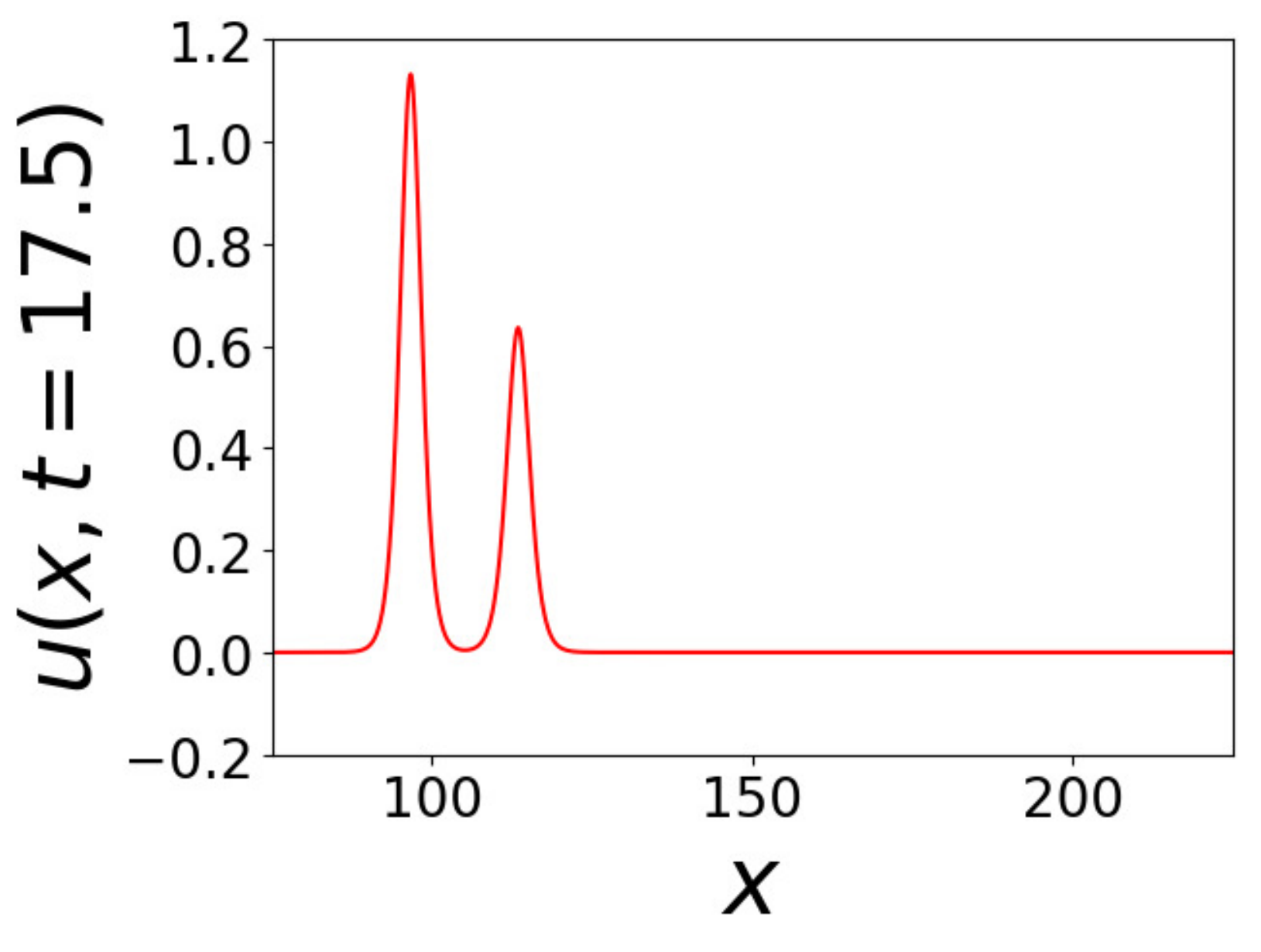}
			\caption{At~$t=17.5$}
			\label{plot0_2}
		\end{subfigure}%
		\begin{subfigure}{.34\textwidth}
			\centering
			\includegraphics[scale=0.34]{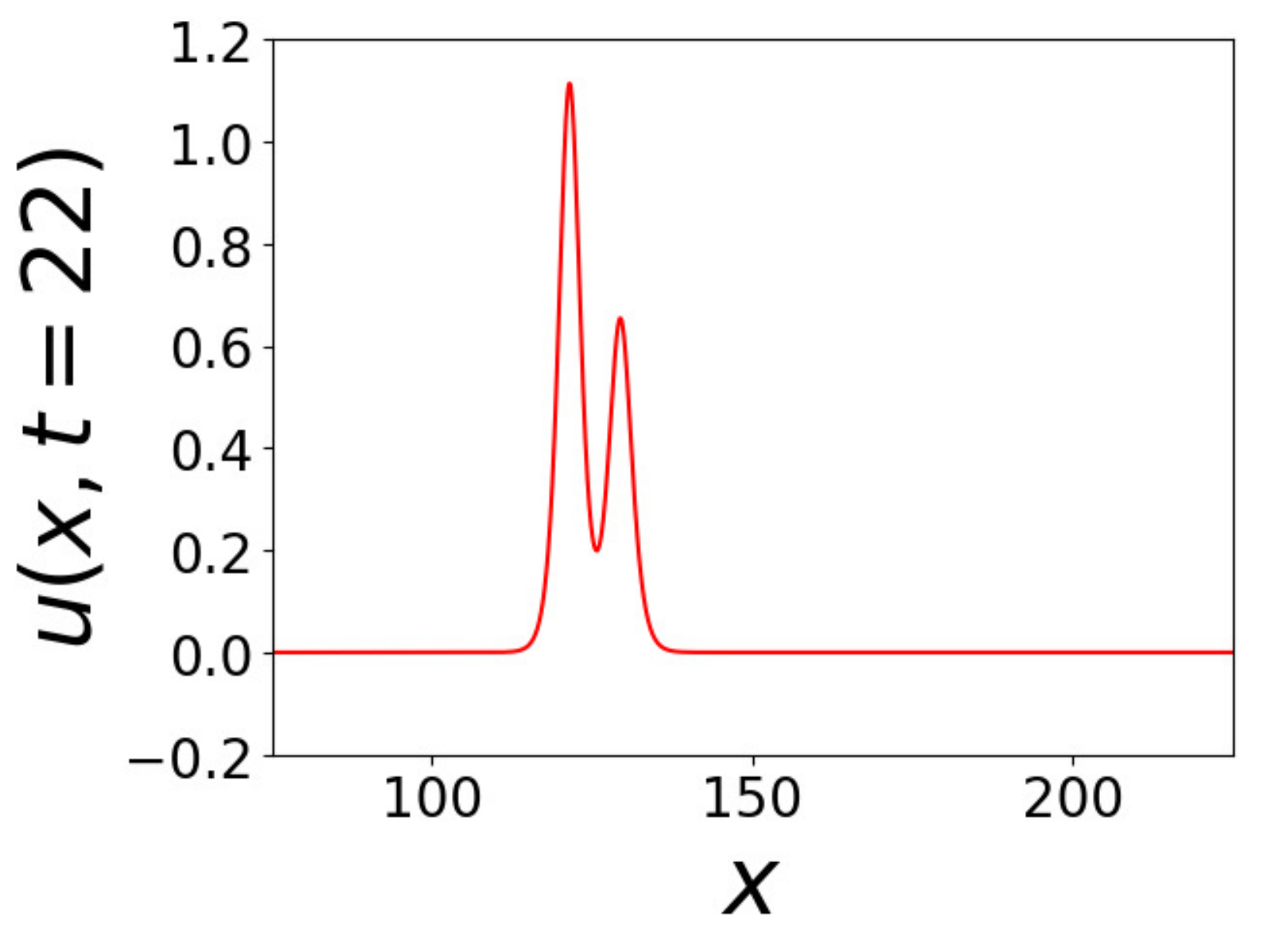}
			\caption{At~$t=22$}
			\label{plot0_3}
		\end{subfigure}%
		\begin{subfigure}{.34\textwidth}
			\centering
			\includegraphics[scale=0.34]{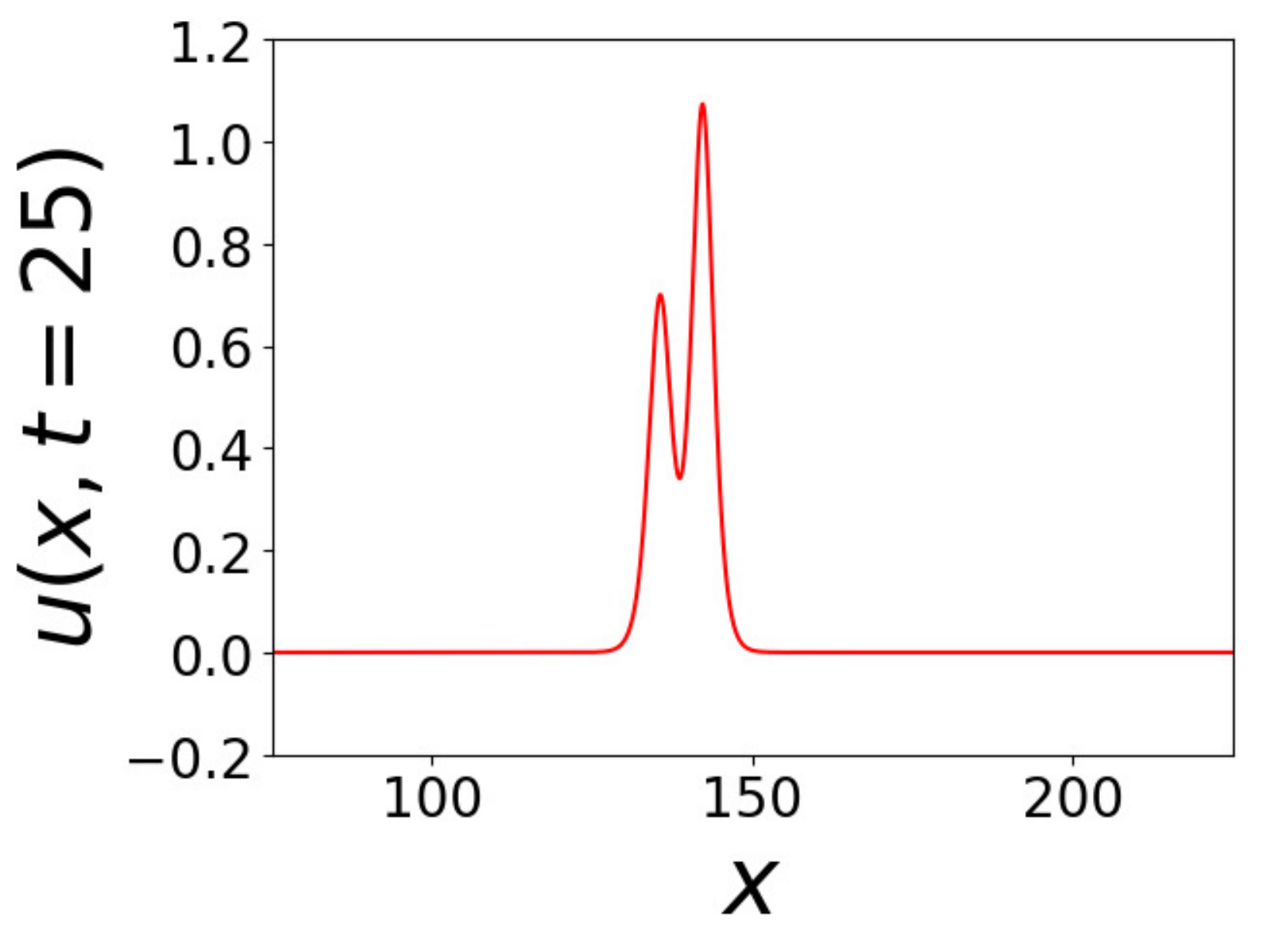}
			\caption{At~$t=25$}
			\label{plot0_4}
		\end{subfigure}
			
		\begin{subfigure}{.34\textwidth}
			\centering
			\includegraphics[scale=0.34]{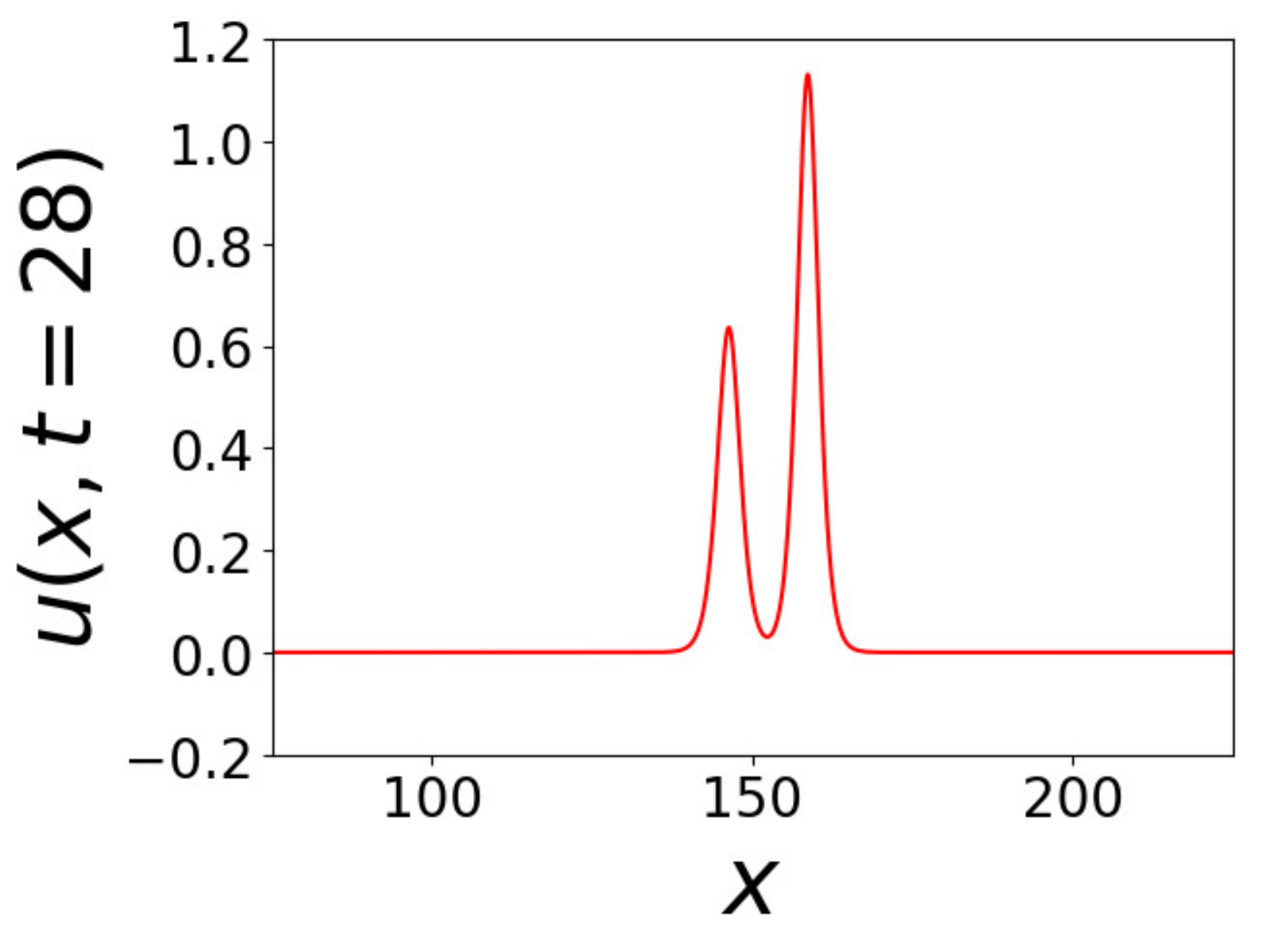}
			\caption{At~$t=28$}
			\label{plot_5}
		\end{subfigure}%
		\begin{subfigure}{.34\textwidth}
			\centering
			\includegraphics[scale=0.34]{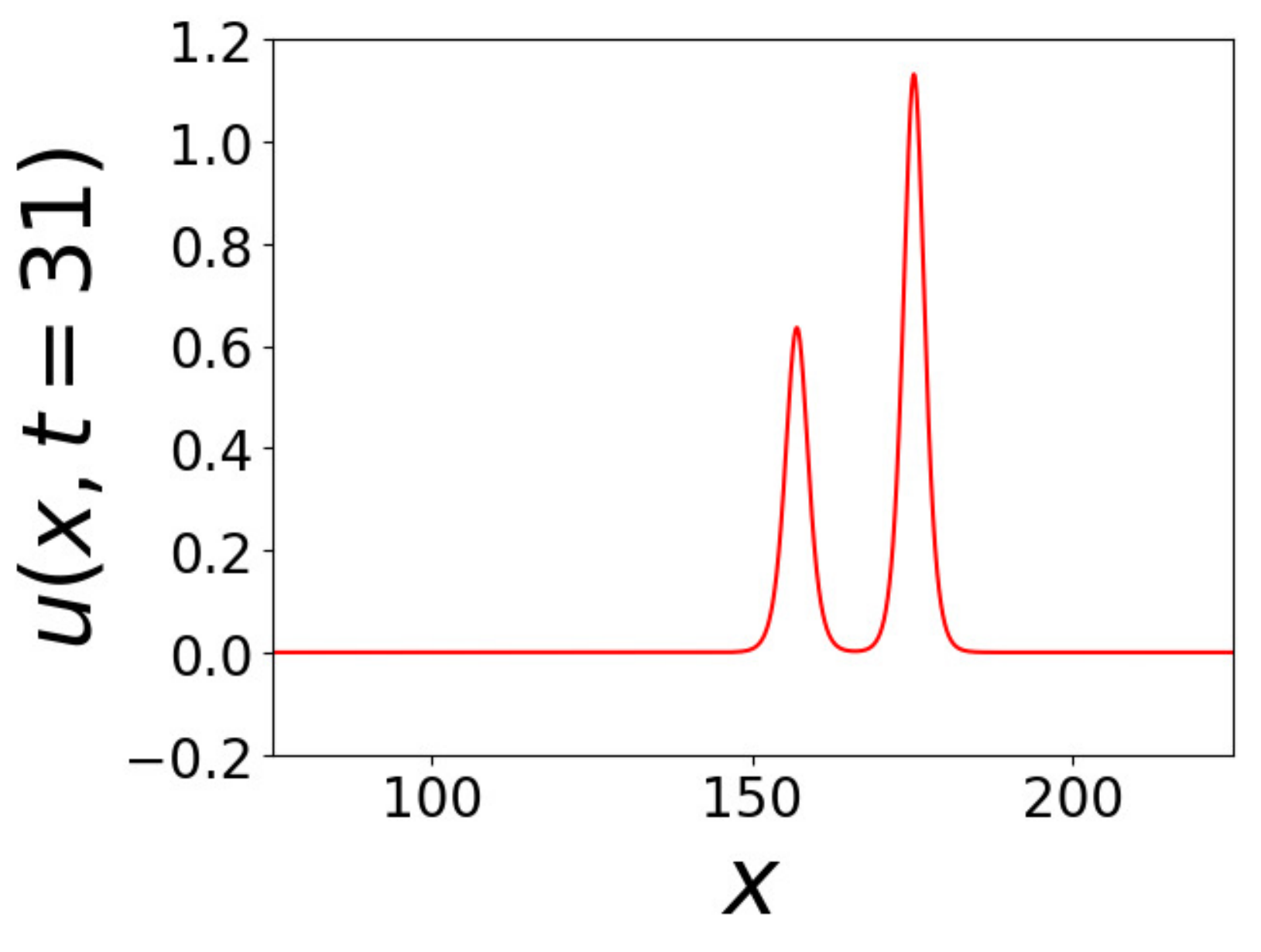}
			\caption{At~$t=31$}
			\label{plot0_6}
		\end{subfigure}%
		\begin{subfigure}{.34\textwidth}
			\centering
			\includegraphics[scale=0.34]{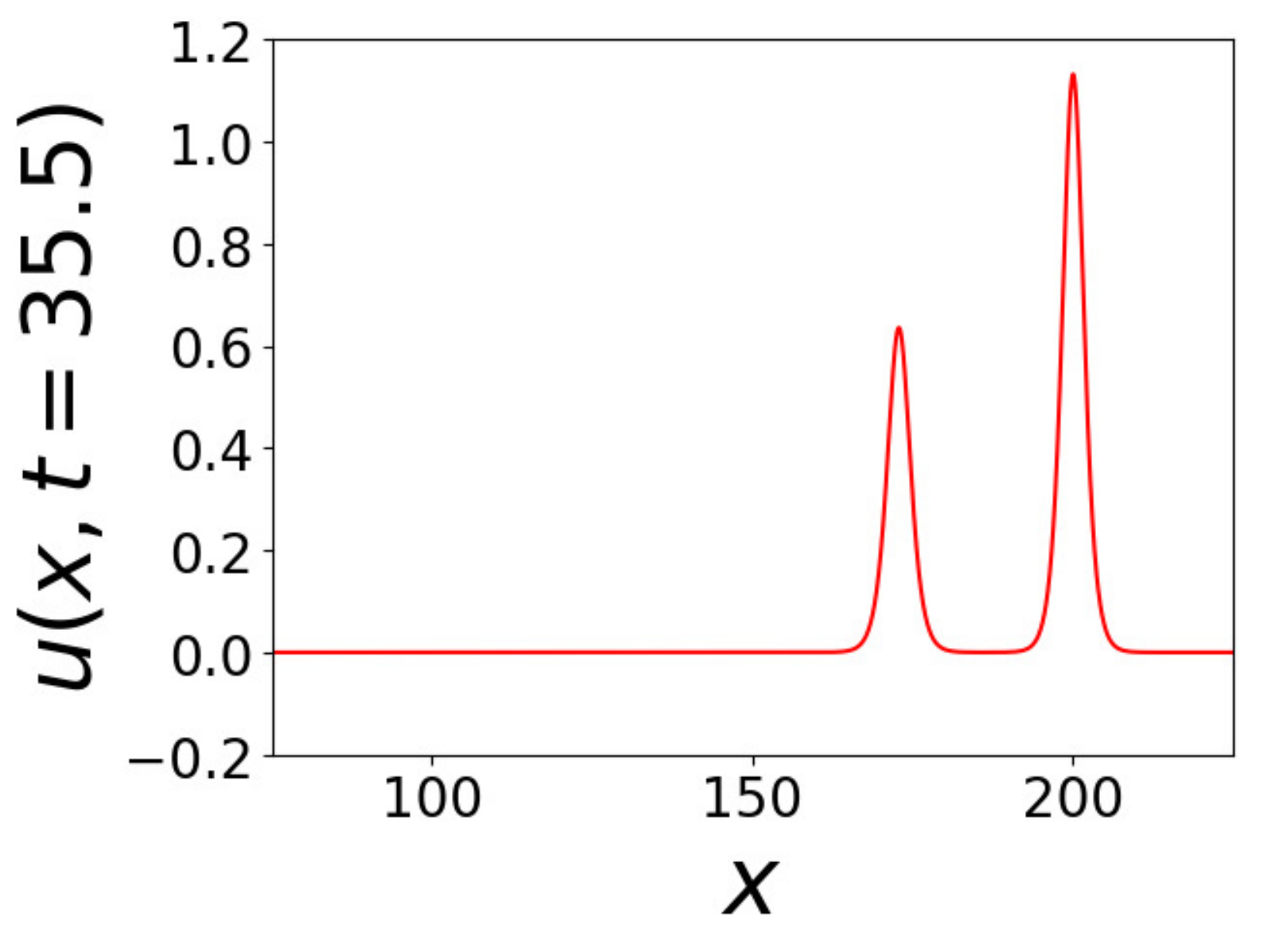}
			\caption{At~$t=35.5$}
			\label{plot0_7}
		\end{subfigure}
		\caption{The spatial-dependence of the $u$ field for two solitons described by mRLW equation, interacting with each other at different points in time, as given in equations~(\ref{fieldqdef}) and~(\ref{1.2}).} 
		\label{plot0_2to0_7}
	\end{figure}

\begin{figure}[b!]
		\centering
		\begin{subfigure}{.5\textwidth}
			\hspace*{-1.2cm}
			\centering
			\includegraphics[scale=0.34]{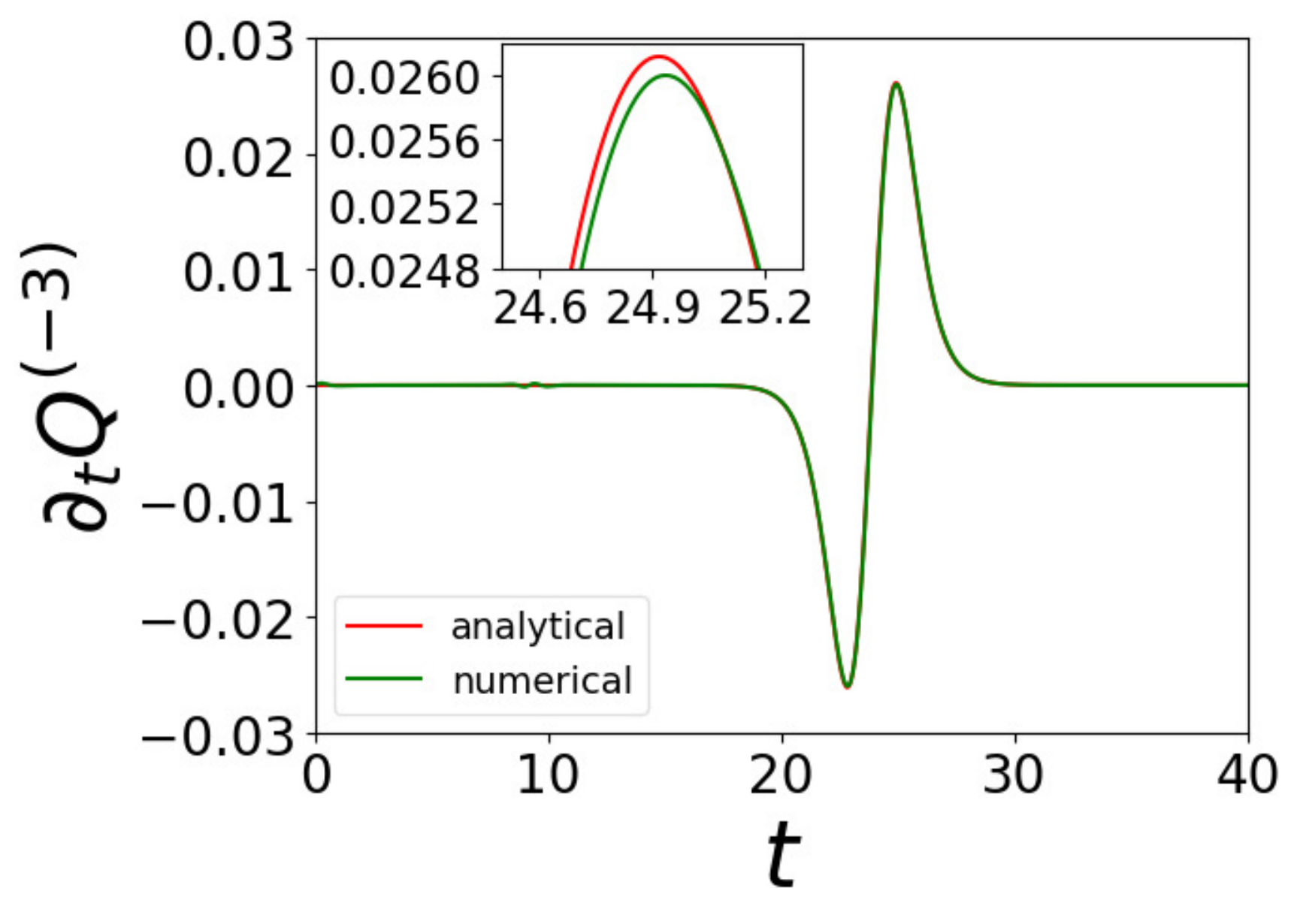}
			\caption{}
			\label{I_1_alpha_8_e1_1_e2_1_e3_1_1_anal_and_num}
		\end{subfigure}%
		\begin{subfigure}{.5\textwidth}
			\hspace*{-1.5cm}
			\centering
			\includegraphics[scale=0.34]{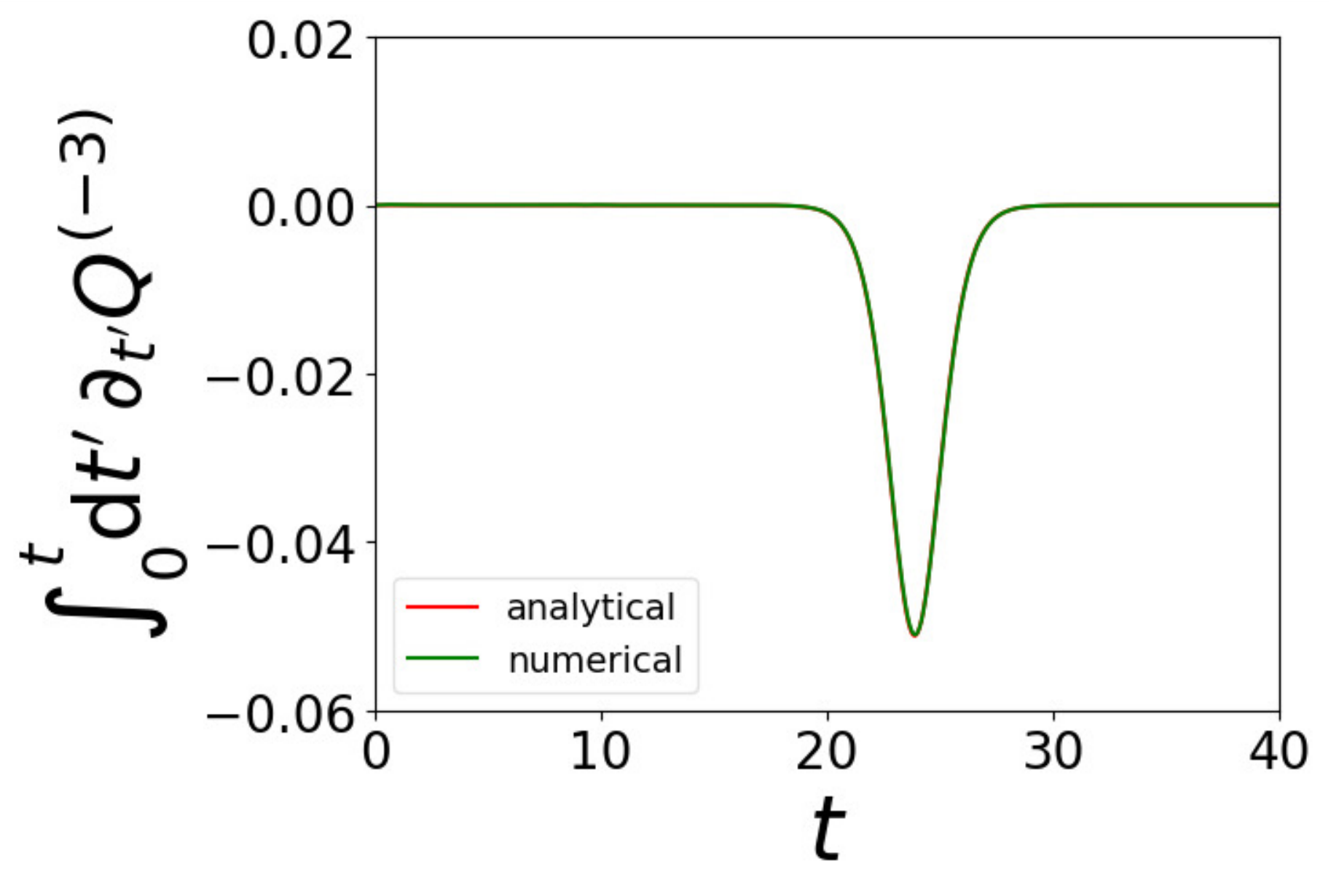}
			\caption{}
			\label{I_1_alpha_8_e1_1_e2_1_e3_1_1_anal_and_num_int_time_space_dQ_dt}
		\end{subfigure}%
			\caption{The time-dependence of the analytical and numerically (represented by 
the red line and the green line, respectively) obtained values of the quantity $\partial_t Q^{(-3)}$ and $\int_{0}^{t} \mathrm{d} t^\prime \, \partial_{t^\prime} Q^{(-3)}$ for the two-soliton simulation of the mRLW equation.}
			\label{I_1_alpha_8_e1_1_e2_1_e3_1_1_analytical_and_numerical}
	\end{figure}

	In figure~\ref{plot0_2to0_7}
	we have plotted the scattering of two solitons described by equation~(\ref{1.2}) at various values of $t$, and in figure~\ref{I_1_alpha_8_e1_1_e2_1_e3_1_1_anal_and_num}
	we have plotted the time dependence of $\partial_t Q^{(-3)}$ for this interaction. Since the mRLW equation possesses an analytical form of the two-soliton solution, in figure~\ref{I_1_alpha_8_e1_1_e2_1_e3_1_1_analytical_and_numerical} we have in fact plotted the values $\partial_t Q^{(-3)}$ for both the analytical and numerical simulation. To better illustrate how small the discrepancy is, we have added an insert of the region near the global maximum on a much smaller scale. We note  that there is hardly any discrepancy between the analytical and numerical values. 
	
	Now, comparing figures~\ref{plot0_2to0_7} and~\ref{I_1_alpha_8_e1_1_e2_1_e3_1_1_anal_and_num}, we see that as $t \to \pm \infty$ the charge is conserved except for during the collision, where it first decreases and then increases back to the same value as before the collision.
	Figure~\ref{I_1_alpha_8_e1_1_e2_1_e3_1_1_anal_and_num_int_time_space_dQ_dt} presents the plots of $\int_{0}^{t} \mathrm{d} t^\prime \, \partial_{t^\prime} Q^{(-3)}$, and we see that  the two lumps exactly cancel each other out, that is,
	\begin{equation}
		\lim_{t \to - \infty} Q^{(-3)} = \lim_{t \to \infty} Q^{(-3)} \,.
	\end{equation}
	Thus, we can conclude that these quantities are indeed quasi-conserved, as has been shown analytically in section \ref{sec:analyticalmrlw}. The agreement of the analytical and numerical results is a good test of the numerical code that we are using.


	\subsubsection{Three-soliton solutions of the mRLW equation} \label{Three_soliton_solutions_of_the_mRLW_equation}
	
\begin{figure}[b!]
		\centering
		\hspace*{-0.1cm}
		\begin{subfigure}{.34\textwidth}
			\centering
			\includegraphics[scale=0.34]{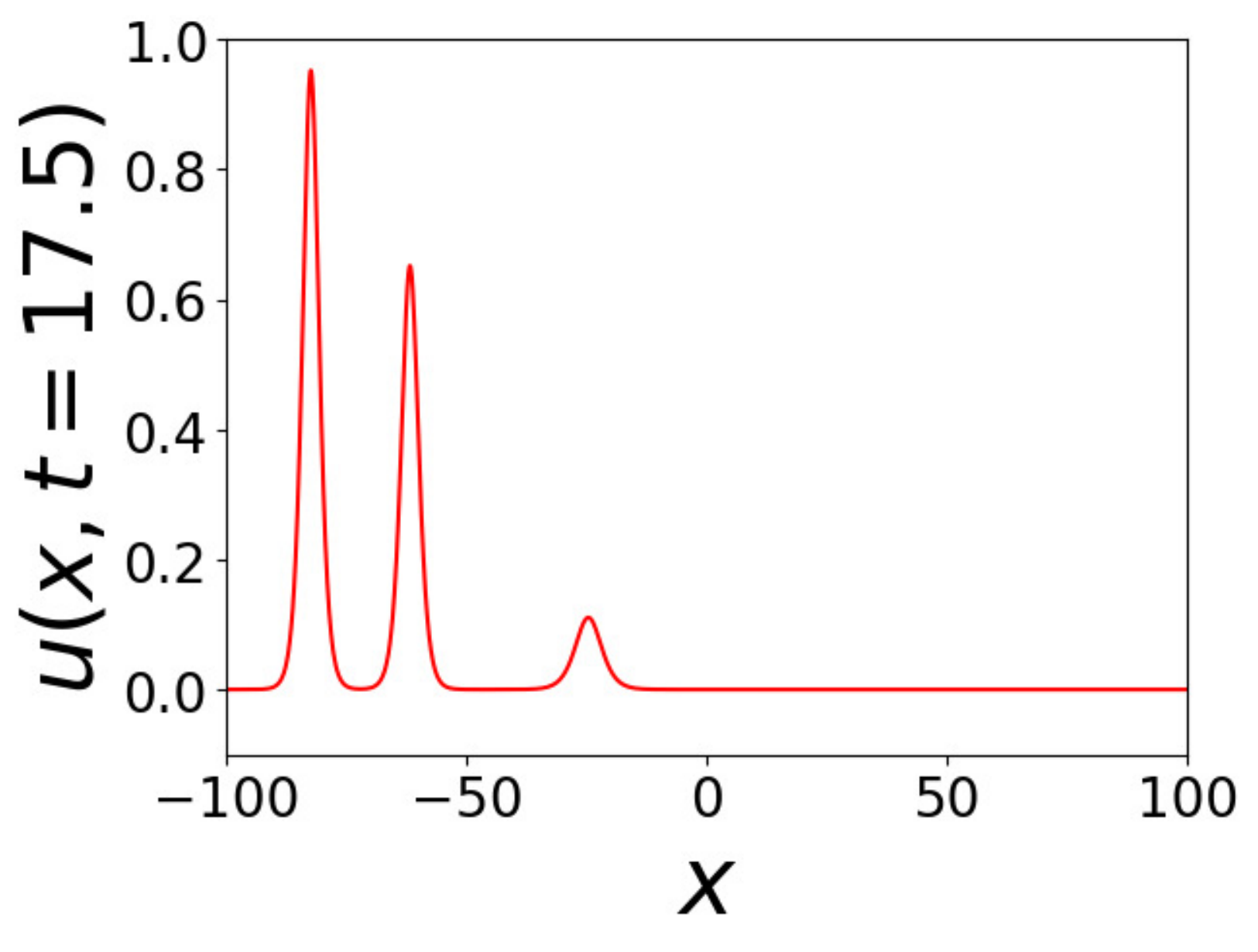}
			\caption{At~$t=17.5$}
			\label{plot1_2}
		\end{subfigure}%
		\begin{subfigure}{.34\textwidth}
			\centering
			\includegraphics[scale=0.34]{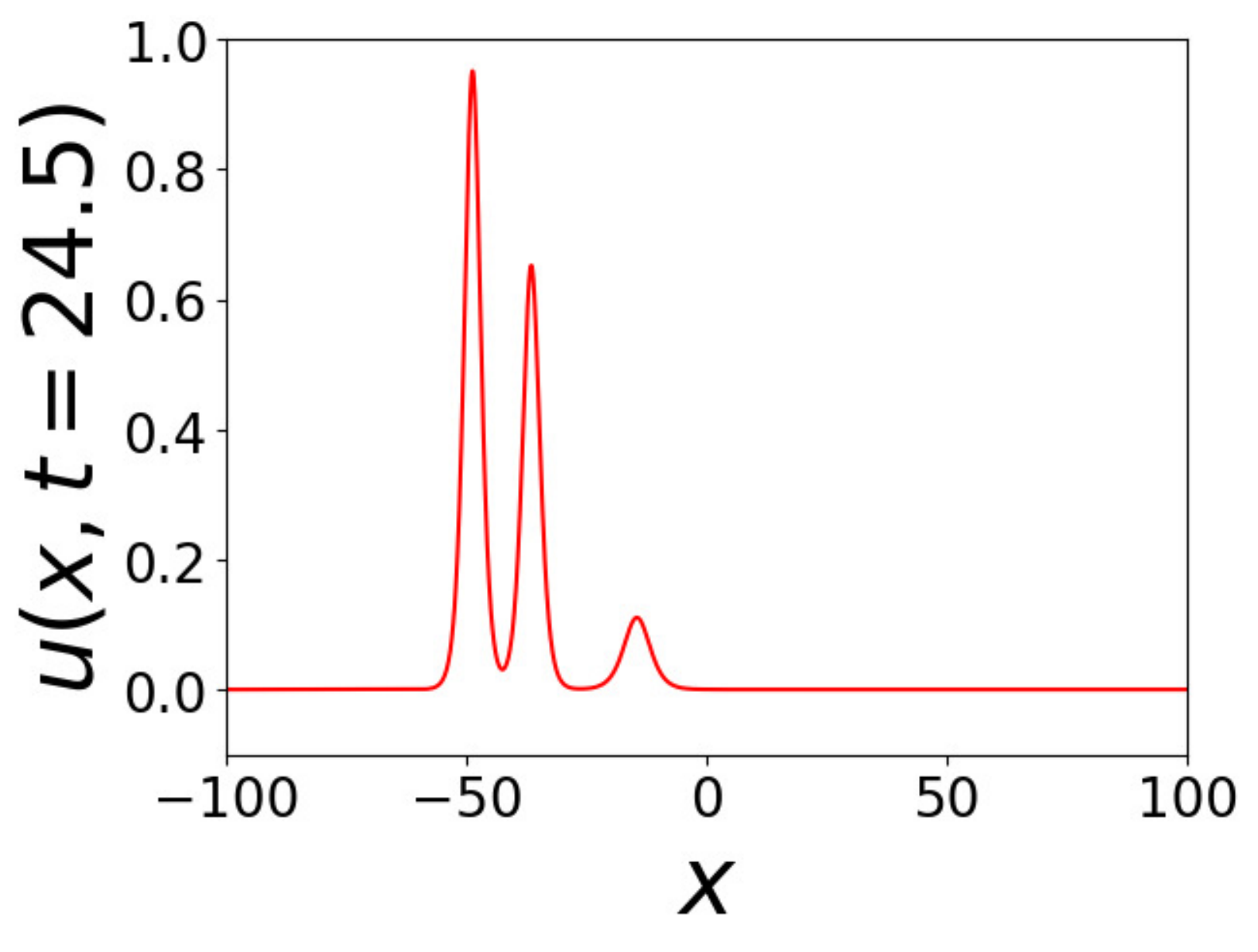}
			\caption{At~$t=24.5$}
			\label{plot1_3}
		\end{subfigure}%
		\begin{subfigure}{.34\textwidth}
			\centering
			\includegraphics[scale=0.34]{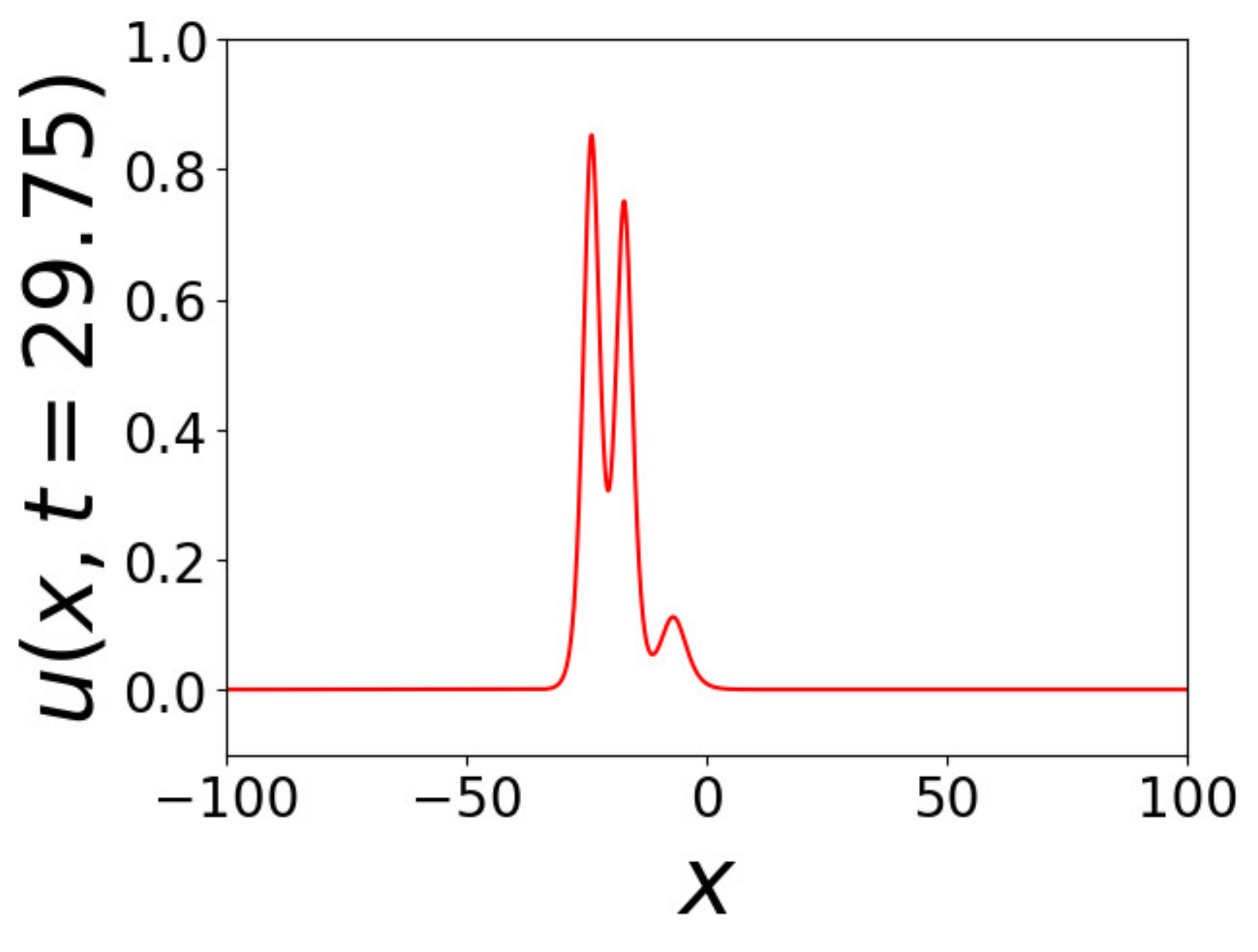}
			\caption{At~$t=29.75$}
			\label{plot1_4}
		\end{subfigure}
		
		\begin{subfigure}{.34\textwidth}
			\centering
			\includegraphics[scale=0.34]{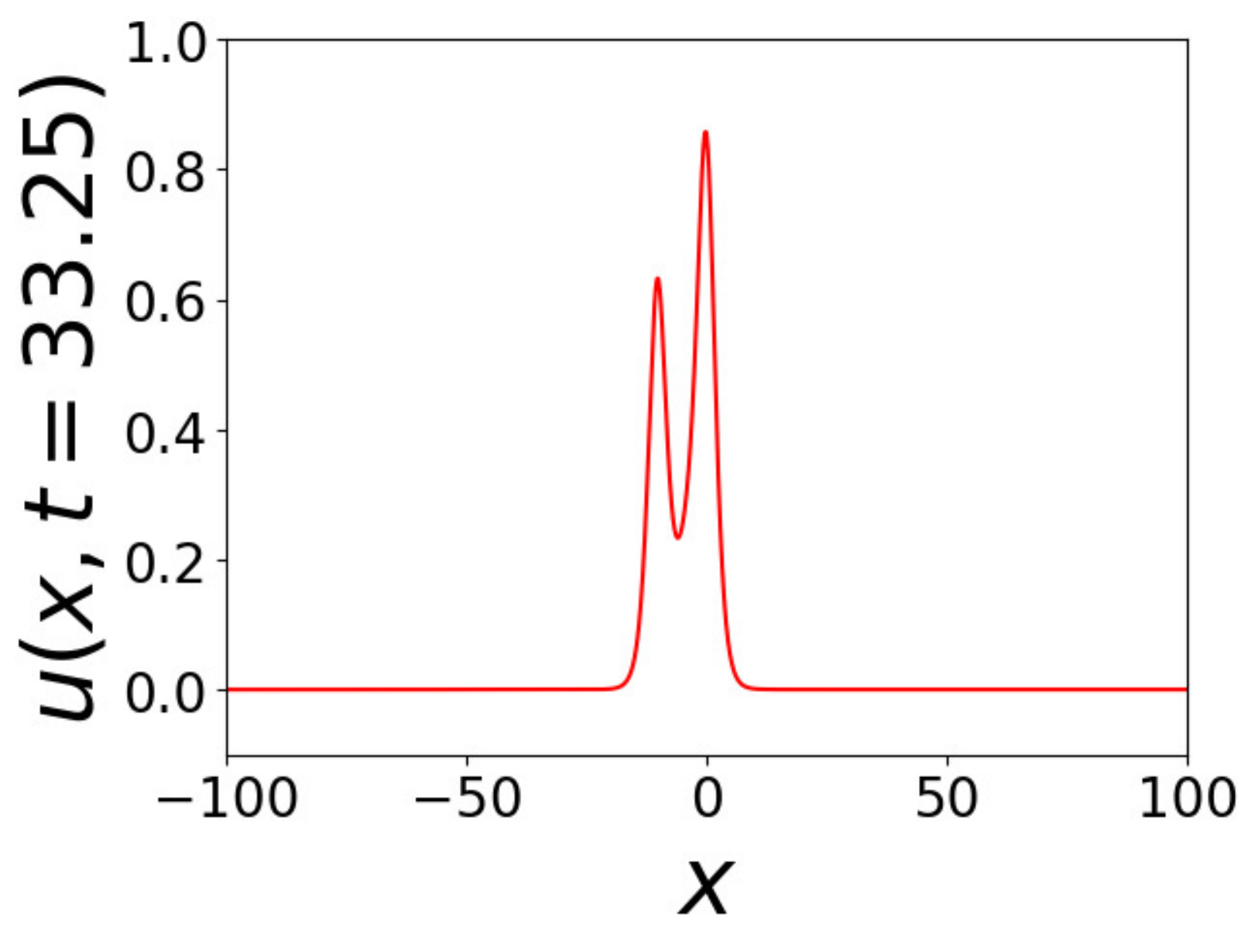}
			\caption{At~$t=33.25$}
			\label{plot1_5}
		\end{subfigure}%
		\begin{subfigure}{.34\textwidth}
			\centering
			\includegraphics[scale=0.34]{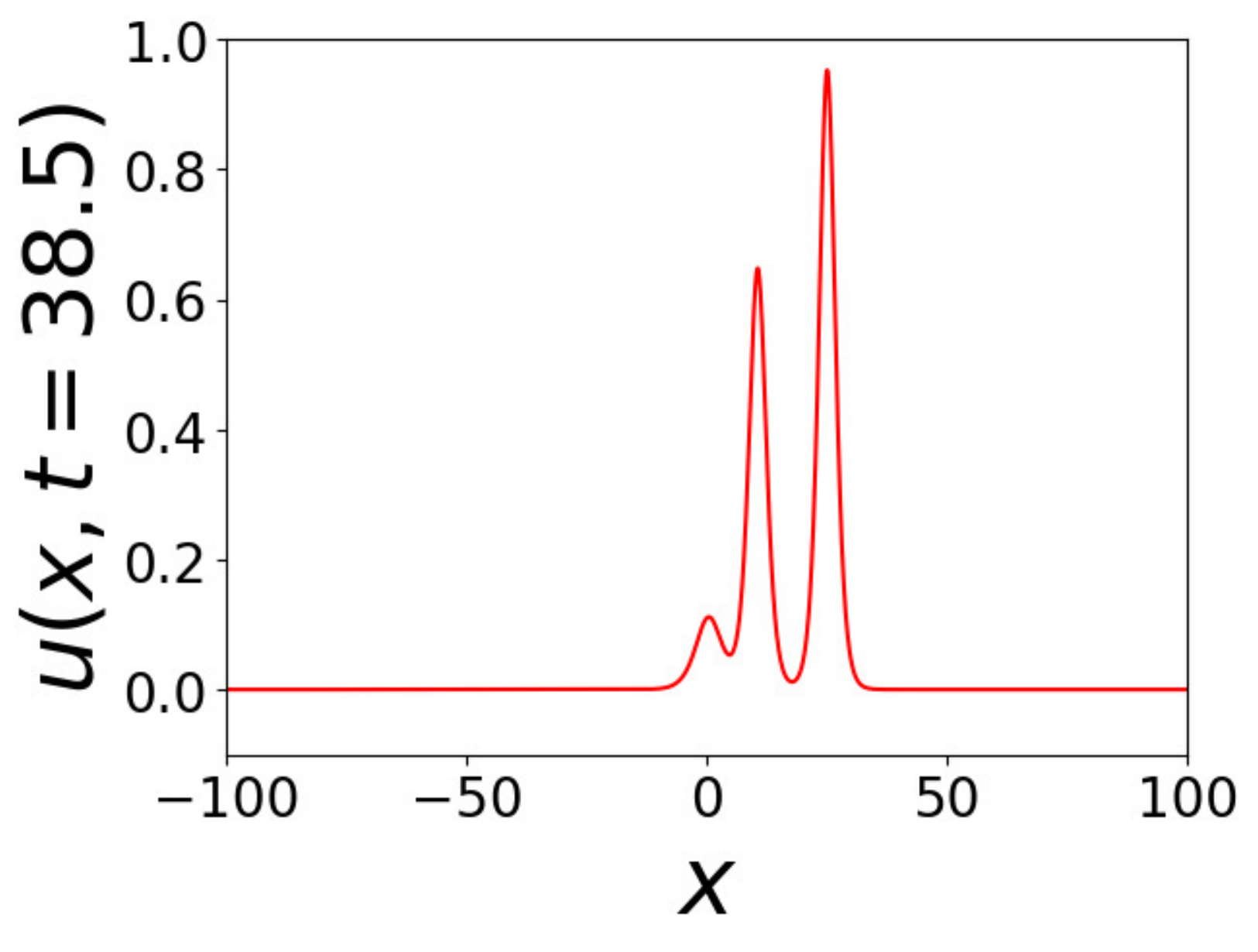}
			\caption{At~$t=38.5$}
			\label{plot1_6}
		\end{subfigure}%
		\begin{subfigure}{.34\textwidth}
			\centering
			\includegraphics[scale=0.34]{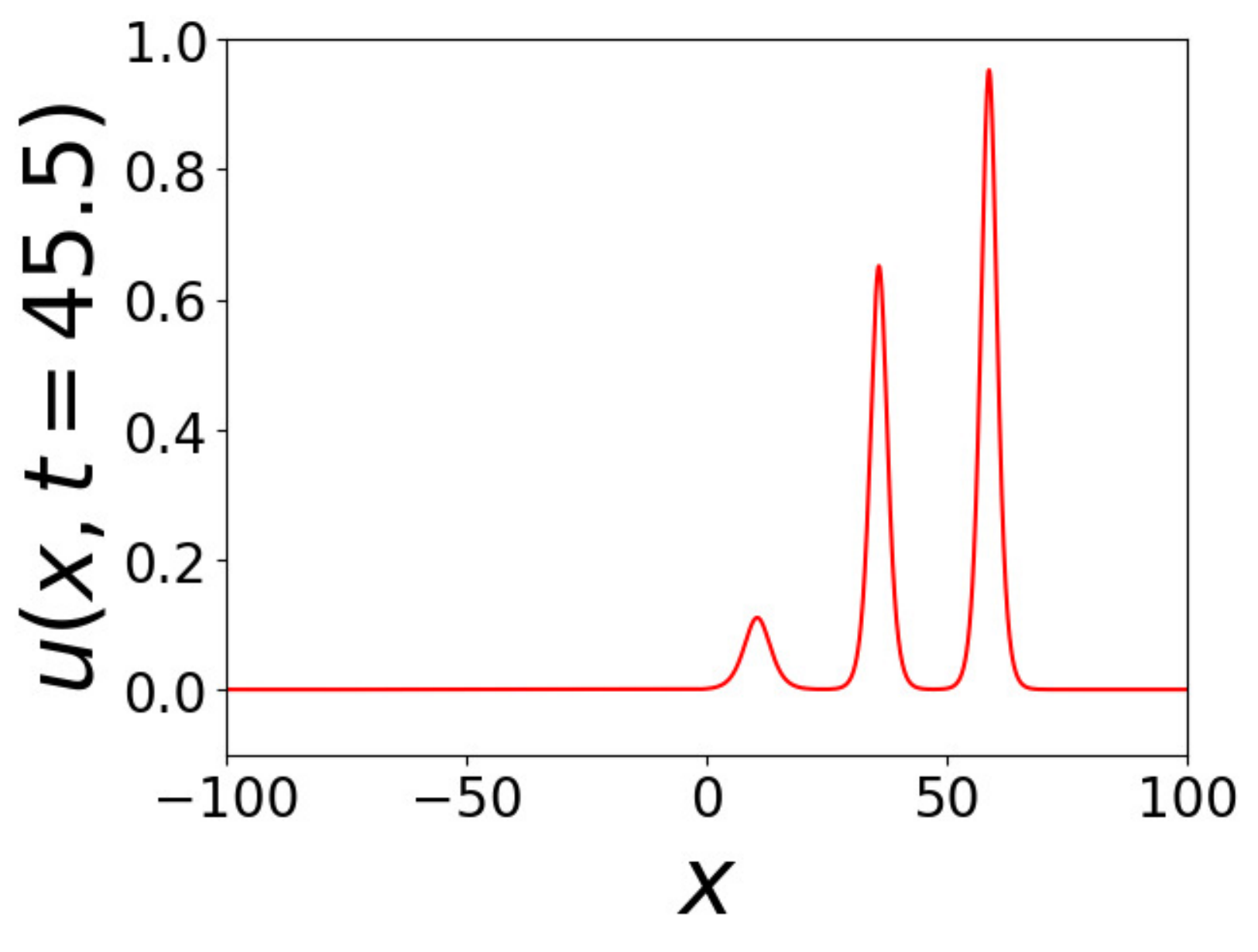}
			\caption{At~$t=45.5$}
			\label{plot1_7}
		\end{subfigure}
		\caption{Snapshots at different points in time of the numerical three-soliton solution of the  mRLW equation (equation (\ref{deformedqkdv}) for $\ve_1=\ve_2=1$), for the case when the solitons collide all around the point $x=0$.} 
		\label{plot1_2to1_7}
	\end{figure}

\begin{figure}[t!]
		\centering
		\begin{subfigure}{.5\textwidth}
			\hspace*{-1.2cm}
			\centering
			\includegraphics[scale=0.34]{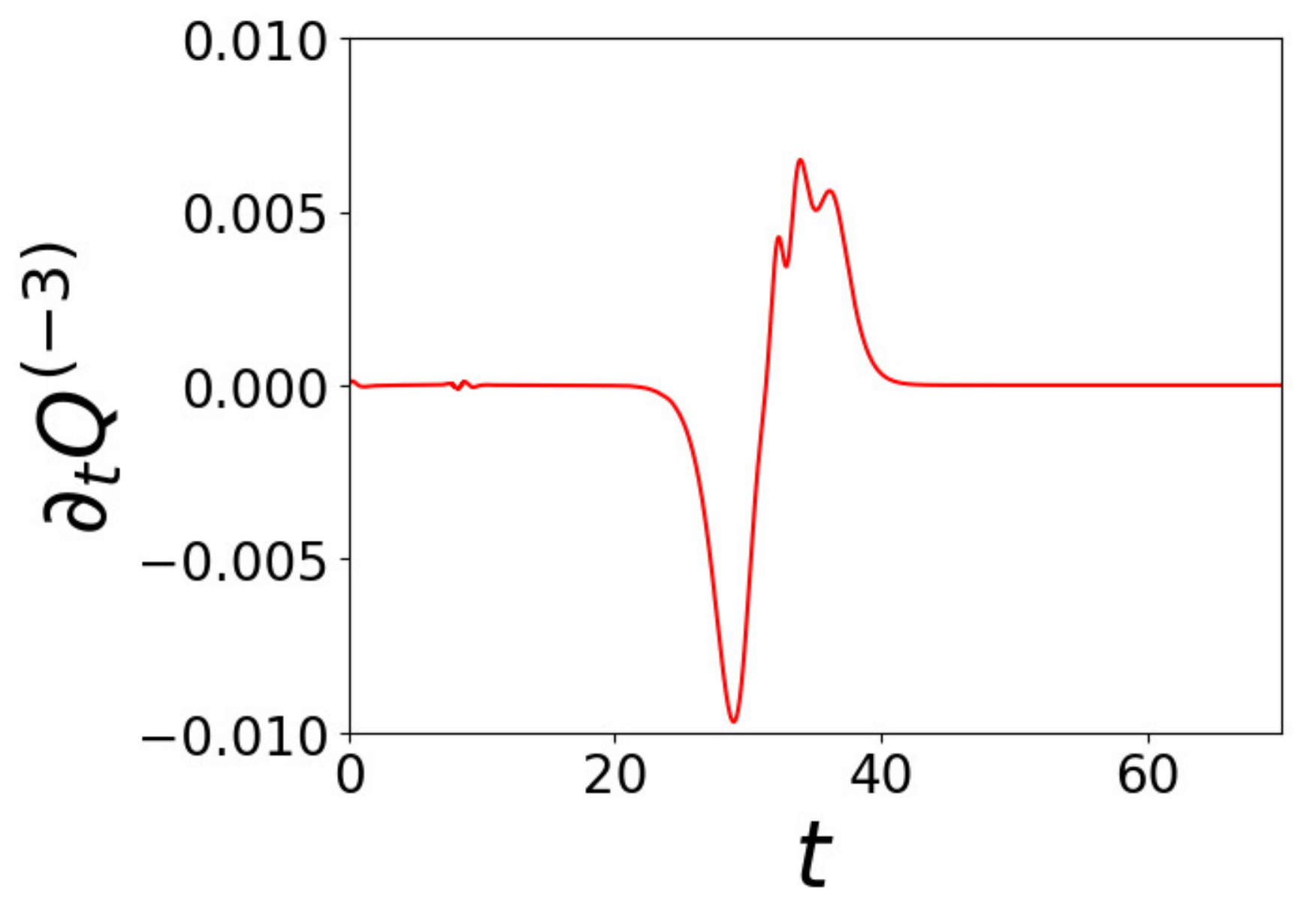}
			\caption{}
			\label{plot1_8}
		\end{subfigure}%
		\begin{subfigure}{.5\textwidth}
			\hspace*{-1.5cm}
			\centering
			\includegraphics[scale=0.34]{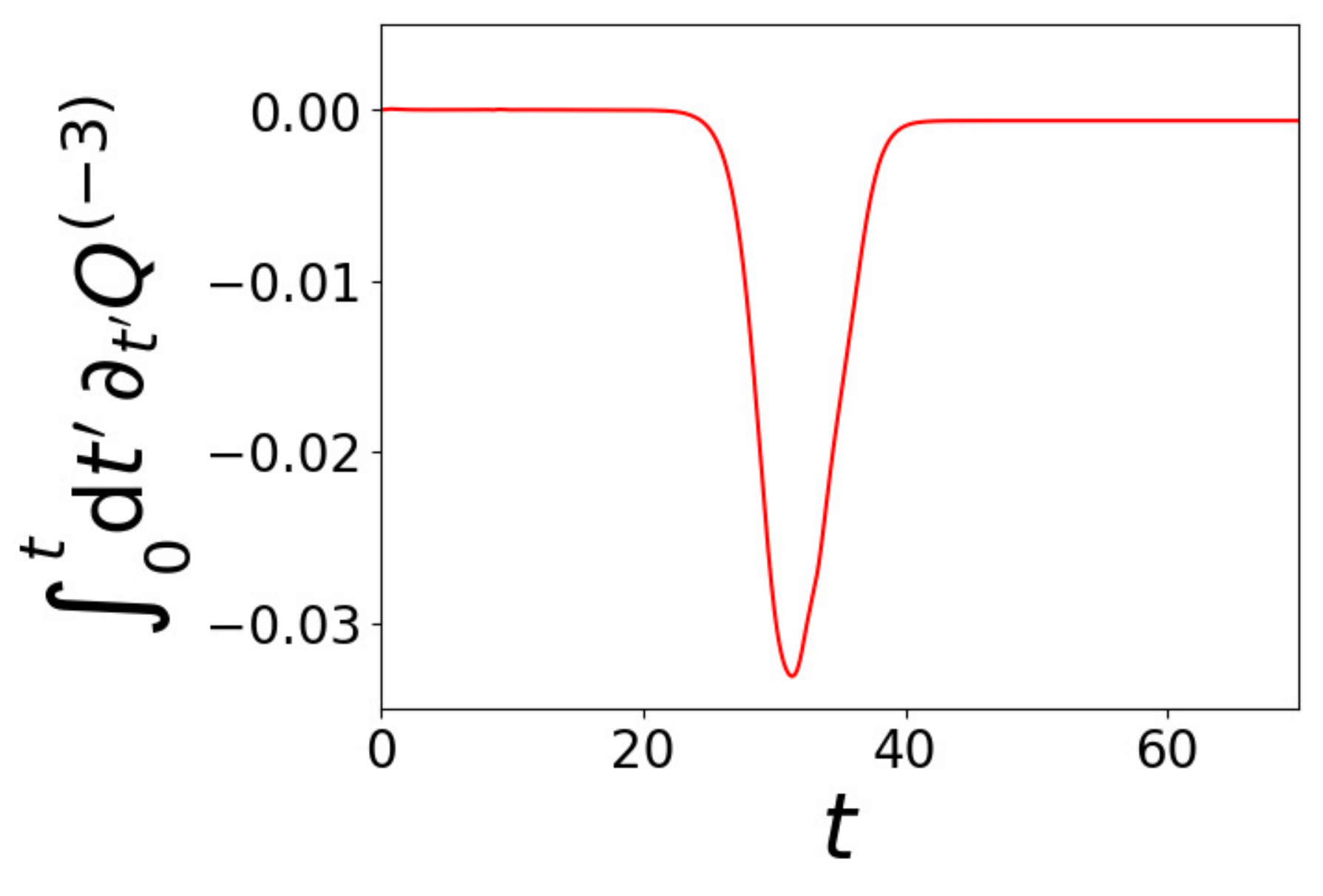}
			\caption{}
			\label{plot1_9}
		\end{subfigure}%
		\caption{The time-dependence of the obtained values of the quantity $\partial_t Q^{(-3)}$ and $\int_{0}^{t} \mathrm{d} t^\prime \, \partial_{t^\prime} Q^{(-3)}$ for the three-soliton simulation of the mRLW equation.}
		\label{plot1_8to1_9}
	\end{figure}

	In figure~\ref{plot1_2to1_7} 
	we present a selection of plots (at different values of $t$) of the $u$ field obtained in a typical numerical simulation of a superposition of three one-soliton solutions. This was done in this way since we do not have analytic three-soliton solutions of the mRLW equation. However, as was discussed in~\cite{first_paper}, such a field is a very good approximation to a solution and the solitons of the numerical three-soliton configuration behave as integrable (or quasi-integrable) solitons. So here we have tested this by looking at the behaviour of corresponding charges. Figure~\ref{plot1_8to1_9} 
	shows the corresponding time dependence of $\partial_t Q^{(-3)}$ and $\int_{0}^{t} \mathrm{d} t^\prime \, \partial_{t^\prime} Q^{(-3)}$. As $Q^{(-3)}$ changes only during the collision, and has the same value in the asymptotic regions (see figure~\ref{plot1_9}), we conclude that our charge is also quasi-conserved for the three-soliton configuration.
	
	In section \ref{sec:kdvsolitons} we have shown that the $u$-field for the exact Hirota three-soliton of the KdV equation is even under the parity transformation (\ref{paritythreesoliton}), when the three solitons collide all at the same point.  As we have discussed in section \ref{sec:paritydynamicsargument}, one should therefore expect that the deformed three-soliton solution should keep that parity property, and so to have an infinity of quasi-conserved charges given by (\ref{chargesdef}) (see section \ref{sec:parityonesol}). The numerical results presented in figure \ref{plot1_8to1_9} show that the charge $Q^{(-3)}$ is indeed quasi-conserved for the mRLW equation, and so it confirms  this expectation.

	\subsection{Quasi-conserved charges of the RLW equation}

	In this subsection we discuss similar topics
 for the 
RLW equation. This model, is described by equation~(\ref{5.02}), in which the conventional re-scaling of the $u$-field used in the literature sets $\alpha=12$.  The model possesses the same exact one-soliton solution with
 the same dispersion relation as the mRLW equation, but it does not possess an analytical two-soliton solutions~\cite{Bryan}. However, since we are 
simulating equation~(\ref{deformedqkdv}) with $\varepsilon_1 = 1$ and $\varepsilon_2 = 0$, we must perform the appropriate rescaling of~$q$  (\textit{i.e.}, $q \to \frac{3}{2}q$) in order to obtain the analytical one-soliton solution.
 Thus, to construct the two- and three numerical systems we have used, as initial conditions for the numerical simulation, a linear superposition of the following analytical single-soliton solutions
\begin{equation}
	q = \frac{3}{2} \sum\limits_{i=1}^n \ln (1 + e^{\Gamma_i})
\end{equation}
where $n=2$ corresponds to the two-soliton simulation and $n=3$ to the three-soliton simulation.

	\subsubsection{Two-soliton solutions of the RLW equation} \label{Two_soliton_solutions_of_the_RLW_equation}
	
\begin{figure}[t!]
		\centering
		\hspace*{-0.1cm}
		\begin{subfigure}{.34\textwidth}
			\centering
			\includegraphics[scale=0.34]{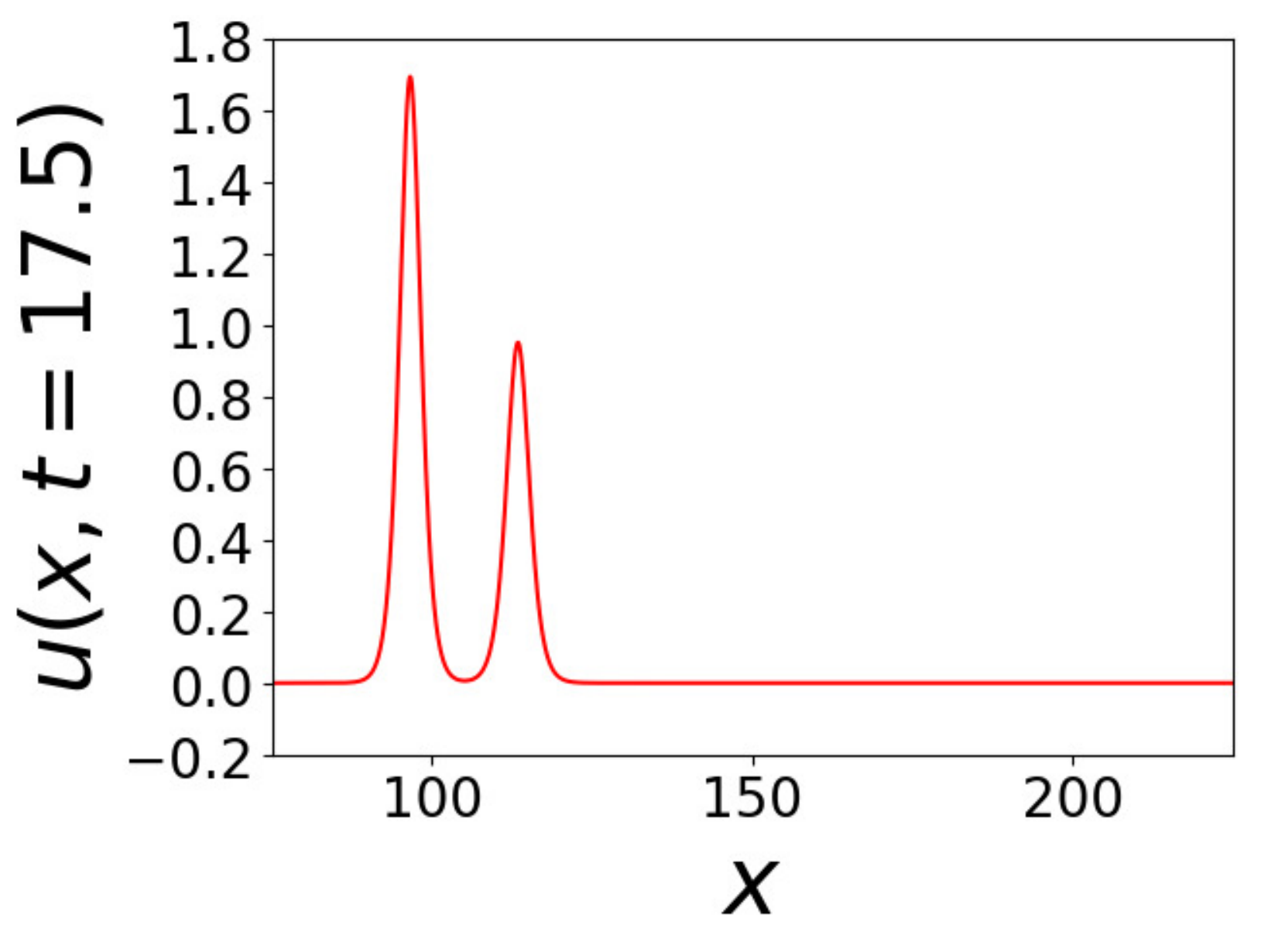}
			\caption{At~$t=17.5$}
			\label{plot2_2}
		\end{subfigure}%
		\begin{subfigure}{.34\textwidth}
			\centering
			\includegraphics[scale=0.34]{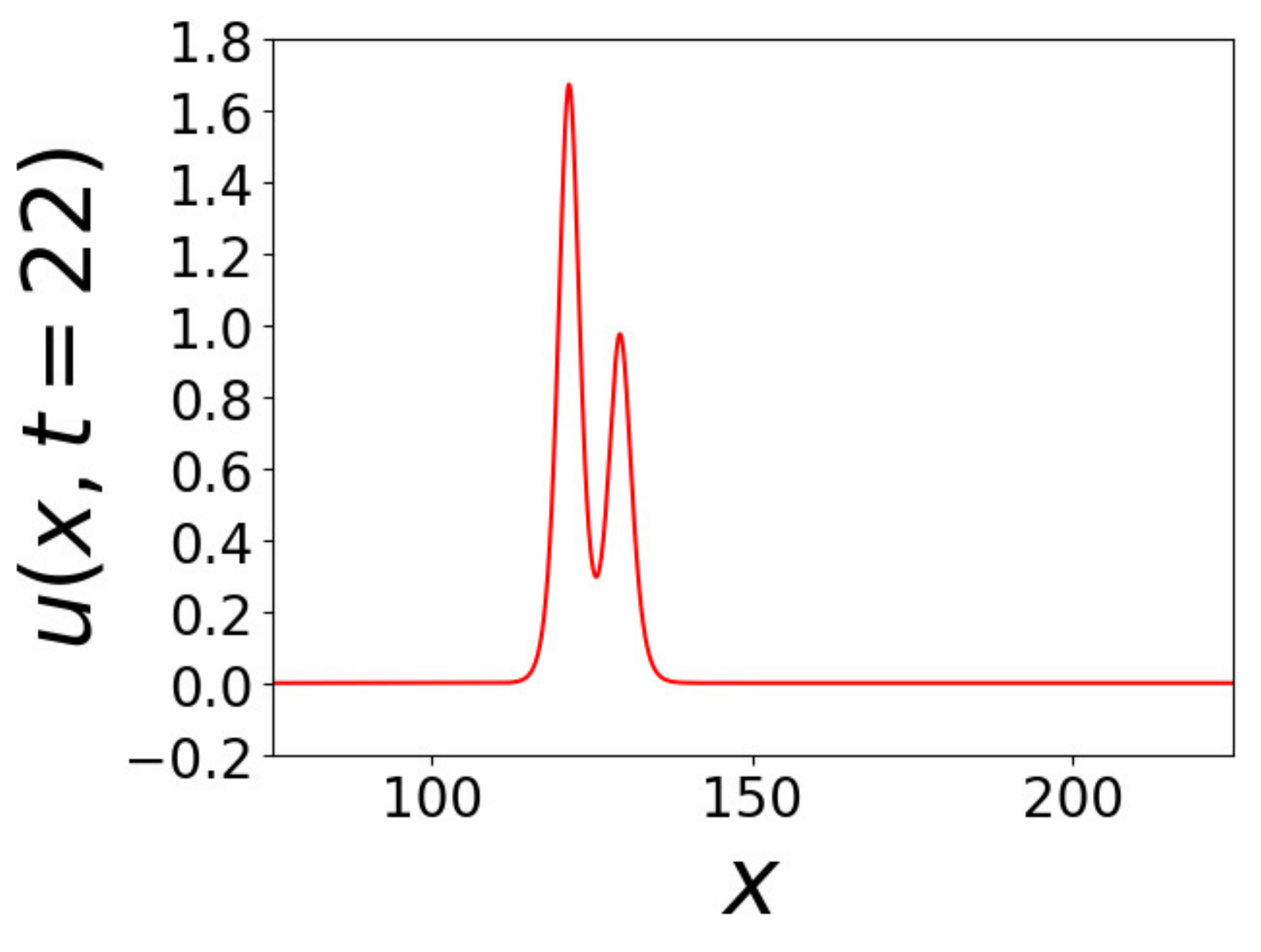}
			\caption{At~$t=22$}
			\label{plot2_3}
		\end{subfigure}%
		\begin{subfigure}{.34\textwidth}
			\centering
			\includegraphics[scale=0.34]{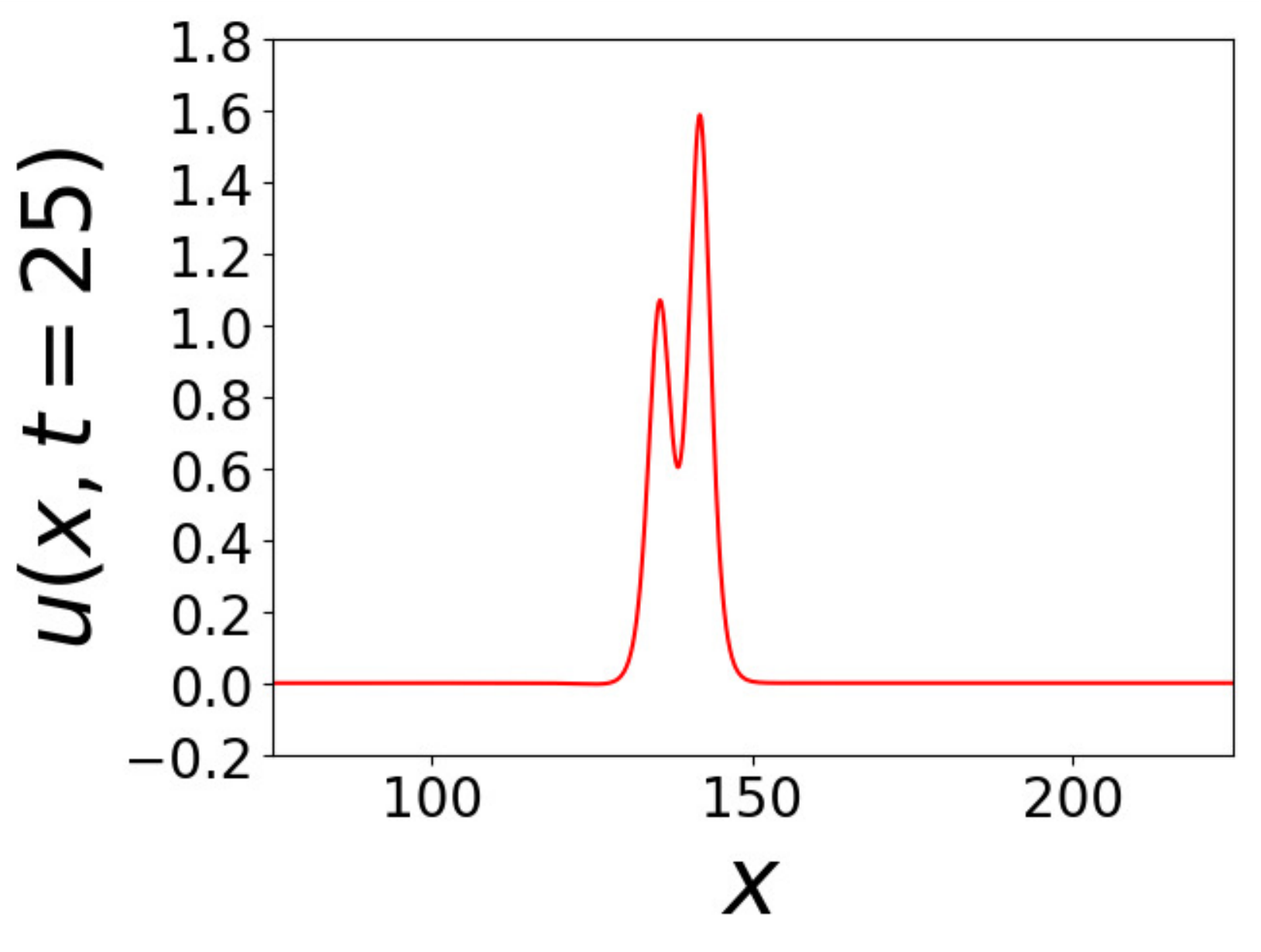}
			\caption{At~$t=25$}
			\label{plot2_4}
		\end{subfigure}
		
		\begin{subfigure}{.34\textwidth}
			\centering
			\includegraphics[scale=0.34]{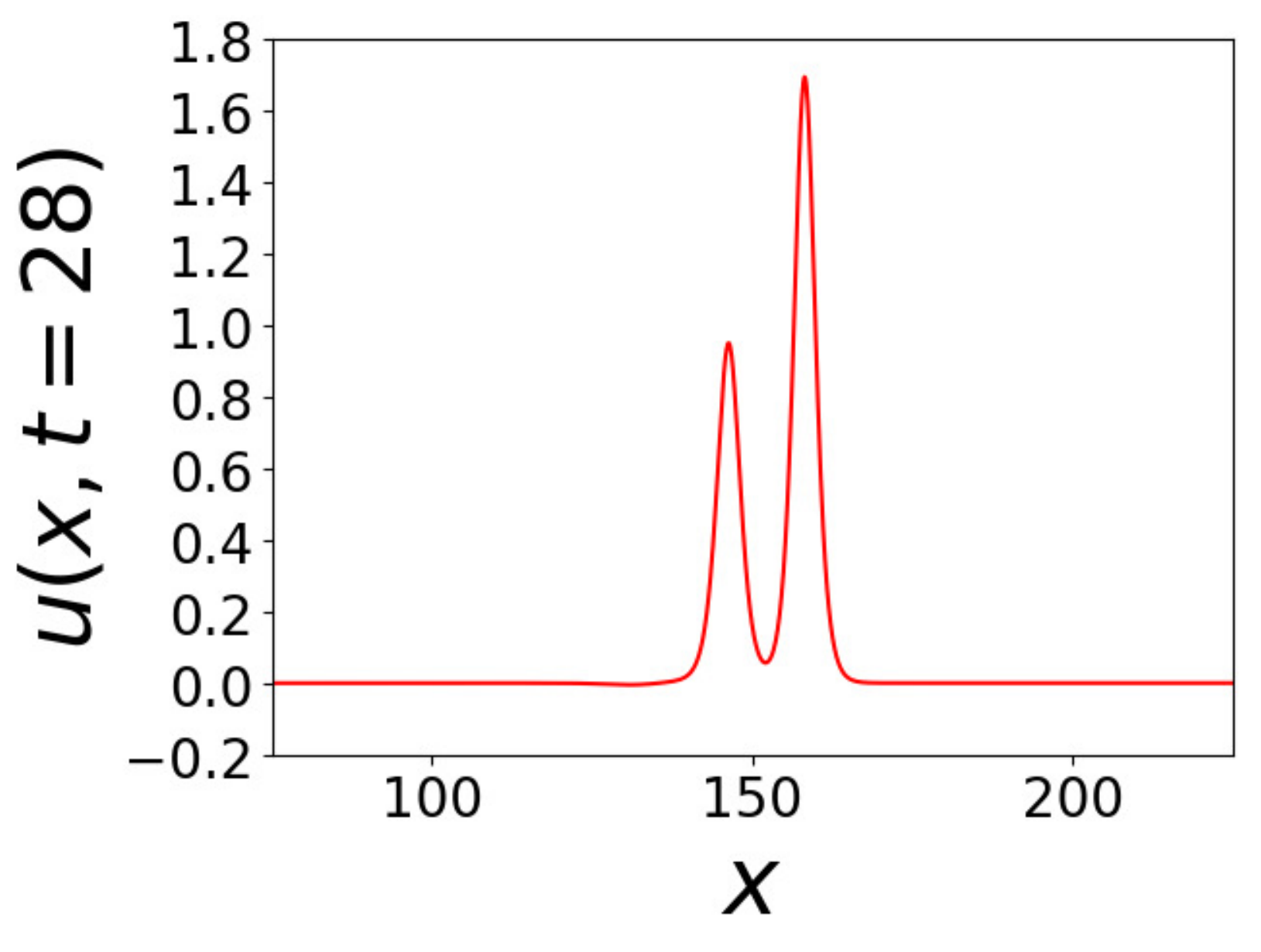}
			\caption{At~$t=28$}
			\label{plot2_5}
		\end{subfigure}%
		\begin{subfigure}{.34\textwidth}
			\centering
			\includegraphics[scale=0.34]{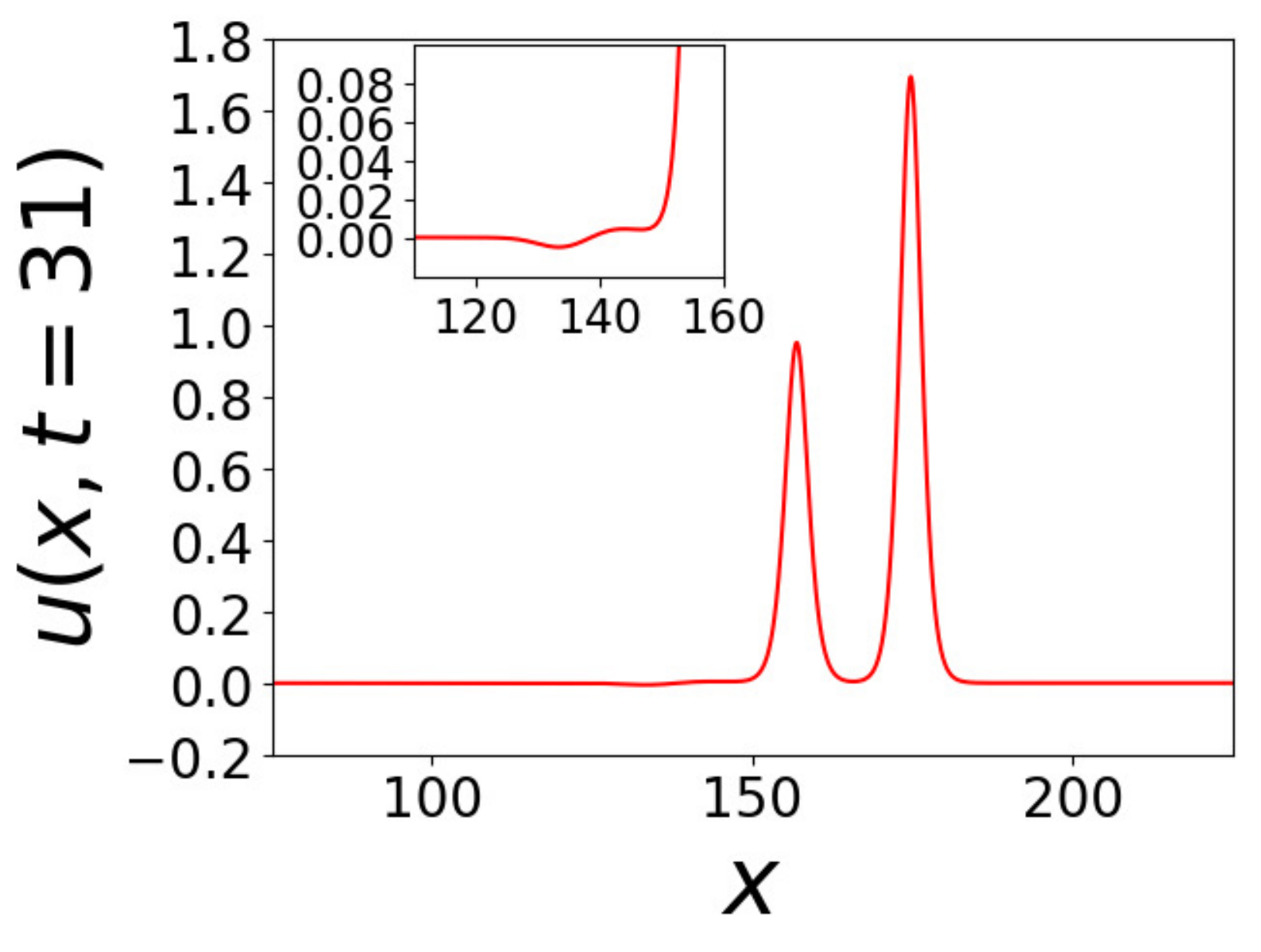}
			\caption{At~$t=31$}
			\label{plot2_6}
		\end{subfigure}%
		\begin{subfigure}{.34\textwidth}
			\centering
			\includegraphics[scale=0.34]{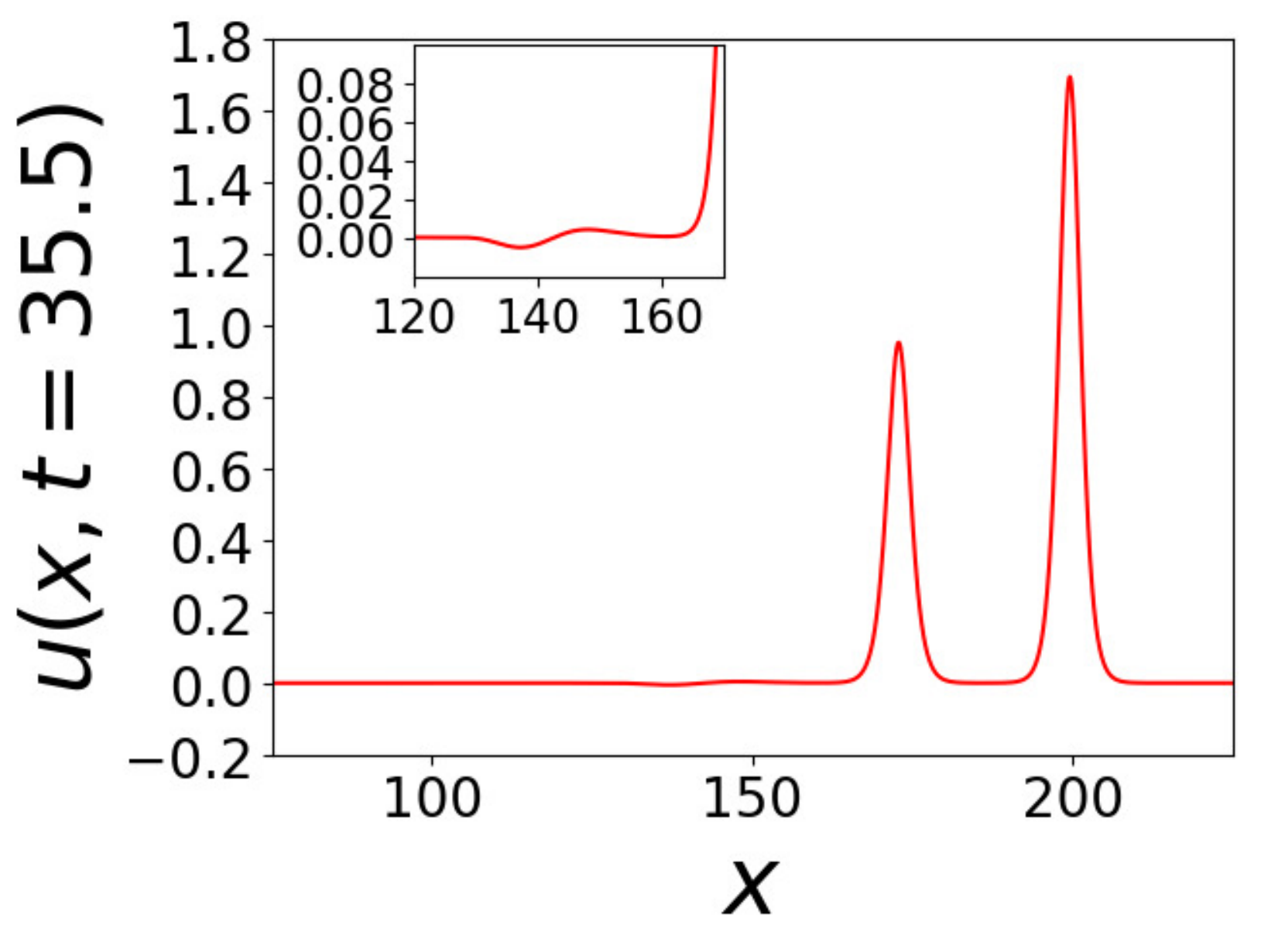}
			\caption{At~$t=35.5$}
			\label{plot2_7}
		\end{subfigure}
		\caption{The field $u$, at selected values of time, seen in our simulation of a two-soliton configuration of the RLW equation.} 
		\label{plot2_2to0_7}
	\end{figure}

\begin{figure}[b!]
		\centering
		\begin{subfigure}{.5\textwidth}
			\hspace*{-1.2cm}
			\centering
			\includegraphics[scale=0.34]{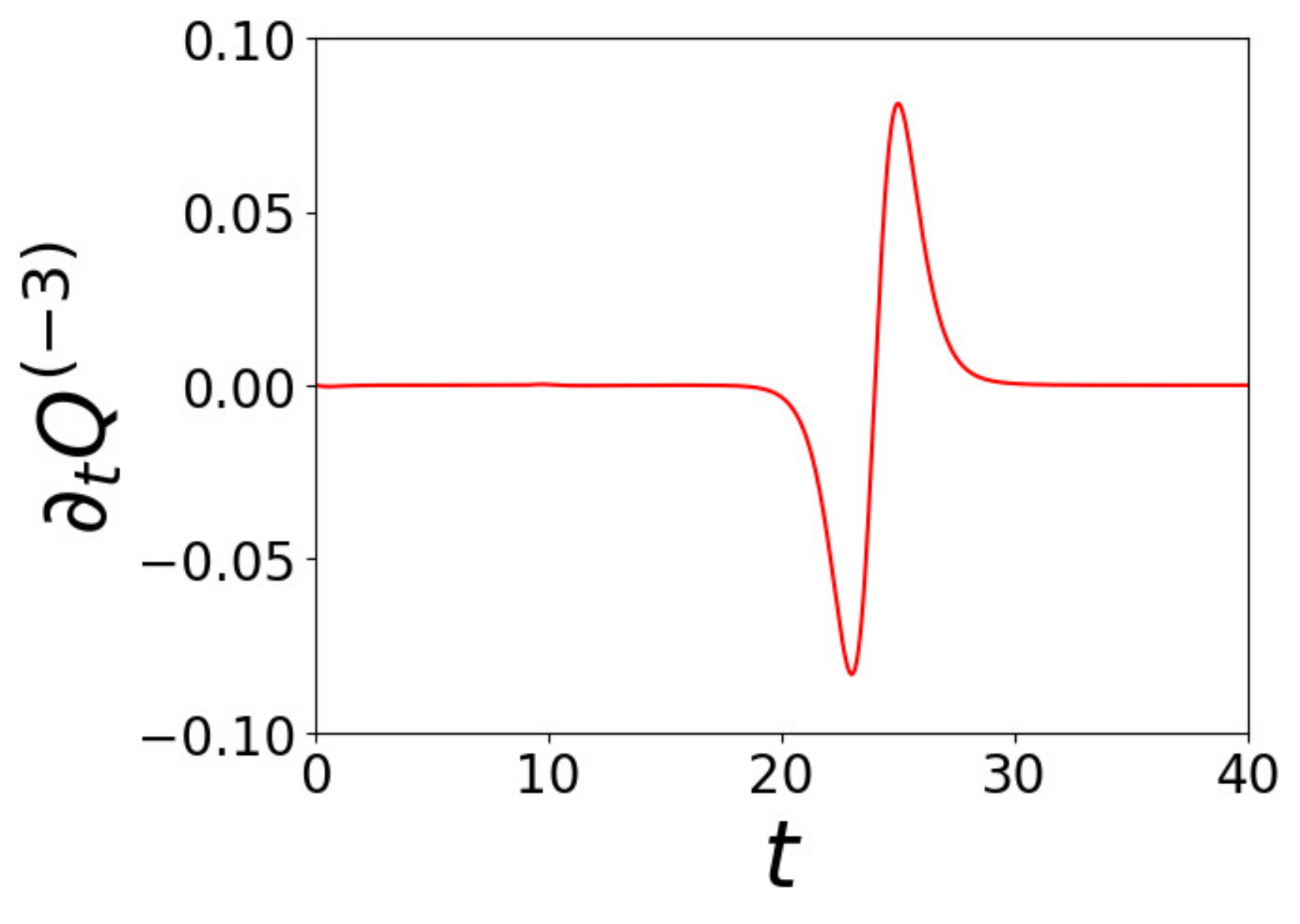}
			\caption{}
			\label{plot2_8}
		\end{subfigure}%
		\begin{subfigure}{.5\textwidth}
			\hspace*{-1.5cm}
			\centering
			\includegraphics[scale=0.34]{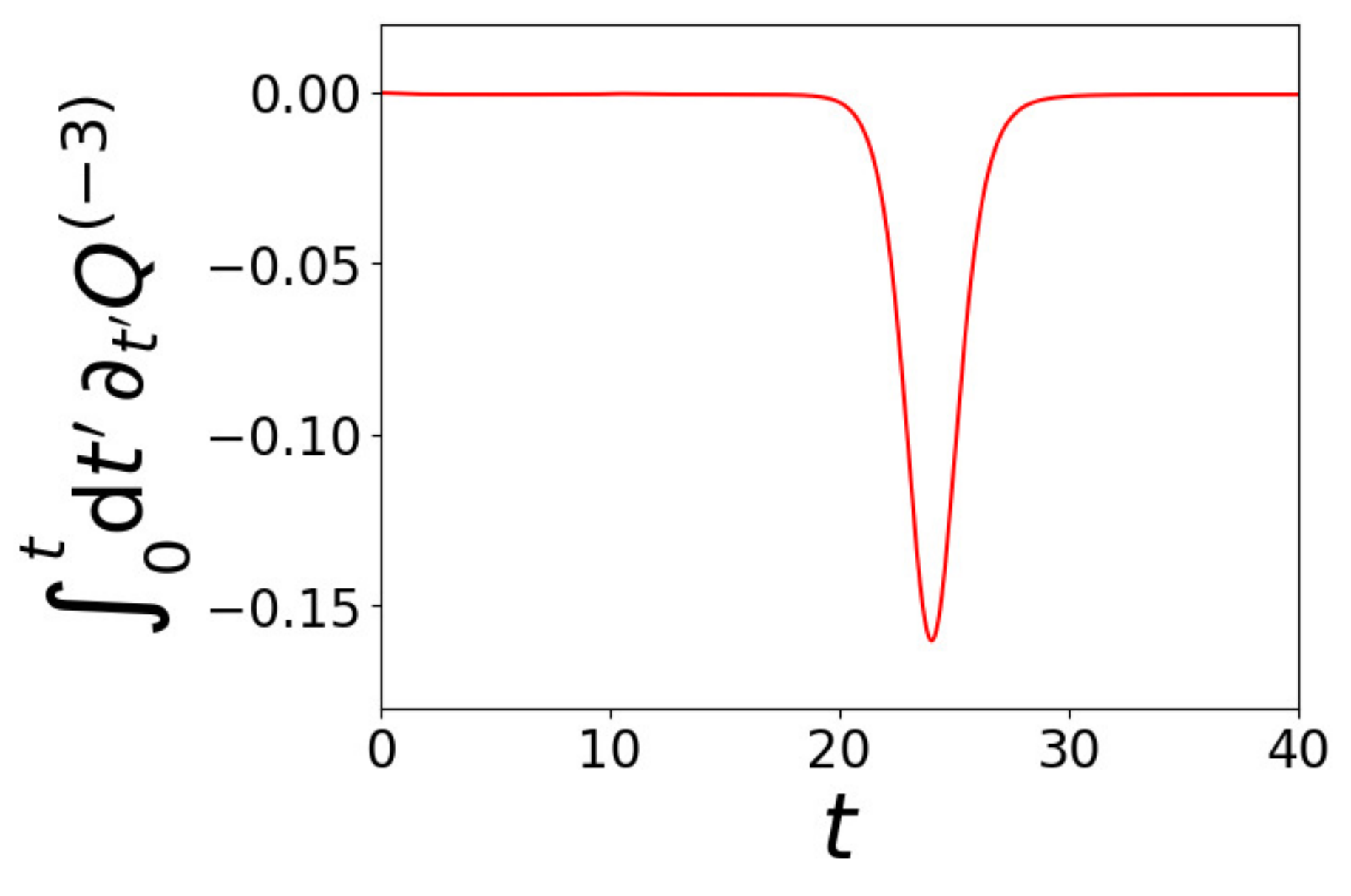}
			\caption{}
			\label{plot2_9}
		\end{subfigure}%
		\caption{The time-dependence of the obtained values of the quantity
 $\partial_t Q^{(-3)}$ and $\int_{0}^{t} \mathrm{d} t^\prime \, \partial_{t^\prime} Q^{(-3)}$  for the two-soliton simulation of the RLW equation.}
		\label{plot2_8to2_9}
	\end{figure}

	The results of the two-soliton simulation for the RLW equation are shown in figure~\ref{plot2_2to0_7}.
	Clearly, the figure shows some radiation (see the inserts in figures~\ref{plot2_6} and~\ref{plot2_7}), and this agrees with
 the results presented in~\cite{Kutluay}. This is expected since the initial conditions used for this simulation do not solve the RLW equation analytically. 
Note that the amplitudes of these solitons are bigger compared to the solitons of the mRLW equation due to the aforementioned rescaling of~$q$.
 Furthermore, in figure~\ref{plot2_8to2_9}	
	we have presented the plots of the time dependence of 
$\partial_t Q^{(-3)}$ and $\int_{0}^{t} \mathrm{d} t^\prime \, \partial_{t^\prime} Q^{(-3)}$ seen
 in this simulation. The shapes of the curves in these plots are very similar to the ones found in figure~\ref{I_1_alpha_8_e1_1_e2_1_e3_1_1_analytical_and_numerical}.
 We note that the small observed emission of radiation does not seem to have a visible effect on the quasi-conservation of the charge.

	
	\subsubsection{Three-soliton solutions of the RLW equation} \label{Three_soliton_solutions_of_the_RLW equation}
	
\begin{figure}[b!]
		\centering
		\hspace*{-0.1cm}
		\begin{subfigure}{.34\textwidth}
			\centering
			\includegraphics[scale=0.34]{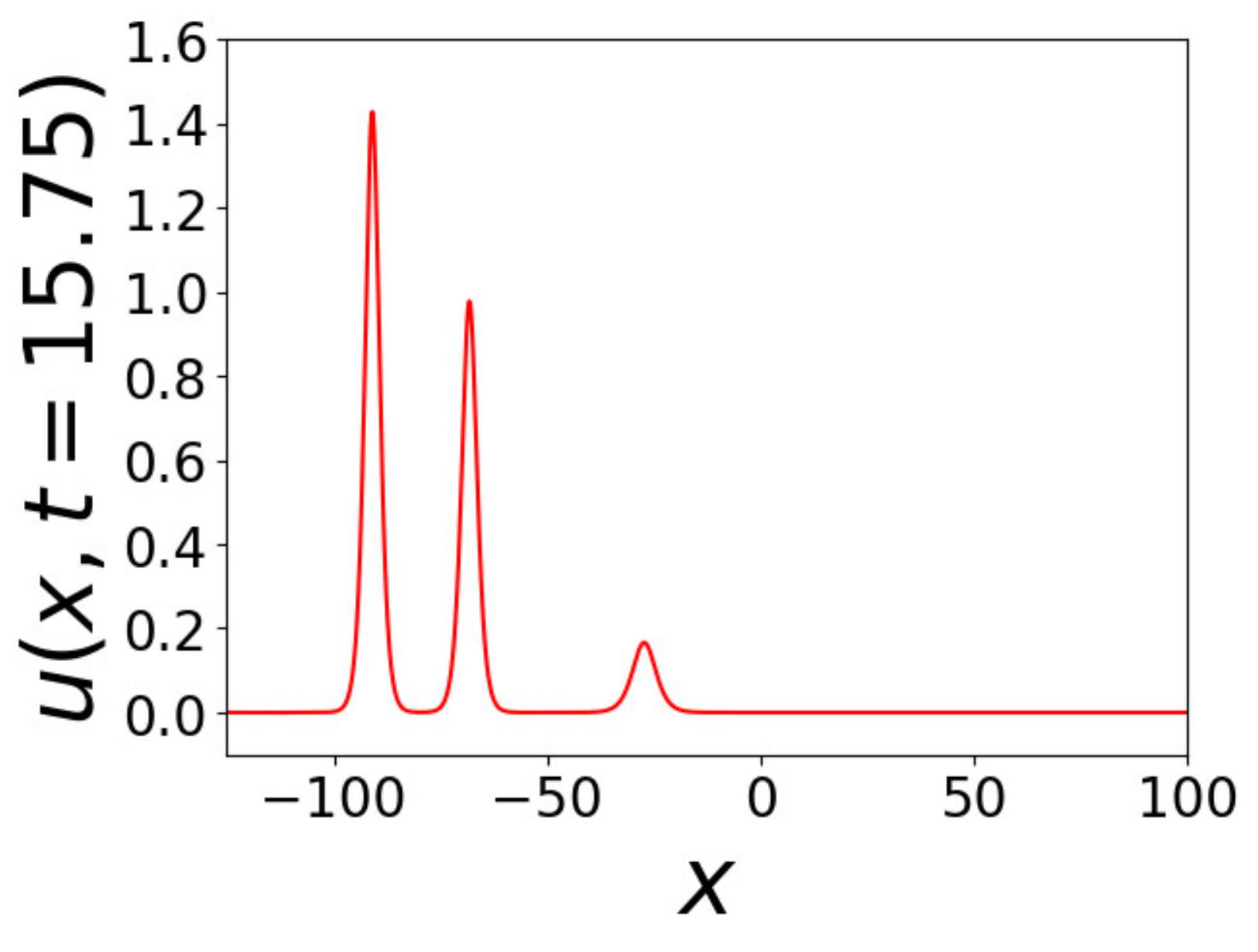}
			\caption{At~$t=15.75$}
			\label{plot3_2}
		\end{subfigure}%
		\begin{subfigure}{.34\textwidth}
			\centering
			\includegraphics[scale=0.34]{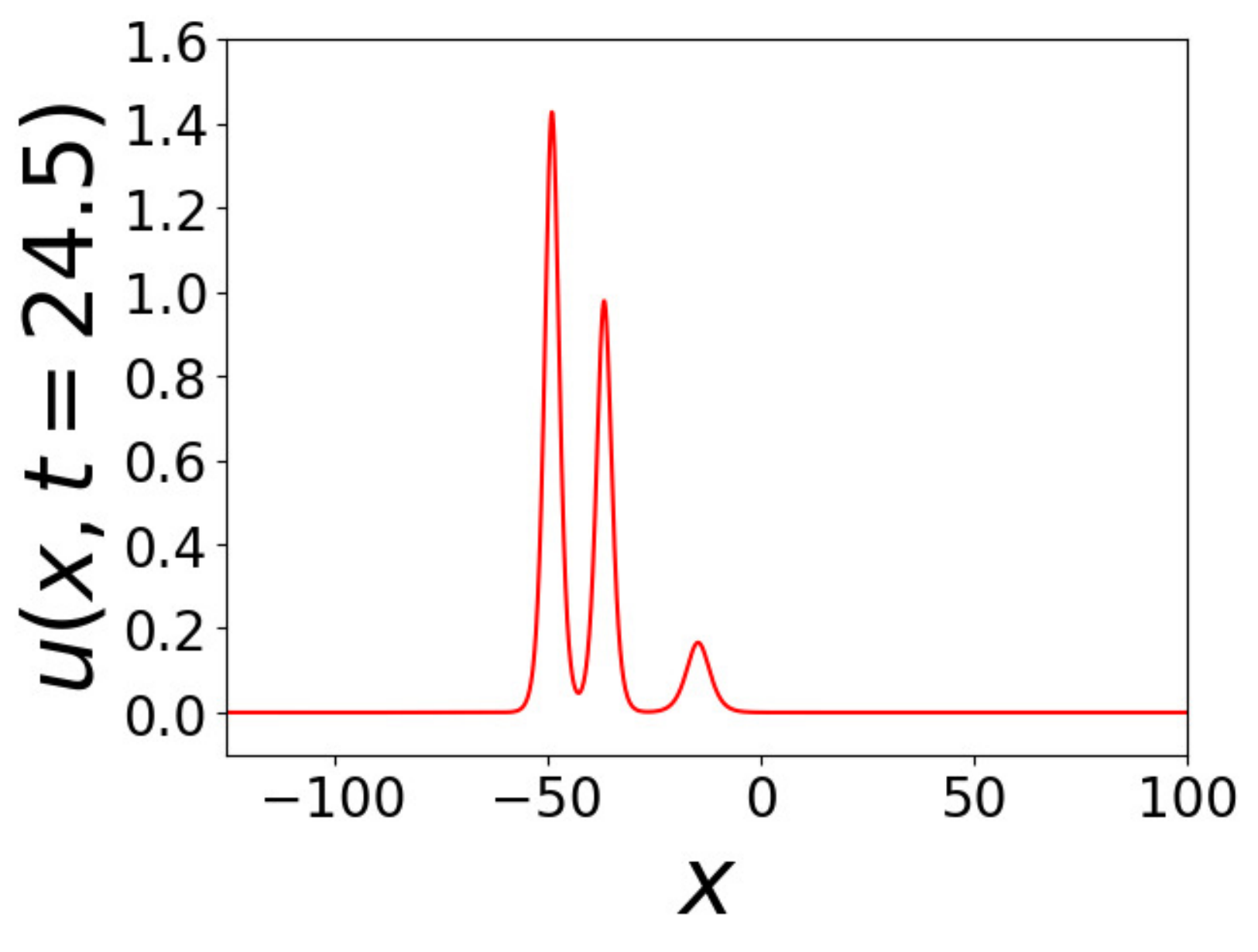}
			\caption{At~$t=24.5$}
			\label{plot3_3}
		\end{subfigure}%
		\begin{subfigure}{.34\textwidth}
			\centering
			\includegraphics[scale=0.34]{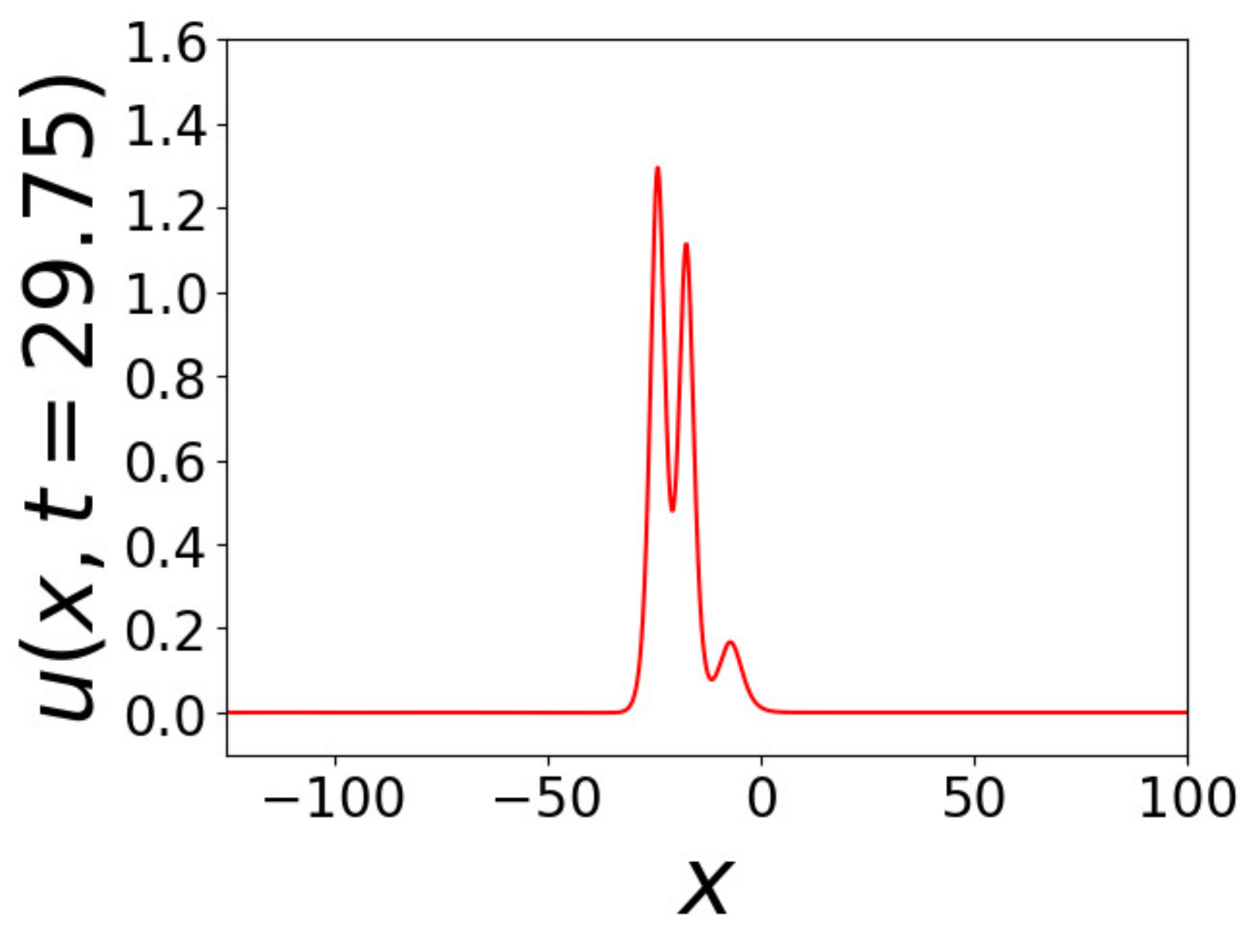}
			\caption{At~$t=29.75$}
			\label{plot3_4}
		\end{subfigure}
		
		\begin{subfigure}{.34\textwidth}
			\centering
			\includegraphics[scale=0.34]{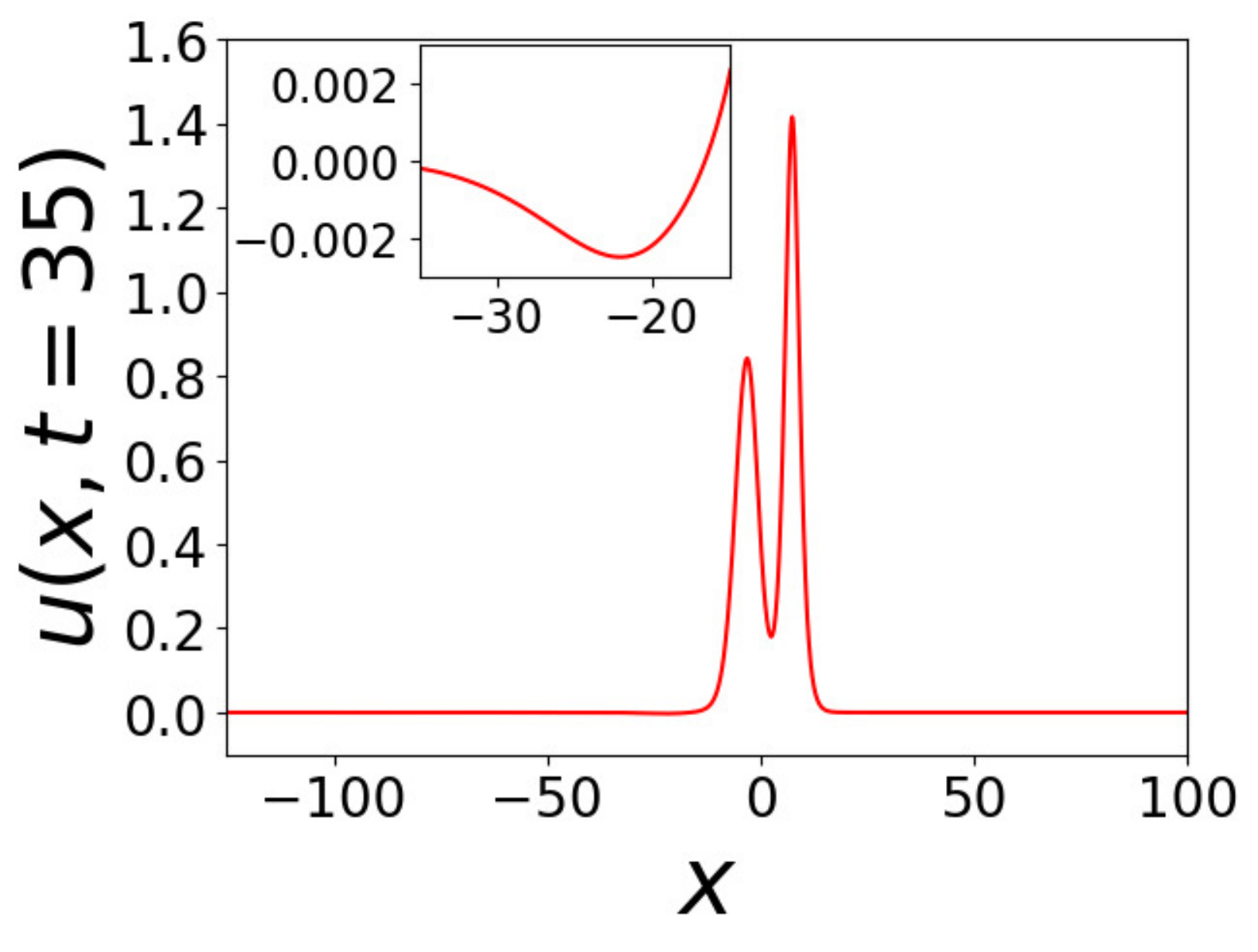}
			\caption{At~$t=35$}
			\label{plot3_5}
		\end{subfigure}%
		\begin{subfigure}{.34\textwidth}
			\centering
			\includegraphics[scale=0.34]{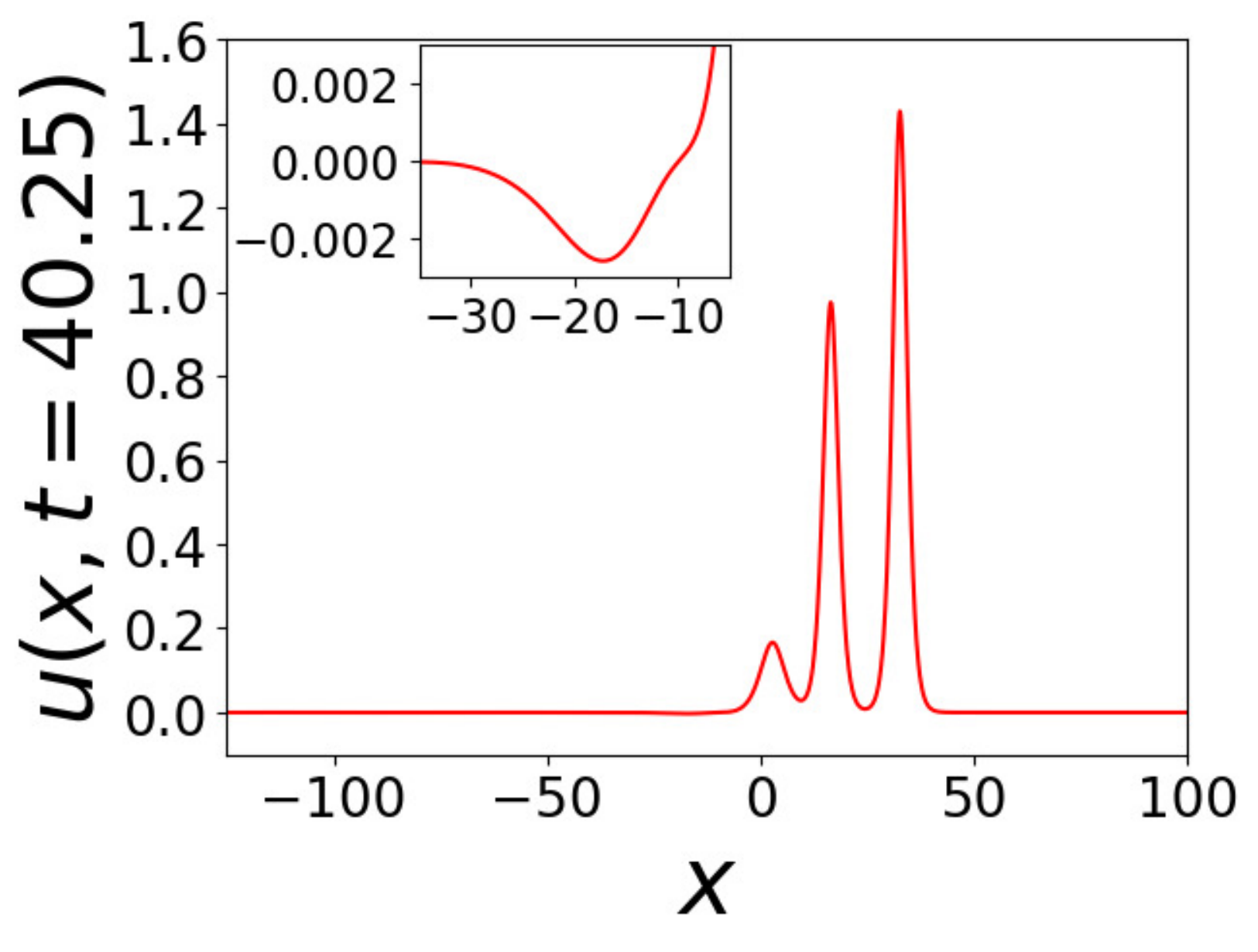}
			\caption{At~$t=40.25$}
			\label{plot3_6}
		\end{subfigure}%
		\begin{subfigure}{.34\textwidth}
			\centering
			\includegraphics[scale=0.34]{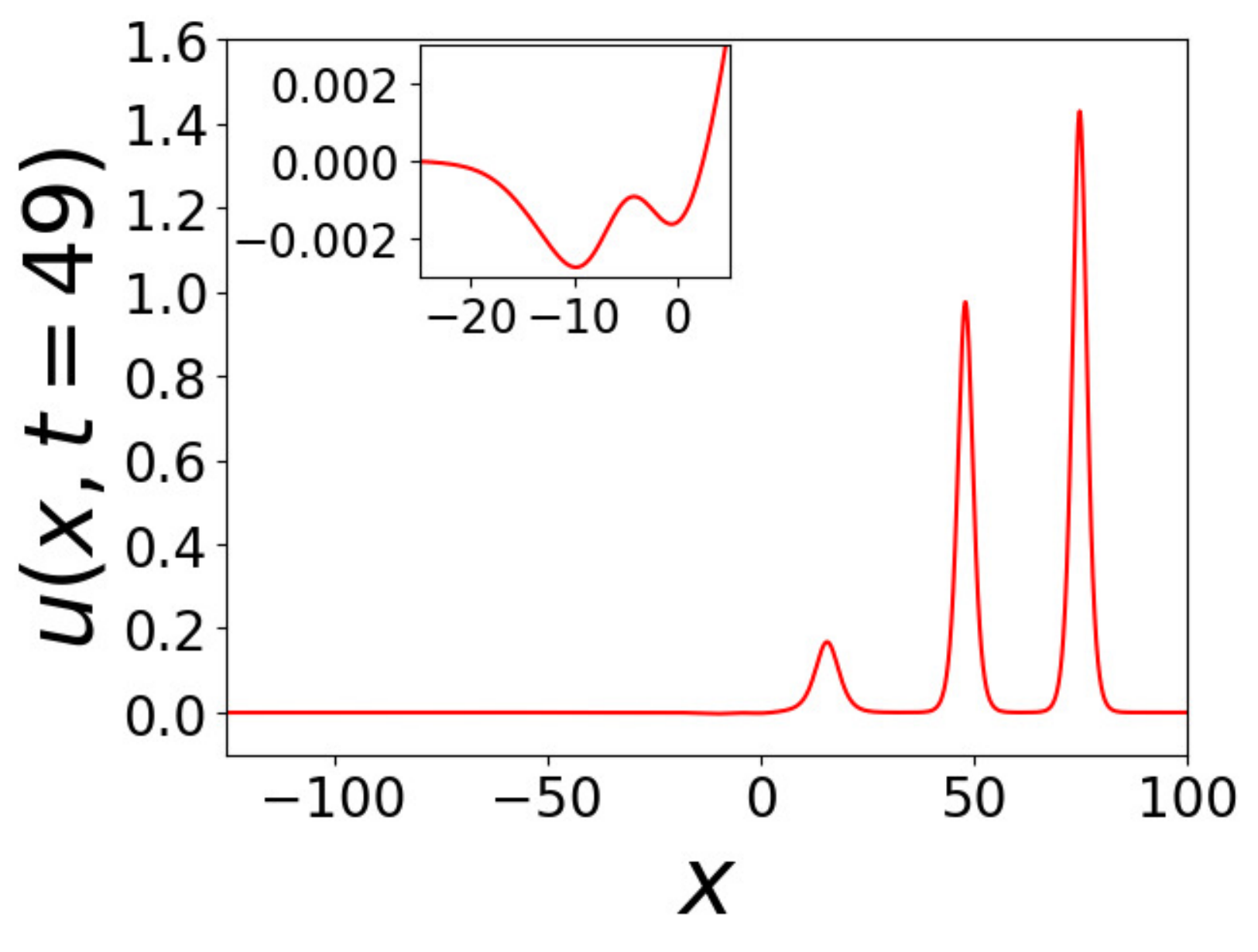}
			\caption{At~$t=49$}
			\label{plot3_7}
		\end{subfigure}
		\caption{The field $u$, for a few values of time, found in our simulation of
the three-soliton system of the RLW equation.} 
		\label{plot3_2to3_7}
	\end{figure}

\begin{figure}[t!]
		\centering
		\begin{subfigure}{.5\textwidth}
			\hspace*{-1.2cm}
			\centering
			\includegraphics[scale=0.34]{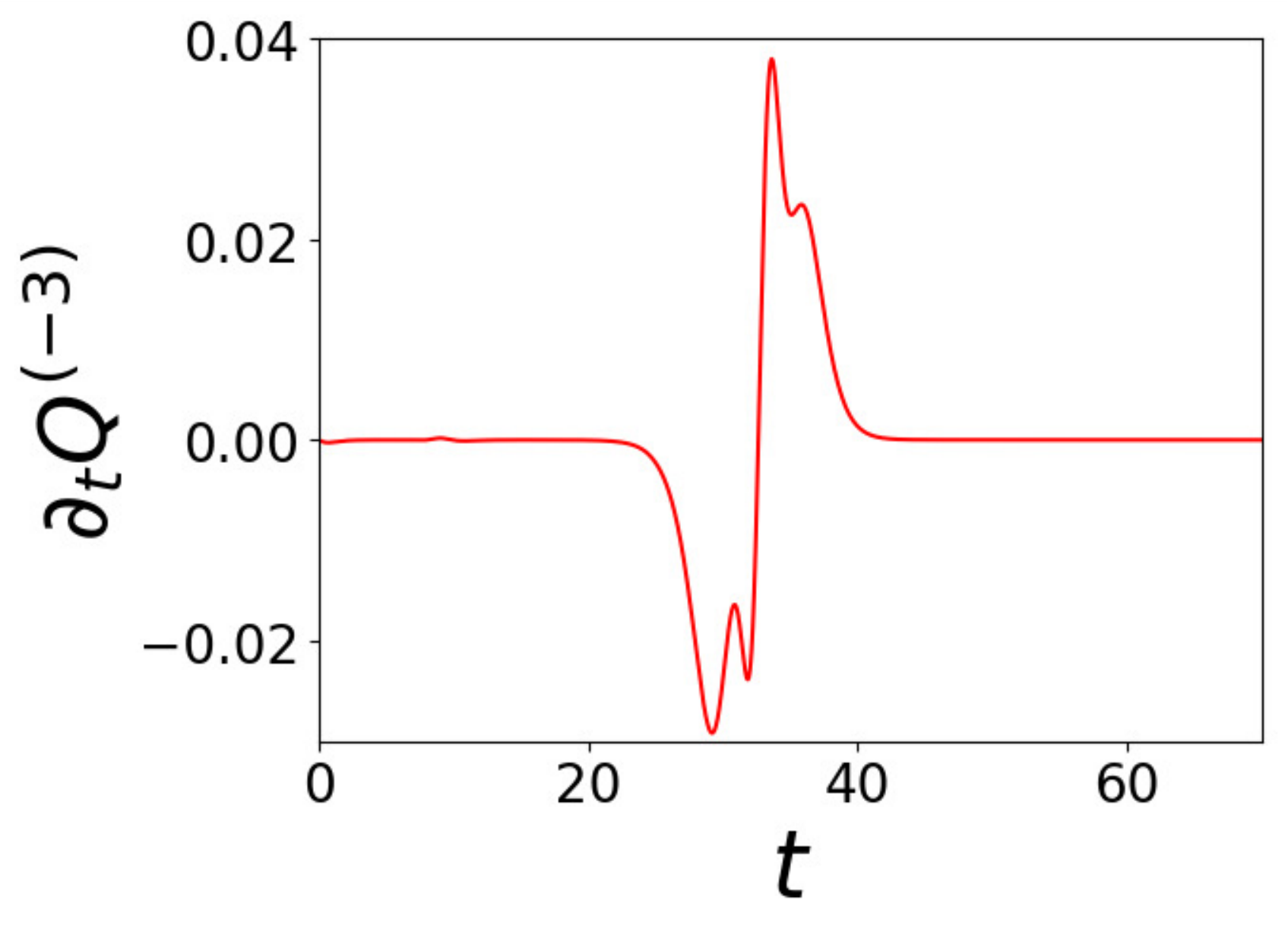}
			\caption{}
			\label{plot3_8}
		\end{subfigure}%
		\begin{subfigure}{.5\textwidth}
			\hspace*{-1.5cm}
			\centering
			\includegraphics[scale=0.34]{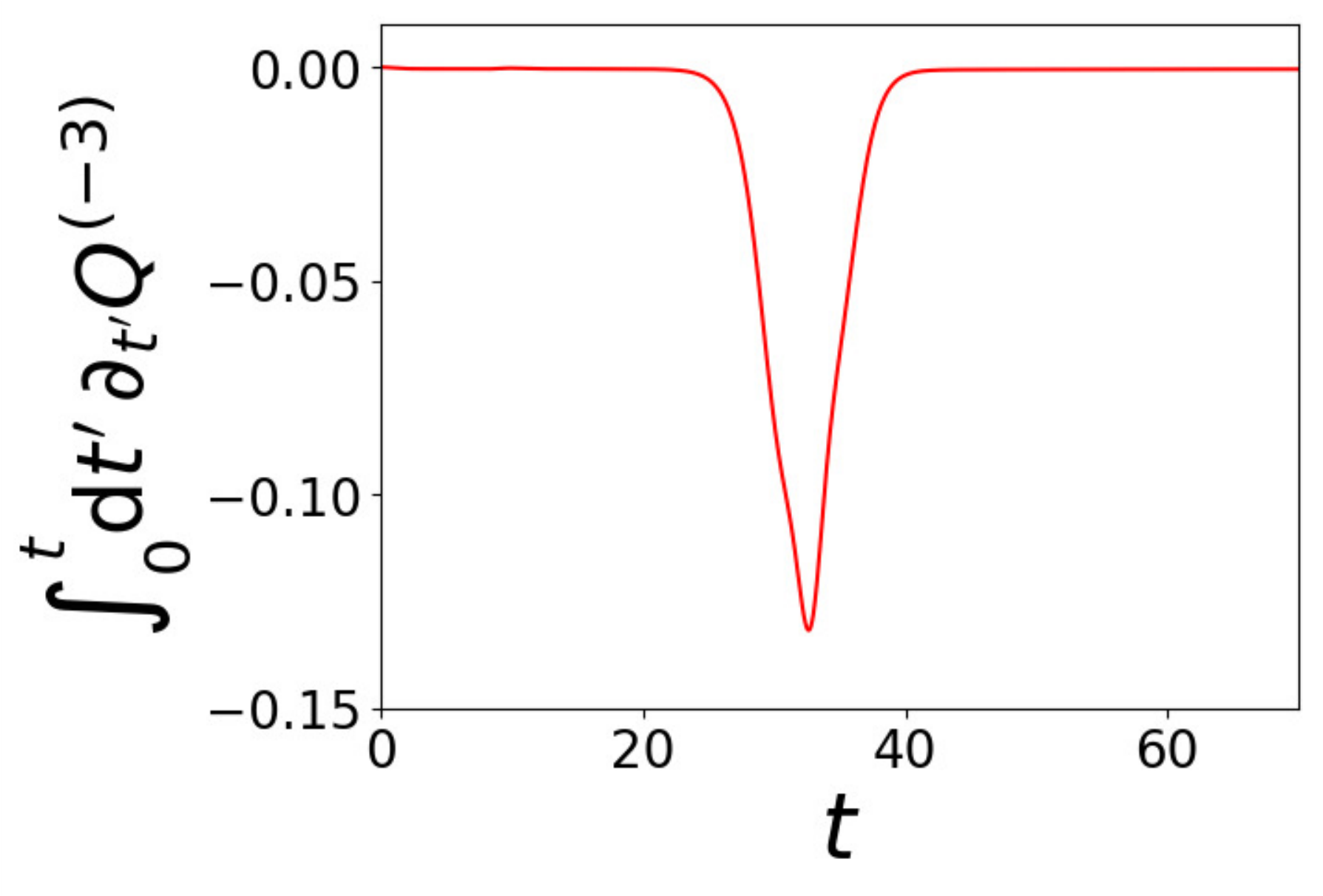}
			\caption{}
			\label{plot3_9}
		\end{subfigure}%
		\caption{The time-dependence of the quantity $\partial_t Q^{(-3)}$ and $\int_{0}^{t} \mathrm{d} t^\prime \, \partial_{t^\prime} Q^{(-3)}$ found
 in our three-soliton simulation of the RLW equation.}
		\label{plot3_8to3_9}
	\end{figure}

	Next we have looked at the three solitons systems of the RLW equation. In figure~\ref{plot3_2to3_7}  
	we present the plots of the fields $u$ seen at various times in their evolution, and in figure~\ref{plot3_8to3_9} the plots of the corresponding 
$\partial_t Q^{(-3)}$ and $\int_{0}^{t} \mathrm{d} t^\prime \, \partial_{t^\prime} Q^{(-3)}$. 
	Note that the amplitudes of the solitons are again larger when compared with the solitons of the mRLW equation 
(see figure~\ref{plot1_2to1_7}). This is due to the factor of $3/2$, mentioned before. The three solitons behave in a very similar way to
 the solitons shown in figure~\ref{plot1_2to1_7} except that, around the time they collide, a small bit of radiation is emitted (see the 
inserts in figures~\ref{plot3_5},~\ref{plot3_6} and~\ref{plot3_7}). Just as for the two-soliton simulation, this emission of radiation does not seem to
 have a visible effect on the quasi-conservation of the charge.

	\subsection{Intermediate models; Quasi-conserved charges for $\varepsilon_1 = 1$ and $\varepsilon_2 \neq 1$} \label{Intermediate_models}
	
	So far we have discussed the properties of solutions of the mRLW and RLW equations. However, as we have shown in section~\ref{The_analytical_Hirota_soliton_solutions}, our basic equation (\ref{deformedqkdv}) possesses one-soliton solutions, whose analytical form is given by equation~(\ref{onegen}), for any real value of $\varepsilon_2$ as long as
$\varepsilon_1 = 1$. So in this section we investigate the question of quasi-integrability of two- and three-soliton systems by performing simulations with the initial conditions taken from linear superpositions of such analytical single-soliton solutions, that is, from
\begin{equation}
q = \frac{3}{\(2+{\ve}_2\)} \sum\limits_{i=1}^n \ln\(1+e^{\Gamma_i}\) \,, \label{ic1}
\end{equation}
where $n=2$ corresponds to the two-soliton simulation and $n=3$ to the three-soliton simulation. We present
 the results of such two- and three-soliton simulations of equation~(\ref{deformedqkdv}), for various values of $\varepsilon_2 \neq 1$ while
 keeping $\varepsilon_1 = 1$ constant.

	\subsubsection{Two-soliton configurations} \label{Two_soliton_configurations2}
	
\begin{figure}[t!]
		\centering
		\hspace*{-0.1cm}
		\begin{subfigure}{.34\textwidth}
			\centering
			\includegraphics[scale=0.34]{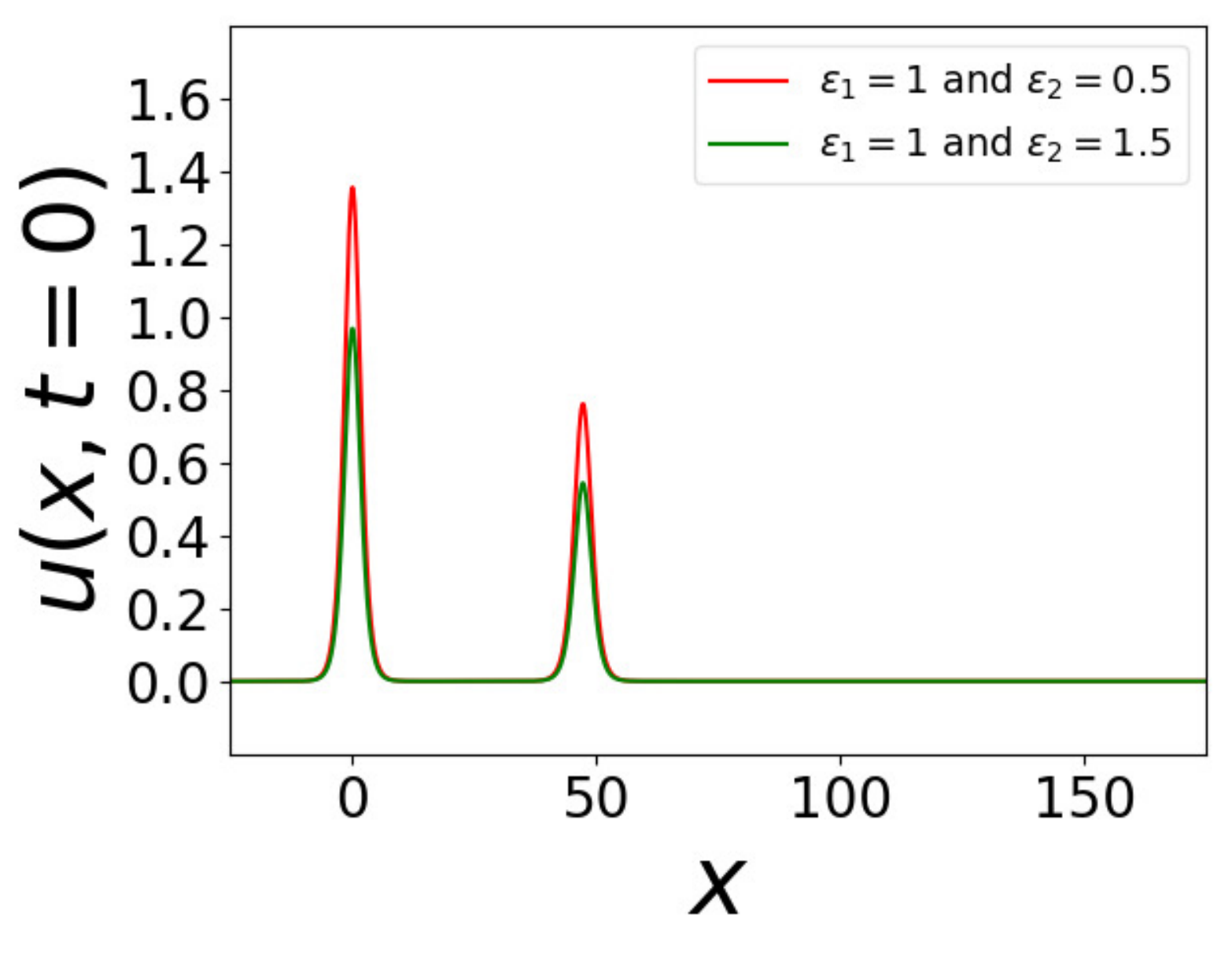}
			\caption{At~$t=0$}
			\label{plot11_2}
		\end{subfigure}%
		\begin{subfigure}{.34\textwidth}
			\centering
			\includegraphics[scale=0.34]{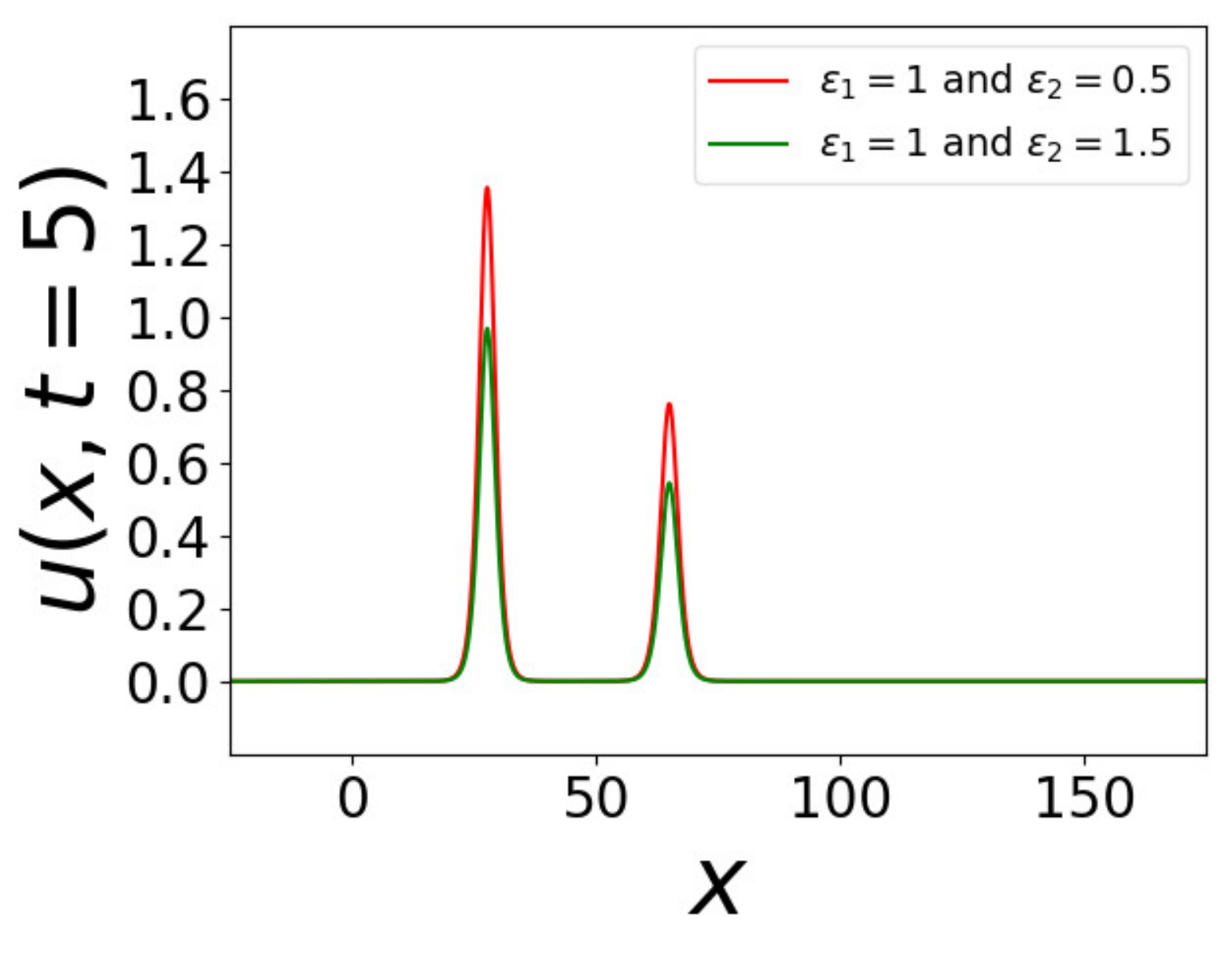}
			\caption{At~$t=5$}
			\label{plot11_3}
		\end{subfigure}%
		\begin{subfigure}{.34\textwidth}
			\centering
			\includegraphics[scale=0.34]{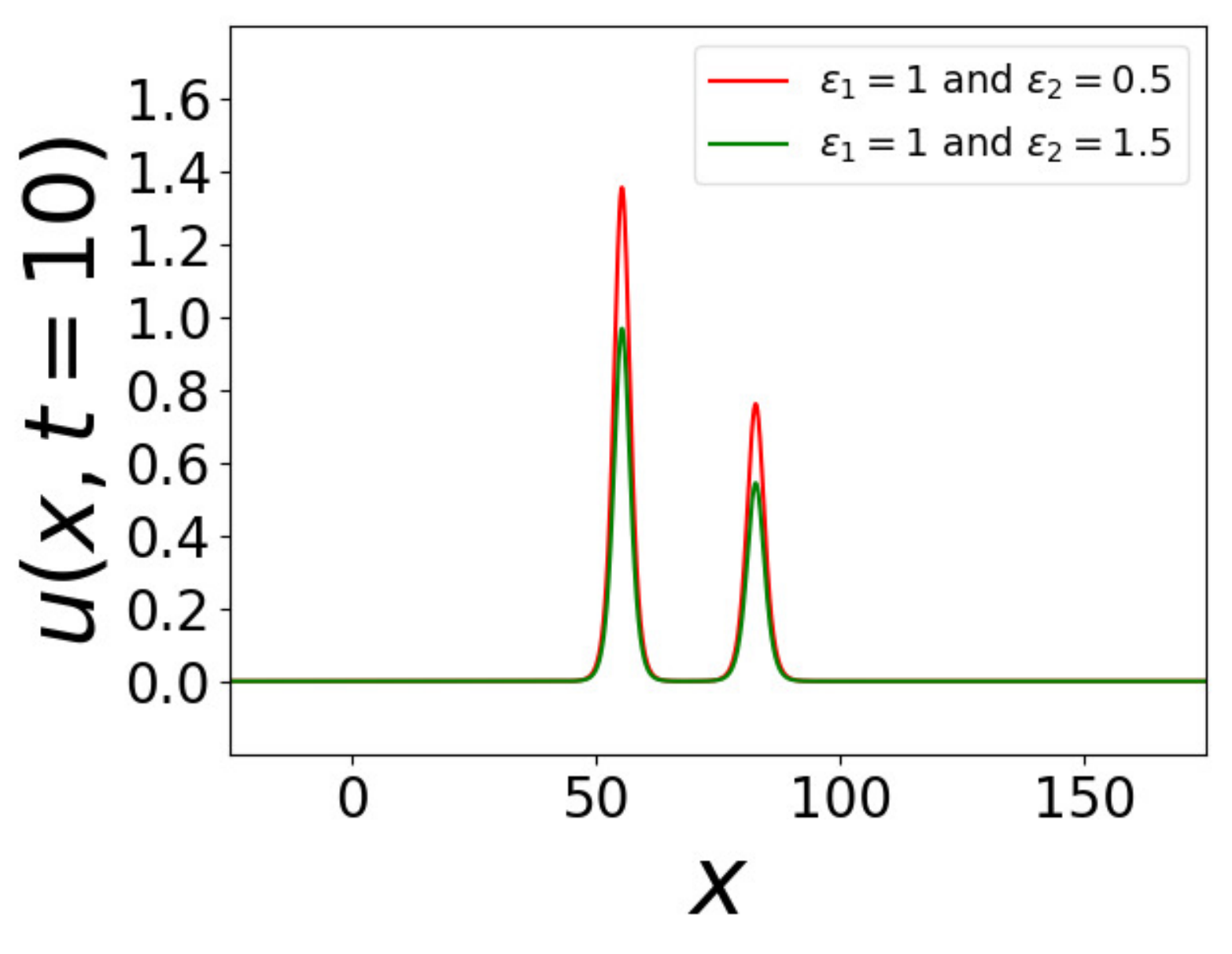}
			\caption{At~$t=10$}
			\label{plot11_4}
		\end{subfigure}
		
		\begin{subfigure}{.34\textwidth}
			\centering
			\includegraphics[scale=0.34]{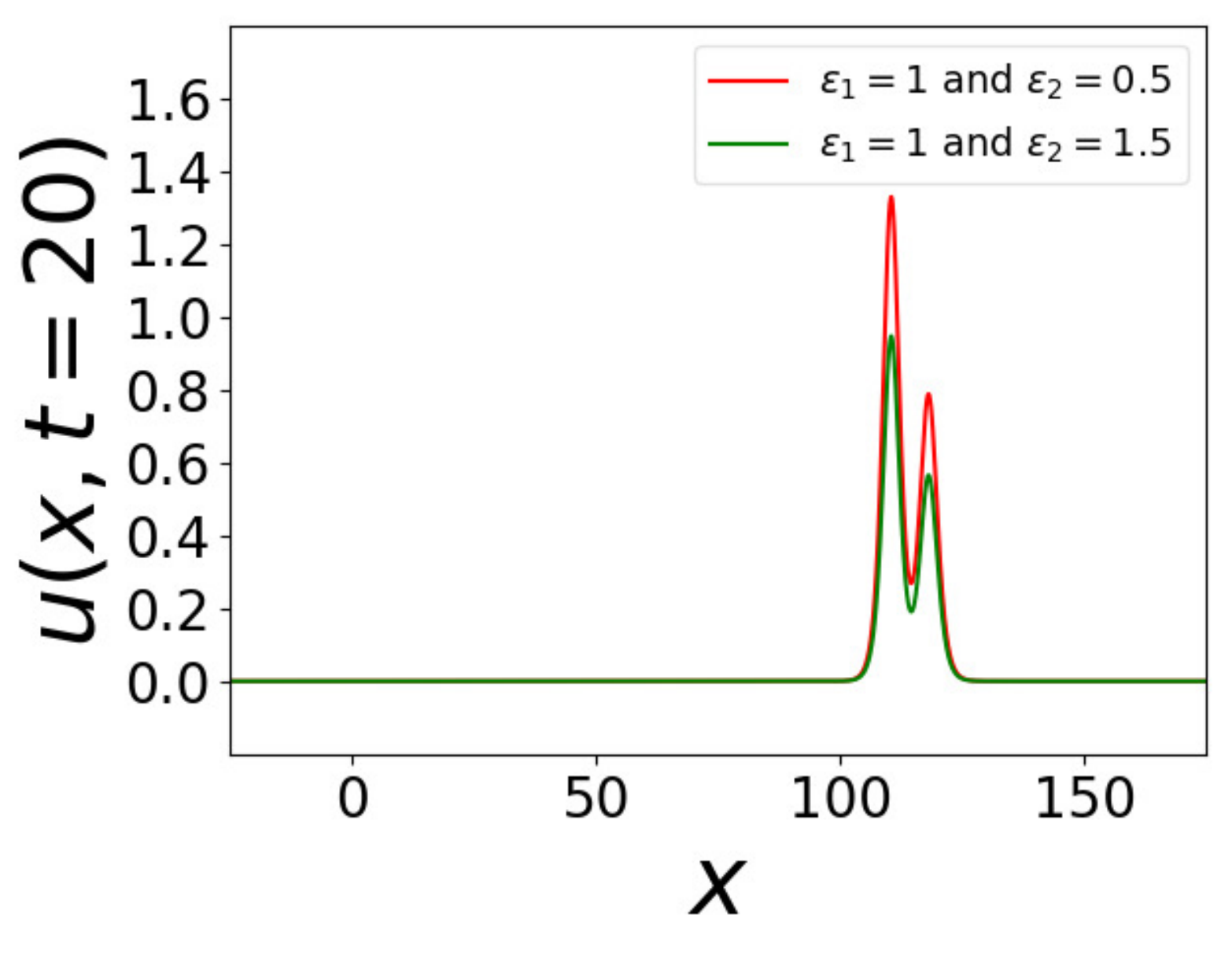}
			\caption{At~$t=20$}
			\label{plot11_5}
		\end{subfigure}%
		\begin{subfigure}{.34\textwidth}
			\centering
			\includegraphics[scale=0.34]{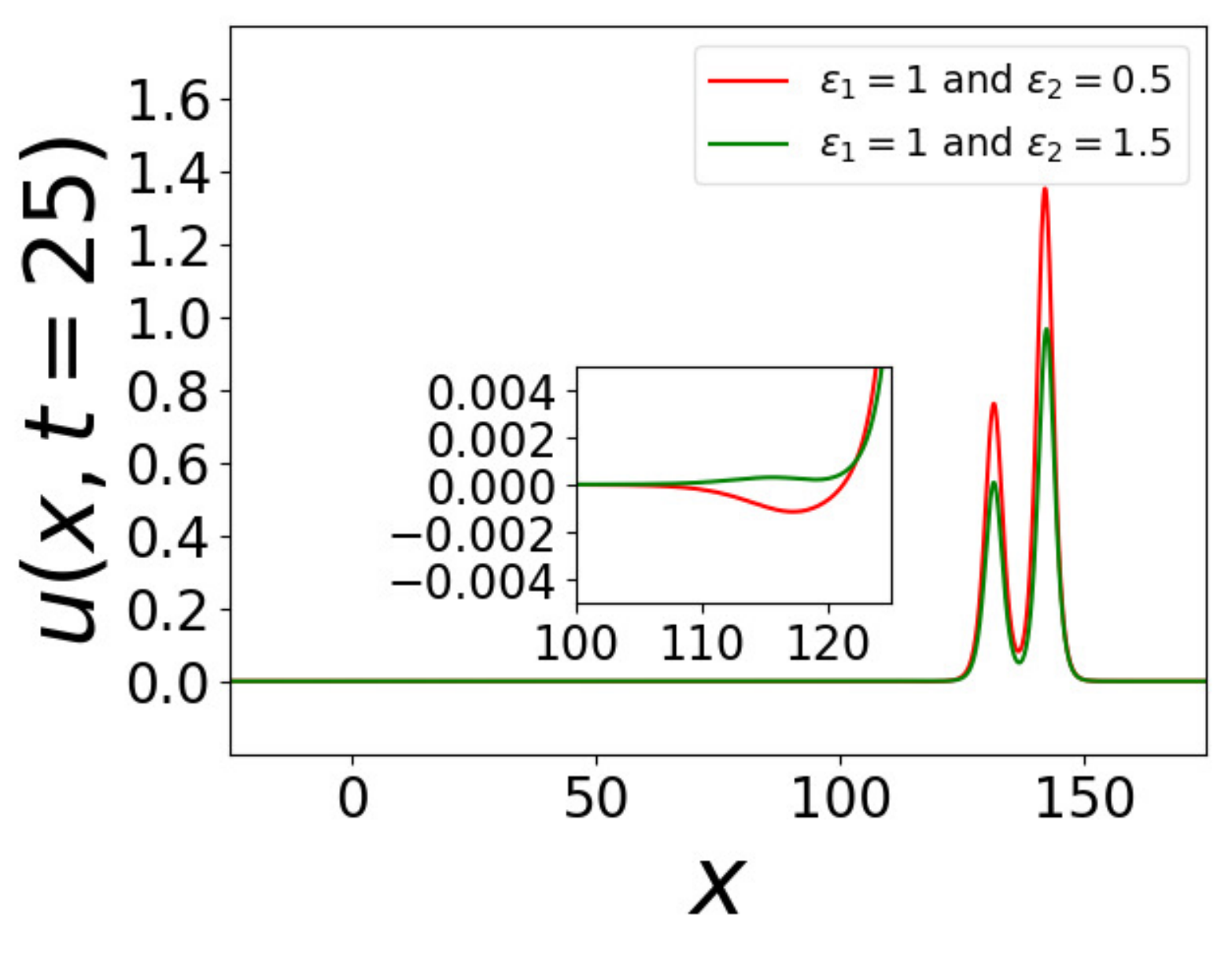}
			\caption{At~$t=25$}
			\label{plot11_6}
		\end{subfigure}%
		\begin{subfigure}{.34\textwidth}
			\centering
			\includegraphics[scale=0.34]{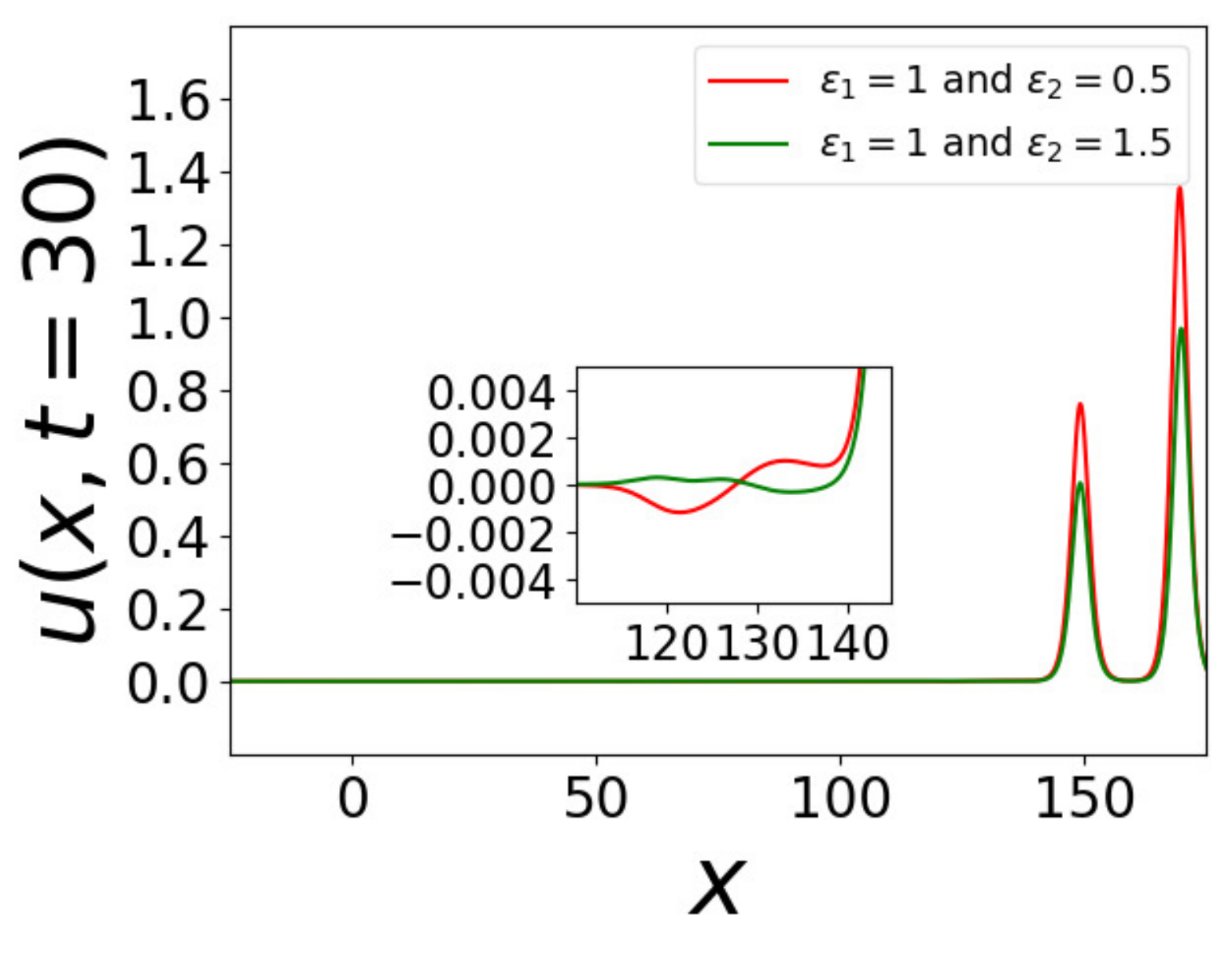}
			\caption{At~$t=30$}
			\label{plot11_7}
		\end{subfigure}
		\caption{The time-evolution of two-soliton systems found in our simulation of equation~(\ref{deformedqkdv}) with $\varepsilon_1 = 1$, $\varepsilon_2 = 0.5$ (red curve) and $\varepsilon_1 = 1$, $\varepsilon_2 = 1.5$ (green curve).} 
		\label{plot11_2to11_7}
	\end{figure}

\begin{figure}[b!]
		\centering
		\hspace*{0.5cm}
		\begin{subfigure}{.34\textwidth}
			\hspace*{-1.2cm}
			\centering
			\includegraphics[scale=0.34]{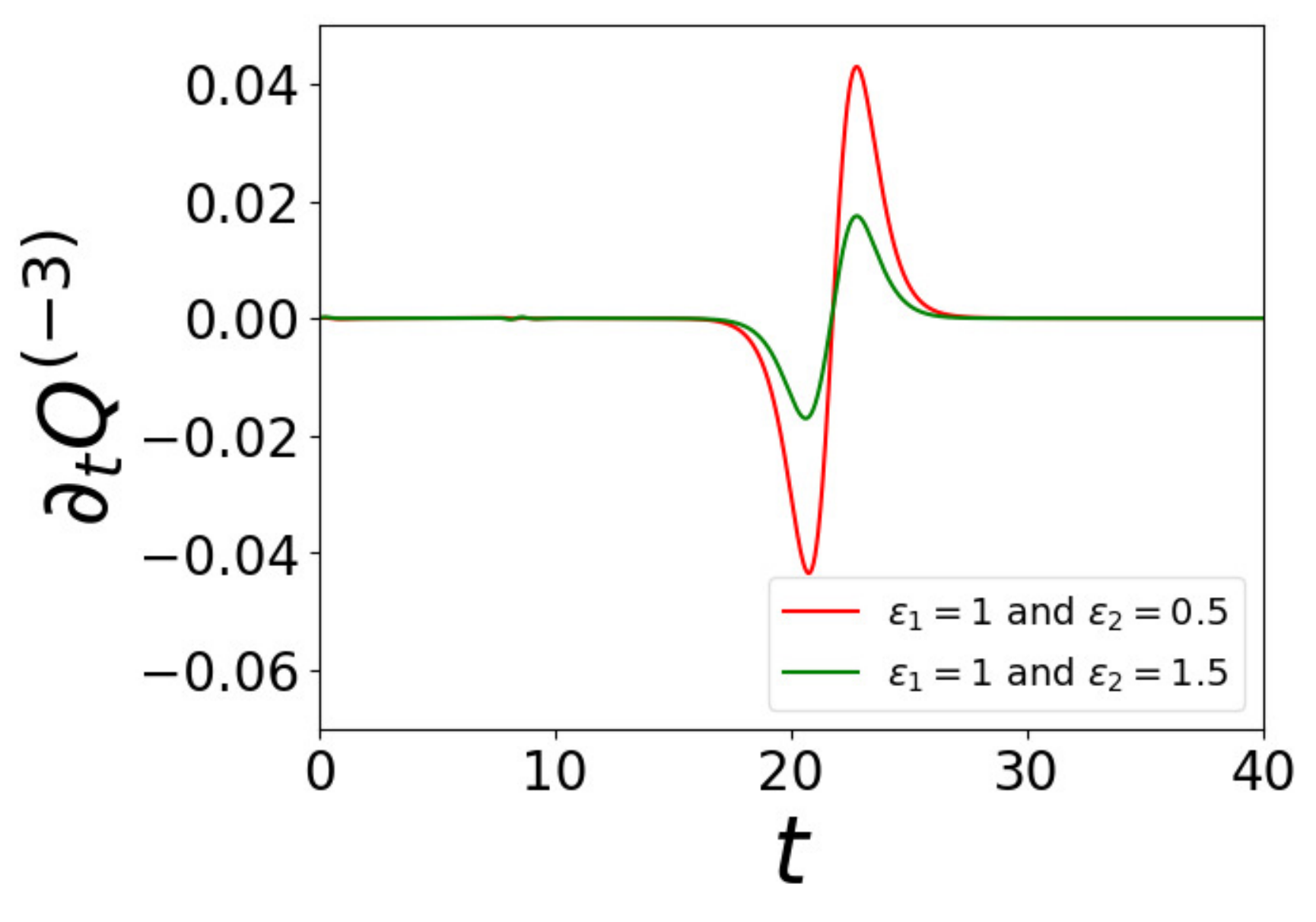}
			\caption{}
			\label{plot11_8}
		\end{subfigure}%
		\begin{subfigure}{.34\textwidth}
			\hspace*{-1.2cm}
			\centering
			\includegraphics[scale=0.34]{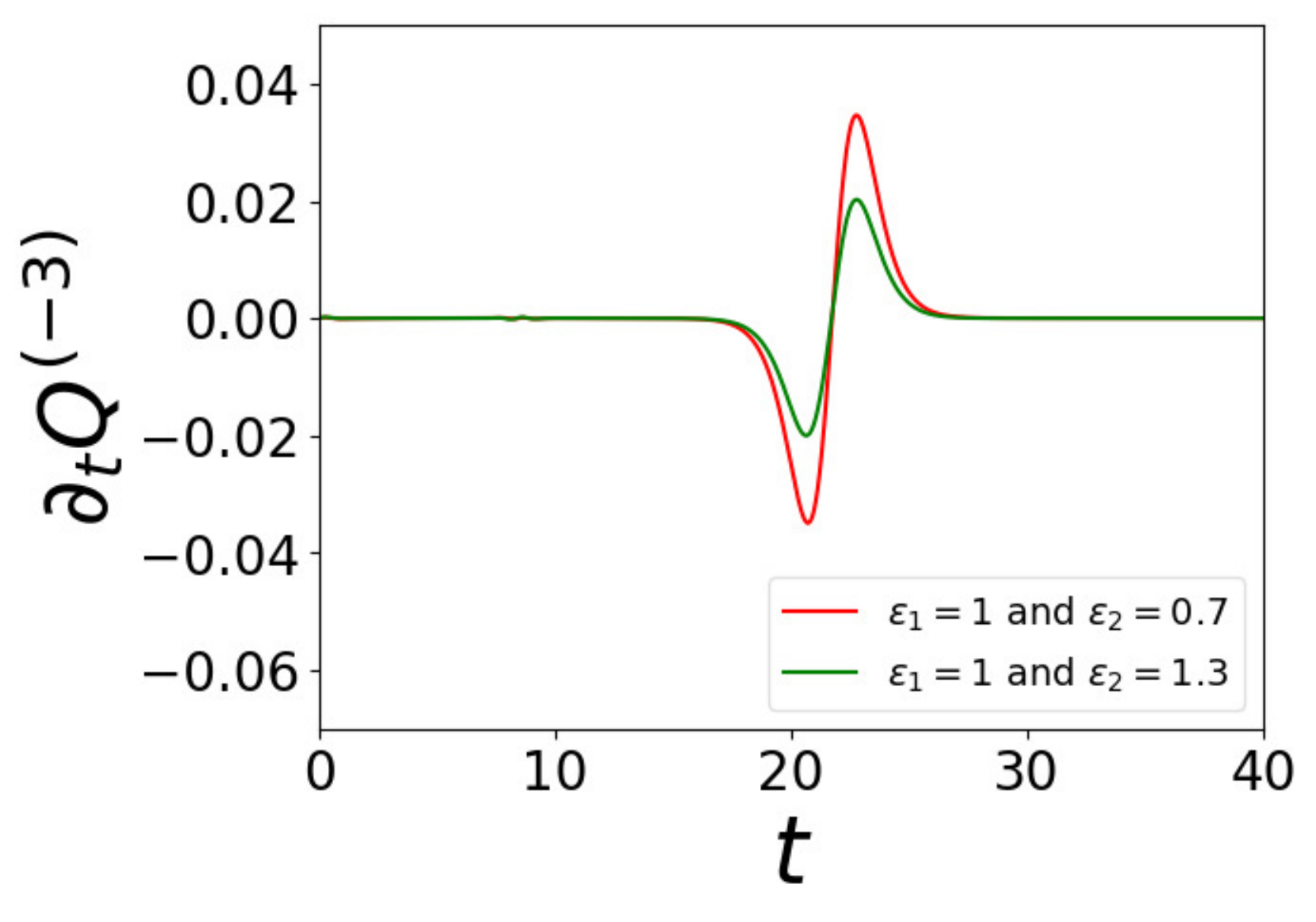}
			\caption{}
			\label{plot11_9}
		\end{subfigure}%
		\begin{subfigure}{.34\textwidth}
			\hspace*{-1.2cm}
			\centering
			\includegraphics[scale=0.34]{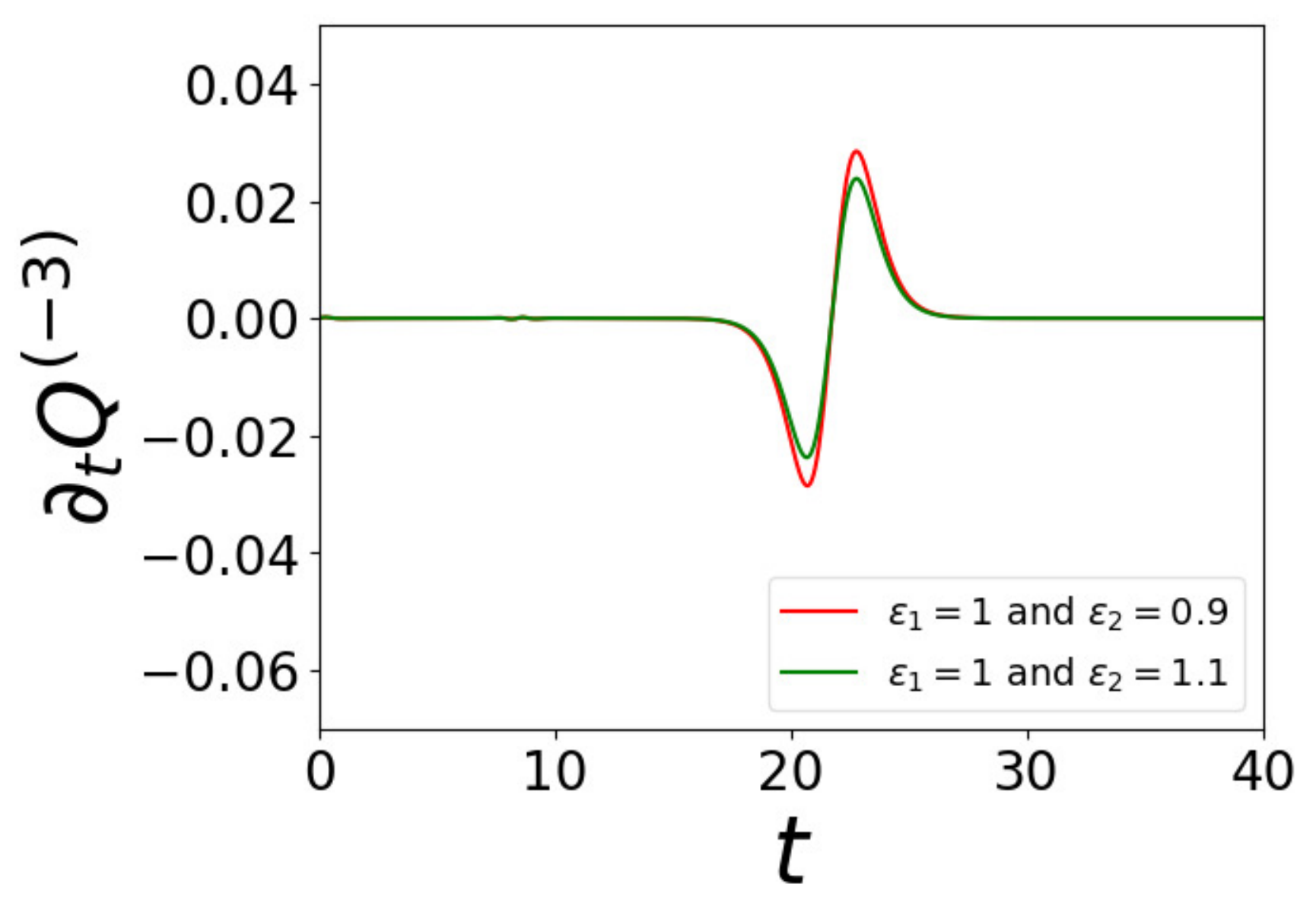}
			\caption{}
			\label{plot11_10}
		\end{subfigure}%
		\caption{The time-dependence of the  obtained values of the quantity $\partial_t Q^{(-3)}$ of the simulation of equation~(\ref{deformedqkdv}), for two-soliton solutions, 
 with different values of $\varepsilon_1 = 1$, $\varepsilon_2 \neq 1$.}
		\label{plot11_8to11_10}
	\end{figure}

\begin{figure}[t!]
		\centering
		\hspace*{0.5cm}
		\begin{subfigure}{.34\textwidth}
			\hspace*{-1.4cm}
			\centering
			\includegraphics[scale=0.32]{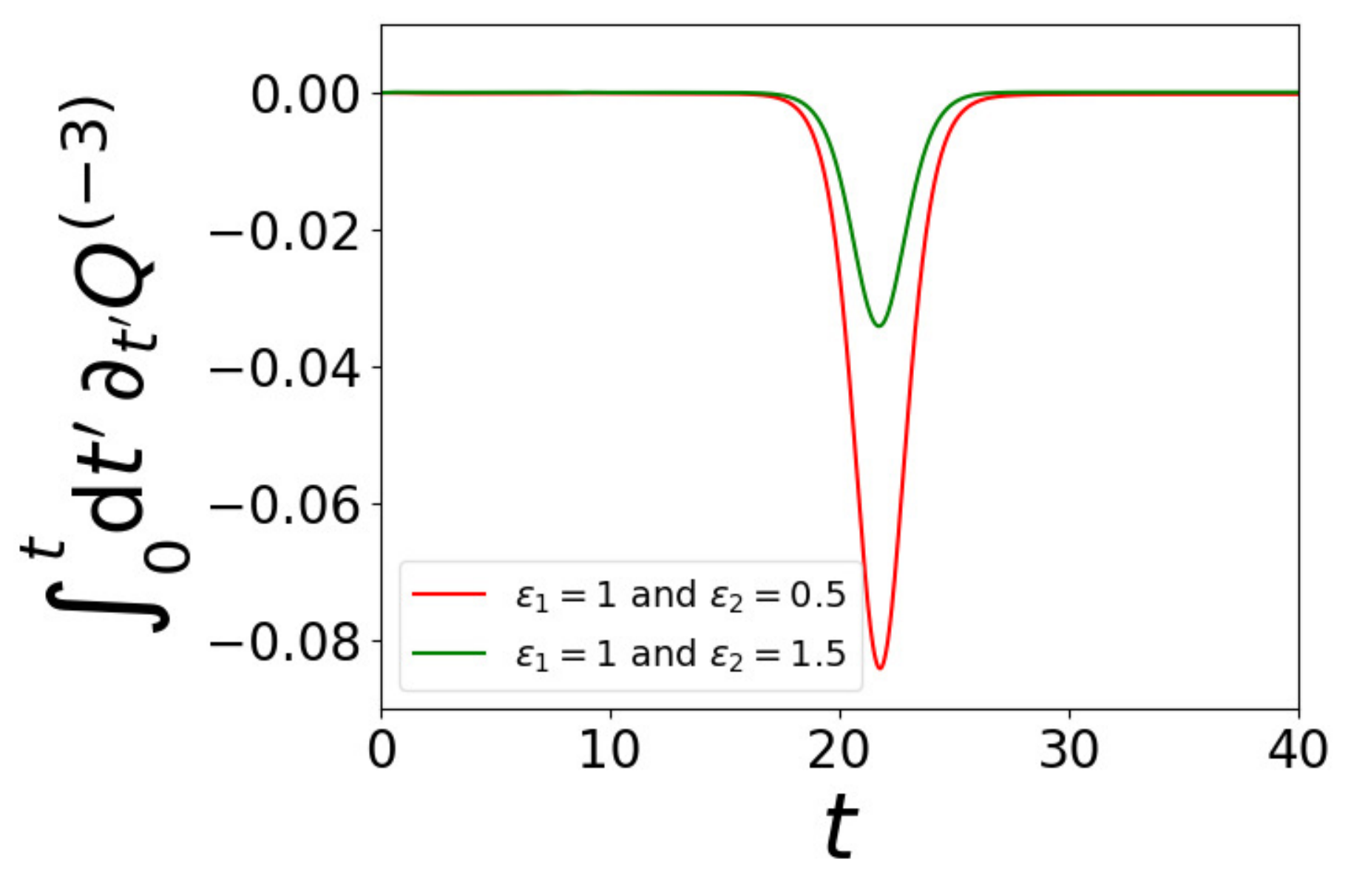}
			\caption{}
			\label{plot11_11}
		\end{subfigure}%
		\begin{subfigure}{.34\textwidth}
			\hspace*{-1.4cm}
			\centering
			\includegraphics[scale=0.32]{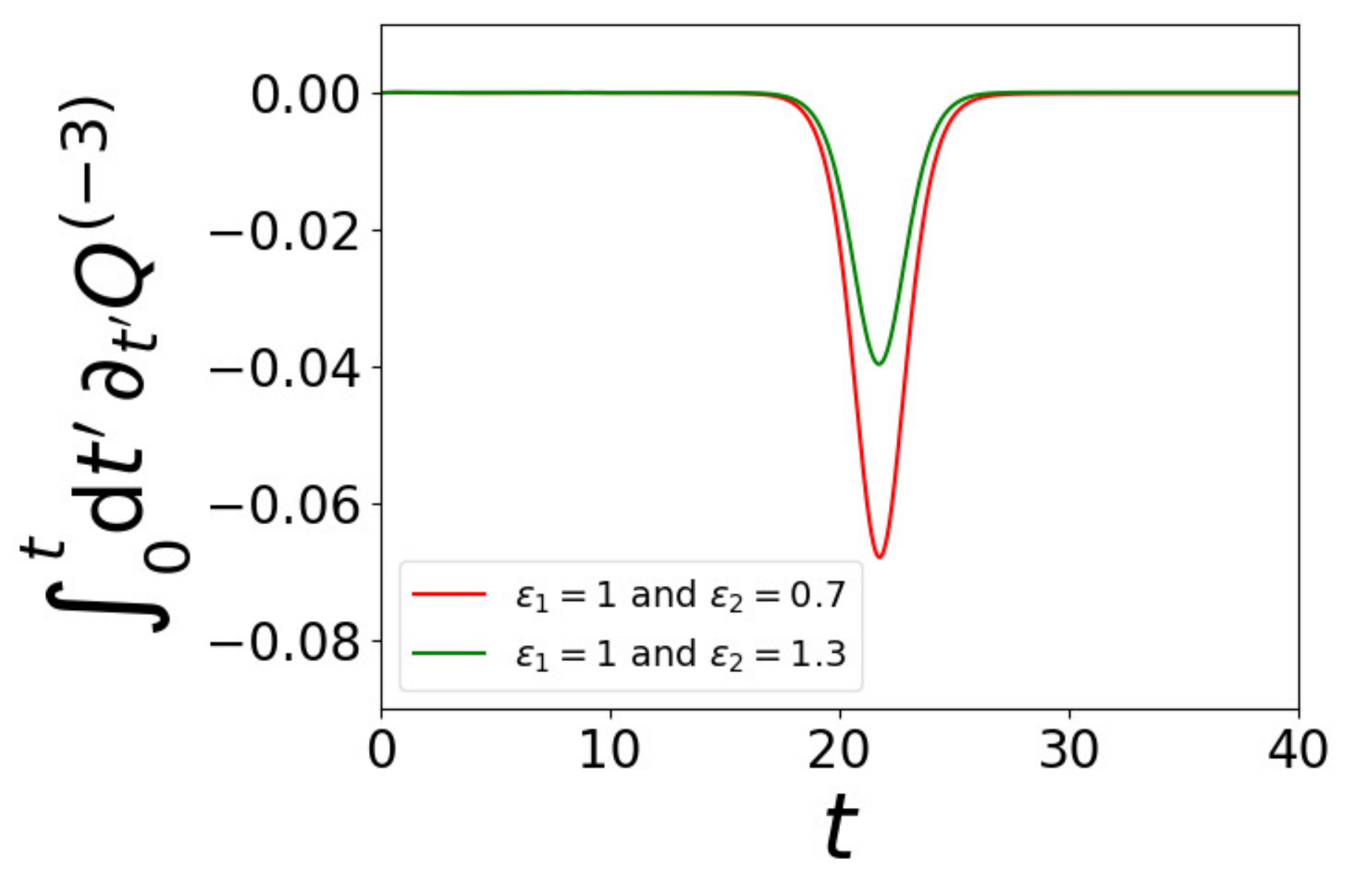}
			\caption{}
			\label{plot11_12}
		\end{subfigure}%
		\begin{subfigure}{.34\textwidth}
			\hspace*{-1.4cm}
			\centering
			\includegraphics[scale=0.32]{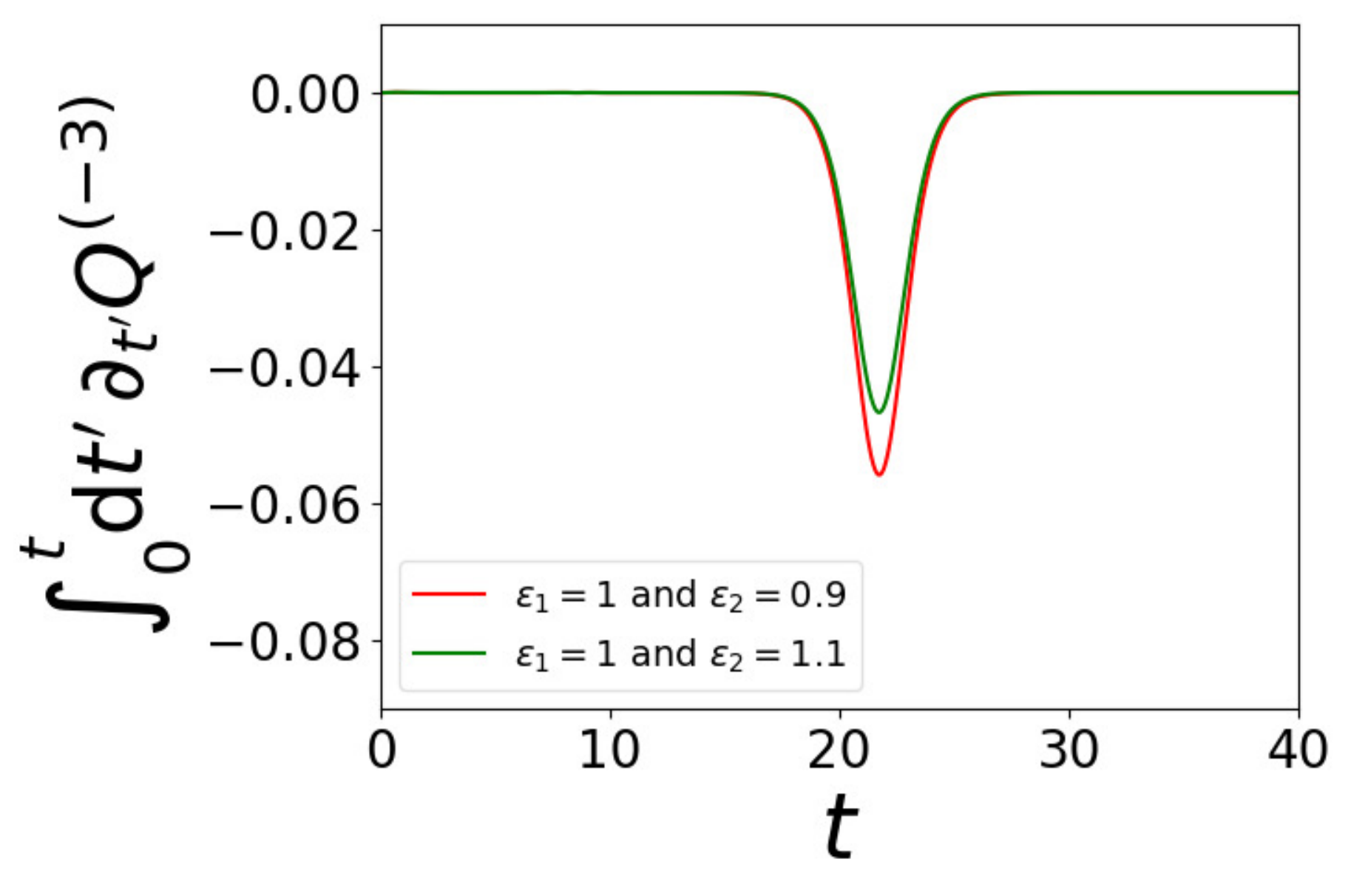}
			\caption{}
			\label{plot11_13}
		\end{subfigure}%
		\caption{The time-dependence of the obtained values of  $\int_{0}^{t} \mathrm{d} t^\prime \, \partial_{t^\prime} Q^{(-3)}$ seen in our simulations of
 equation~(\ref{deformedqkdv}), for two-soliton solutions, for   $\varepsilon_1 = 1$, and $\varepsilon_2 = 0.7$ and $\varepsilon_2 = 1.3$.}
		\label{plot11_11to11_13}
	\end{figure}

	As before, first we present our results for two solitons. Figure~\ref{plot11_2to11_7}	
	shows the time-evolution of the $u$ field for a simulation with $\varepsilon_2 = 0.5$ 
and a simulation with $\varepsilon_2 = 1.5$. Figure~\ref{plot11_8}
	and~\ref{plot11_11}
	show the time dependence of $\partial_t Q^{(-3)}$ and $\int_{0}^{t} \mathrm{d} t^\prime \, \partial_{t^\prime} Q^{(-3)}$ seen these simulations. 
The system for $\varepsilon_2 = 0.5$ emits a small amount of (visible) radiation during the interaction of the soliton fields (similarly to the two-soliton 
simulation for the RLW equation (see subsubsection~\ref{Two_soliton_solutions_of_the_RLW_equation})). This radiation is difficult to spot on the scale shown 
in figure~\ref{plot11_2to11_7}, and so we have added inserts in plots~\ref{plot11_6} and~\ref{plot11_7} to show more clearly this radiation emitted during the interaction.
	Looking at figure~\ref{plot11_8}, we see that $Q^{(-3)}$ changes significantly only during the interaction of 
the two soliton fields. Furthermore, figure~\ref{plot11_11} shows that the quantity $Q^{(-3)}$ satisfies
	\begin{equation}
	\lim_{t \to - \infty} Q^{(-3)} = \lim_{t \to \infty} Q^{(-3)} \,,
	\end{equation}
	which is again what we saw for the RLW and mRLW equation. 
	
	The other plots in figures~\ref{plot11_8to11_10} and~\ref{plot11_11to11_13} show the the time dependence of $\partial_t Q^{(-3)}$ and $\int_{0}^{t} \mathrm{d} t^\prime \, \partial_{t^\prime} Q^{(-3)}$ for simulations with various other values of $\varepsilon_2 \neq 1$. They show similar patterns to the ones described for $\varepsilon_2 = 0.5$ and $\varepsilon_2 = 1.5$. Note that as $|1-\varepsilon_2|$ becomes smaller, the values of $\partial_t Q^{(-3)}$ and $\int_{0}^{t} \mathrm{d} t^\prime \, \partial_{t^\prime} Q^{(-3)}$ converge closer to the values displayed in figure~\ref{I_1_alpha_8_e1_1_e2_1_e3_1_1_analytical_and_numerical}.
	
	\subsubsection{Three-soliton configurations}
	
\begin{figure}[t!]
		\centering
		\hspace*{-0.1cm}
		\begin{subfigure}{.34\textwidth}
			\centering
			\includegraphics[scale=0.34]{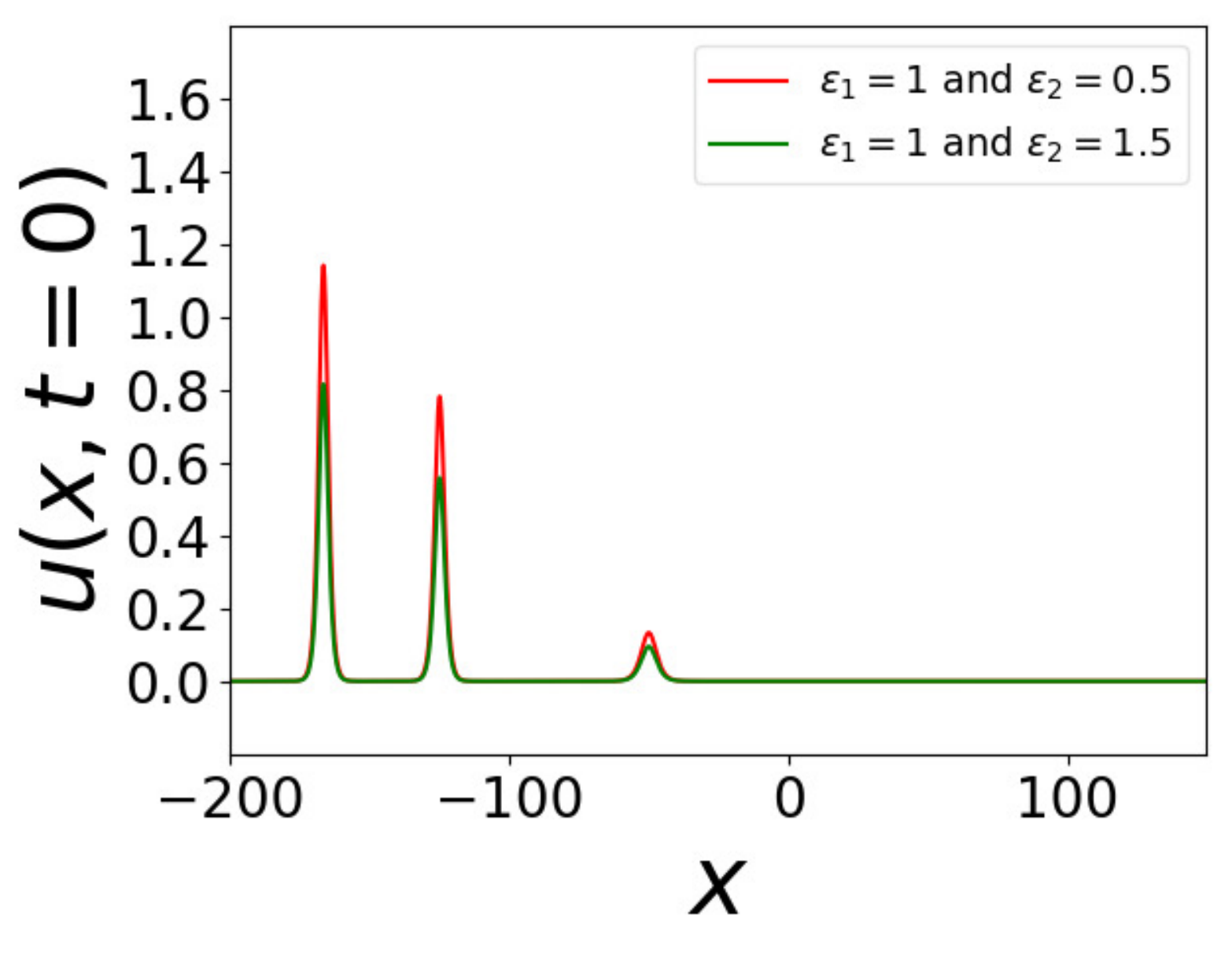}
			\caption{At~$t=0$}
			\label{plot12_2}
		\end{subfigure}%
		\begin{subfigure}{.34\textwidth}
			\centering
			\includegraphics[scale=0.34]{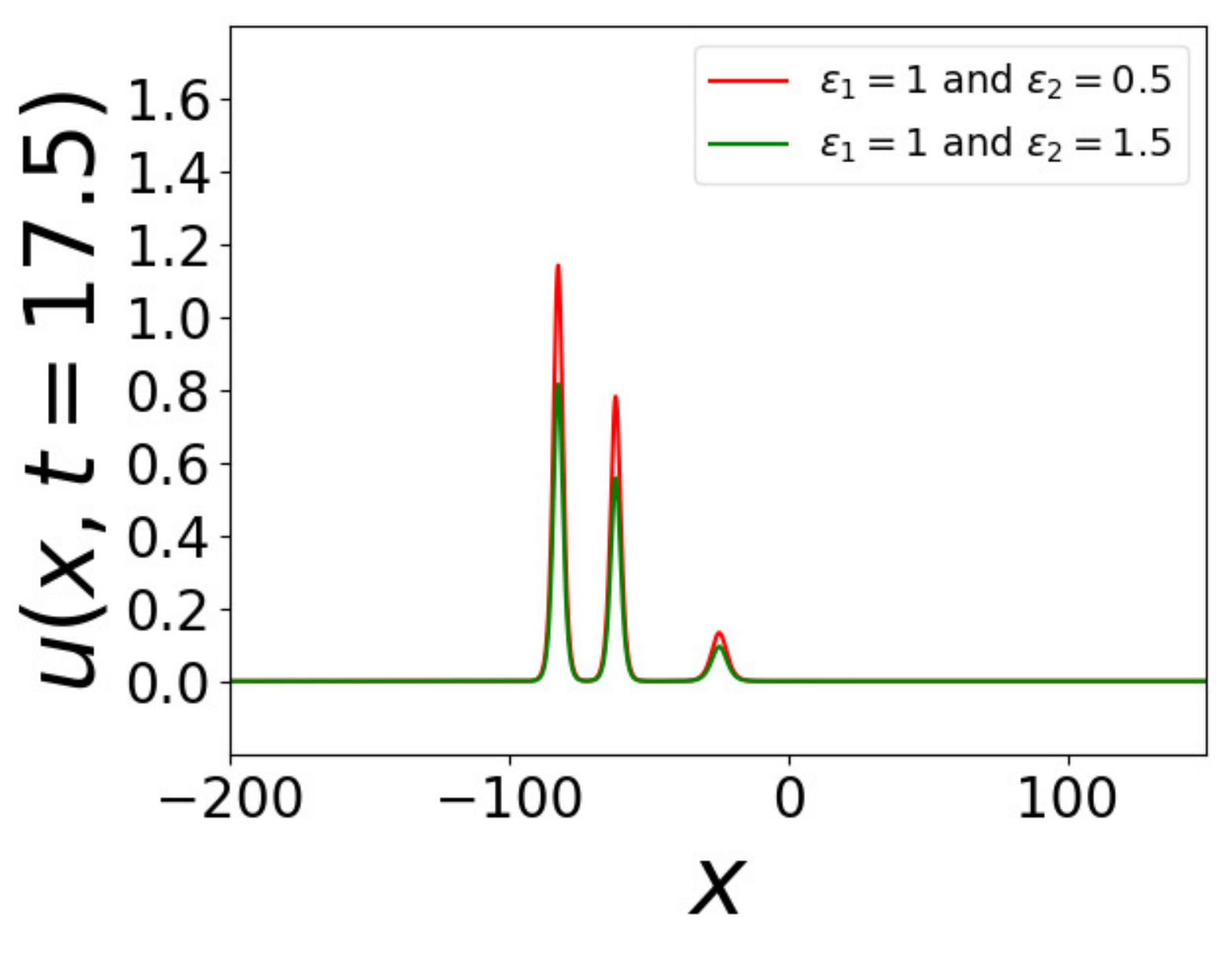}
			\caption{At~$t=17.5$}
			\label{plot12_3}
		\end{subfigure}%
		\begin{subfigure}{.34\textwidth}
			\centering
			\includegraphics[scale=0.34]{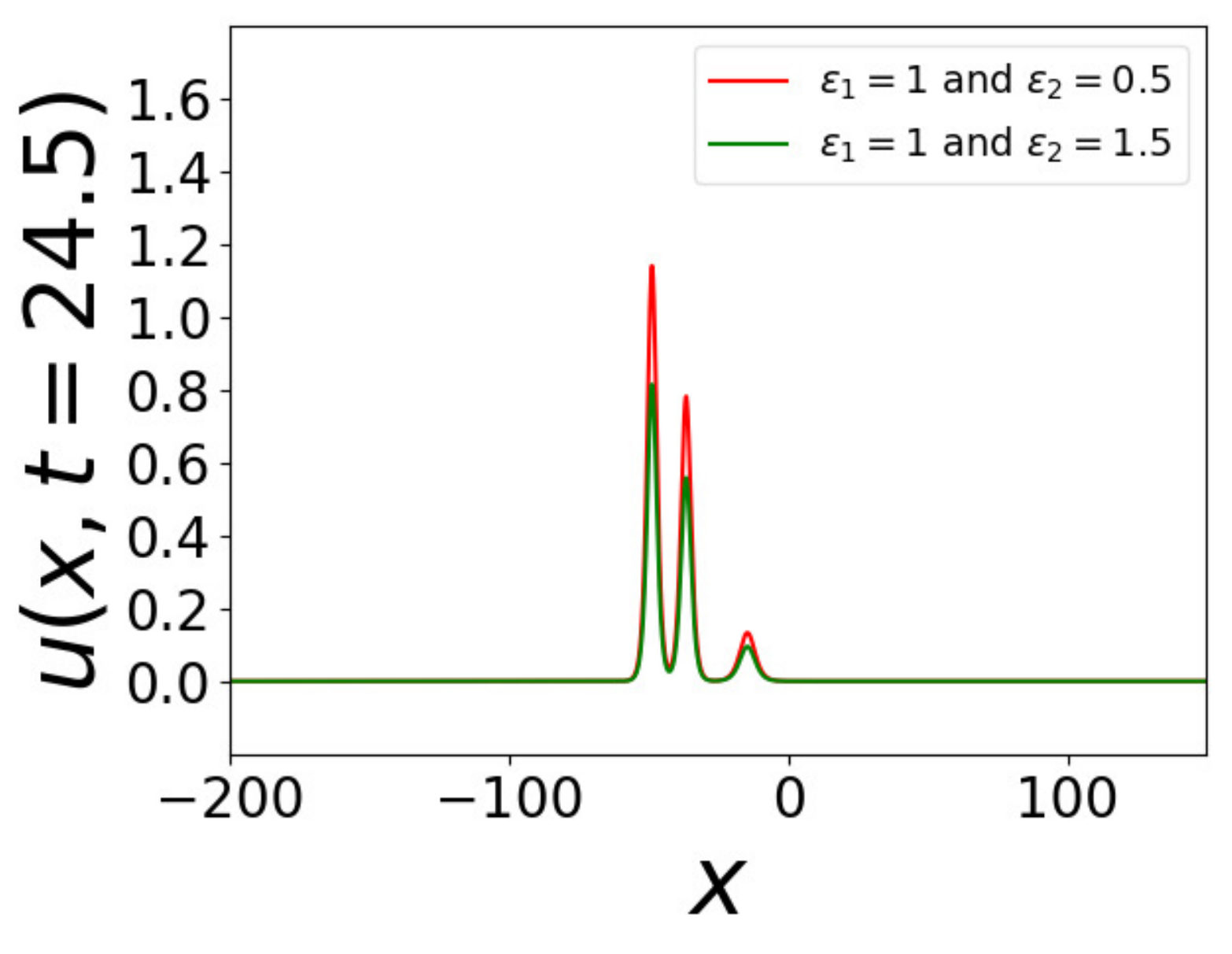}
			\caption{At~$t=24.5$}
			\label{plot12_4}
		\end{subfigure}
		
		\begin{subfigure}{.34\textwidth}
			\centering
			\includegraphics[scale=0.34]{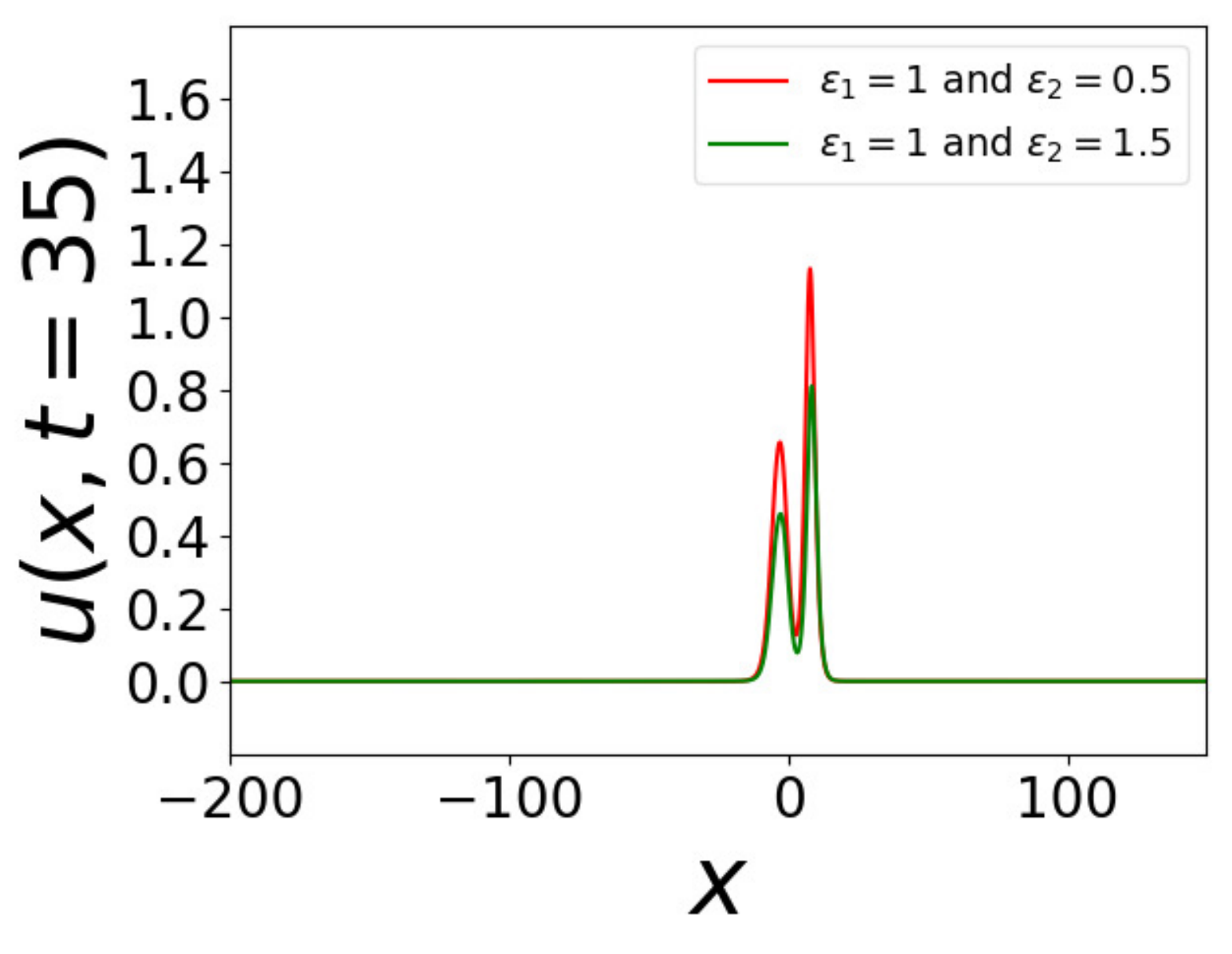}
			\caption{At~$t=35$}
			\label{plot12_5}
		\end{subfigure}%
		\begin{subfigure}{.34\textwidth}
			\centering
			\includegraphics[scale=0.34]{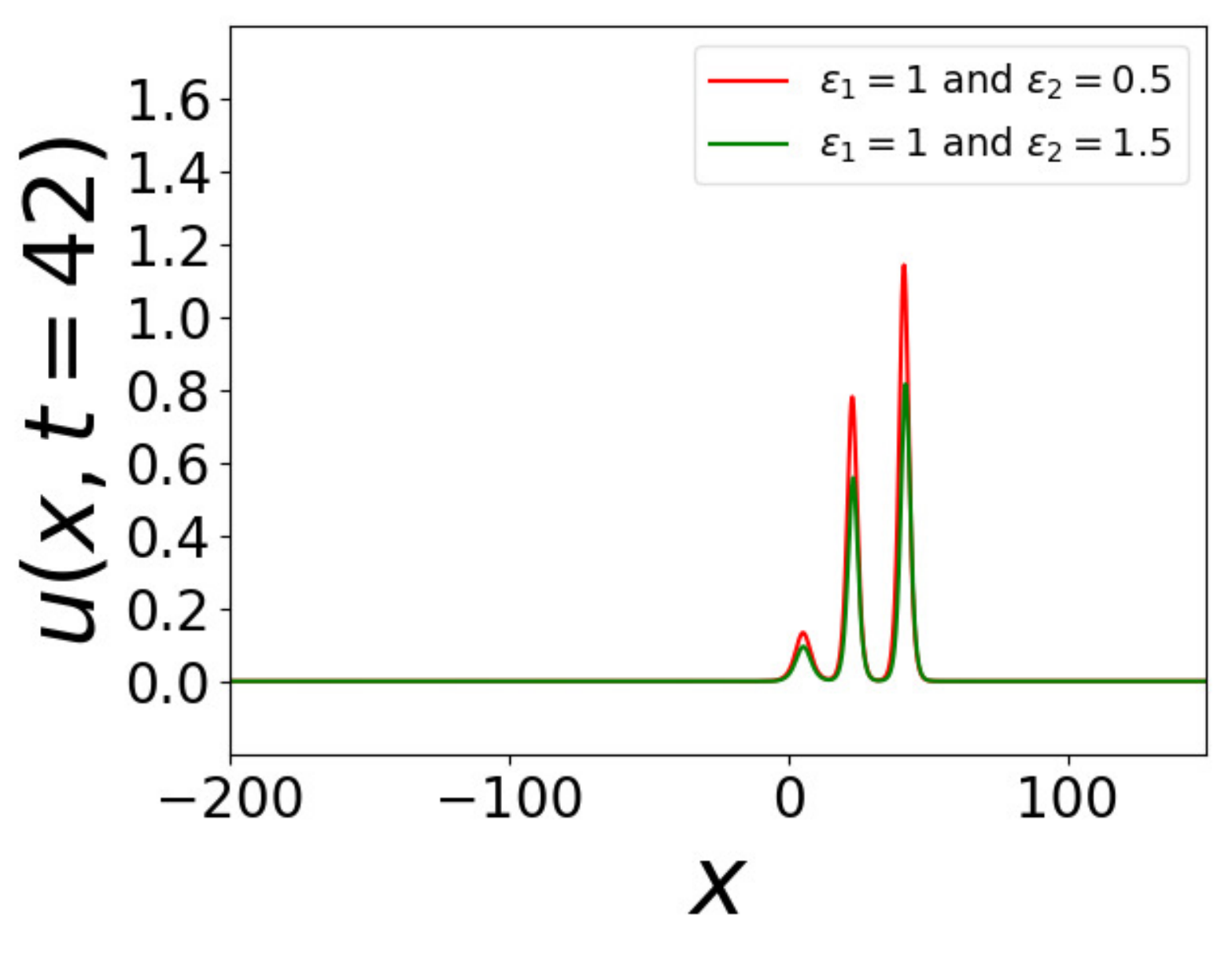}
			\caption{At~$t=42$}
			\label{plot12_6}
		\end{subfigure}%
		\begin{subfigure}{.34\textwidth}
			\centering
			\includegraphics[scale=0.34]{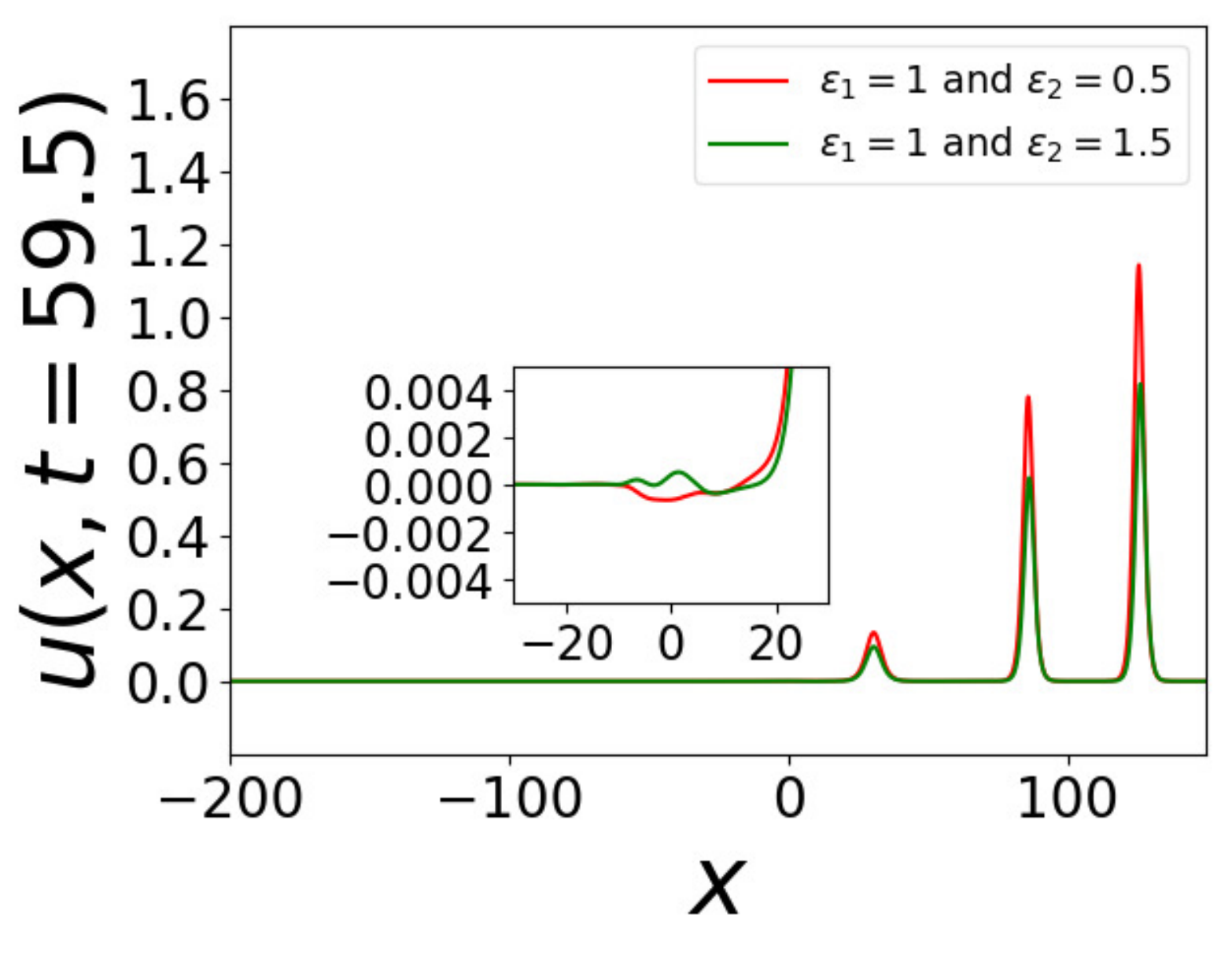}
			\caption{At~$t=59.5$}
			\label{plot12_7}
		\end{subfigure}
		\caption{The time-evolution of two three-soliton systems seen in in our simulations of the equation~(\ref{deformedqkdv}) with
 $\varepsilon_1 = 1$, $\varepsilon_2 = 0.5$ (red curve) and $\varepsilon_1 = 1$, $\varepsilon_2 = 1.5$ (green curve).} 
		\label{plot12_2to12_7}
	\end{figure}

\begin{figure}[b!]
		\centering
		\hspace*{0.5cm}
		\begin{subfigure}{.34\textwidth}
			\hspace*{-1.2cm}
			\centering
			\includegraphics[scale=0.34]{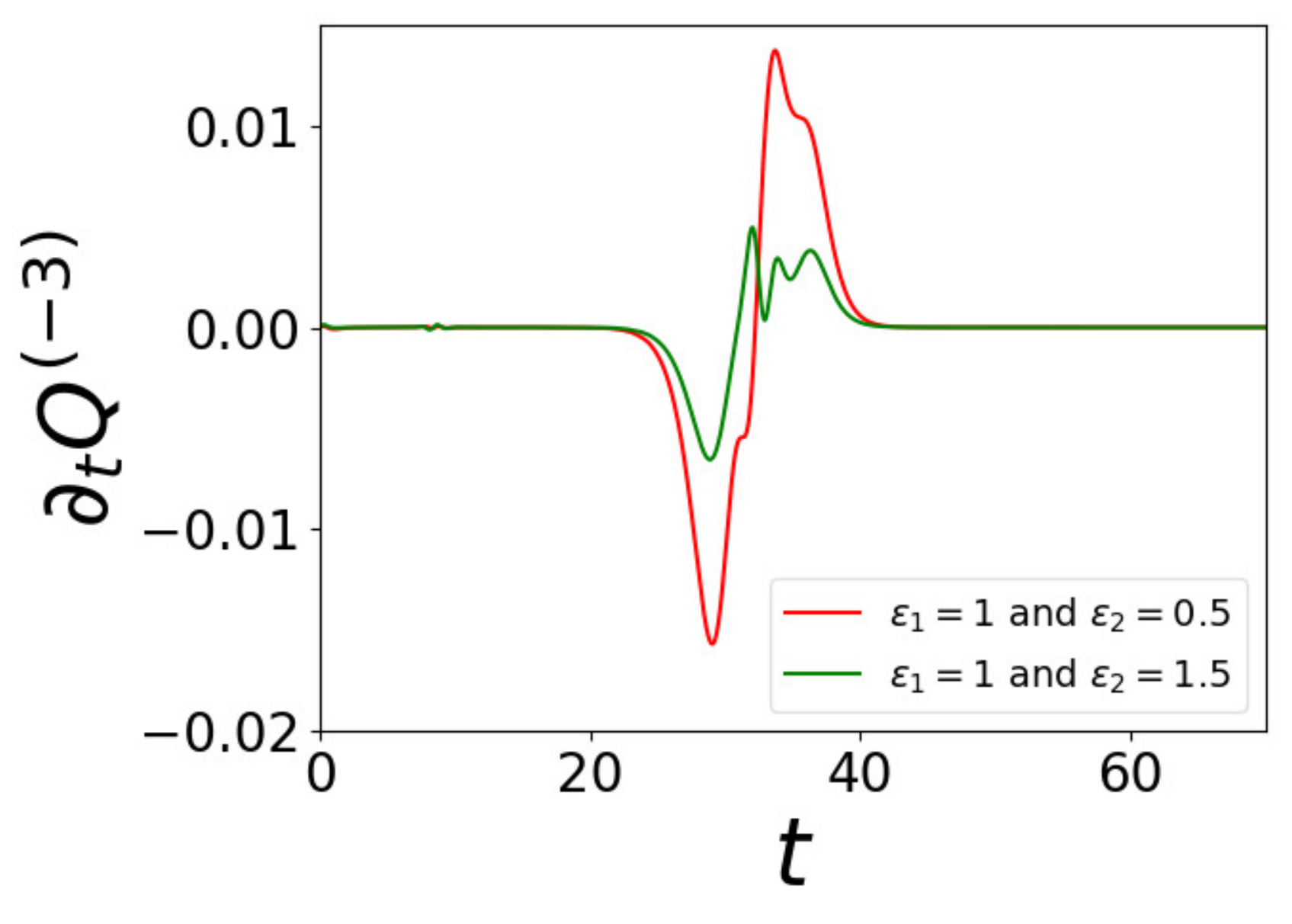}
			\caption{}
			\label{plot12_8}
		\end{subfigure}%
		\begin{subfigure}{.34\textwidth}
			\hspace*{-1.2cm}
			\centering
			\includegraphics[scale=0.34]{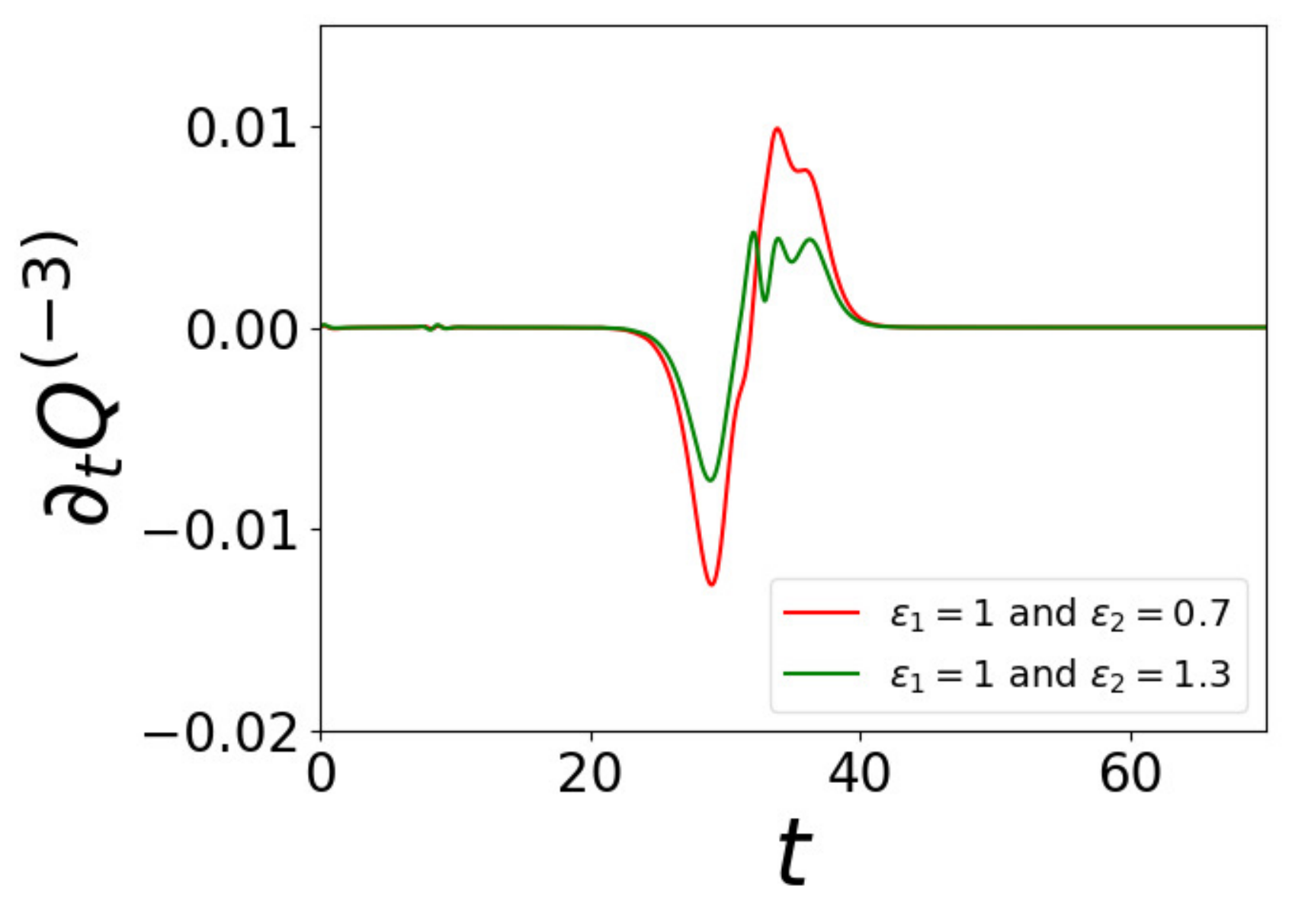}
			\caption{}
			\label{plot12_9}
		\end{subfigure}%
		\begin{subfigure}{.34\textwidth}
			\hspace*{-1.2cm}
			\centering
			\includegraphics[scale=0.34]{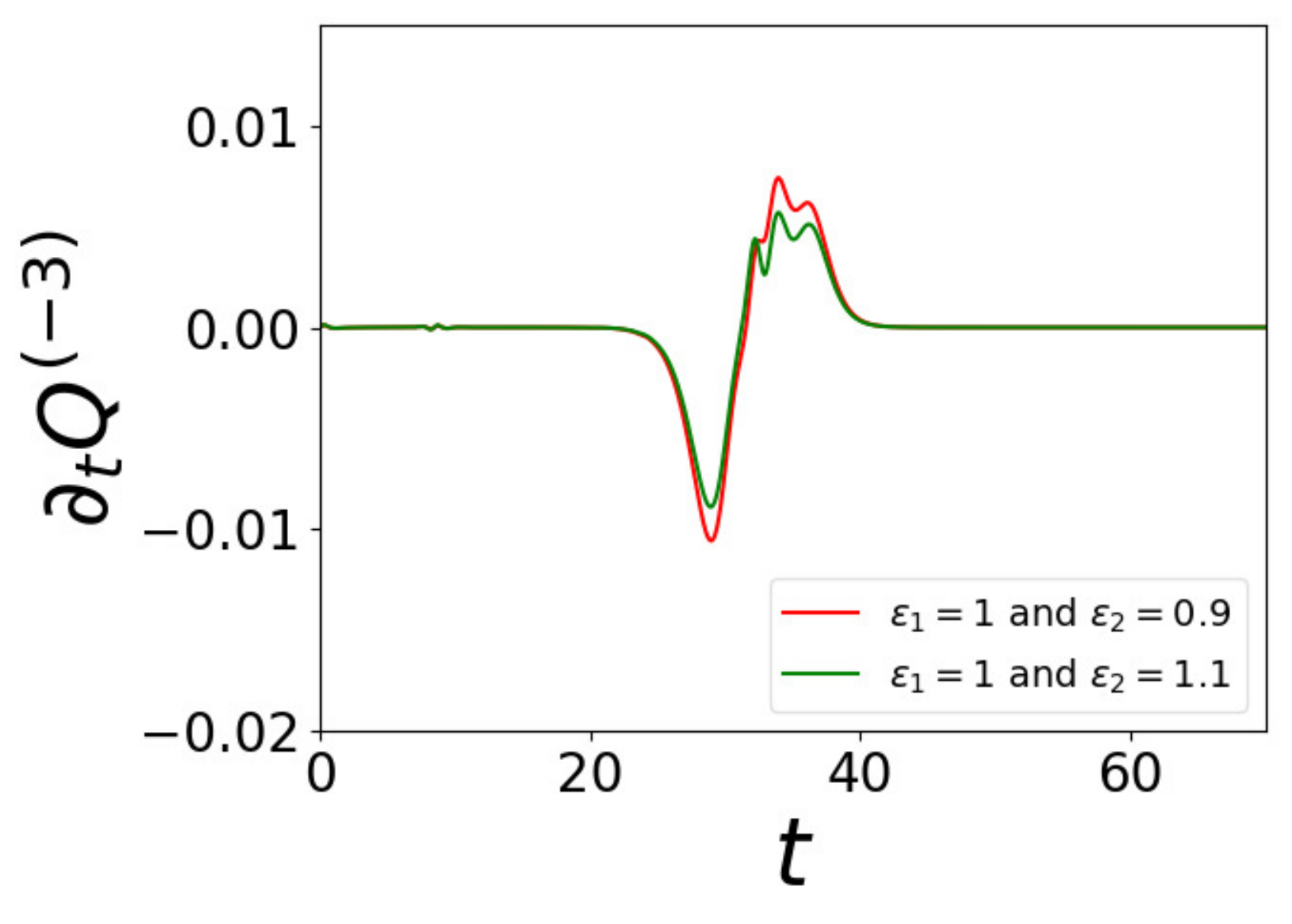}
			\caption{}
			\label{plot12_10}
		\end{subfigure}%
		\caption{The time-dependence of $\partial_t Q^{(-3)}$ for a three-soliton system as determined by the simulation of equation~(\ref{deformedqkdv}) with 
  $\varepsilon_1 = 1$ and different values of $\varepsilon_2 \neq 1$.}
		\label{plot12_8to12_10}
	\end{figure}

\begin{figure}[t!]
		\centering
		\hspace*{0.5cm}
		\begin{subfigure}{.34\textwidth}
			\hspace*{-1.4cm}
			\centering
			\includegraphics[scale=0.32]{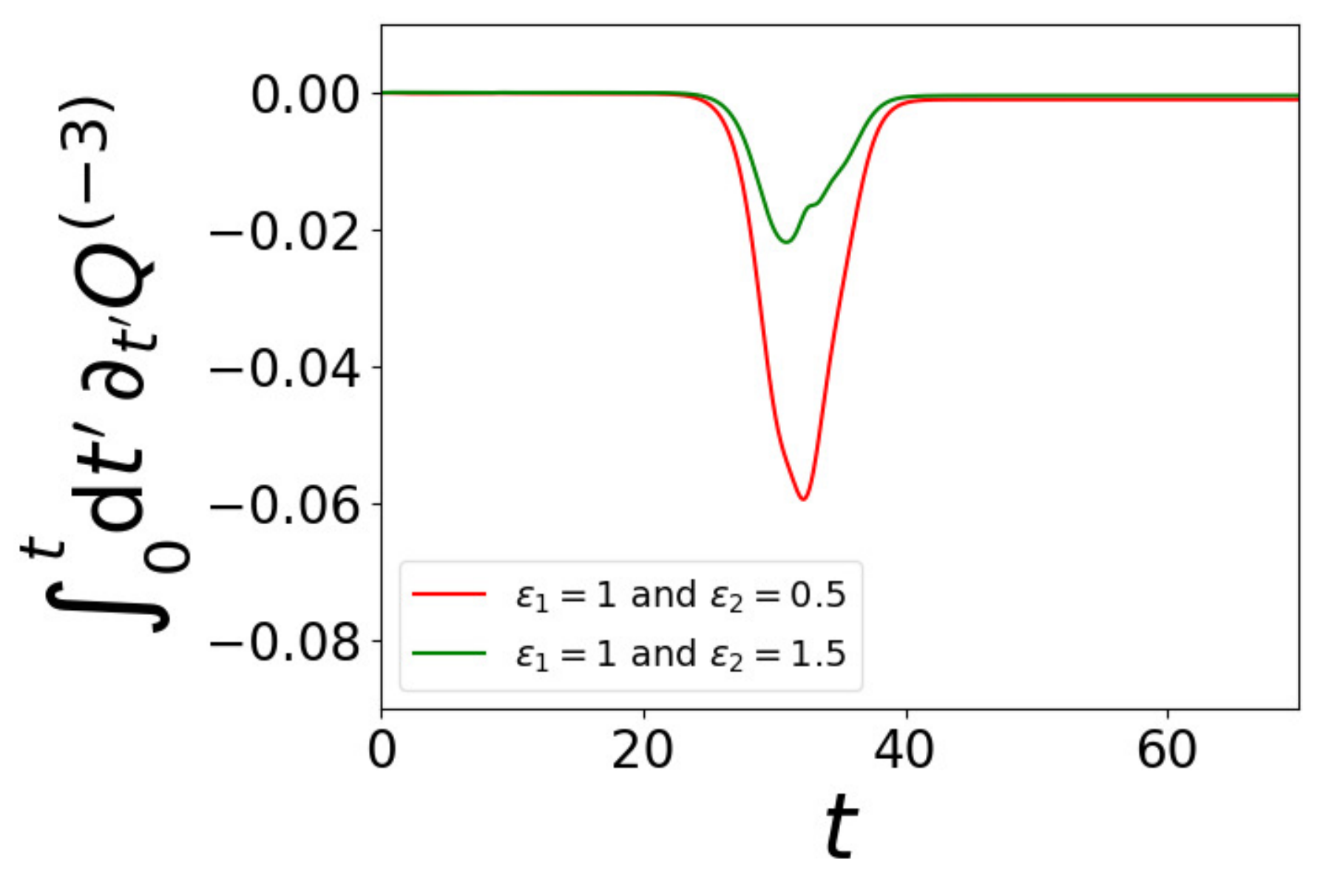}
			\caption{}
			\label{plot12_11}
		\end{subfigure}%
		\begin{subfigure}{.34\textwidth}
			\hspace*{-1.4cm}
			\centering
			\includegraphics[scale=0.32]{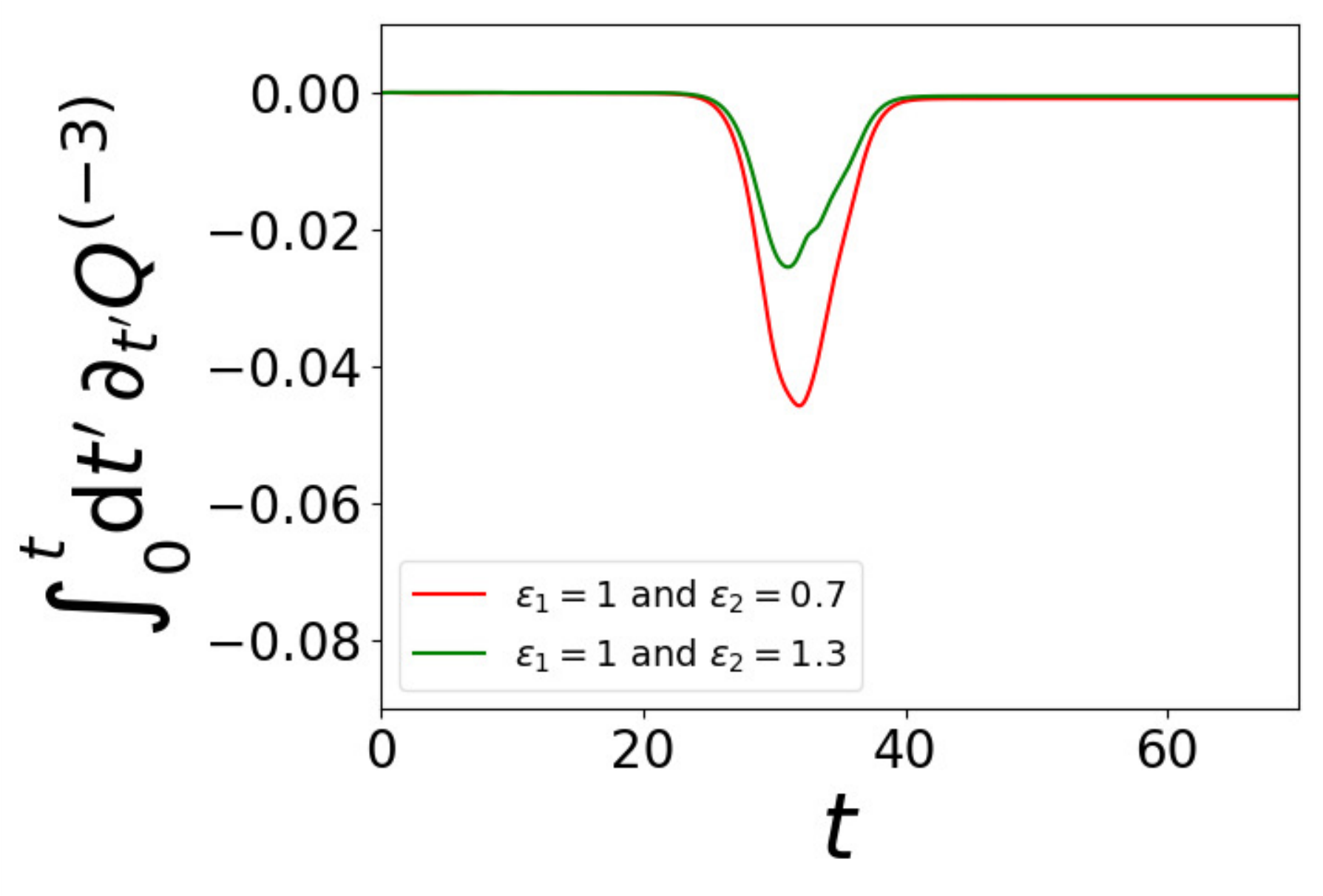}
			\caption{}
			\label{plot12_12}
		\end{subfigure}%
		\begin{subfigure}{.34\textwidth}
			\hspace*{-1.4cm}
			\centering
			\includegraphics[scale=0.32]{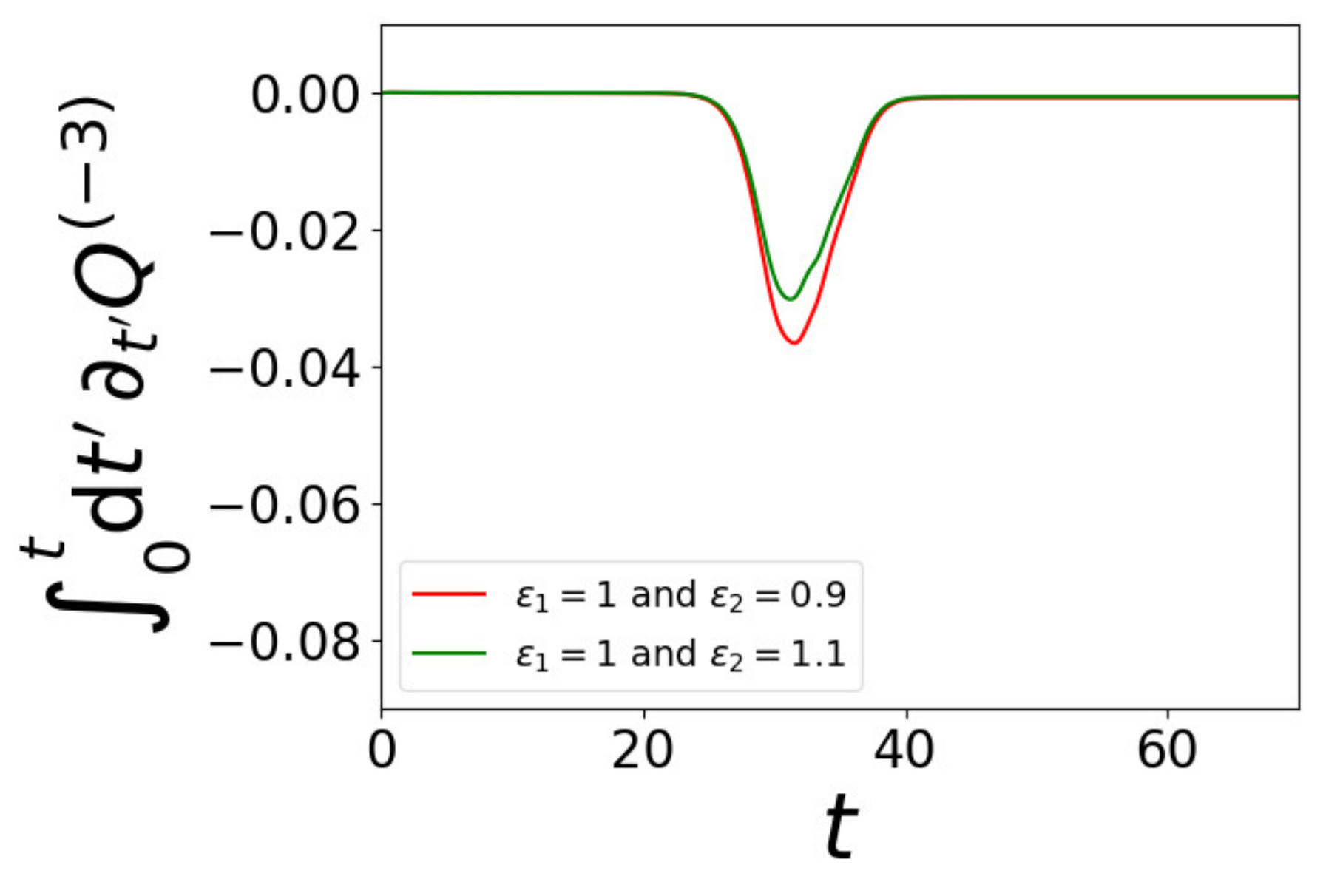}
			\caption{}
			\label{plot12_13}
		\end{subfigure}%
		\caption{The time-dependence of the quantity $\int_{0}^{t} \mathrm{d} t^\prime \, \partial_{t^\prime} Q^{(-3)}$ as determined in the simulation
 of equation~(\ref{deformedqkdv}), for three-soliton solutions, with 
  $\varepsilon_1 = 1$ and different values of $\varepsilon_2 \neq 1$.} 
		\label{plot12_11to12_13}
	\end{figure}

	Next, we also looked at some three solitons systems which were constructed using equation~(\ref{ic1}), with $n=3$, as the initial conditions. Figure~\ref{plot12_2to12_7}
	shows the time evolution seen in two simulations for the values $\varepsilon_1=1$,
 $\varepsilon_2=0.5$ and $\varepsilon_1=1$, $\varepsilon_2=1.5$. These plots show that the three solitons interact with each other in a way very similar to 
the mRLW and RLW equation (see subsections~\ref{Three_soliton_solutions_of_the_mRLW_equation} and~\ref{Three_soliton_solutions_of_the_RLW equation}).
 Unlike for the two-soliton simulation, we now see slightly more radiation emitted in the $\varepsilon_2=1.5$ case than in the $\varepsilon_2=0.5$ case (see the insert in figure~\ref{plot12_7}). 
Figures~\ref{plot12_8to12_10}
	and~\ref{plot12_11to12_13}
	present the plots of time dependence of $\partial_t Q^{(-3)}$ and $\int_{0}^{t} \mathrm{d} t^\prime \, \partial_{t^\prime} Q^{(-3)}$ seen in our simulations.
 Figure~\ref{plot12_8to12_10} clearly shows that $Q^{(-3)}$  only changes during the interaction of the soliton fields, and figure~\ref{plot12_11to12_13} demonstrates 
 that this quantity is indeed truly quasi-conserved in the sense that
	\begin{equation}
	\lim_{t \to - \infty} Q^{(-3)} = \lim_{t \to \infty} Q^{(-3)} \,.
	\end{equation}

	\subsection{Intermediate models; Quasi-conserved charges for various values of $\varepsilon_1$ and $\varepsilon_2$} 
        
We also would like to consider other values of $\varepsilon_1$ and $\varepsilon_2$. However, we have no analytical expressions of one-soliton solutions when $\varepsilon_1 \neq 1$. Therefore, we have used the analytical two-soliton solutions of the mRLW equation as initial conditions for the two-soliton configurations, and the linear superposition of three single-soliton solutions of the mRLW equation for the three-soliton simulations. 

In the next subsection we use the same values as those used in subsection~\ref{Intermediate_models} (for $\varepsilon_1=1$ and $\varepsilon_2 \neq 1$) to simulate the two-soliton configurations. We have repeated these simulations to study the effect of using the `wrong' initial conditions. Then in the subsequent subsubsections we investigate the quasi-integrability for various values of $\varepsilon_1 \neq 1$ while keeping $\varepsilon_2 = 1$.

	\subsubsection{Two-soliton configurations for $\varepsilon_1=1$ and $\varepsilon_2 \neq 1$} \label{Two_soliton_configurations1}

\begin{figure}[b!]
		\centering
		\hspace*{-0.1cm}
		\begin{subfigure}{.34\textwidth}
			\centering
			\includegraphics[scale=0.34]{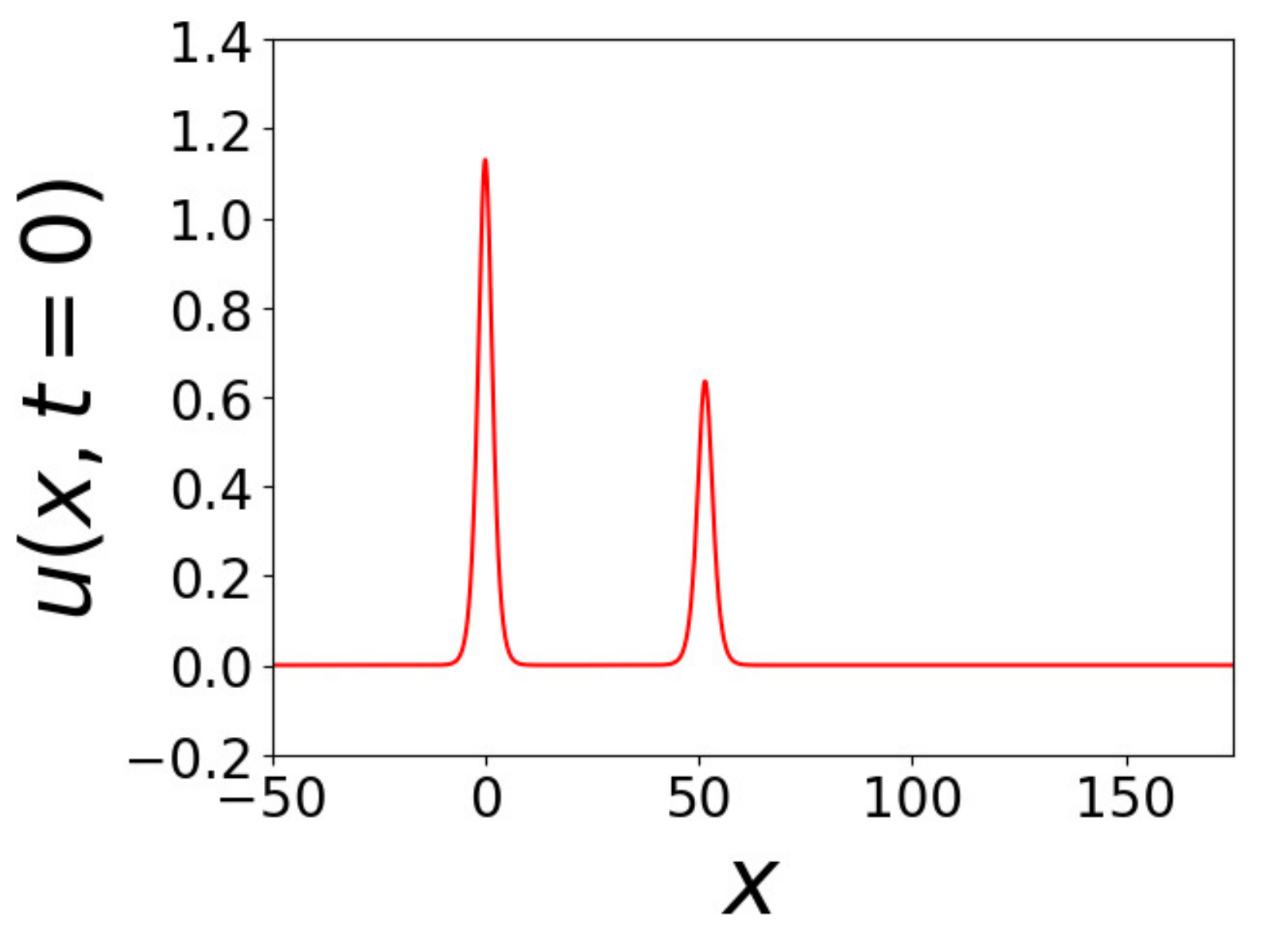}
			\caption{At~$t=0$}
			\label{plot4_2}
		\end{subfigure}%
		\begin{subfigure}{.34\textwidth}
			\centering
			\includegraphics[scale=0.34]{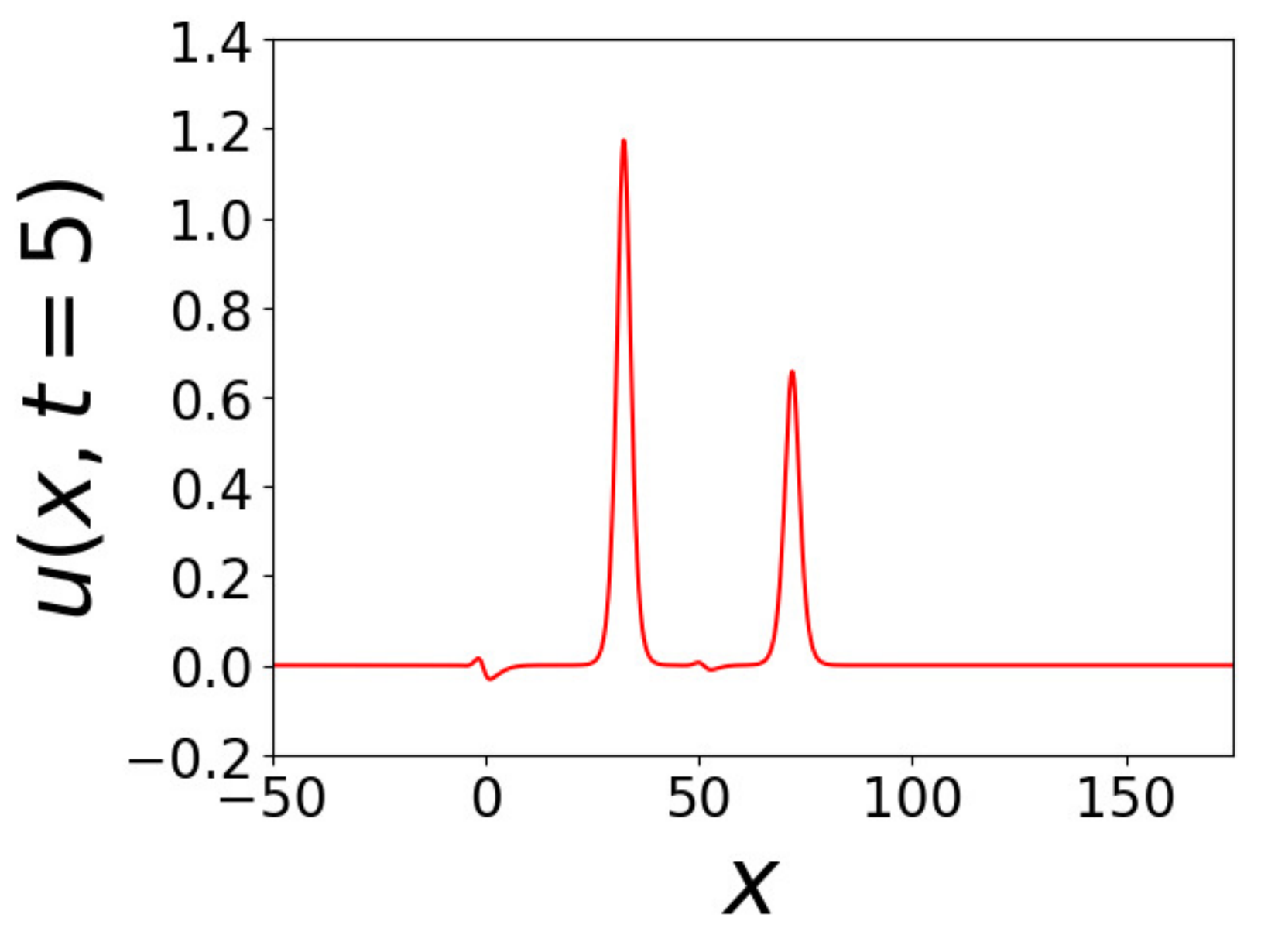}
			\caption{At~$t=5$}
			\label{plot4_3}
		\end{subfigure}%
		\begin{subfigure}{.34\textwidth}
			\centering
			\includegraphics[scale=0.34]{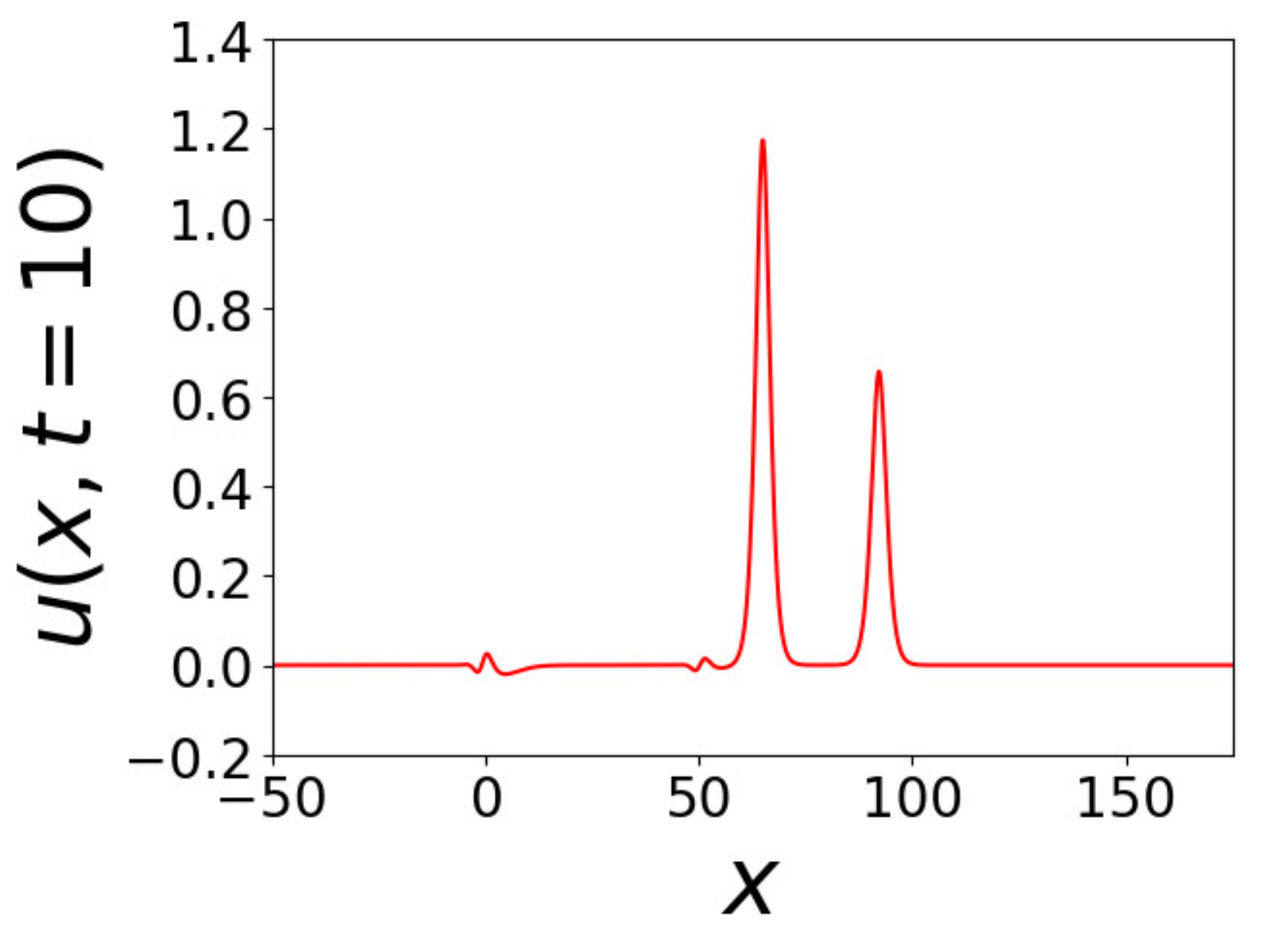}
			\caption{At~$t=10$}
			\label{plot4_4}
		\end{subfigure}
		
		\begin{subfigure}{.34\textwidth}
			\centering
			\includegraphics[scale=0.34]{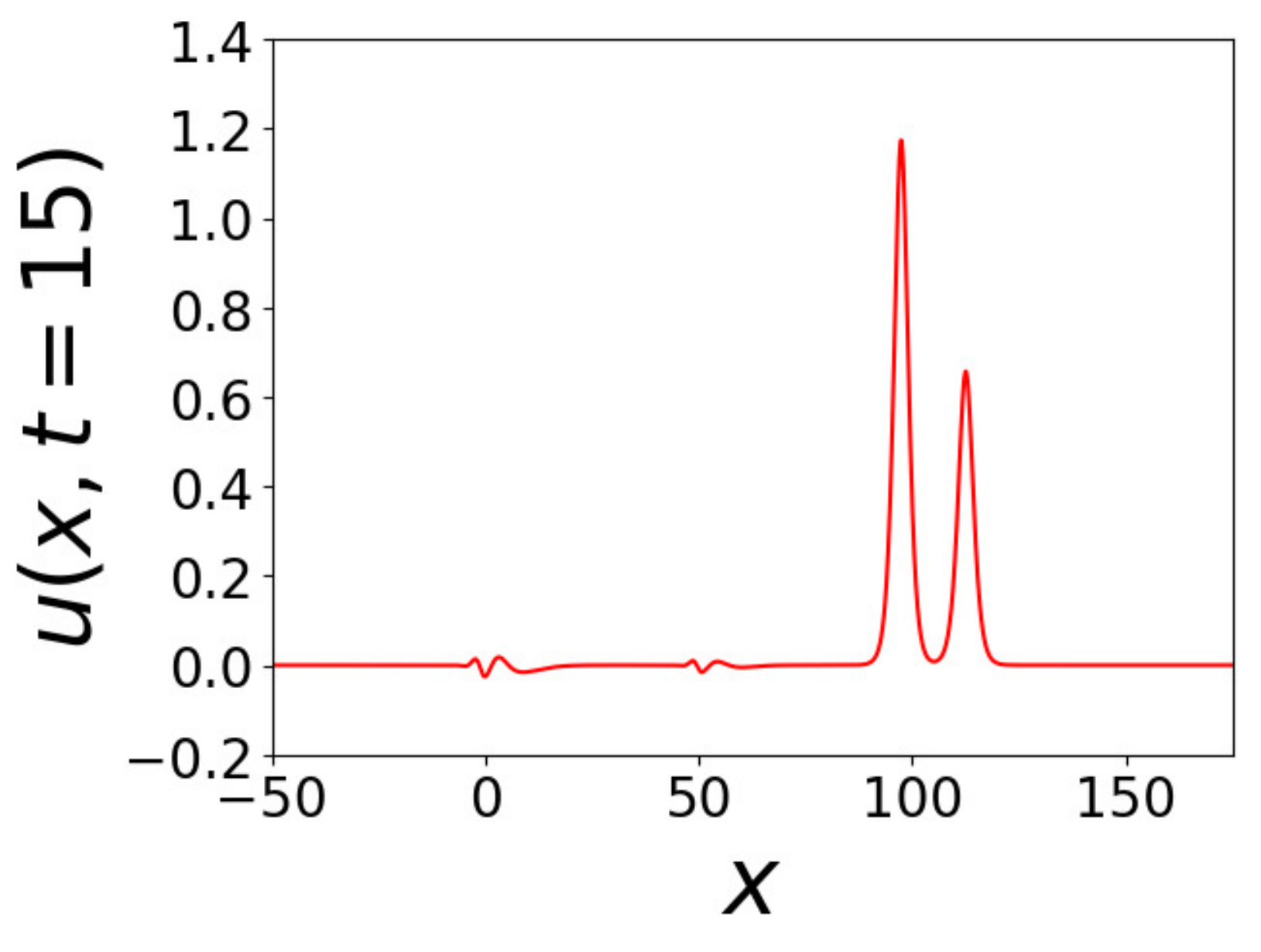}
			\caption{At~$t=15$}
			\label{plot4_5}
		\end{subfigure}%
		\begin{subfigure}{.34\textwidth}
			\centering
			\includegraphics[scale=0.34]{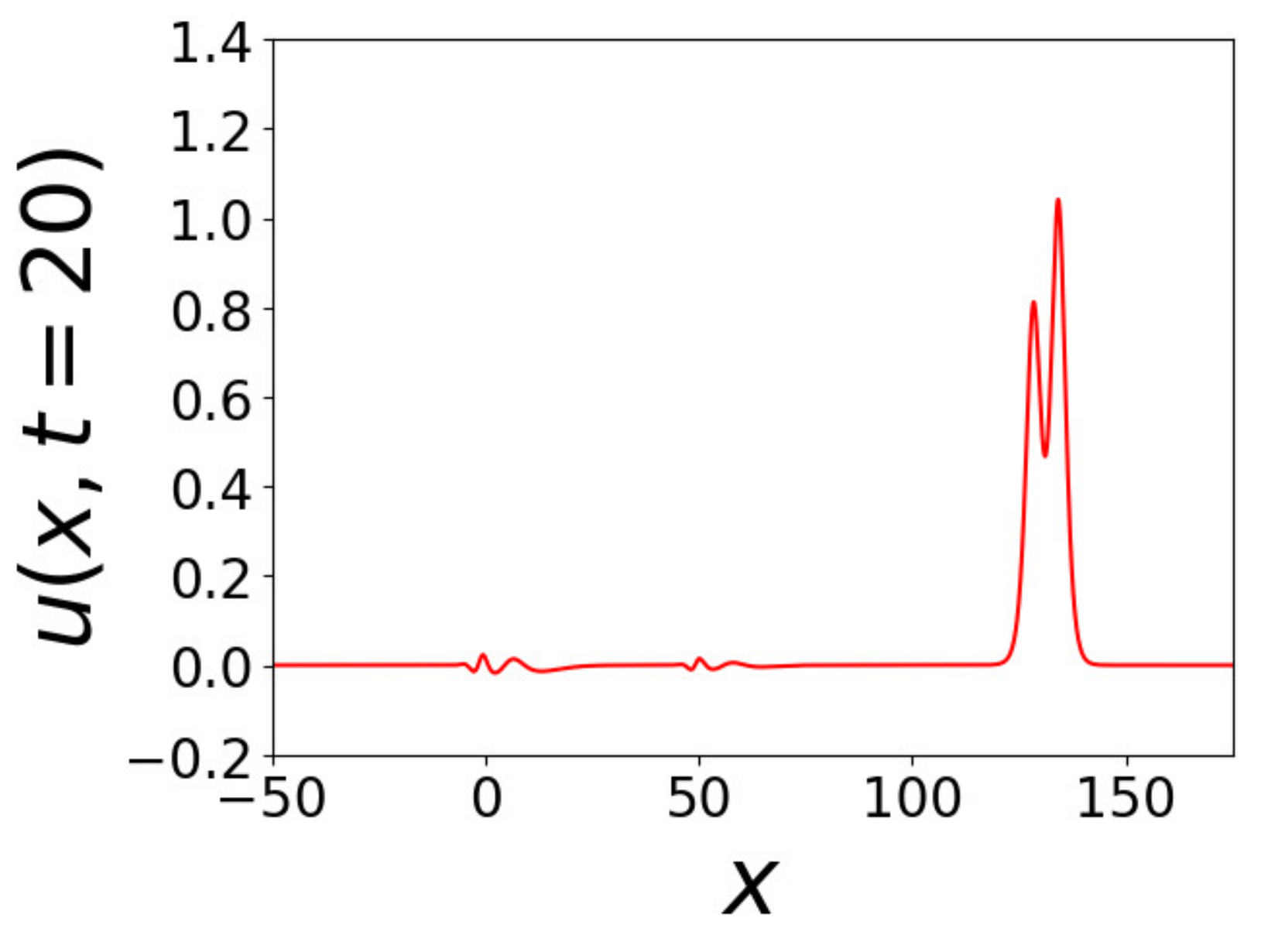}
			\caption{At~$t=20$}
			\label{plot4_6}
		\end{subfigure}%
		\begin{subfigure}{.34\textwidth}
			\centering
			\includegraphics[scale=0.34]{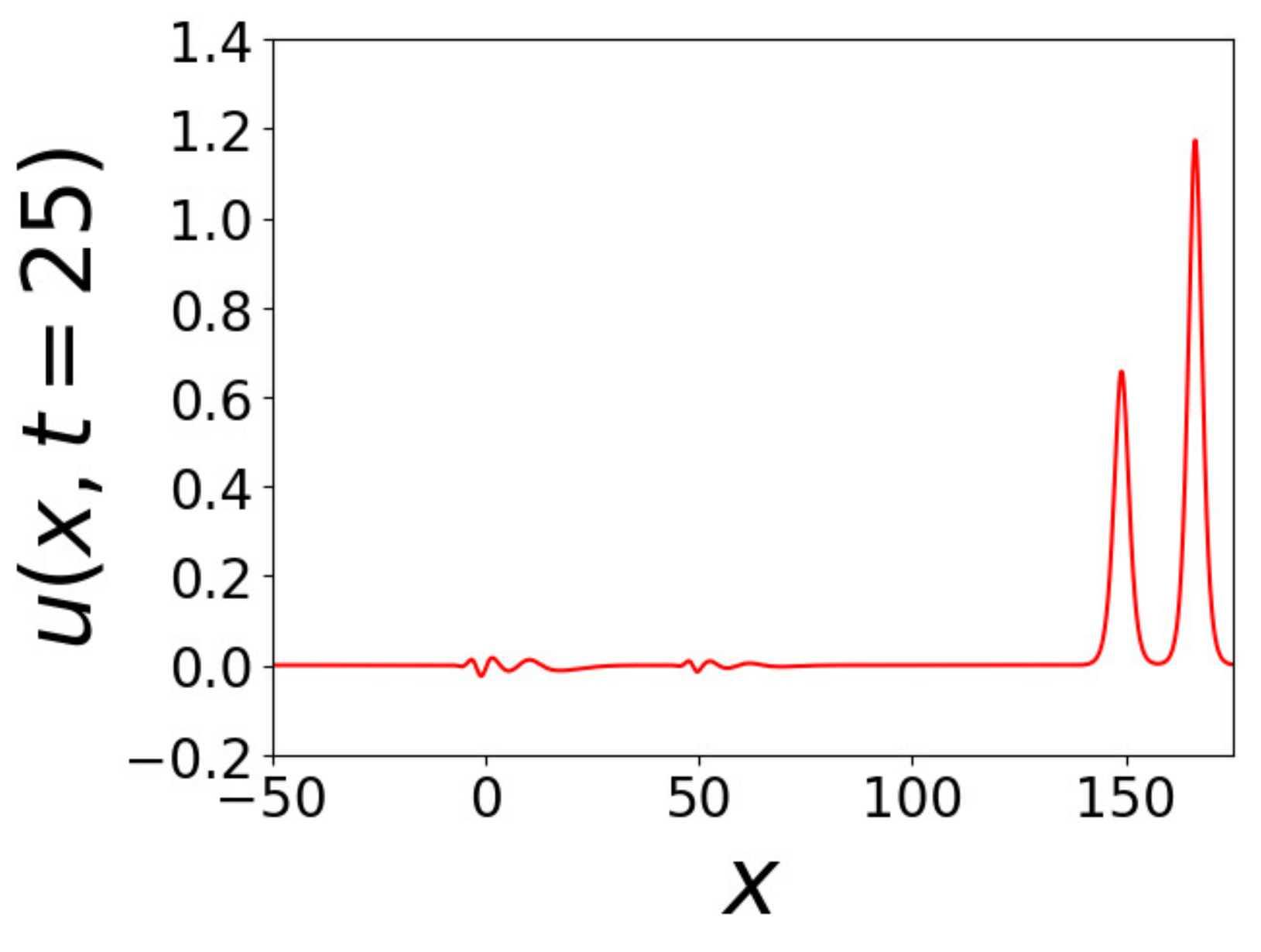}
			\caption{At~$t=25$}
			\label{plot4_7}
		\end{subfigure}
		\caption{The $u$ field configurations seen in our time-evolution of the numerical time-evolution  of the equation~(\ref{deformedqkdv}) for $\varepsilon_1 = 1$ and $\varepsilon_2 = 1.5$. at several values of time $t$} 
		\label{plot4_2to4_7}
	\end{figure}

\begin{figure}[t!]
		\centering
		\begin{subfigure}{.5\textwidth}
			\hspace*{-1.2cm}
			\centering
			\includegraphics[scale=0.34]{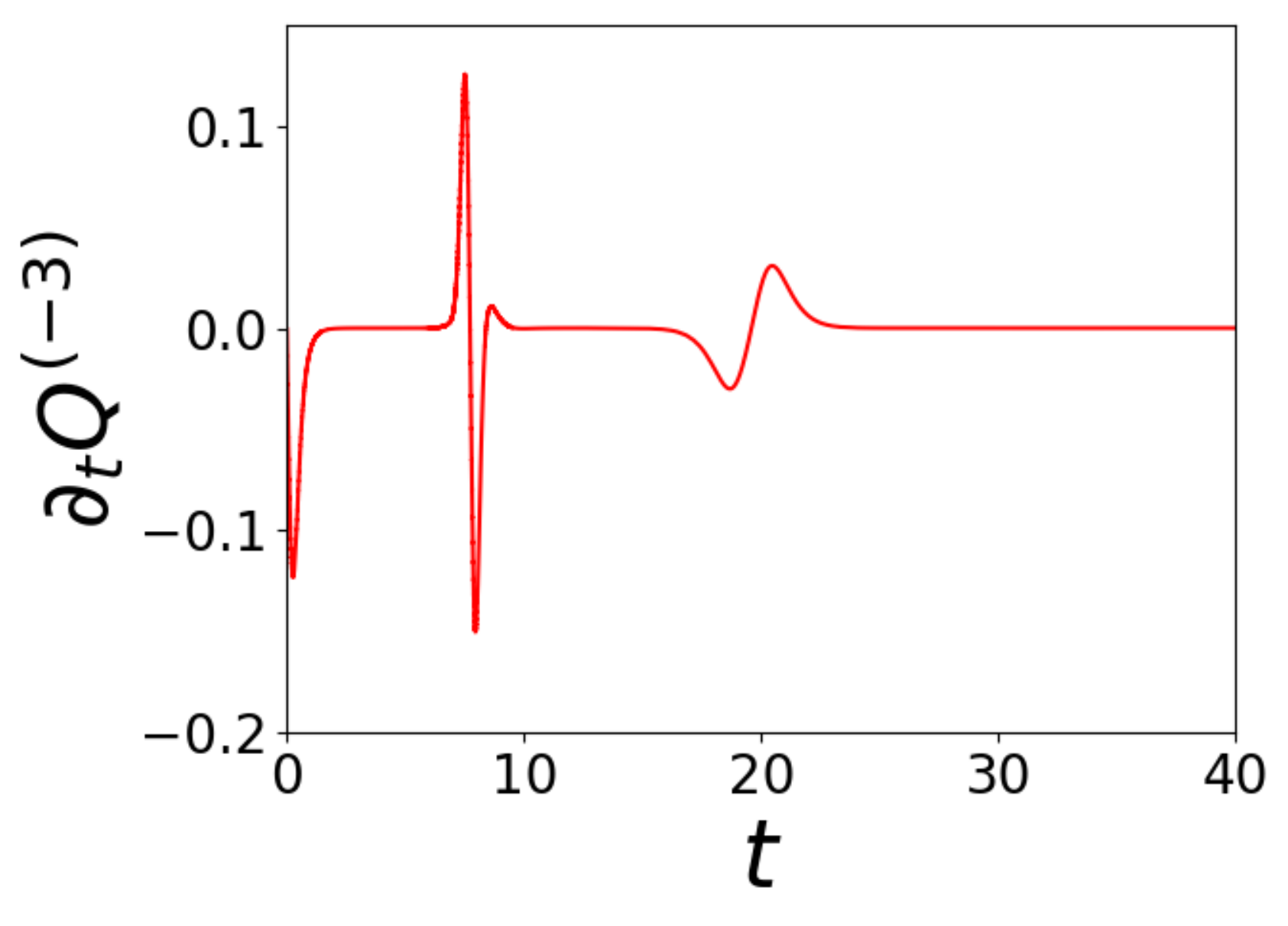}
			\caption{}
			\label{plot4_8}
		\end{subfigure}%
		\begin{subfigure}{.5\textwidth}
			\hspace*{-1.5cm}
			\centering
			\includegraphics[scale=0.34]{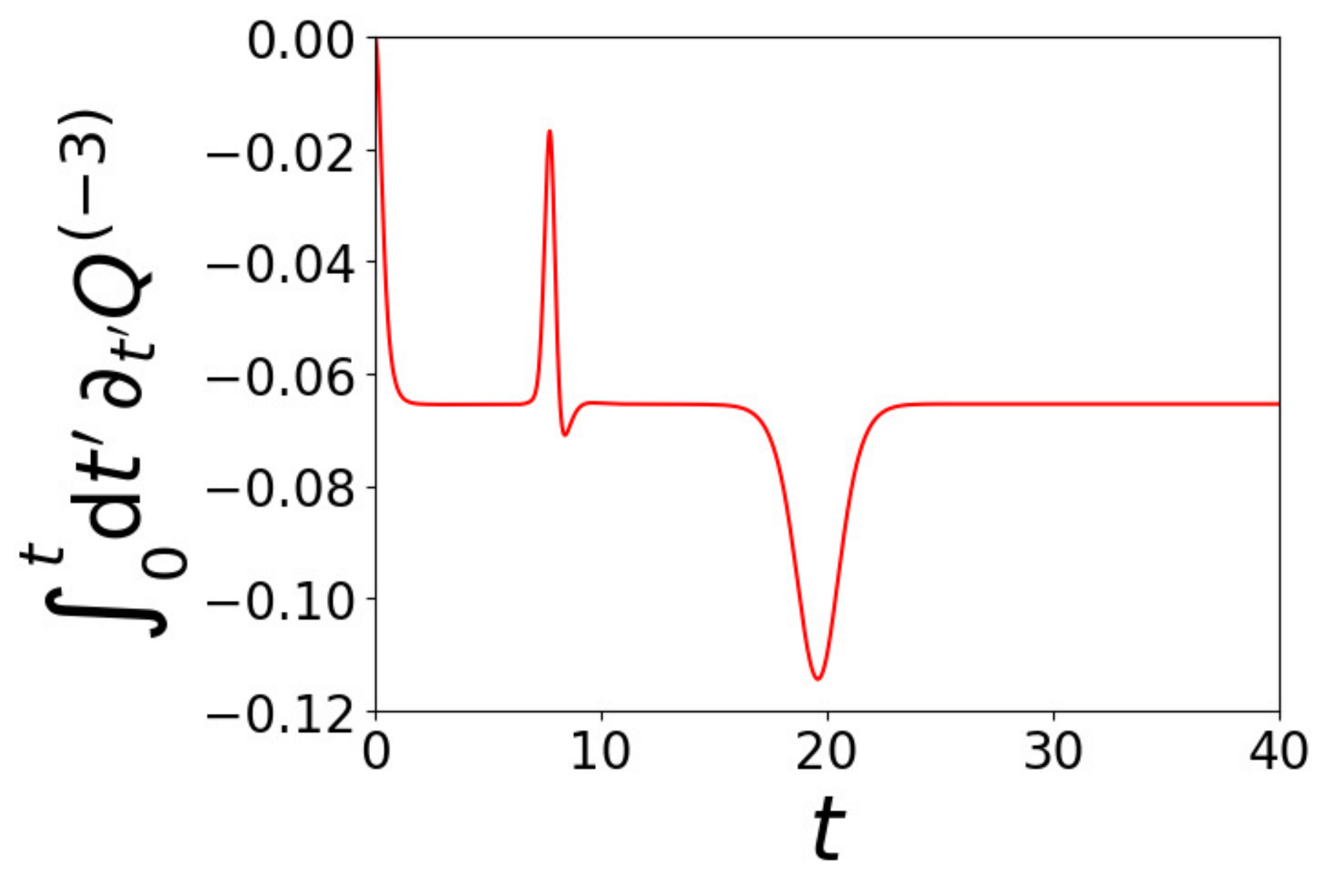}
			\caption{}
			\label{plot4_9}
		\end{subfigure}%
		\caption{The time dependence of the quantity $\partial_t Q^{(-3)}$ and $\int_{0}^{t} \mathrm{d} t^\prime \, \partial_{t^\prime} Q^{(-3)}$ obtained in two-soliton simulation of equation~(\ref{deformedqkdv}) with $\varepsilon_1 = 1$ and $\varepsilon_2 = 1.5$.}
		\label{plot4_8to4_9}
	\end{figure}

	Figure~\ref{plot4_2to4_7}
	presents various plots of the simulation with $\varepsilon_1 = 1$ and $\varepsilon_2 = 1.5$, and figure~\ref{plot4_8to4_9}
	shows the plots of the time-dependence of the  corresponding quantities 
$\partial_t Q^{(-3)}$ and $\int_{0}^{t} \mathrm{d} t^\prime \, \partial_{t^\prime} Q^{(-3)}$. Figure~\ref{plot4_2} shows the initial
 condition at~$t=0$. Figure~\ref{plot4_3} shows these two soliton fields after they have evolved for $5$ units of time. Looking at
 that figure, we note that both solitons have moved to the right and have left some radiation behind immediately after the start of the simulation, while
 their amplitudes have increased slightly. This is because the true soliton of such a system is probably significantly different the
one described by the initial conditions used in this simulation. This implies that the field $u$ changes from the form given by the initial conditions and settles to a true solution, and in doing so the two solitons leave behind some radiation. Then figure~\ref{plot4_4} 
shows the largest soliton field after it has interacted with the radiation emitted by the smallest soliton field.
 Figure~\ref{plot4_5} shows the solution at $t=15$ when the largest soliton starts to interact with the smallest soliton.
 Figure~\ref{plot4_6} shows the actual soliton fields interacting with each other. Finally, figure~\ref{plot4_7} shows the field when the solitons have moved 
away from each other again. These changes are all seen in figure~\ref{plot4_8}; from the beginning (at $t=0$ to $t \simeq 2$) the initial
 conditions settle down to the true solution of the equation leaving behind some radiation, from about $t \simeq 6$ to $t \simeq 10$ 
the largest soliton interacts with the radiation left behind by the smallest soliton. Finally, from $t \simeq 16$ to $t \simeq 23$ the two
 soliton fields interact, after which the solitons keep moving unhindered. Notice also that figure~\ref{plot4_9} shows that, after the the
 initial conditions settle down to the true solution, all the lumps cancel each other out.
	
\begin{figure}[t!]
		\centering
		\hspace*{0.5cm}
		\begin{subfigure}{.34\textwidth}
			\hspace*{-1.2cm}
			\centering
			\includegraphics[scale=0.34]{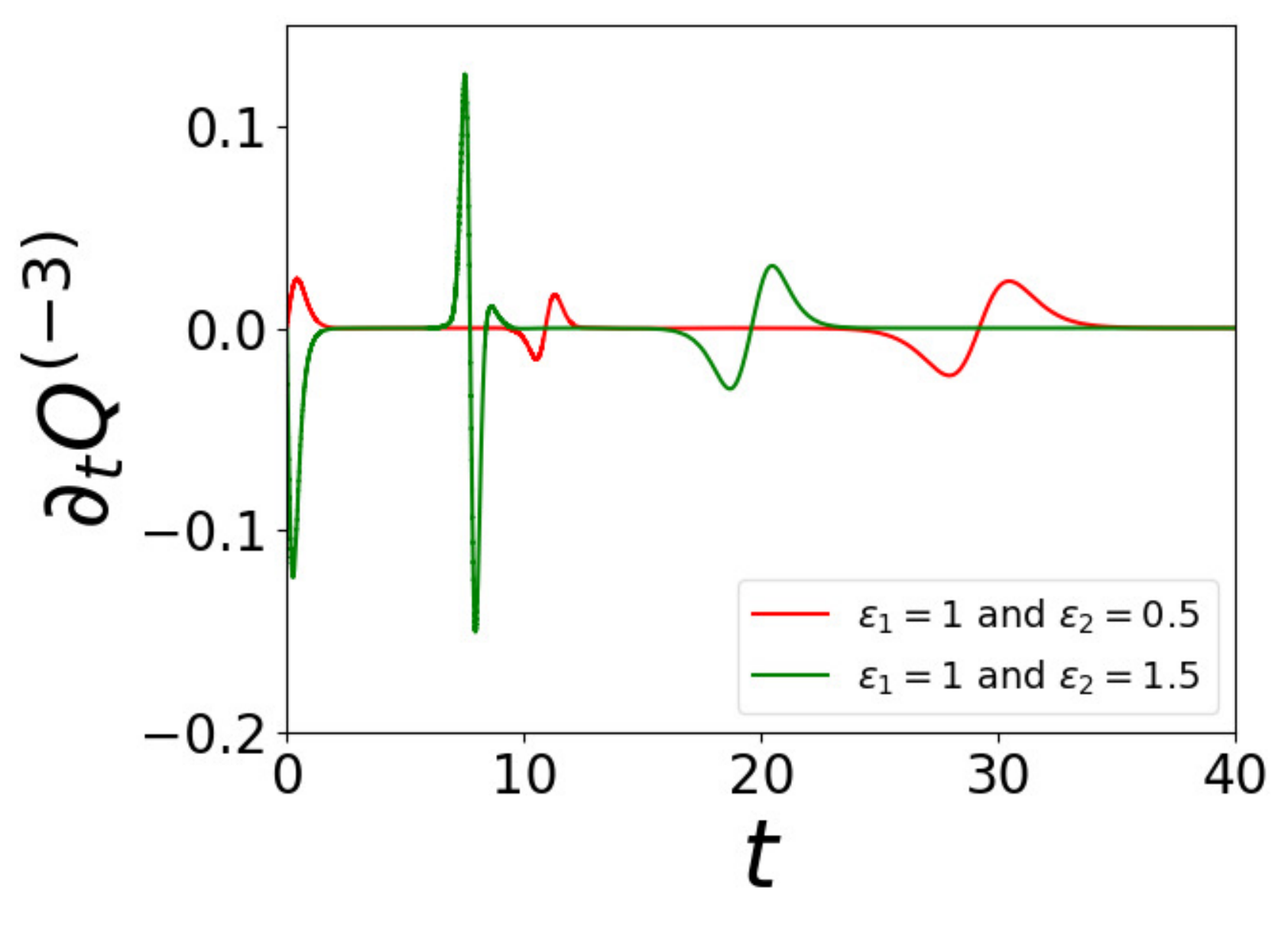}
			\caption{}
			\label{plot5_8}
		\end{subfigure}%
		\begin{subfigure}{.34\textwidth}L
			\hspace*{-1.2cm}
			\centering
			\includegraphics[scale=0.34]{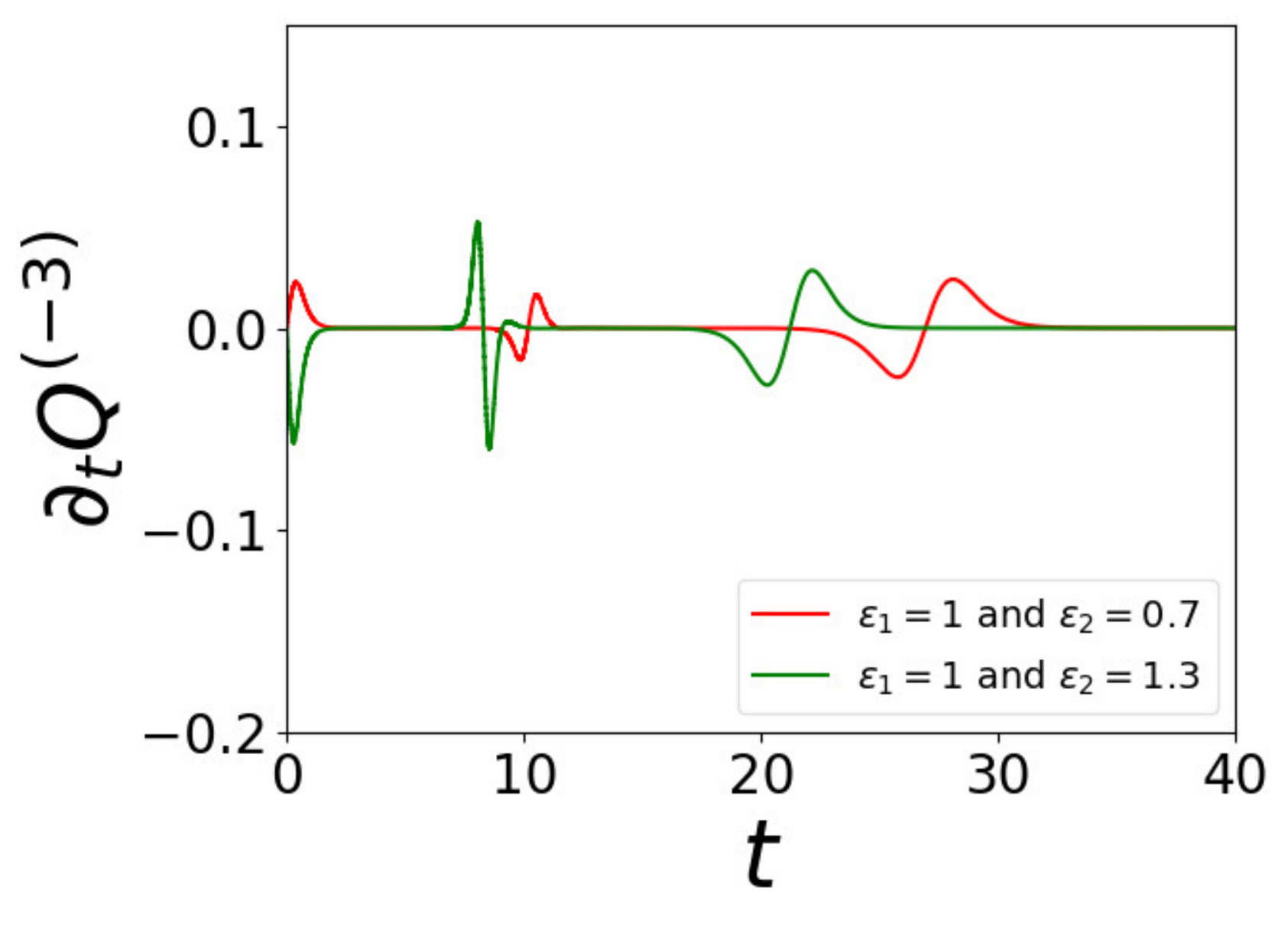}
			\caption{}
			\label{plot5_9}
		\end{subfigure}%
		\begin{subfigure}{.34\textwidth}
			\hspace*{-1.2cm}
			\centering
			\includegraphics[scale=0.34]{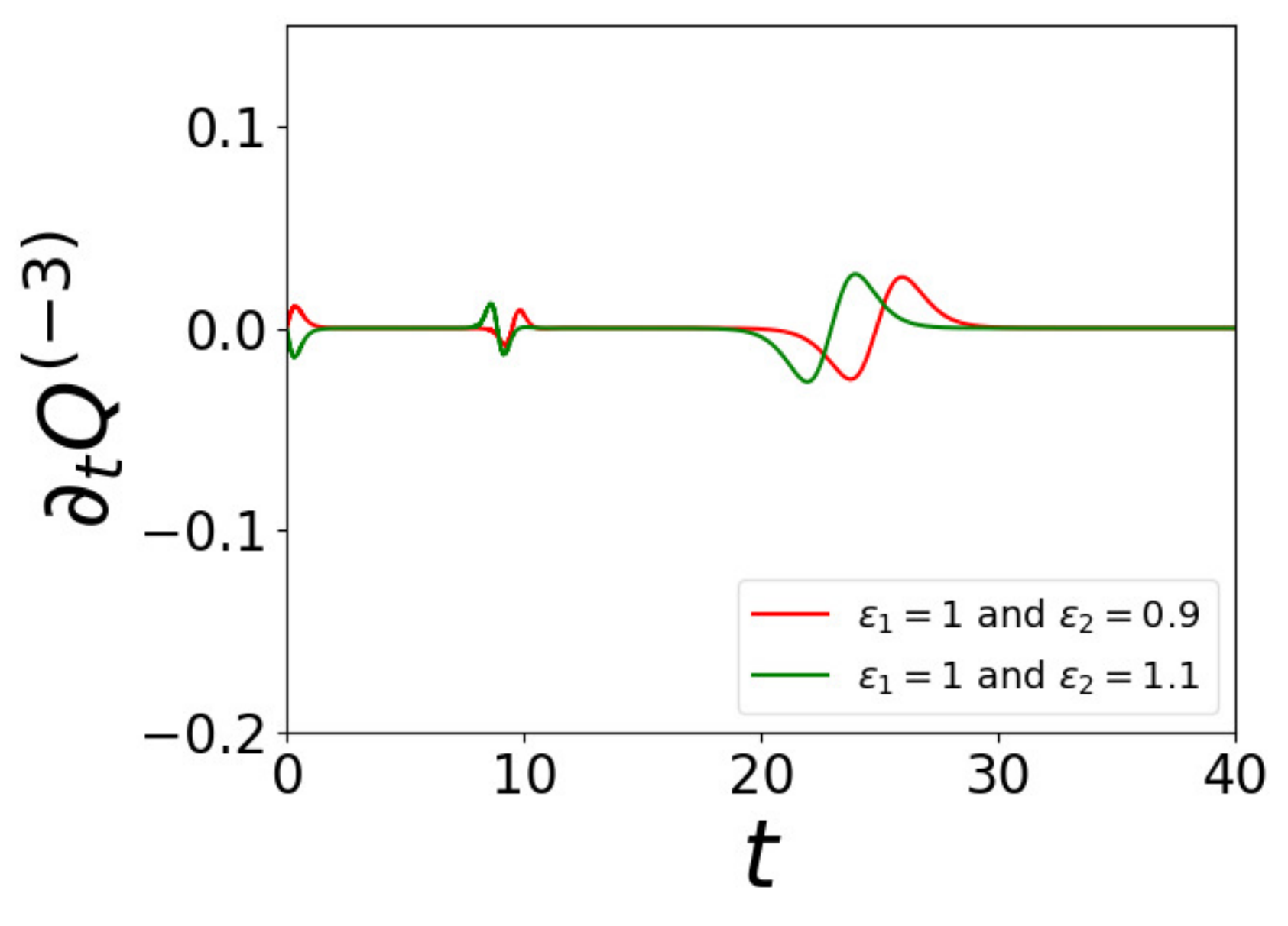}
			\caption{}
			\label{plot5_10}
		\end{subfigure}%
		\caption{The time-dependence of the obtained values of  $\partial_t Q^{(-3)}$ seen in the simulation of equation~(\ref{deformedqkdv})
 for $\varepsilon_1 = 1$ and $\varepsilon_2 \neq 1$, with the analytical two-soliton solutions of the mRLW equation used as initial conditions.}
		\label{plot5_8to5_10}
	\end{figure}

	For other values of $\varepsilon_1 = 1$ and $\varepsilon_2 \neq 1$ we have observed a very similar behaviour to the one described above. 
To illustrate this better, figure~\ref{plot5_8to5_10}
	presents three plots comparing  $\partial_t Q^{(-3)}$ for $\varepsilon_2 = 0.5$ and $\varepsilon_2 = 1.5$ (see figure~\ref{plot5_8}), 
$\varepsilon_2 = 0.7$ and $\varepsilon_2 = 1.3$ (see figure~\ref{plot5_9}), and $\varepsilon_2 = 0.9$ and $\varepsilon_2 = 1.1$ (see figure~\ref{plot5_10}). 
We see that the larger $|1-\varepsilon_2|$ becomes, the more intense are the changes of $Q^{(-3)}$ when the solitons readjust their form from the one described by
the initial conditions and when the larger soliton interacts with the radiation left behind by the smaller soliton. 
This is expected since the larger the value of $|1-\varepsilon_2|$ is, the larger is the difference between
 the initial conditions and the true solutions of the  equation, which results in a faster rate of change of the charge $Q^{(-3)}$.
 Furthermore, this implies that the radiation left behind by the soliton is larger, and so the interaction between the larger soliton and the 
radiation due to the smaller soliton is  more intense. However, when the two actual soliton fields collide, we see that as $|1-\varepsilon_2|$ becomes smaller, 
the peak of $\partial_t Q^{(-3)}$ converges closer to the peak shown in figure~\ref{I_1_alpha_8_e1_1_e2_1_e3_1_1_anal_and_num}, both 
in terms of amplitude as well as in terms of its location on the horizontal axis. The horizontal movement is due to the fact that when the solitons emit radiation,
 their amplitudes increase/decrease (depending on whether $\varepsilon_2<1$ or $\varepsilon_2>1$) which causes their velocities to increase/decrease as well. 
	
\begin{figure}[t!]
		\centering
		\hspace*{-0.1cm}
		\begin{subfigure}{.34\textwidth}
			\centering
			\includegraphics[scale=0.34]{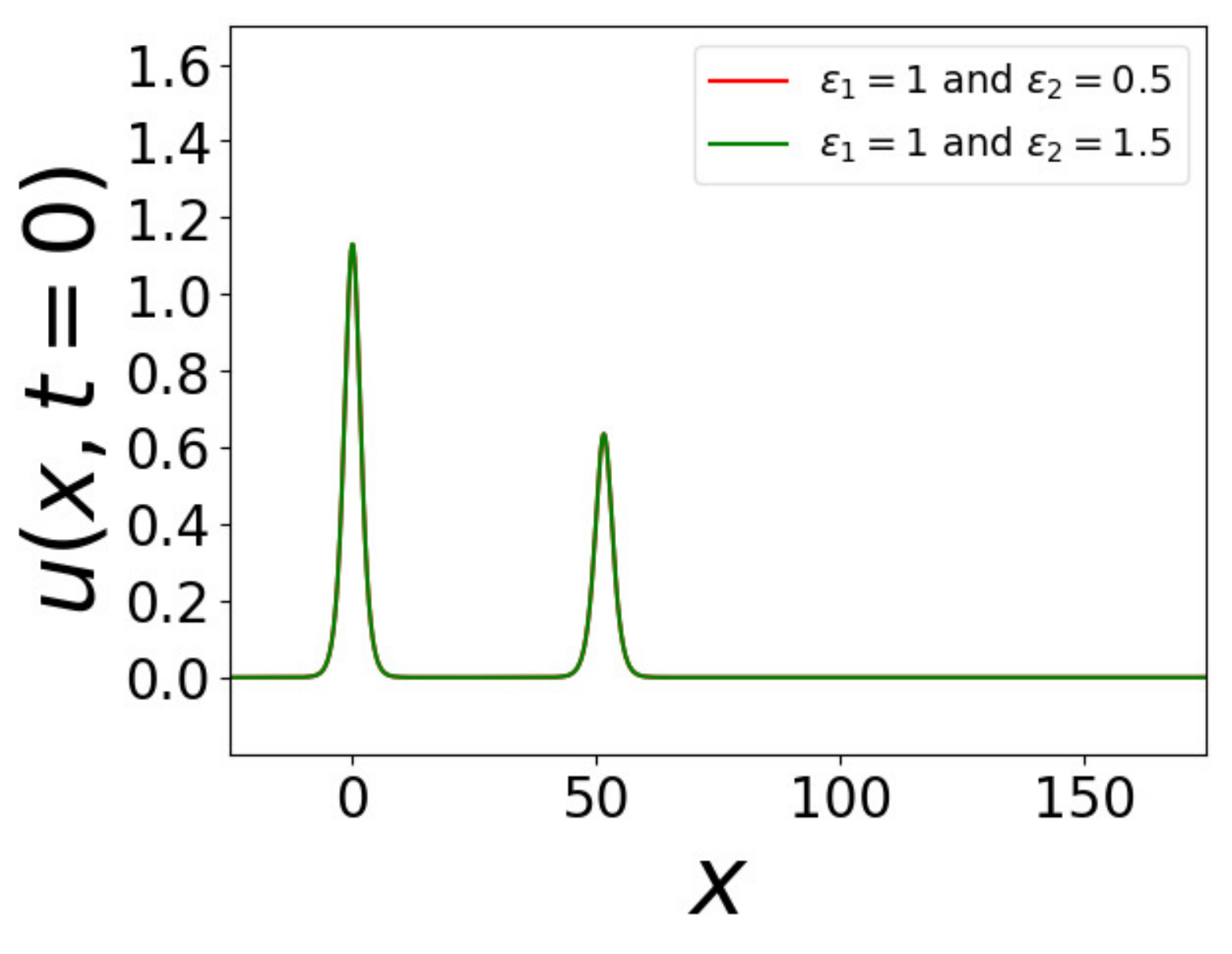}
			\caption{At~$t=0$}
			\label{plot5_2}
		\end{subfigure}%
		\begin{subfigure}{.34\textwidth}
			\centering
			\includegraphics[scale=0.34]{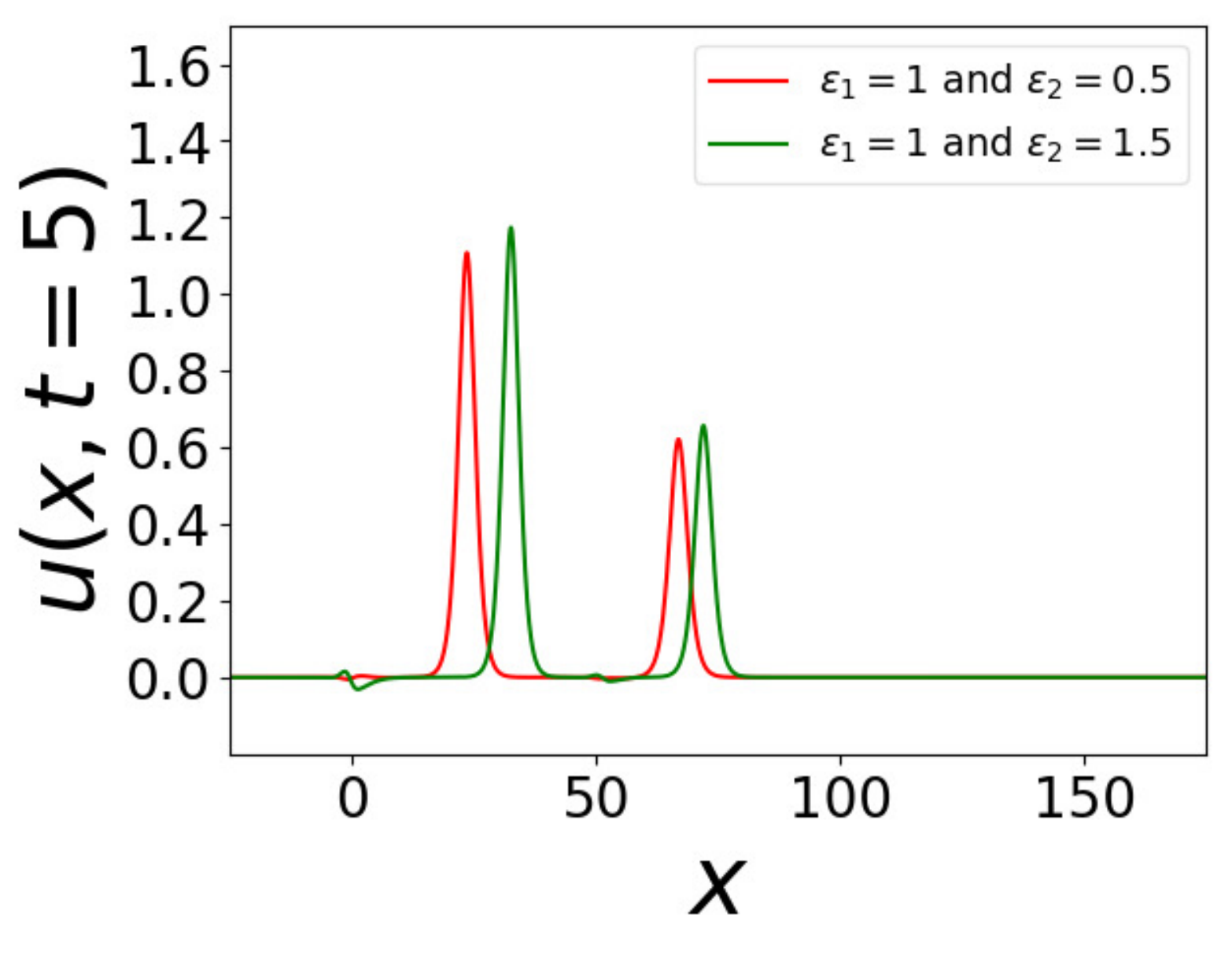}
			\caption{At~$t=5$}
			\label{plot5_3}
		\end{subfigure}%
		\begin{subfigure}{.34\textwidth}
			\centering
			\includegraphics[scale=0.34]{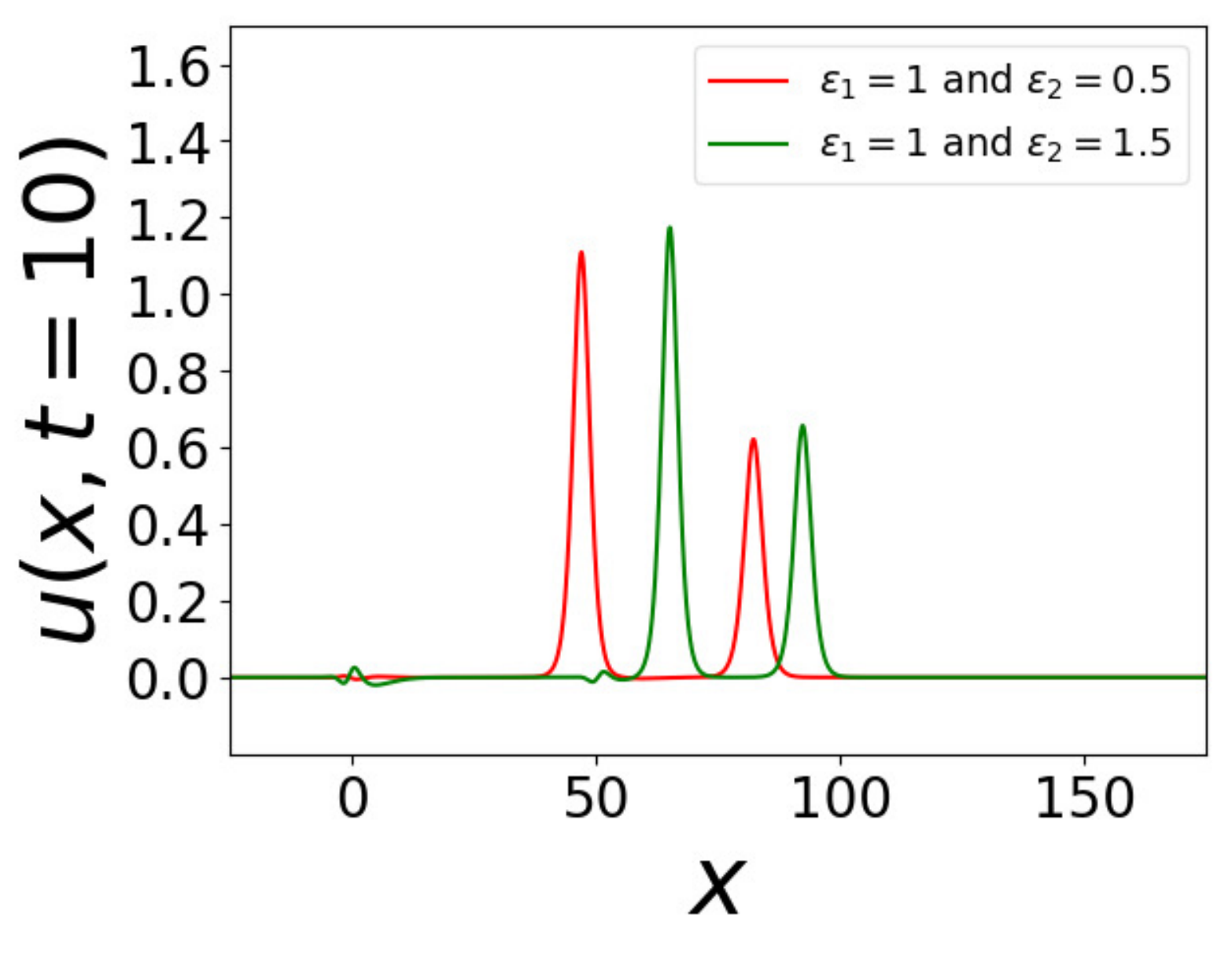}
			\caption{At~$t=10$}
			\label{plot5_4}
		\end{subfigure}
		
		\begin{subfigure}{.34\textwidth}
			\centering
			\includegraphics[scale=0.34]{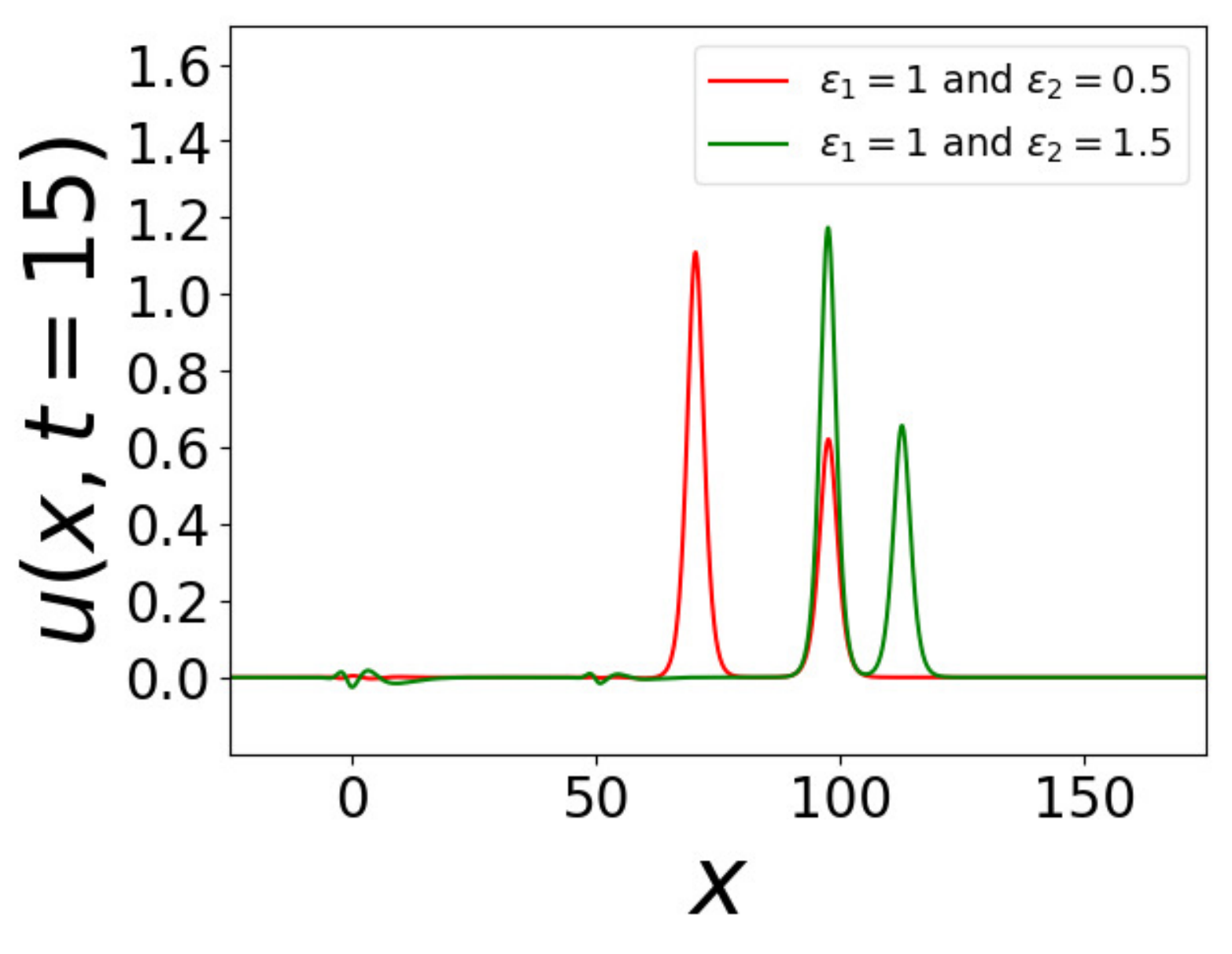}
			\caption{At~$t=15$}
			\label{plot5_5}
		\end{subfigure}%
		\begin{subfigure}{.34\textwidth}
			\centering
			\includegraphics[scale=0.34]{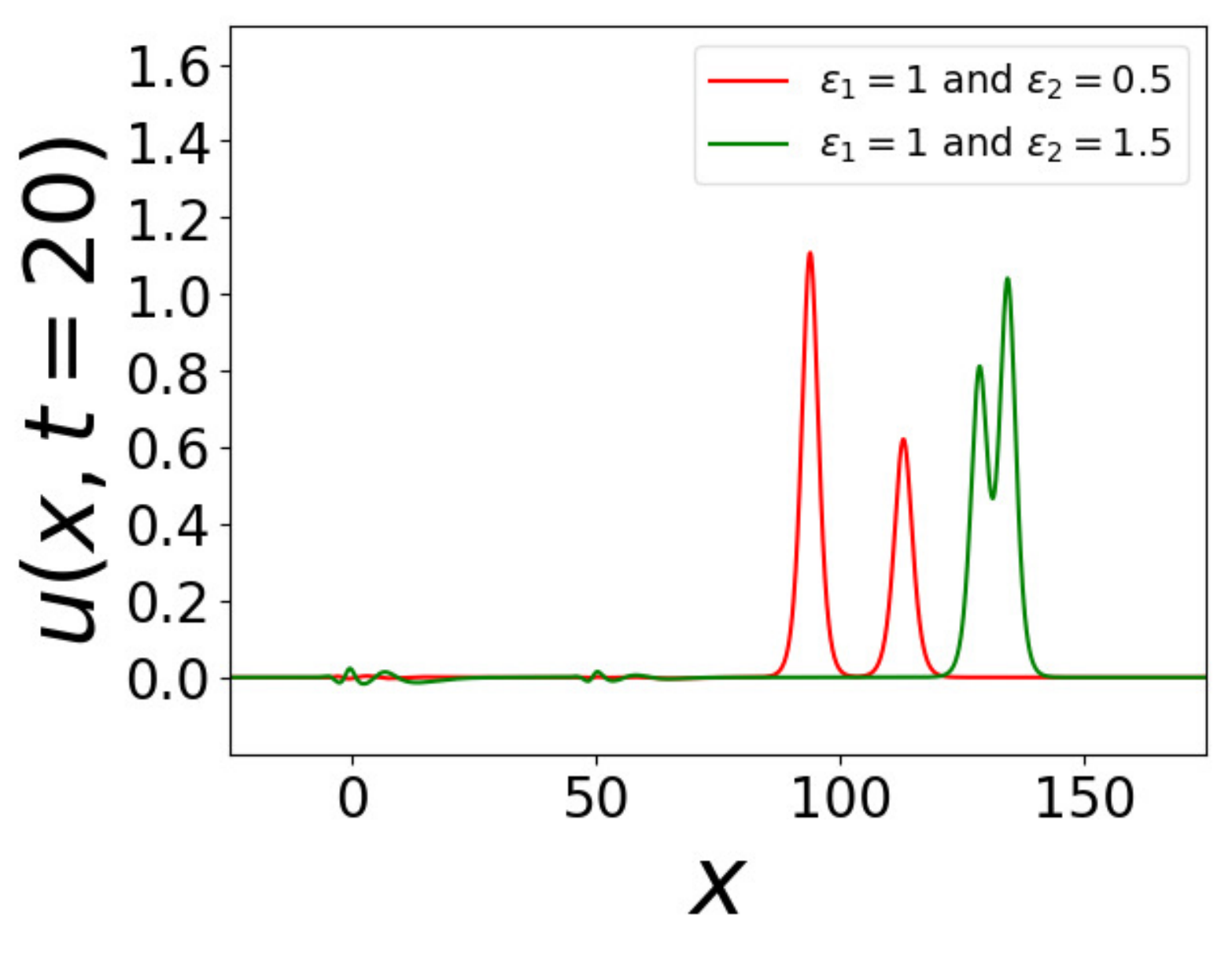}
			\caption{At~$t=20$}
			\label{plot5_6}
		\end{subfigure}%
		\begin{subfigure}{.34\textwidth}
			\centering
			\includegraphics[scale=0.34]{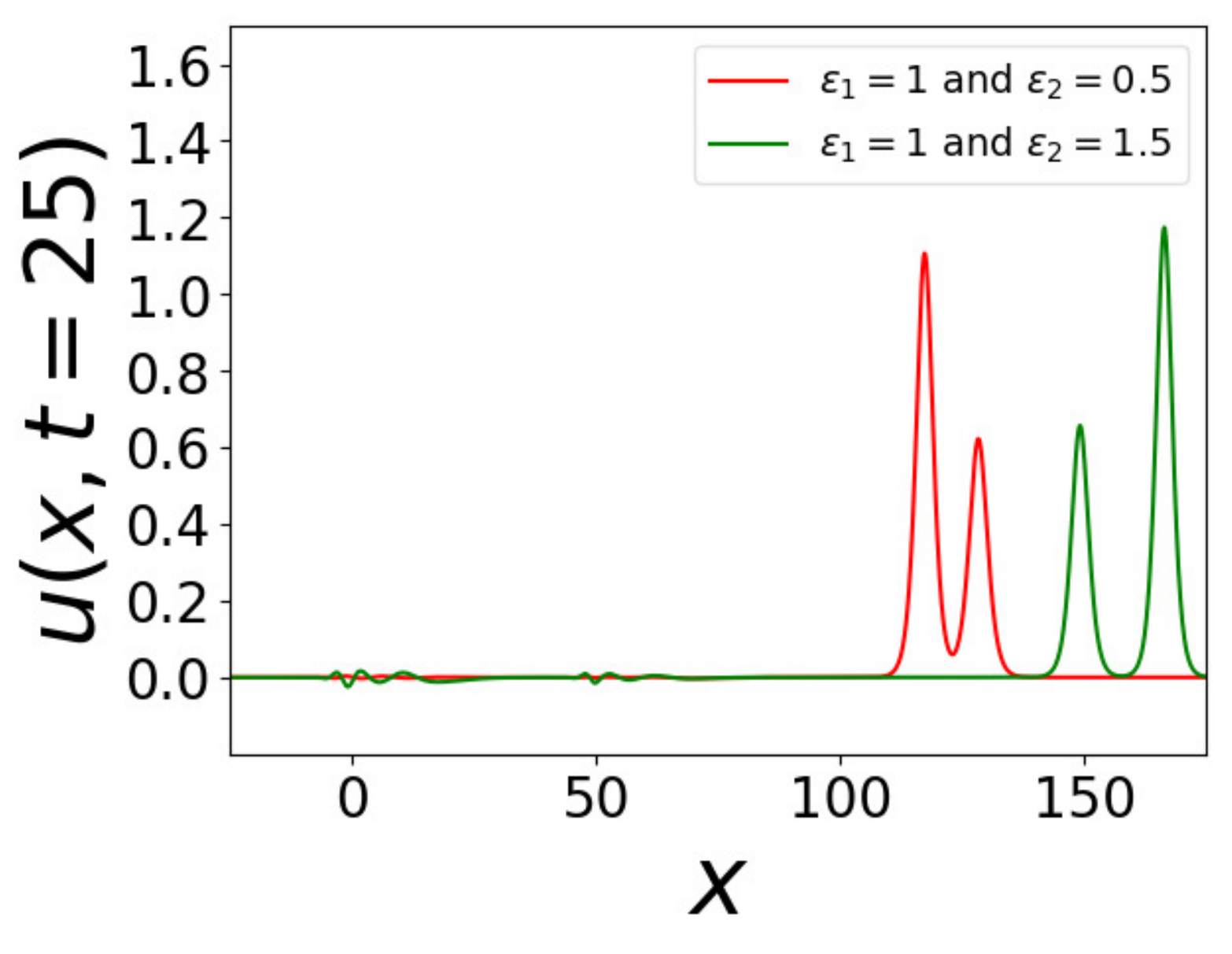}
			\caption{At~$t=25$}
			\label{plot5_7}
		\end{subfigure}
		\caption{The time-evolution of two two-soliton systems studied in the equation~(\ref{deformedqkdv}) with $\varepsilon_1 = 1$ 
and $\varepsilon_2 = 0.5$ (red curve), and $\varepsilon_1 = 1$ and $\varepsilon_2 = 1.5$ (green curve).} 
		\label{plot5_2to5_7}
	\end{figure}

\begin{figure}[b!]
		\centering
		\hspace*{0.5cm}
		\begin{subfigure}{.34\textwidth}
			\hspace*{-1.4cm}
			\centering
			\includegraphics[scale=0.32]{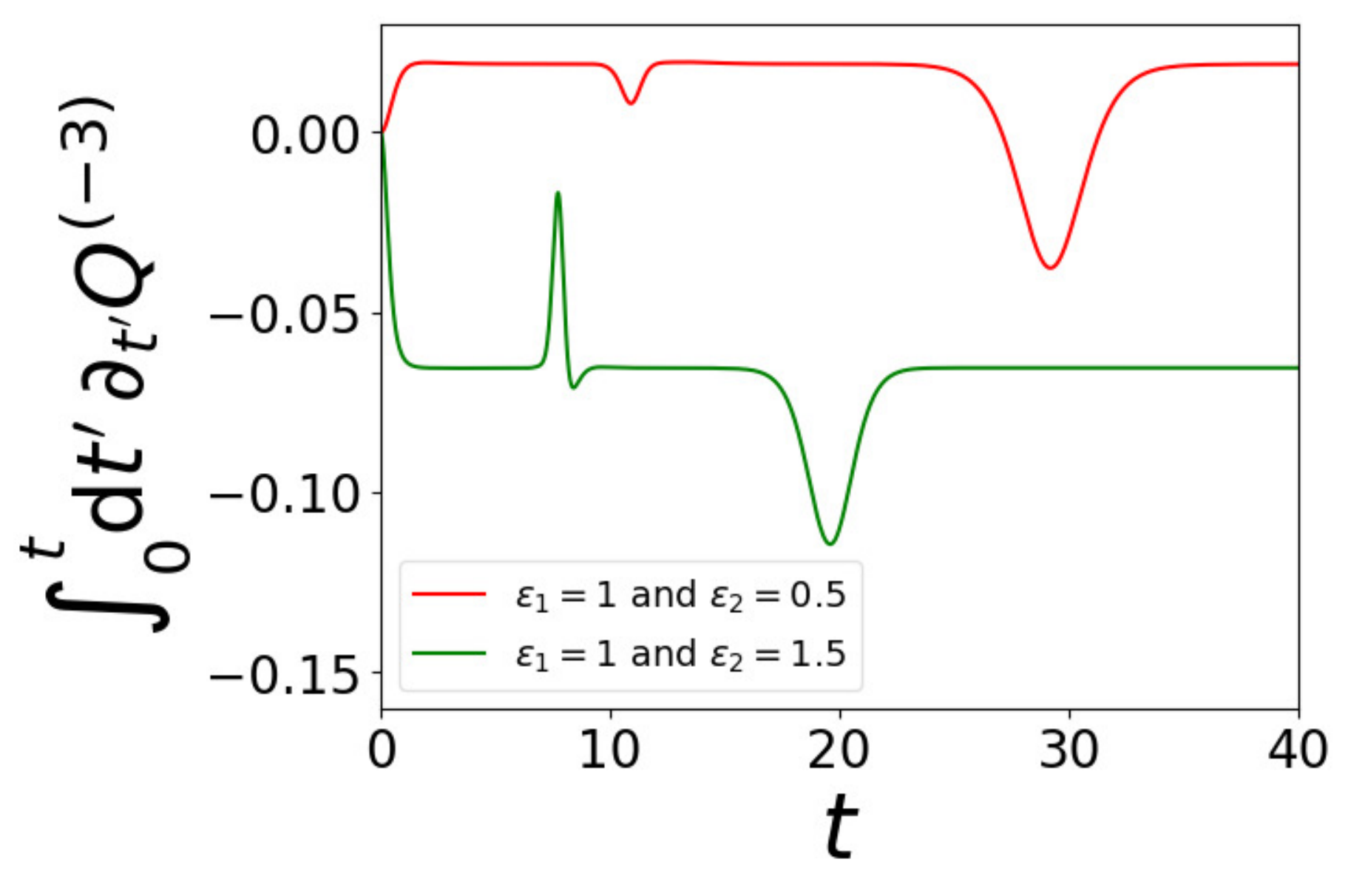}
			\caption{}
			\label{plot5_11}
		\end{subfigure}%
		\begin{subfigure}{.34\textwidth}
			\hspace*{-1.4cm}
			\centering
			\includegraphics[scale=0.32]{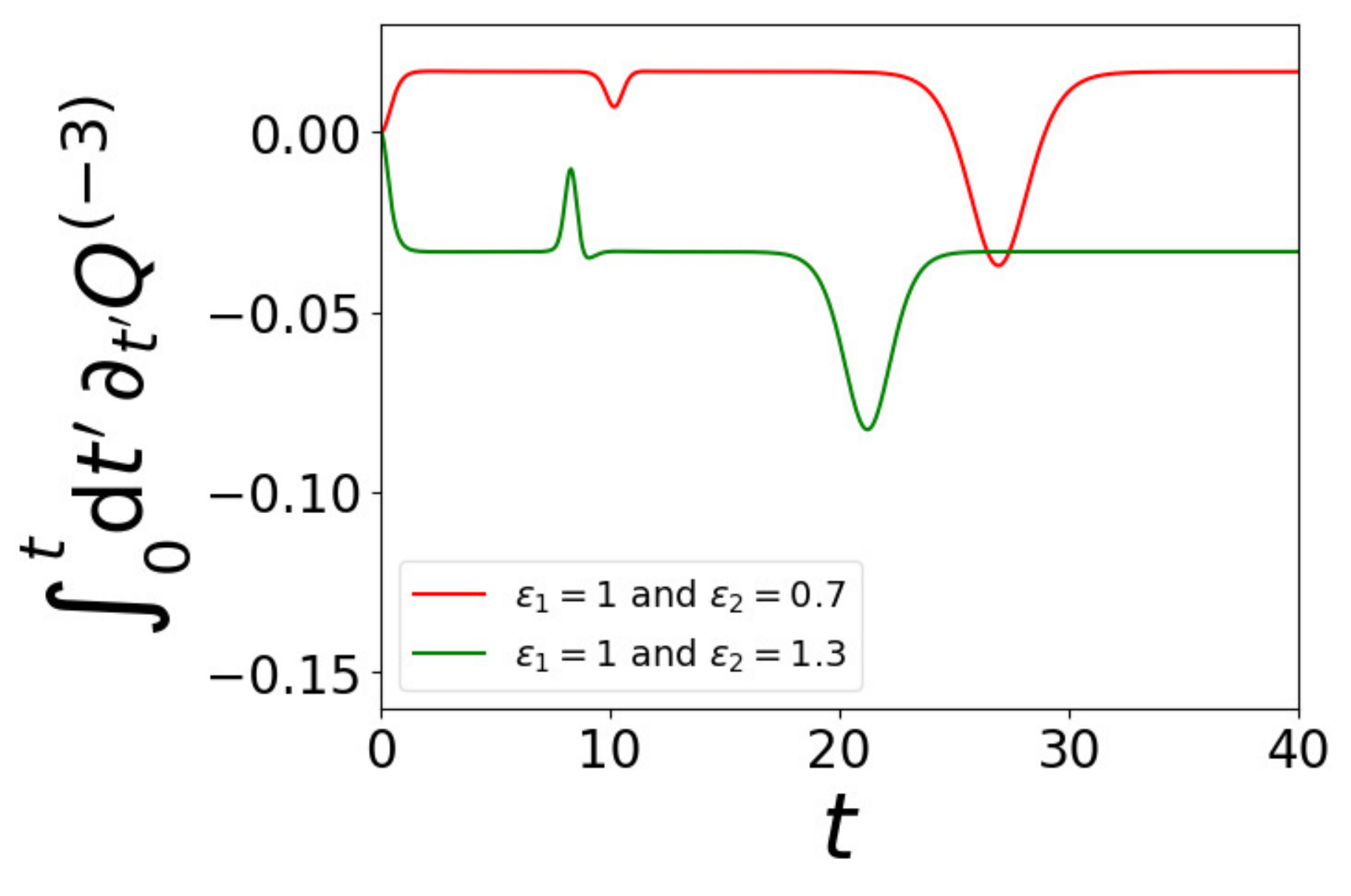}
			\caption{}
			\label{plot5_12}
		\end{subfigure}%
		\begin{subfigure}{.34\textwidth}
			\hspace*{-1.4cm}
			\centering
			\includegraphics[scale=0.32]{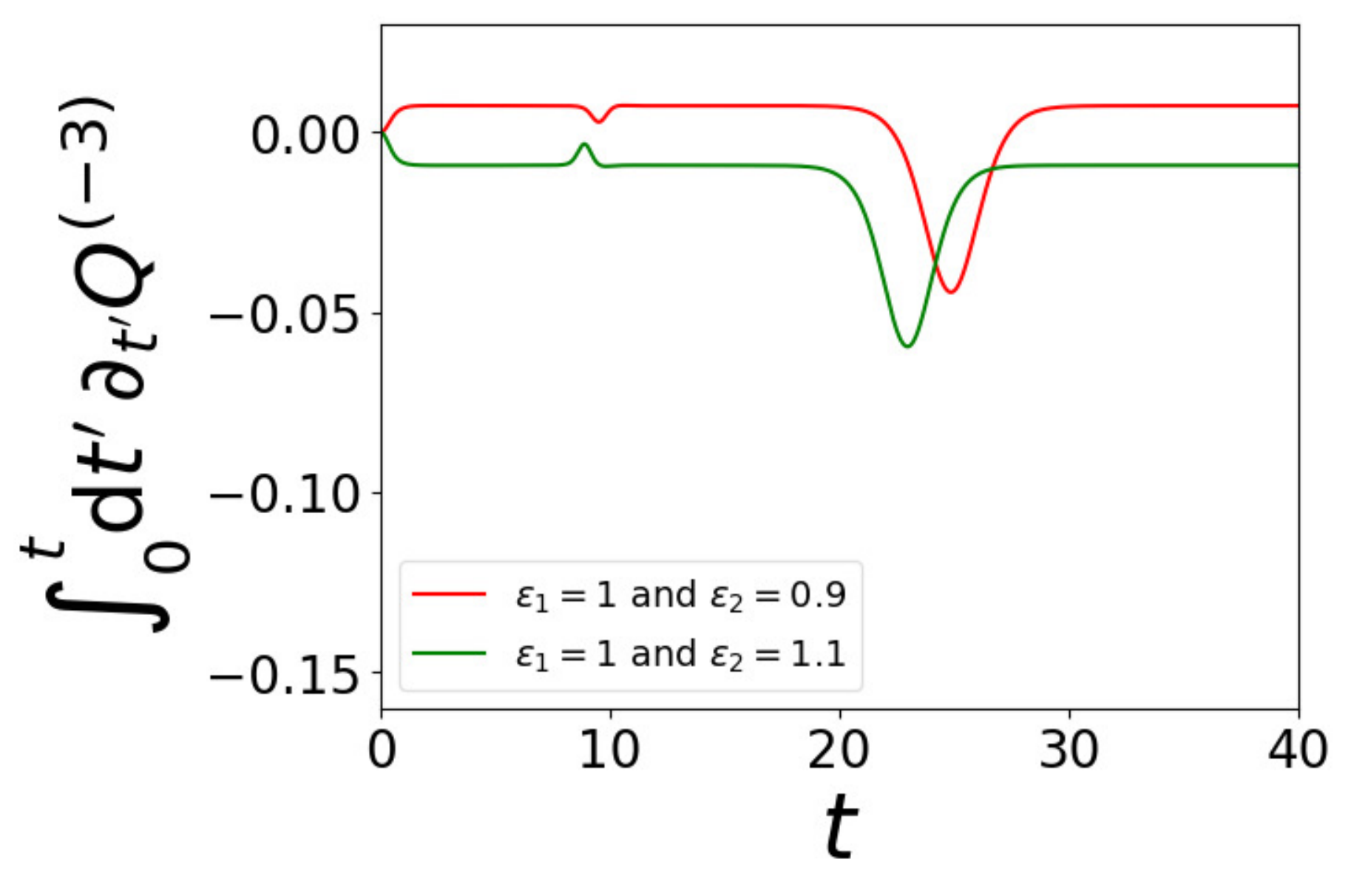}
			\caption{}
			\label{plot5_13}
		\end{subfigure}%
		\caption{The time-dependence of  $\int_{0}^{t} \mathrm{d} t^\prime \, \partial_{t^\prime} Q^{(-3)}$ seen in the simulation of
 equation~(\ref{deformedqkdv}) with different values of $\varepsilon_1 = 1$ and $\varepsilon_2 \neq 1$.}
		\label{plot5_11to5_13}
	\end{figure}

	To illustrate better these observed properties, figure~\ref{plot5_2to5_7} 
	presents again the various plots shown in figure~\ref{plot4_2to4_7} corresponding to $\varepsilon_1 = 1$ and $\varepsilon_2 = 1.5$
 but it  also shows the results of the simulation corresponding to $\varepsilon_1 = 1$ and $\varepsilon_2 = 0.5$, in order to compare the two simulations. This figure also
 illustrates the fact that if $\varepsilon_2 > 1$, then the solitons' amplitudes and velocities increase as a result of the emission of this radiation. On the other hand, 
if $\varepsilon_2 < 1$, the solitons' amplitudes and velocities decrease. Figure~\ref{plot5_11to5_13} 
	presents the corresponding values of $\int_{0}^{t} \mathrm{d} t^\prime \, \partial_{t^\prime} Q^{(-3)}$.  Overall, we see that after the fields have settled from their form given by the initial conditions to the values corresponding to a solution, 
all the lumps in figure~\ref{plot5_8to5_10} cancel each other out.

	\subsubsection{Two-soliton configurations for $\varepsilon_1 \neq 1$ and $\varepsilon_2 = 1$}
	Armed with the observation of the last subsection  we present here results of some simulations for $\varepsilon_1 \neq 1$ and $\varepsilon_2 = 1$ where, just as in the previous subsubsection, the initial conditions were constructed from the analytical two-soliton solution of the mRLW equation.

\begin{figure}[t!]
		\centering
		\hspace*{-0.1cm}
		\begin{subfigure}{.34\textwidth}
			\centering
			\includegraphics[scale=0.34]{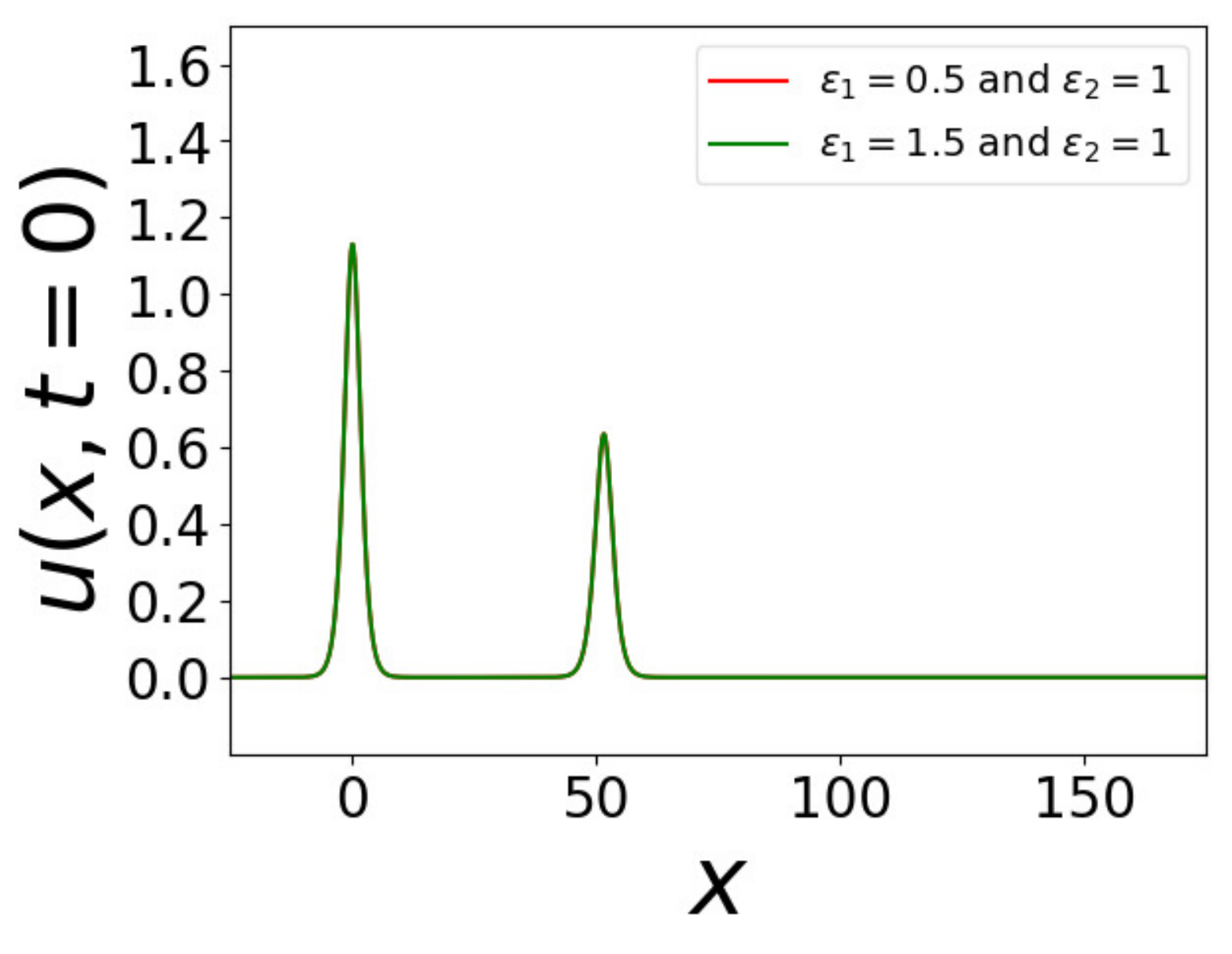}
			\caption{At~$t=0$}
			\label{plot7_2}
		\end{subfigure}%
		\begin{subfigure}{.34\textwidth}
			\centering
			\includegraphics[scale=0.34]{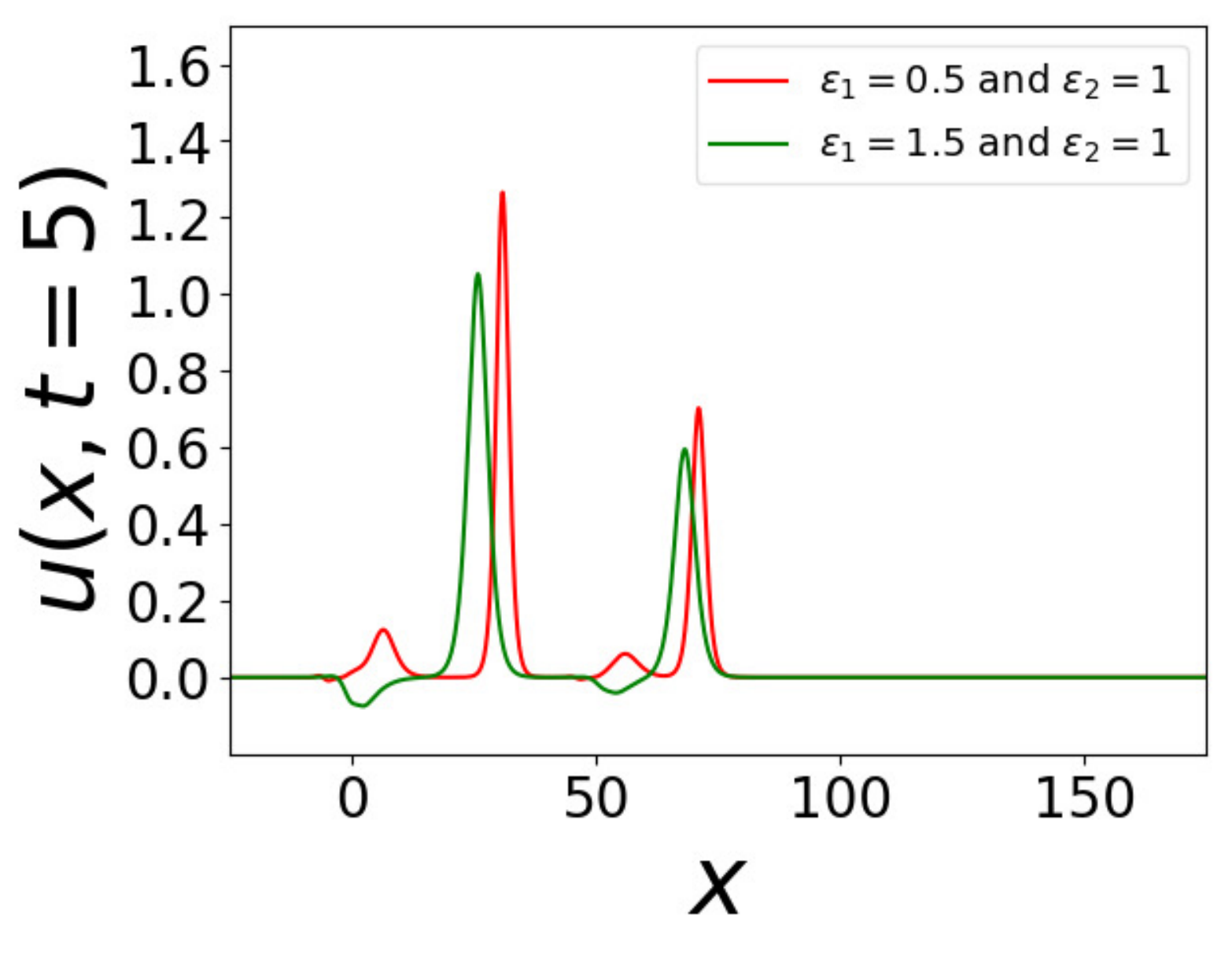}
			\caption{At~$t=5$}
			\label{plot7_3}
		\end{subfigure}%
		\begin{subfigure}{.34\textwidth}
			\centering
			\includegraphics[scale=0.34]{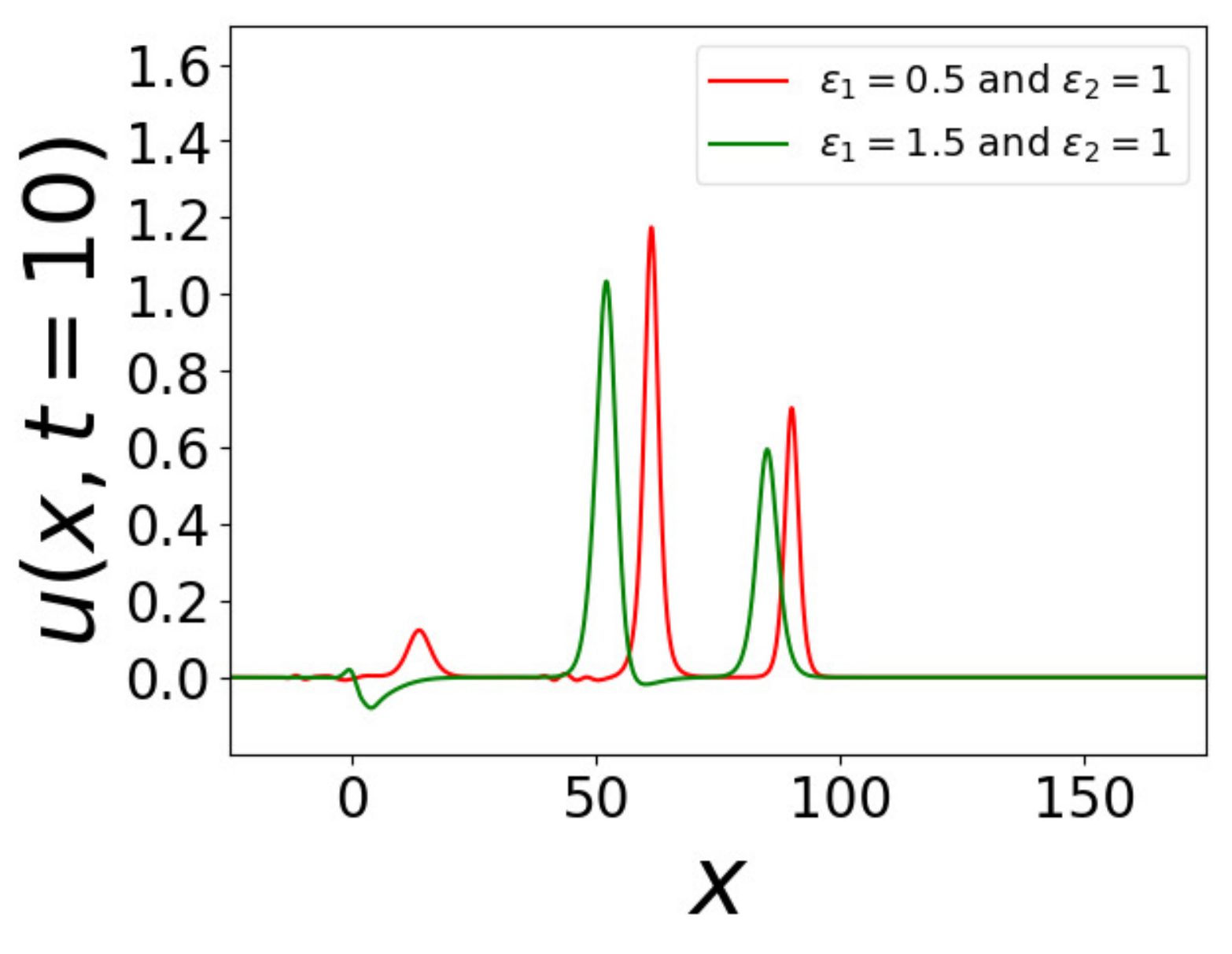}
			\caption{At~$t=10$}
			\label{plot7_4}
		\end{subfigure}
		
		\begin{subfigure}{.34\textwidth}
			\centering
			\includegraphics[scale=0.34]{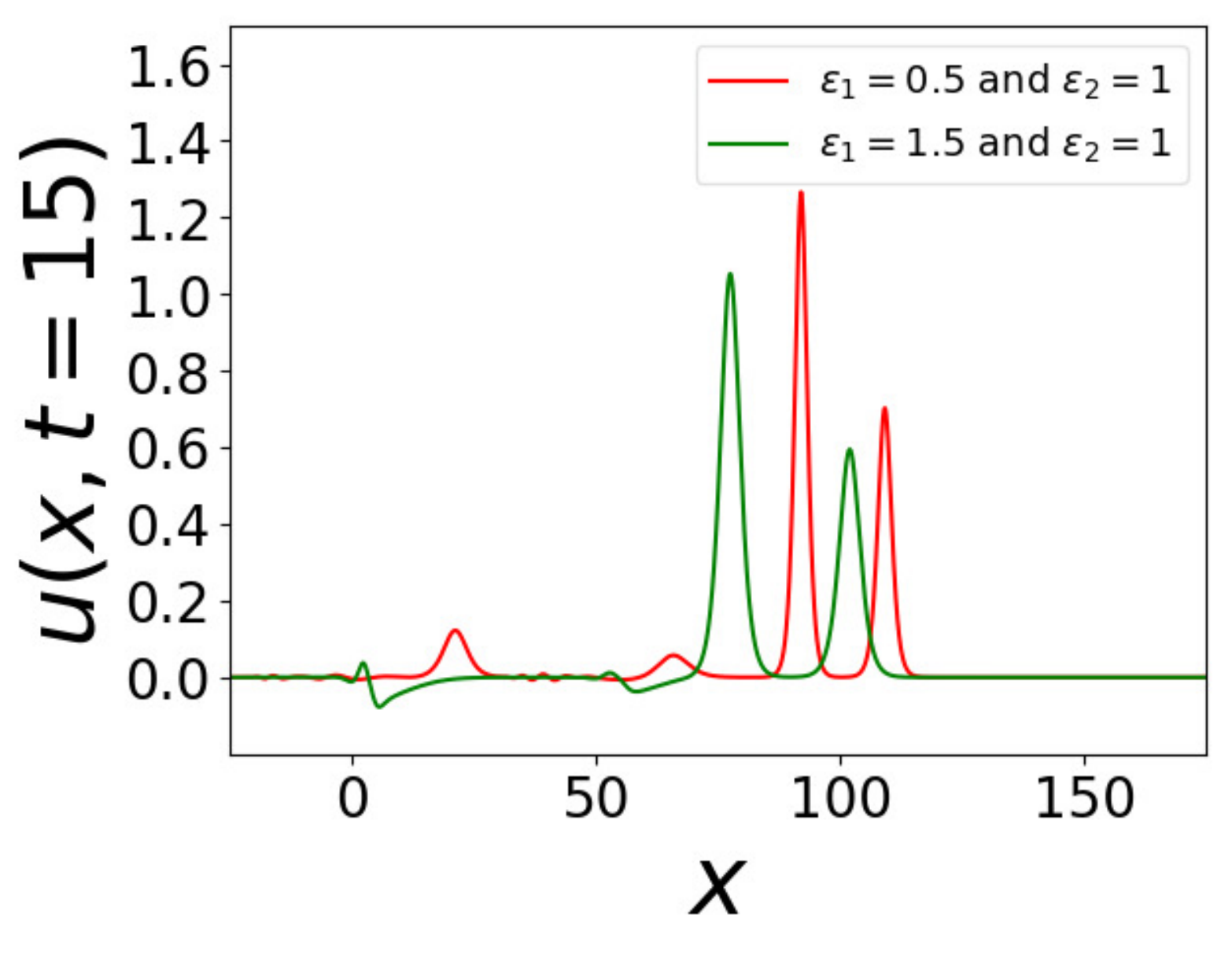}
			\caption{At~$t=15$}
			\label{plot7_5}
		\end{subfigure}%
		\begin{subfigure}{.34\textwidth}
			\centering
			\includegraphics[scale=0.34]{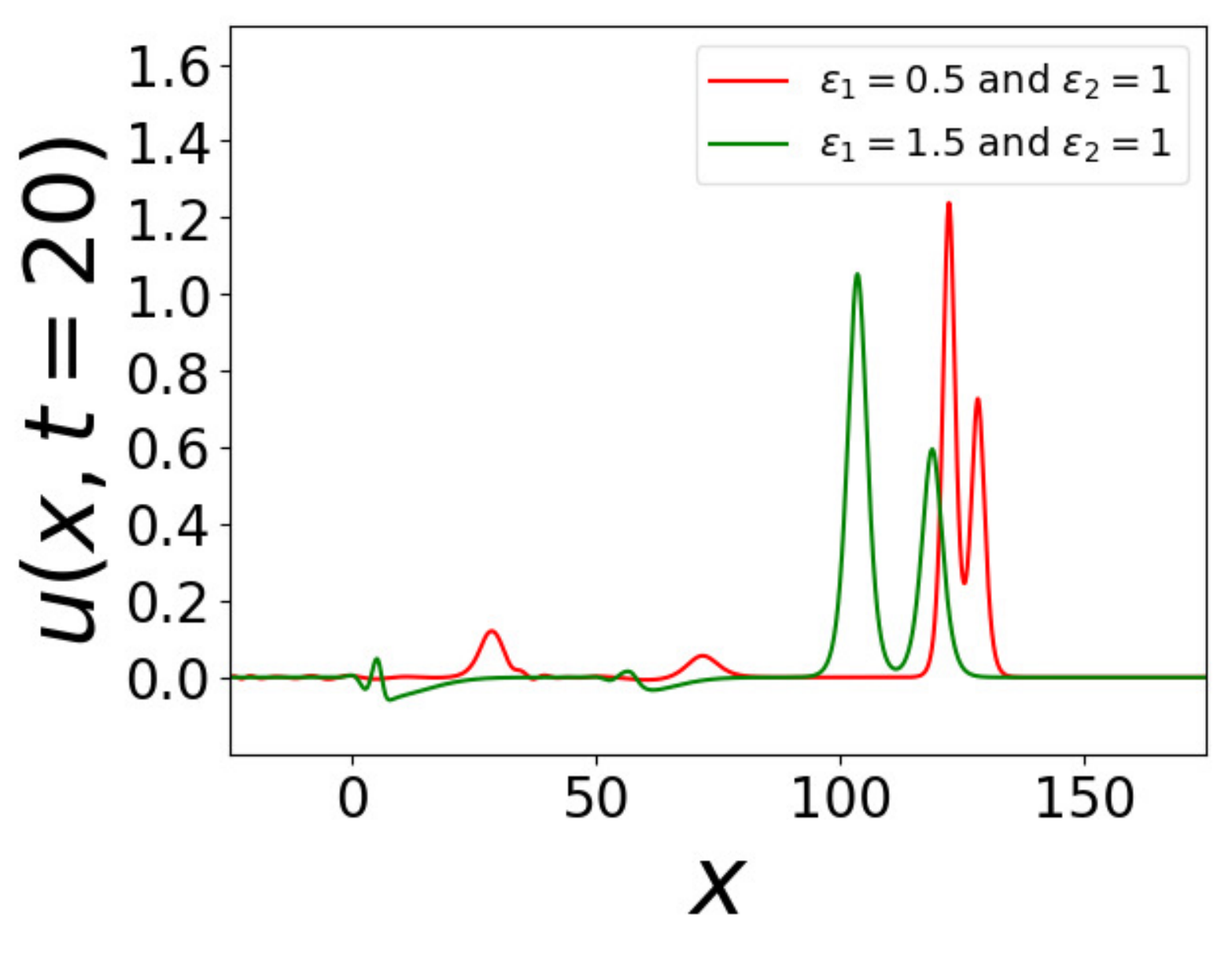}
			\caption{At~$t=20$}
			\label{plot7_6}
		\end{subfigure}%
		\begin{subfigure}{.34\textwidth}
			\centering
			\includegraphics[scale=0.34]{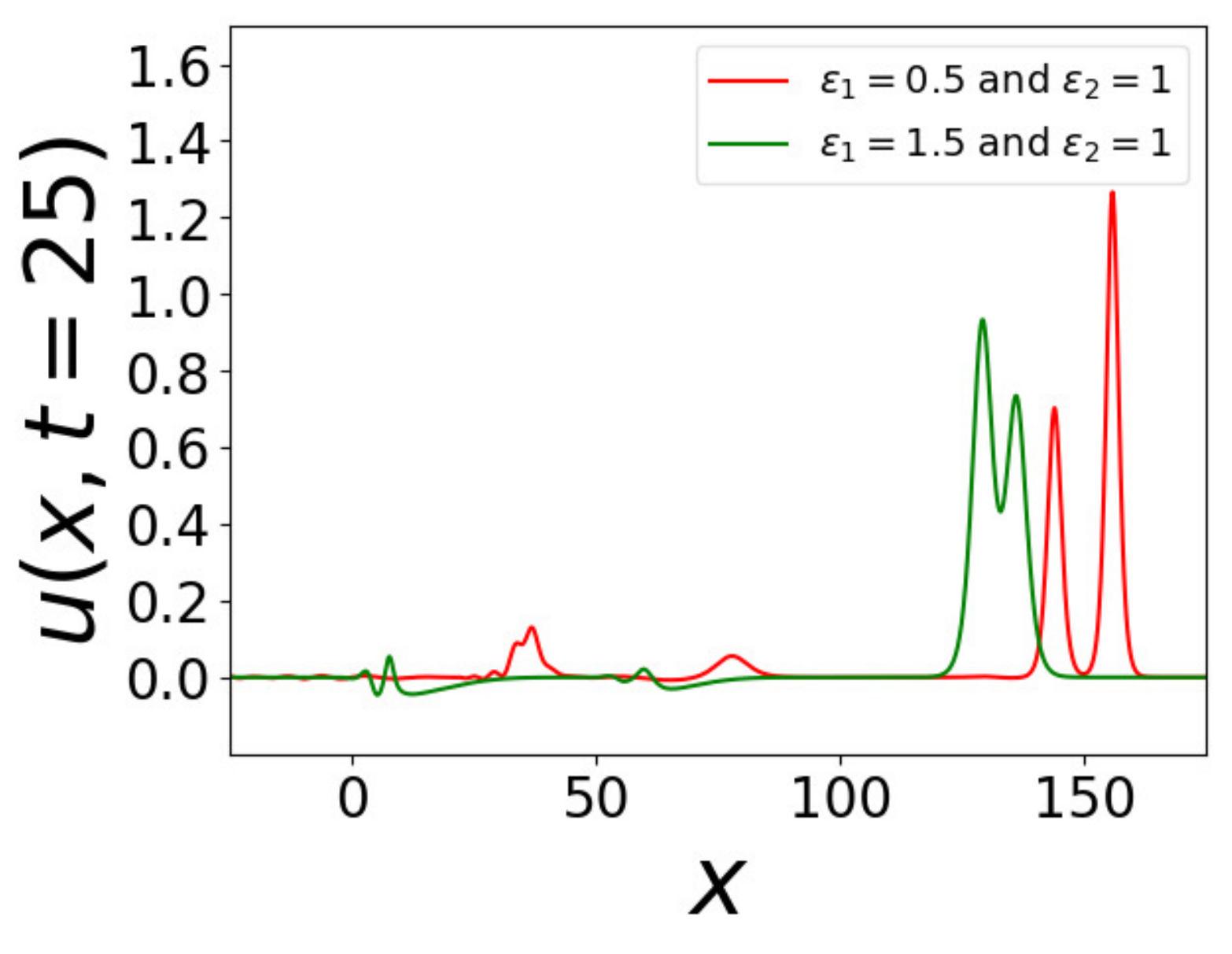}
			\caption{At~$t=25$}
			\label{plot7_7}
		\end{subfigure}
		\caption{The time-evolution of two two-soliton systems of equation~(\ref{deformedqkdv}) for $\varepsilon_1 = 0.5$, $\varepsilon_2 = 1$  (red curve)
and $\varepsilon_1 = 1.5$, $\varepsilon_2 = 1$, (green curve) in which the initial conditions were taken from solitons of the mRLW equation.} 
		\label{plot7_2to7_7}
	\end{figure}

\begin{figure}[b!]
		\centering
		\hspace*{0.5cm}
		\begin{subfigure}{.34\textwidth}
			\hspace*{-1.2cm}
			\centering
			\includegraphics[scale=0.34]{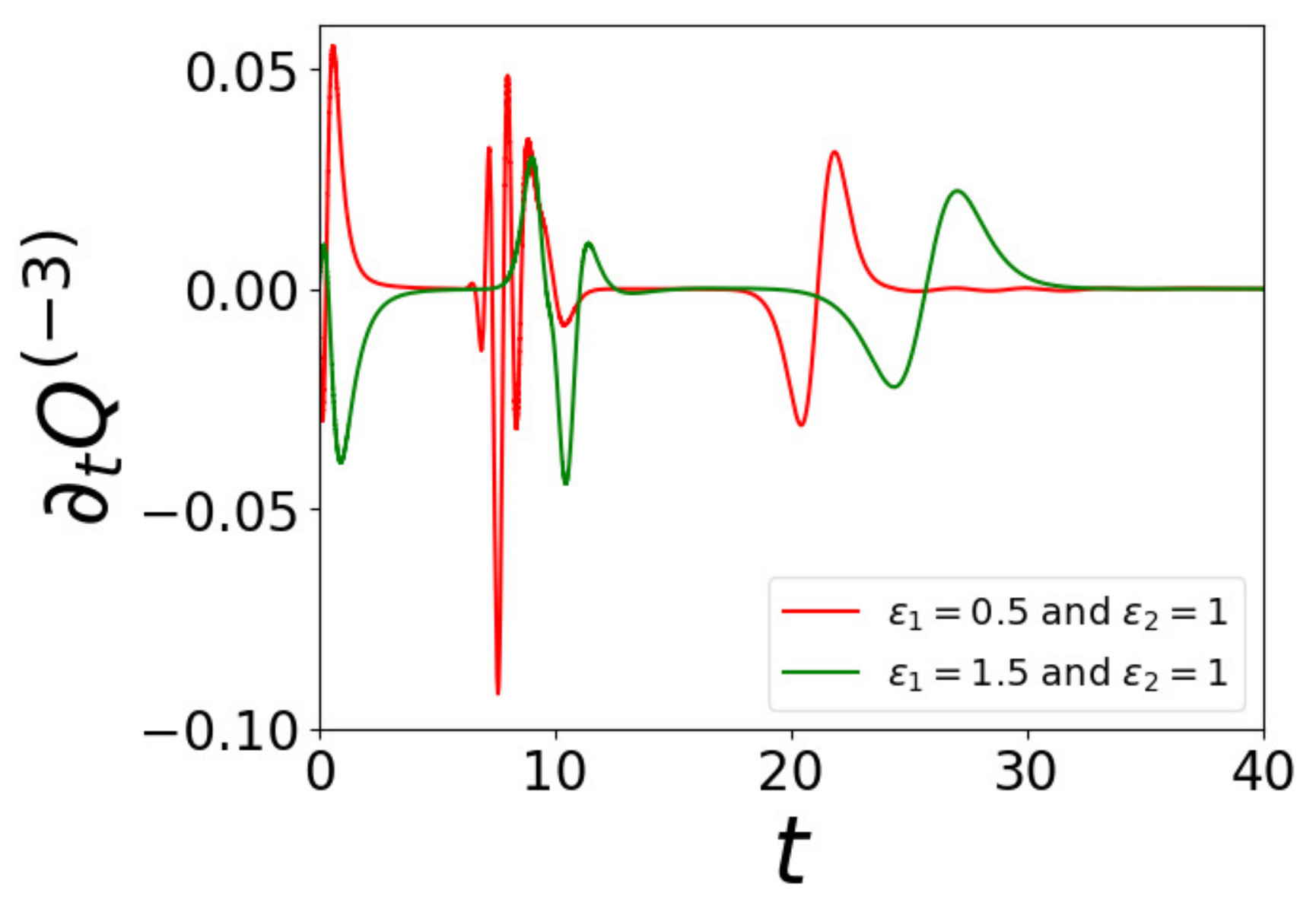}
			\caption{}
			\label{plot7_8}
		\end{subfigure}%
		\begin{subfigure}{.34\textwidth}
			\hspace*{-1.2cm}
			\centering
			\includegraphics[scale=0.34]{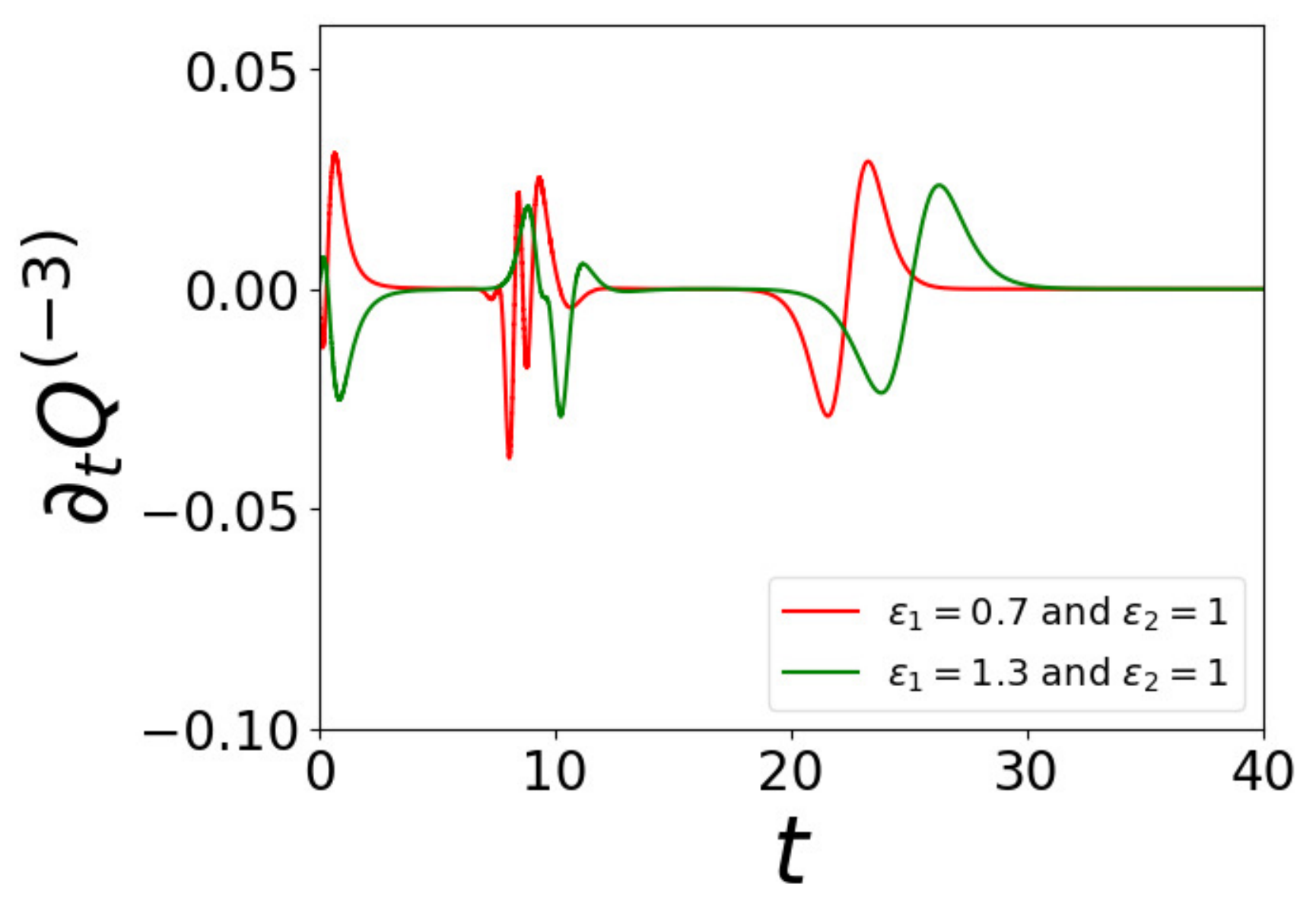}
			\caption{}
			\label{plot7_9}
		\end{subfigure}%
		\begin{subfigure}{.34\textwidth}
			\hspace*{-1.2cm}
			\centering
			\includegraphics[scale=0.34]{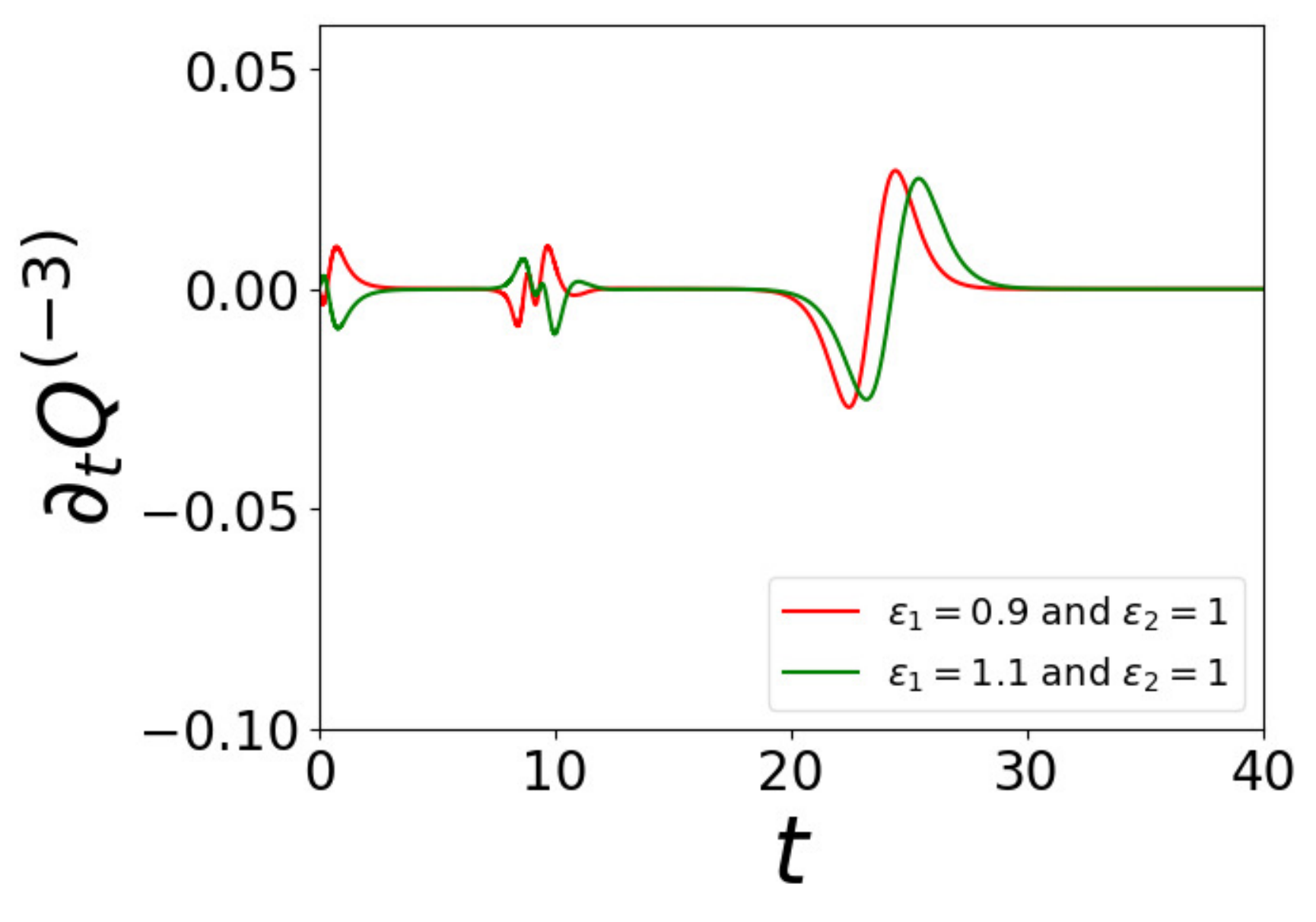}
			\caption{}
			\label{plot7_10}
		\end{subfigure}%
		\caption{The time-dependence of the determined values of the quantity $\partial_t Q^{(-3)}$ see in the simulations of equation~(\ref{deformedqkdv}) 
for different values of $\varepsilon_1 \neq 1$, $\varepsilon_2 = 1$.}
		\label{plot7_8to7_10}
	\end{figure}

\begin{figure}[t!]
		\centering
		\hspace*{0.5cm}
		\begin{subfigure}{.34\textwidth}
			\hspace*{-1.4cm}
			\centering
			\includegraphics[scale=0.32]{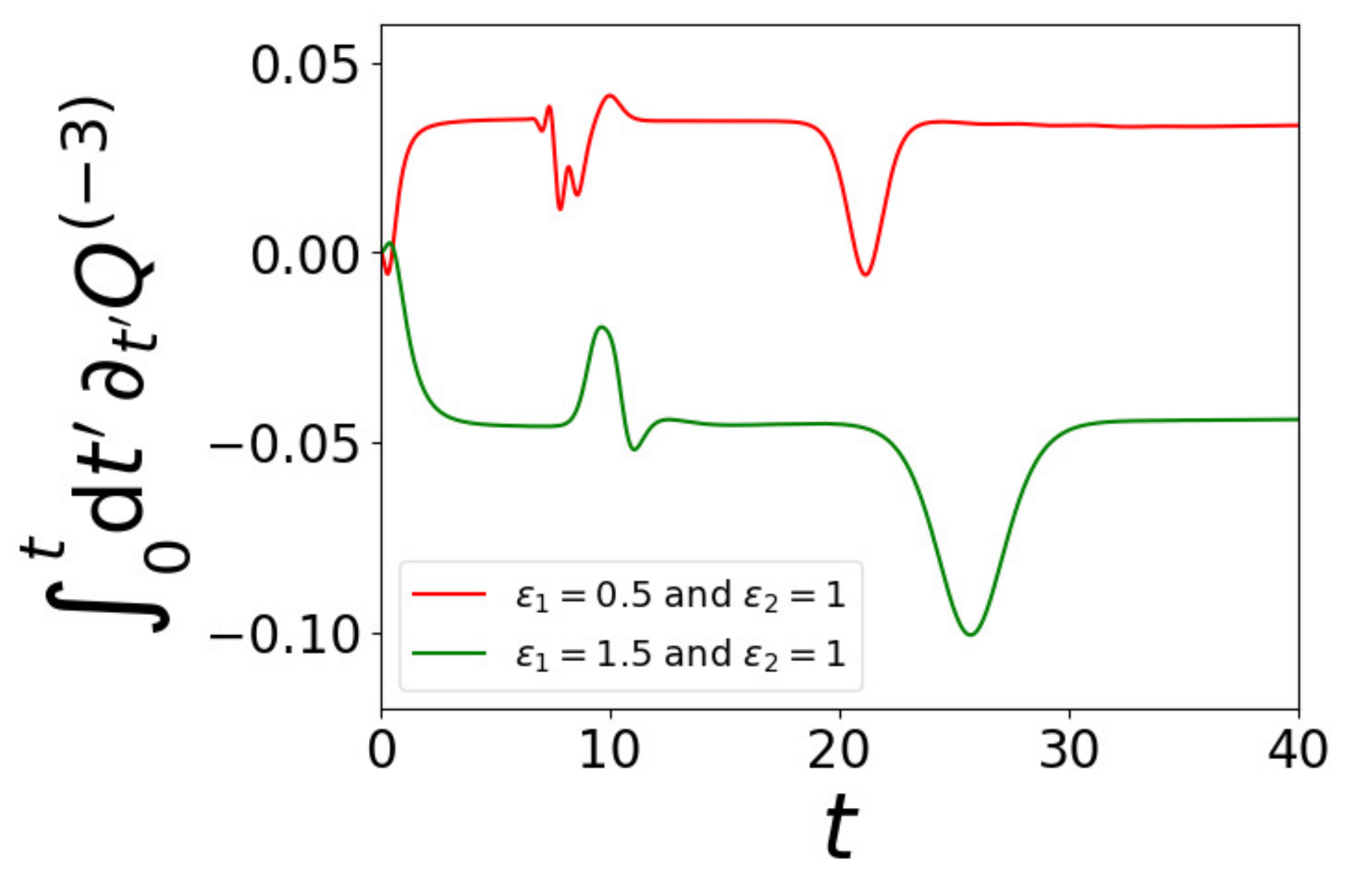}
			\caption{}
			\label{plot7_11}
		\end{subfigure}%
		\begin{subfigure}{.34\textwidth}
			\hspace*{-1.4cm}
			\centering
			\includegraphics[scale=0.32]{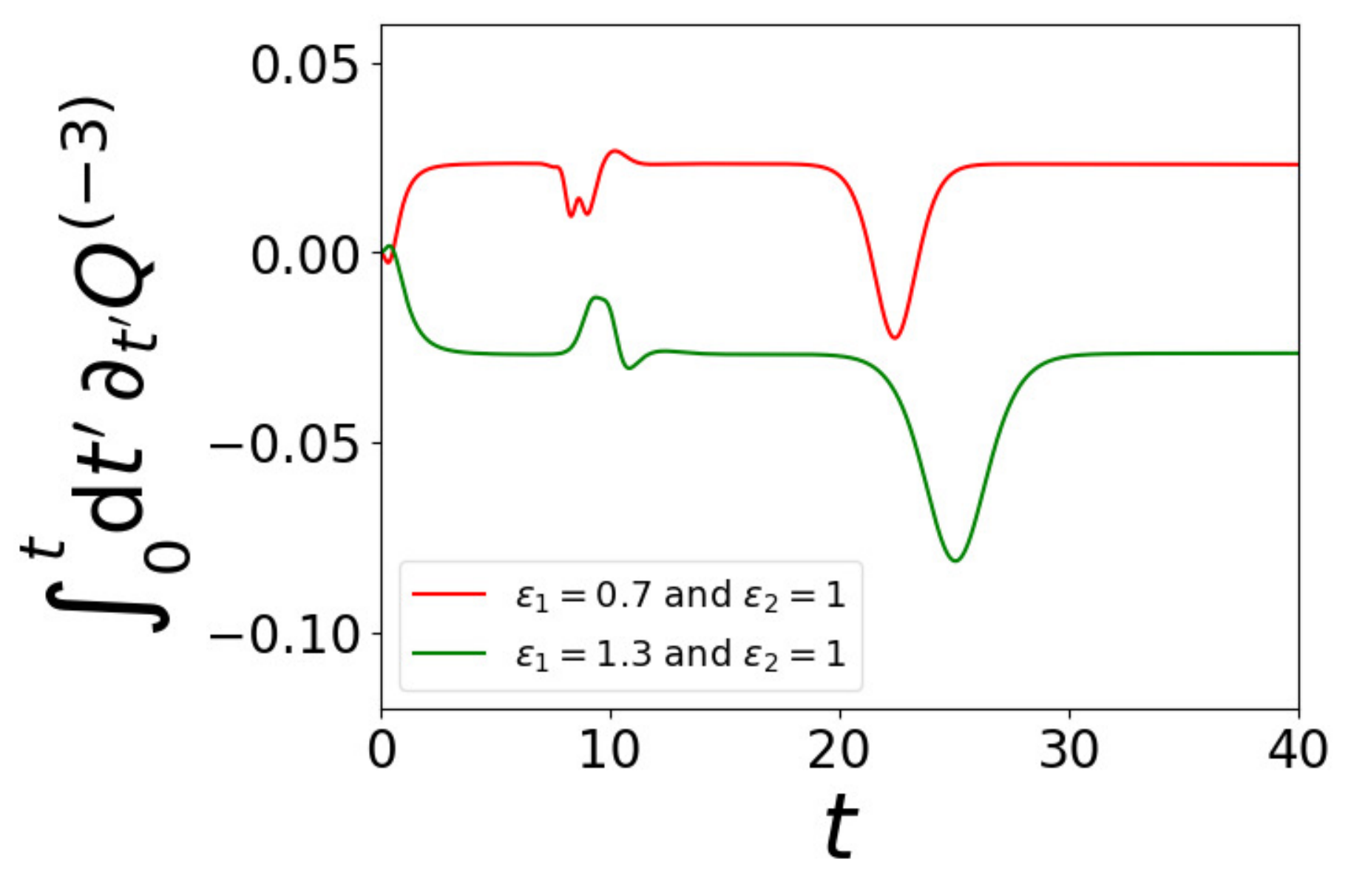}
			\caption{}
			\label{plot7_12}
		\end{subfigure}%
		\begin{subfigure}{.34\textwidth}
			\hspace*{-1.4cm}
			\centering
			\includegraphics[scale=0.32]{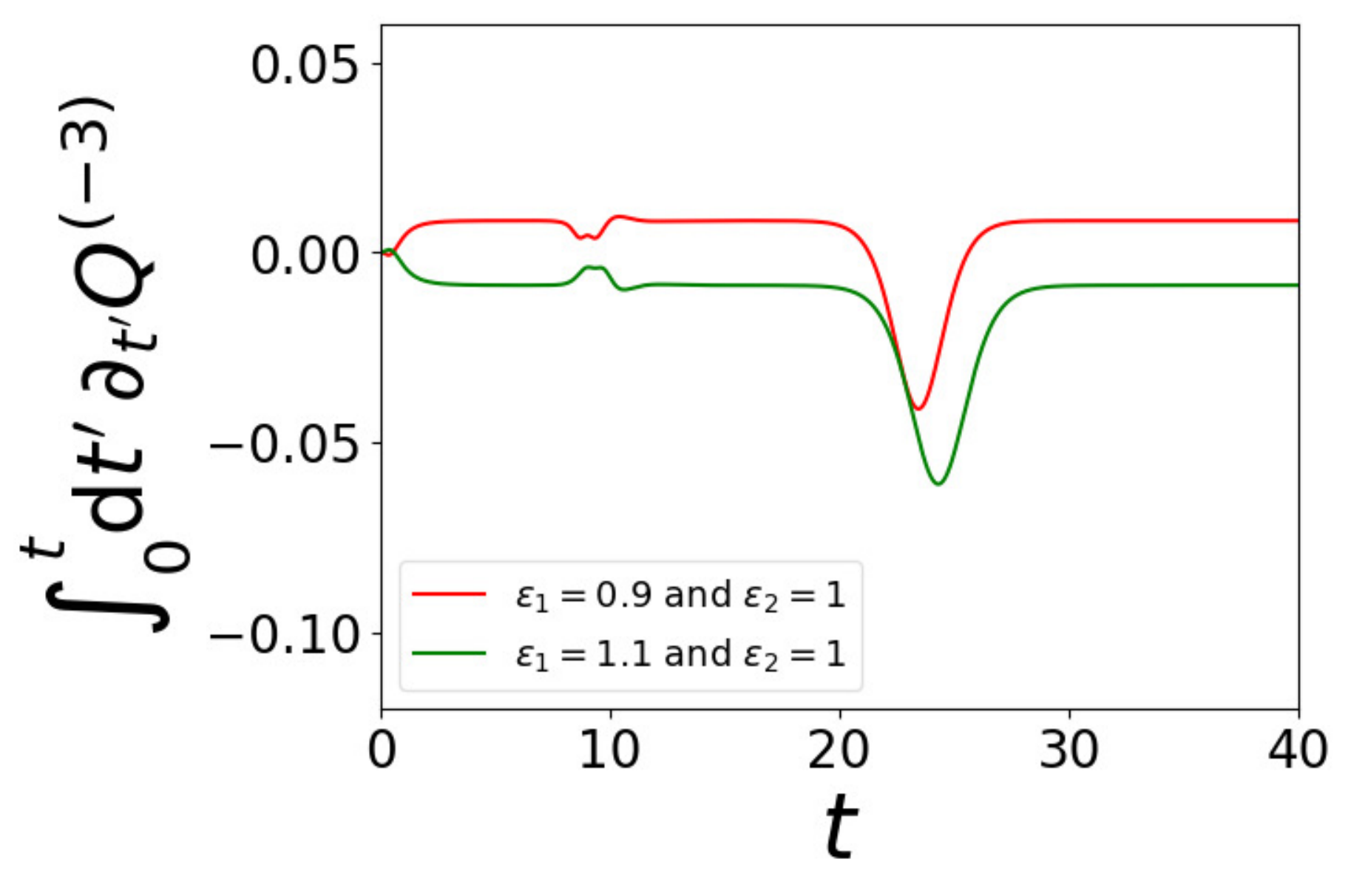}
			\caption{}
			\label{plot7_13}
		\end{subfigure}%
		\caption{The time-dependence of  $\int_{0}^{t} \mathrm{d} t^\prime \, \partial_{t^\prime} Q^{(-3)}$ seen in the simulations of
 equation~(\ref{deformedqkdv}) for different values of $\varepsilon_1 \neq 1$, $\varepsilon_2 = 1$.}
		\label{plot7_11to7_13}
	\end{figure}

	Figure~\ref{plot7_2to7_7}
	shows the $u$ fields seen in the two-soliton simulation corresponding to $\varepsilon_1 = 0.5$ (respresented by the red curve) and $\varepsilon_1 = 1.5$ (represented by the green curve).
 We observe that when  $\varepsilon_1 < 1$  the radiation emitted by the solitons has initially a positive amplitude, and the solitons' amplitudes and velocities 
increases as a result of emission of  this radiation. On the other hand, if $\varepsilon_1 > 1$, the radiation emitted by the solitons has initially a negative amplitude,
 and the solitons' amplitudes and velocities decrease. Comparing figure~\ref{plot7_2to7_7} with figure~\ref{plot5_2to5_7}, we see that changing $\varepsilon_1$ (while keeping $\varepsilon_2 = 1$) results in more radiation emitted than when changing $\varepsilon_2$ (while keeping $\varepsilon_1 = 1$) by the same amount.  Figures~\ref{plot7_8to7_10}
	and~\ref{plot7_11to7_13}
	present the time dependence of  $\partial_t Q^{(-3)}$ and $\int_{0}^{t} \mathrm{d} t^\prime \, \partial_{t^\prime} Q^{(-3)}$ observed in these simulations, and for various other simulations with other values of $\varepsilon_1 \neq 1$.

	\subsubsection{Three-soliton configurations for $\varepsilon_1 \neq 1$ and $\varepsilon_2 = 1$}
	
\begin{figure}[b!]
		\centering
		\hspace*{-0.1cm}
		\begin{subfigure}{.34\textwidth}
			\centering
			\includegraphics[scale=0.34]{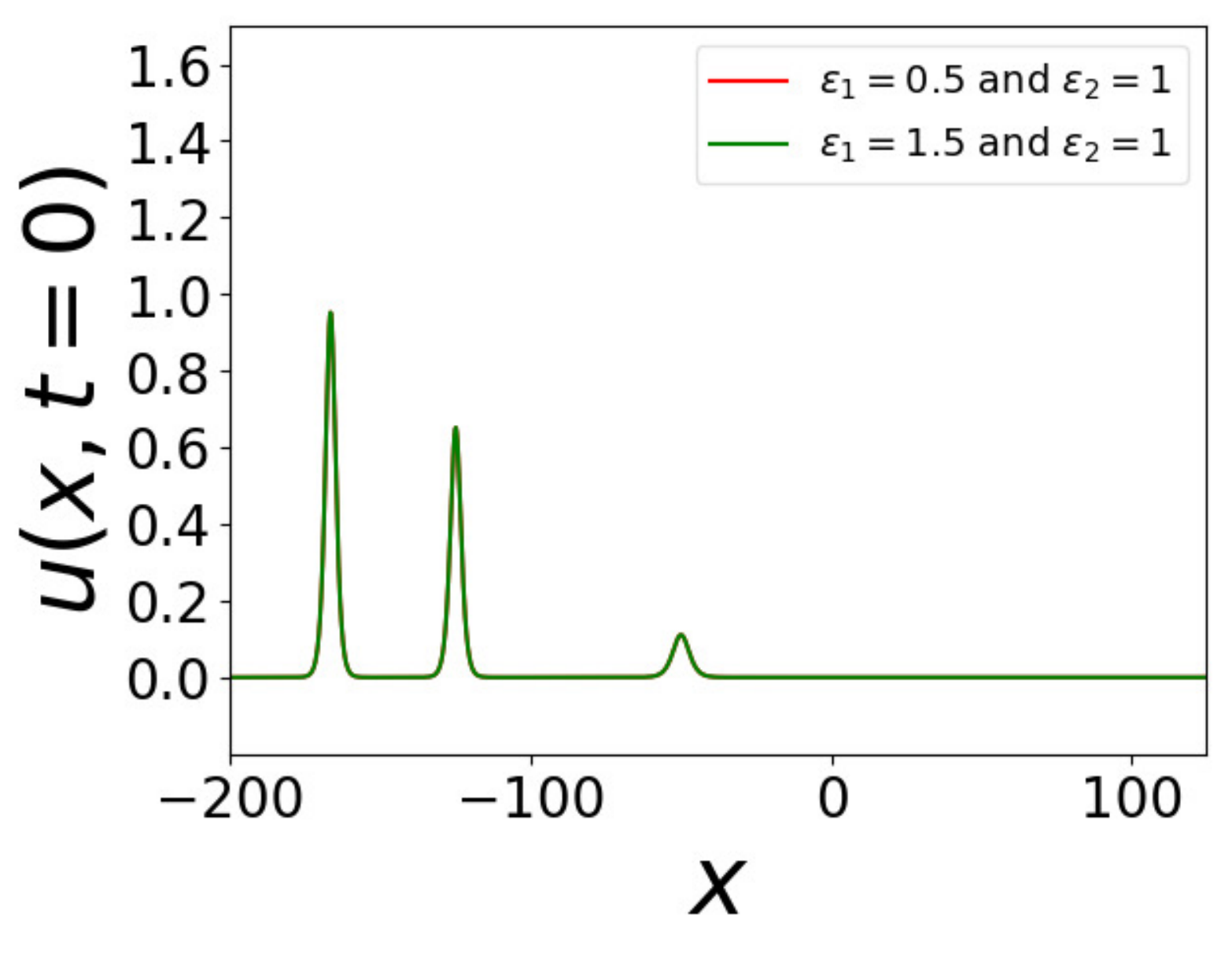}
			\caption{At~$t=0$}
			\label{plot8_2}
		\end{subfigure}%
		\begin{subfigure}{.34\textwidth}
			\centering
			\includegraphics[scale=0.34]{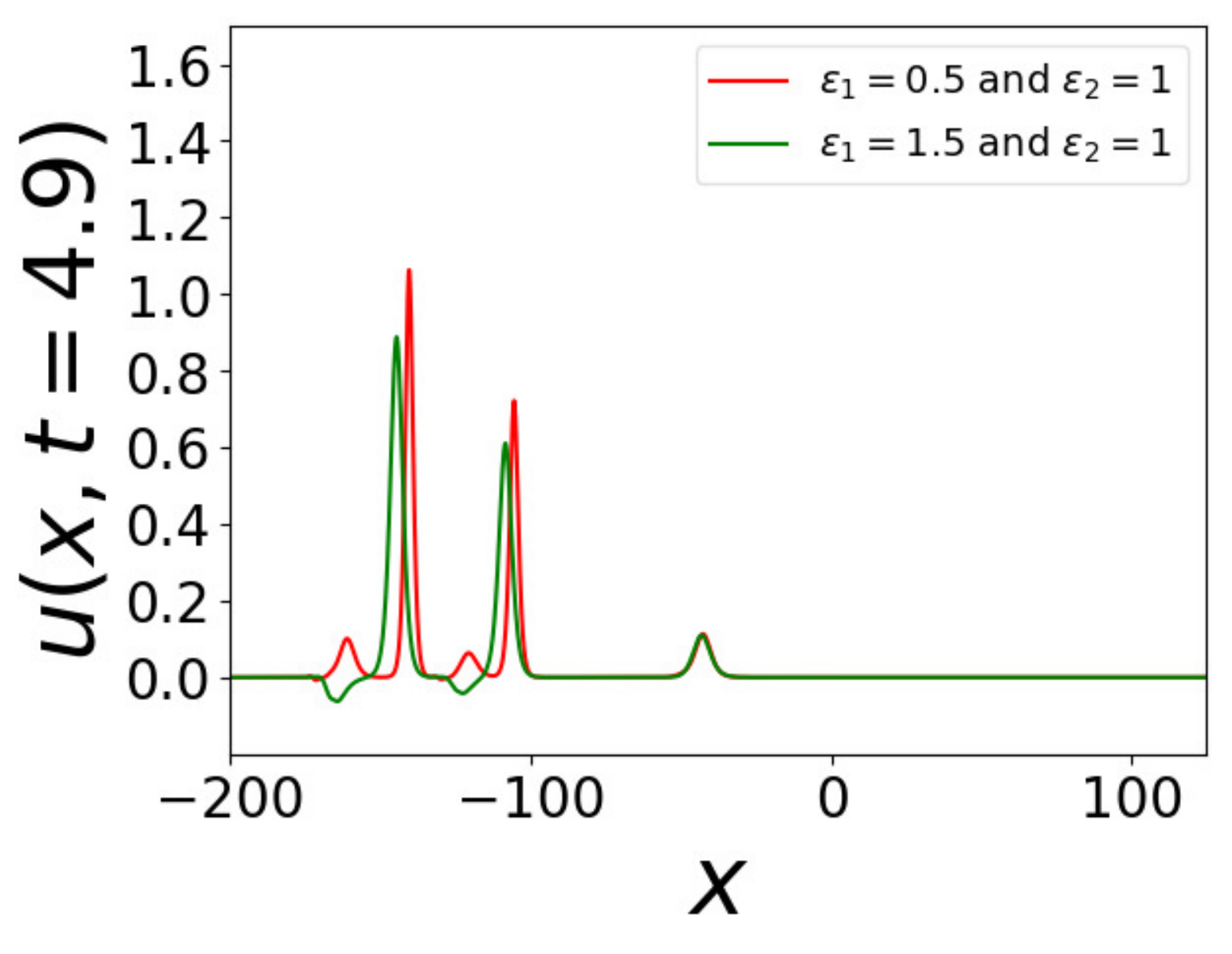}
			\caption{At~$t=4.9$}
			\label{plot8_3}
		\end{subfigure}%
		\begin{subfigure}{.34\textwidth}
			\centering
			\includegraphics[scale=0.34]{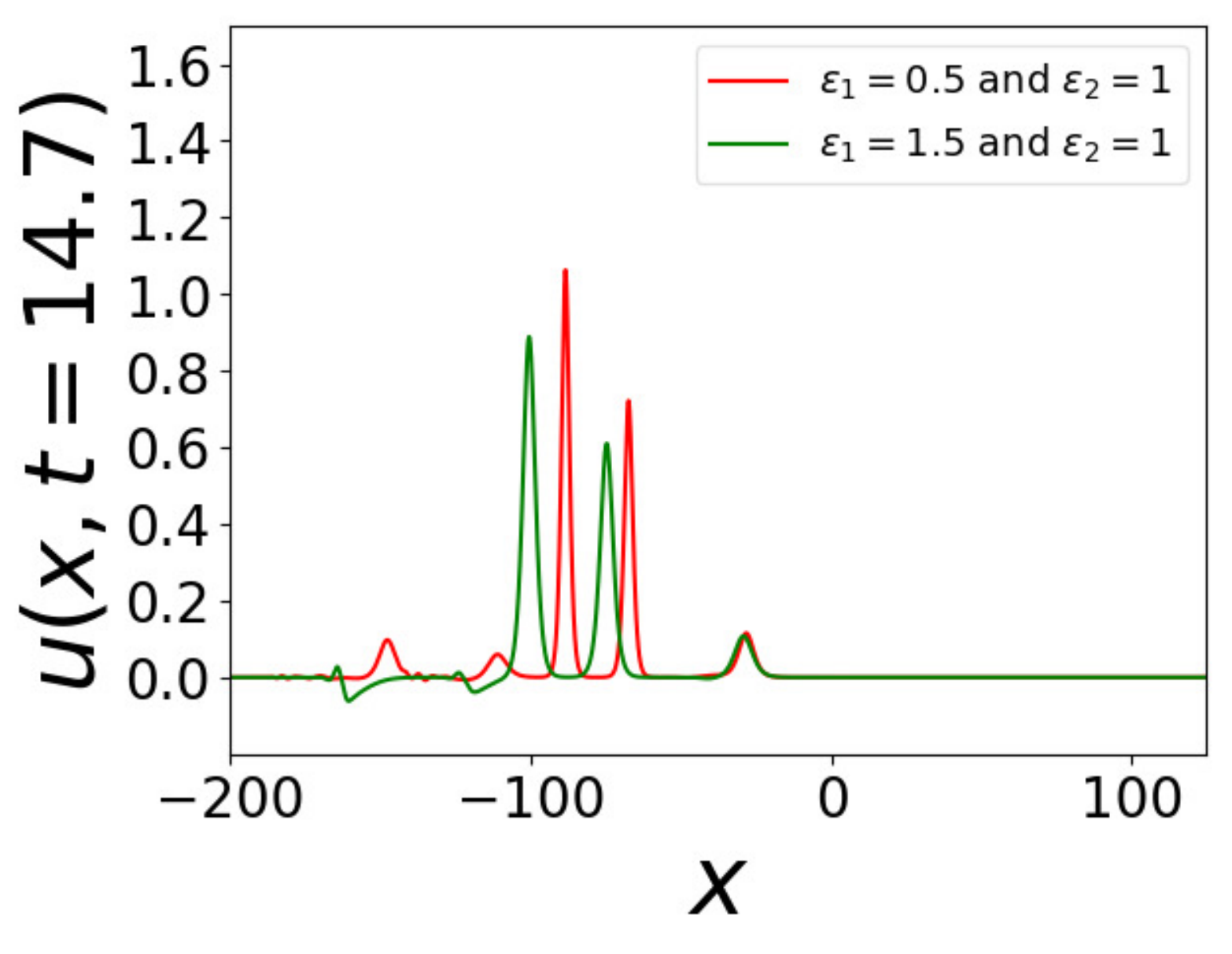}
			\caption{At~$t=14.7$}
			\label{plot8_4}
		\end{subfigure}
		
		\begin{subfigure}{.34\textwidth}
			\centering
			\includegraphics[scale=0.34]{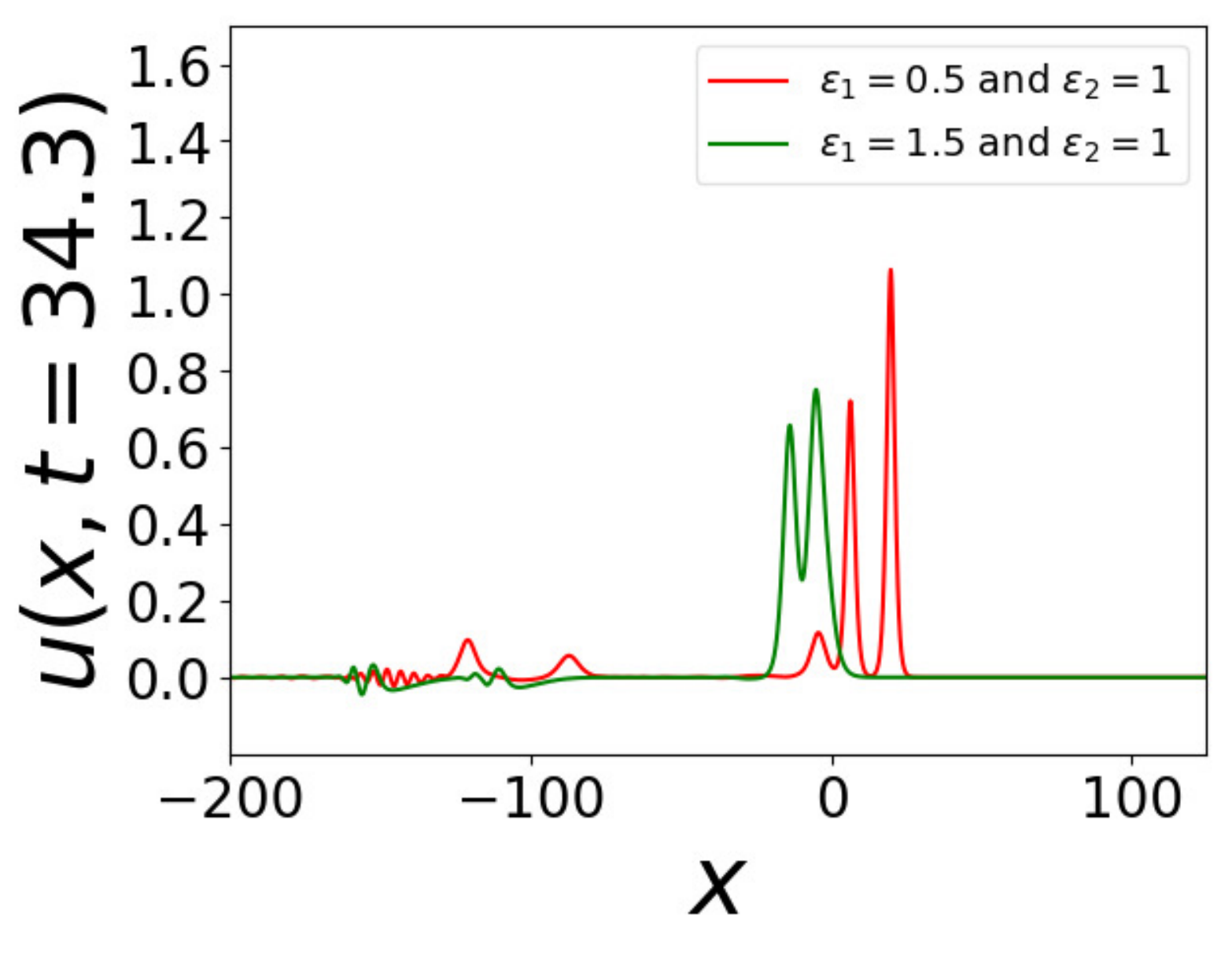}
			\caption{At~$t=34.3$}
			\label{plot8_5}
		\end{subfigure}%
		\begin{subfigure}{.34\textwidth}
			\centering
			\includegraphics[scale=0.34]{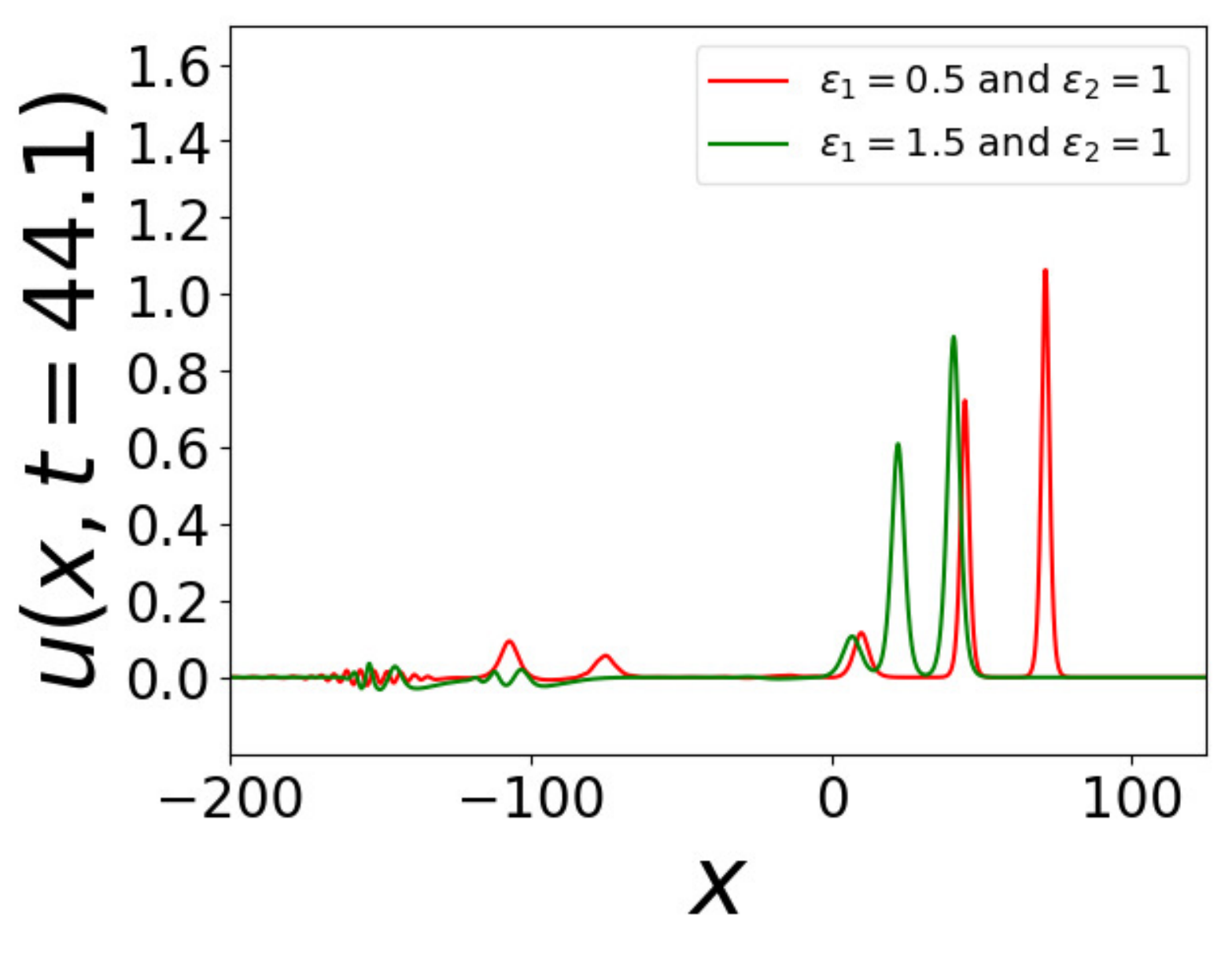}
			\caption{At~$t=44.1$}
			\label{plot8_6}
		\end{subfigure}%
		\begin{subfigure}{.34\textwidth}
			\centering
			\includegraphics[scale=0.34]{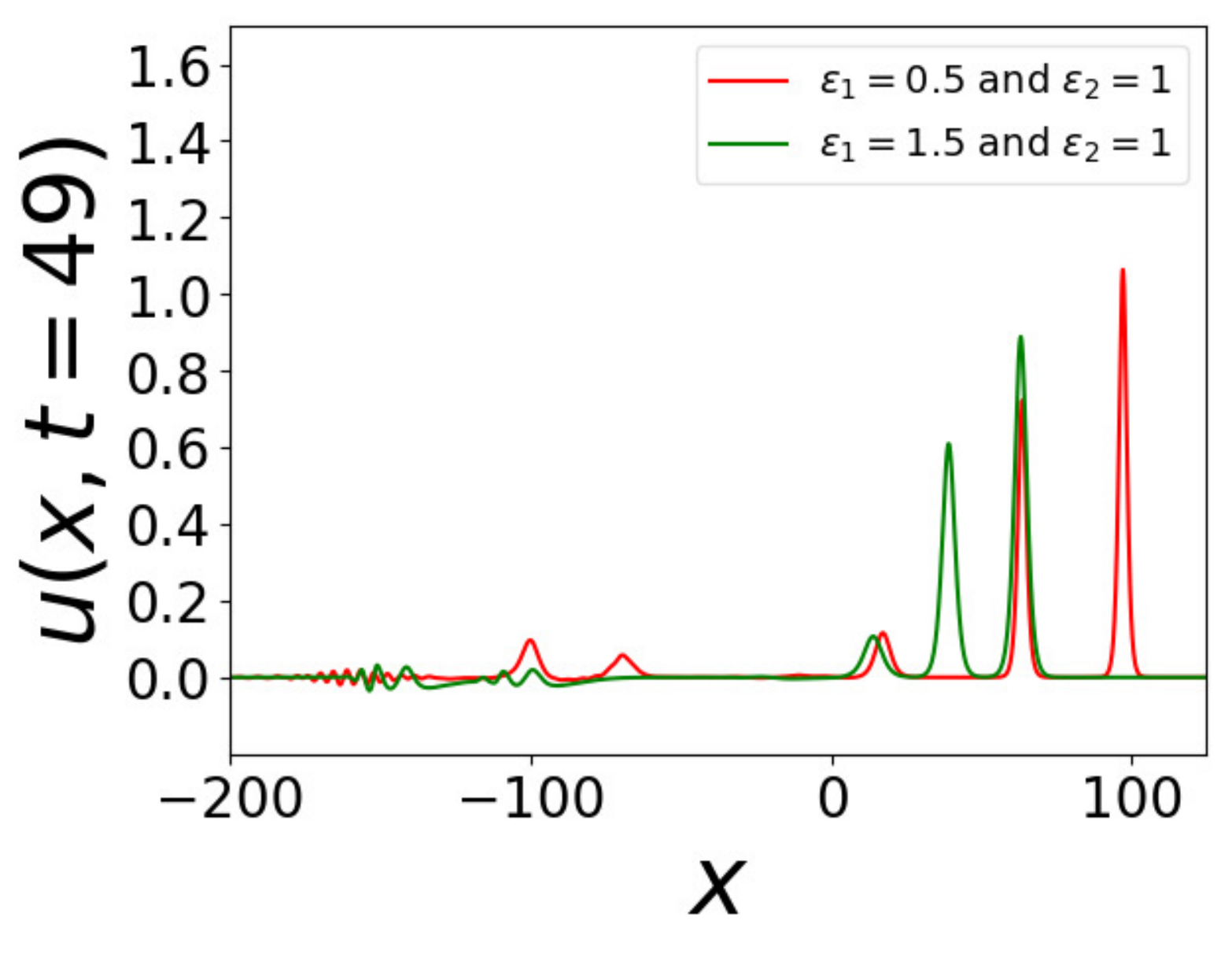}
			\caption{At~$t=49$}
			\label{plot8_7}
		\end{subfigure}
		\caption{The time-evolution of two three-soliton systems seen in our simulations of equation~(\ref{deformedqkdv}) for
 $\varepsilon_1 = 0.5$ and $\varepsilon_2 = 1$ (red curve), and $\varepsilon_1 = 1.5$ and $\varepsilon_2 = 1$ green curve).} 
		\label{plot8_2to8_7}
	\end{figure}

\begin{figure}[t!]
		\centering
		\hspace*{0.5cm}
		\begin{subfigure}{.34\textwidth}
			\hspace*{-1.2cm}
			\centering
			\includegraphics[scale=0.34]{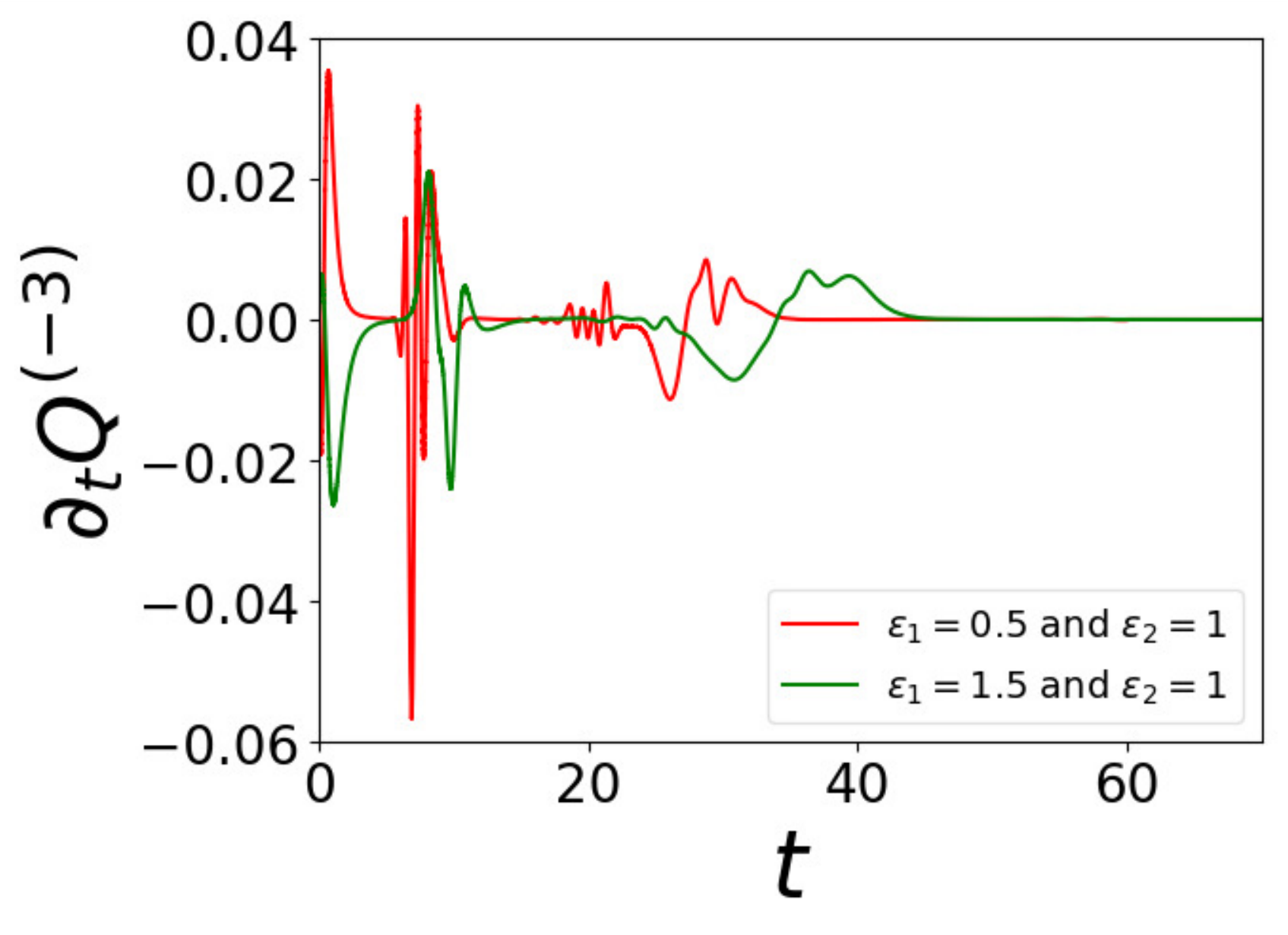}
			\caption{}
			\label{plot8_8}
		\end{subfigure}%
		\begin{subfigure}{.34\textwidth}
			\hspace*{-1.2cm}
			\centering
			\includegraphics[scale=0.34]{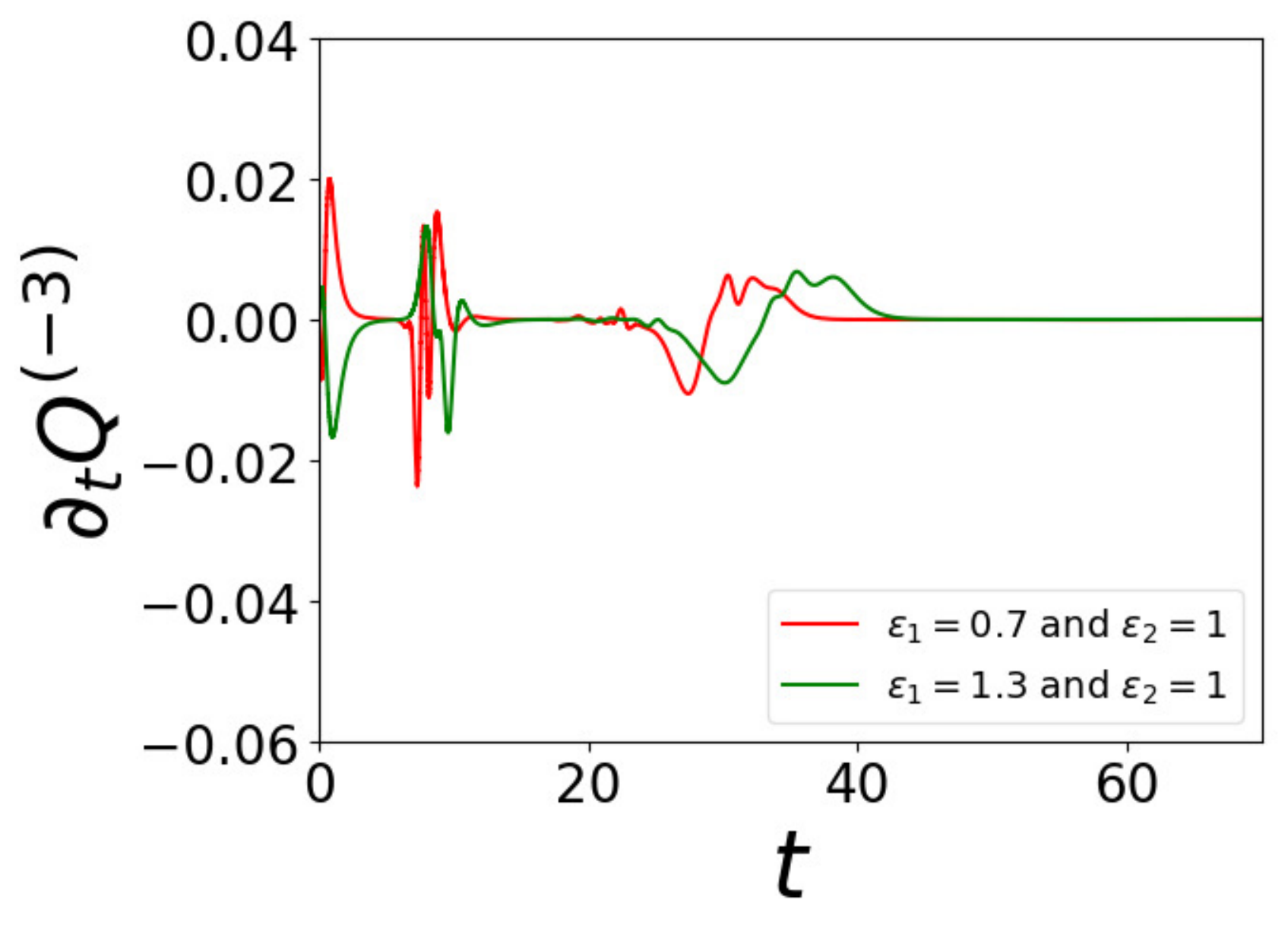}
			\caption{}
			\label{plot8_9}
		\end{subfigure}%
		\begin{subfigure}{.34\textwidth}
			\hspace*{-1.2cm}
			\centering
			\includegraphics[scale=0.34]{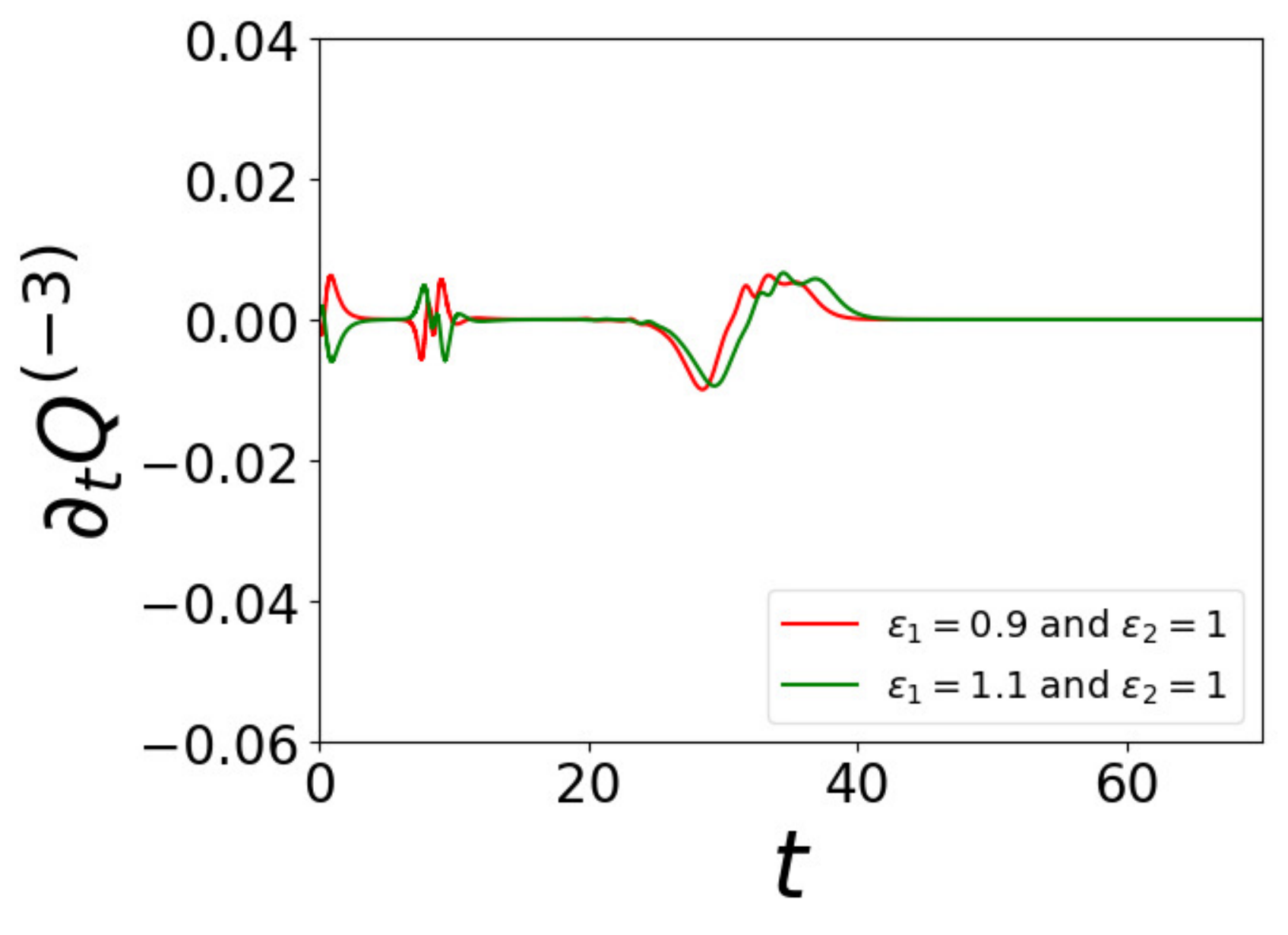}
			\caption{}
			\label{plot8_10}
		\end{subfigure}%
		\caption{The time-dependence of  $\partial_t Q^{(-3)}$ seen in the simulations of equation~(\ref{deformedqkdv}) for two values
 of $\varepsilon_1 \neq 1$ and $\varepsilon_2 = 1$.}
		\label{plot8_8to8_10}
	\end{figure}

\begin{figure}[b!]
		\centering
		\hspace*{0.5cm}
		\begin{subfigure}{.34\textwidth}
			\hspace*{-1.4cm}
			\centering
			\includegraphics[scale=0.32]{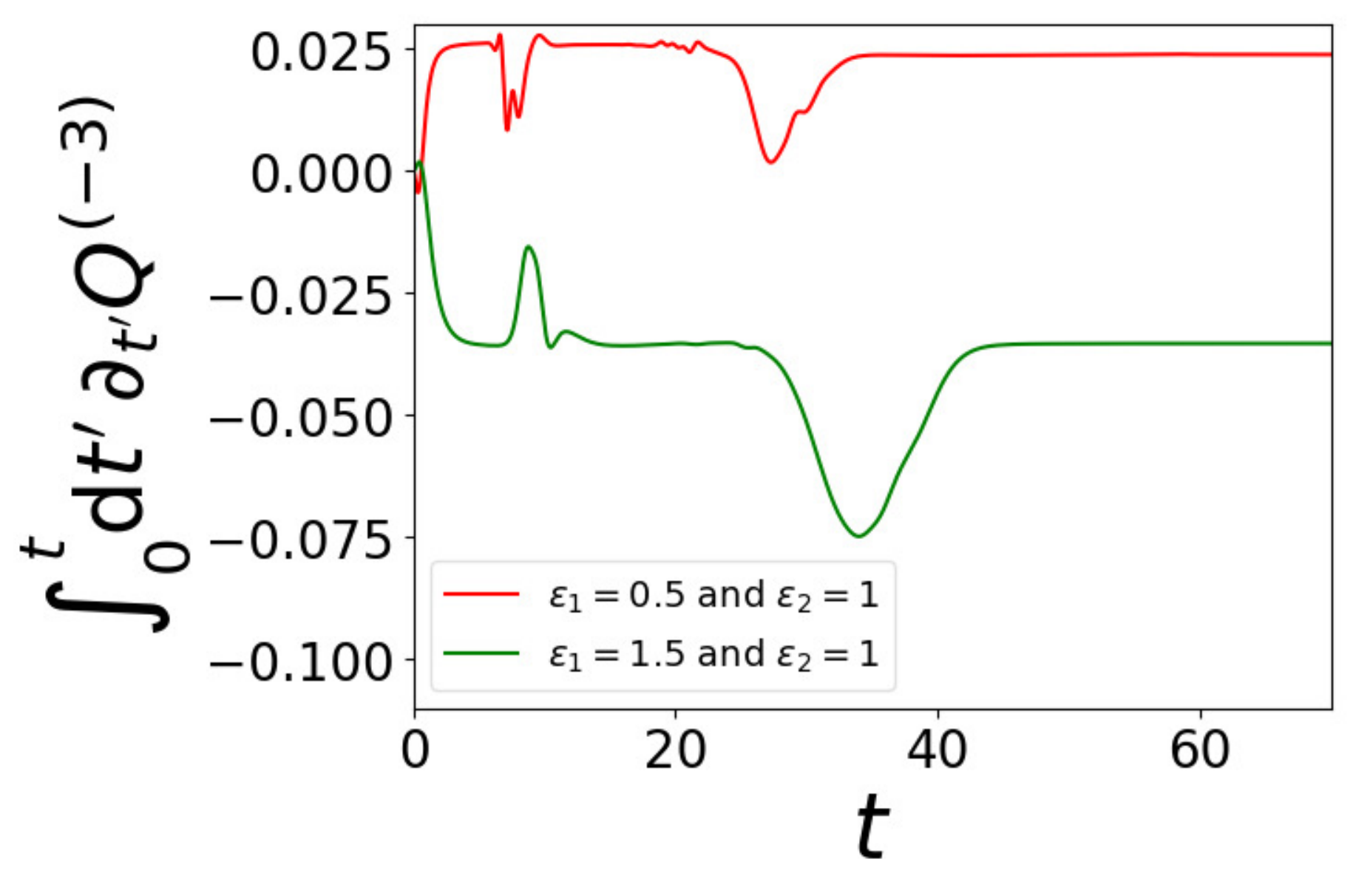}
			\caption{}
			\label{plot8_11}
		\end{subfigure}%
		\begin{subfigure}{.34\textwidth}
			\hspace*{-1.4cm}
			\centering
			\includegraphics[scale=0.32]{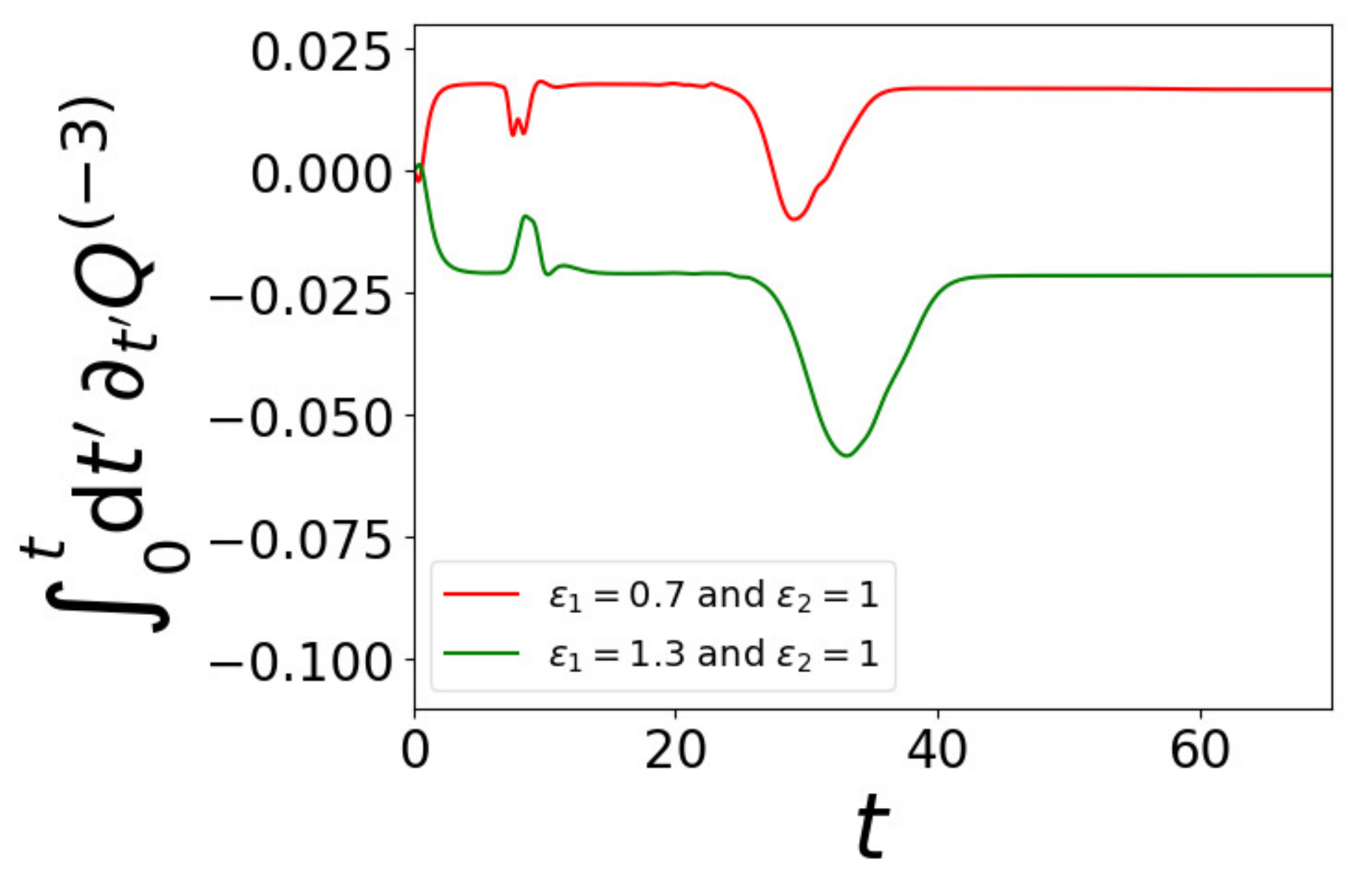}
			\caption{}
			\label{plot8_12}
		\end{subfigure}%
		\begin{subfigure}{.34\textwidth}
			\hspace*{-1.4cm}
			\centering
			\includegraphics[scale=0.32]{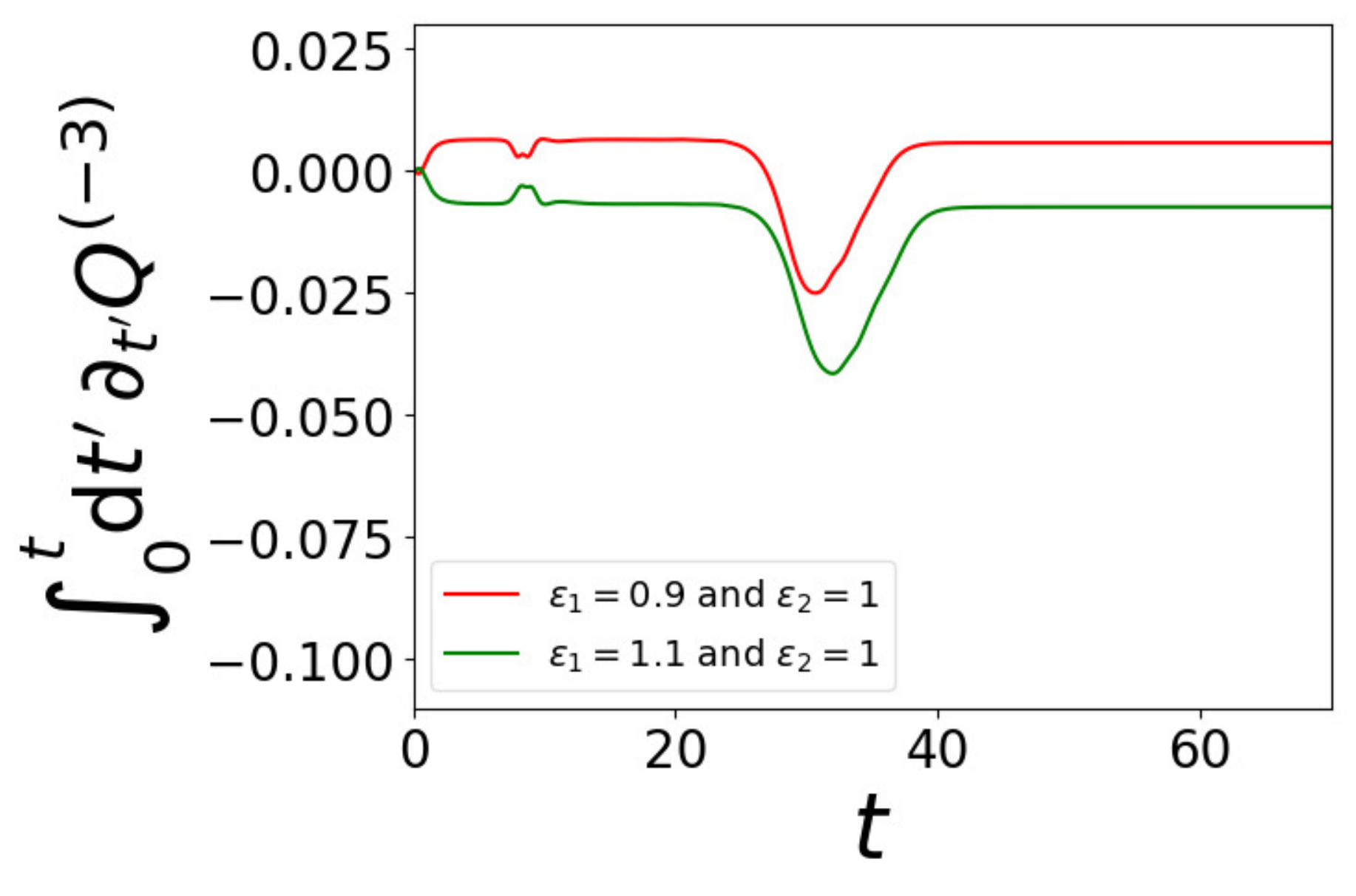}
			\caption{}
			\label{plot8_13}
		\end{subfigure}%
		\caption{The time-dependence of the determined values of $\int_{0}^{t} \mathrm{d} t^\prime \, \partial_{t^\prime} Q^{(-3)}$ 
seen in the simulations of equation~(\ref{deformedqkdv}) for 3 sets of values of $\varepsilon_1 \neq 1$ and $\varepsilon_2 = 1$.}
		\label{plot8_11to8_13}
	\end{figure}

	The  simulations of three soliton systems corresponding to $\varepsilon_1 = 0.5$, $\varepsilon_2 = 1$ and $\varepsilon_1 = 1.5$, $\varepsilon_2 = 1$ are shown
 in figure~\ref{plot8_2to8_7},
	and figures~\ref{plot8_8to8_10} 
	and~\ref{plot8_11to8_13}
	present the time dependence of the corresponding $\partial_t Q^{(-3)}$ and $\int_{0}^{t} \mathrm{d} t^\prime \, \partial_{t^\prime} Q^{(-3)}$.  These simulations are run with the linear superposition of three analytical one-soliton solutions of the mRLW equation as initial conditions, just as in subsubsection~\ref{Three_soliton_solutions_of_the_mRLW_equation}. These simulations show similar
 patterns to those seen for the two-soliton simulations.

	\section{Conclusions}		
	
	We have performed a detailed study of the quasi-integrability properties of various deformations of the KdV equation which include among them the RLW and mRLW equations. The charges were constructed by introducing an anomalous zero-curvature condition as explained in section~\ref{sec:anomalouszc}. Note that only the time component $A_t$ of the Lax pair was deformed away from the KdV potentials. The space component $A_x$ was not deformed and so the charges $Q^{(-2n-1)}$, that we presented in  (\ref{chargesdef}), are the same as the exact conserved charges of the KdV equation. The difference lies in the anomalies $\alpha^{(-2n-1)}$, also introduced in (\ref{chargesdef}), which vanish for KdV but not for its deformations. However, we have shown that the deformations of the KdV theory considered in this paper present a very  interesting property. Even though the charges are not conserved and do vary during the scatterings  of solitons, they all return, after the scattering, to the values they had before the scattering. It is this property, the asymptotic conservation of the charges,  that defines what we call a  quasi-integrability of the theory.   The mechanisms underlying  such  asymptotic conservation of the charges is still not fully understood. What we have learned so far is that the quasi-integrability
	goes hand in hand with some special properties of the soliton solutions under a specific space-time parity transformation, as explained in section  \ref{sec:parity}, and that these properties lead to the asymptotic vanishing of the anomalies. In addition, another intriguing property that correlates with quasi-integrability is that if a soliton solution of the exactly integrable theory possesses the correct behaviour under this parity, then the dynamics of the deformed theories seem to preserve this behaviour under a perturbative expansion around the integrable theory. In section \ref{sec:paritydynamicsargument} we have presented the details of such an expansion and showed	its compatibility with the expected parity behaviour of the solutions.  
	
	Our analysis  of deformations of the KdV theory present two novel ideas in the study of the quasi-integrability. First, we believe it is the first time that the quasi-integrability ideas were tested for the scattering of three solitons, where the implementation of the parity transformation is much more involved. When the three solitons scatter in separated pairs, it is expected that the quasi-integrability of two-soliton scattering would be preserved, since the third soliton is away and not interacting with the other two. However, when all three solitons collide together at the same time and in a given position in space, the three body interactions come into play, and one cannot rely on the two body interactions to analyze the quasi-integrability. We have shown in section \ref{sec:kdvsolitons} that the exact Hirota three-soliton solutions of the KdV equation do have the correct parity properties when their solitons all collide together at the same point in space. Using the results of section \ref{sec:paritydynamicsargument} one could therefore expect that deformed three-soliton solutions would possess the same property. That is exactly what our numerical simulations have shown: the charges are indeed asymptotically conserved when the three solitons of various deformations of the  KdV theory collide together. This provides a strong support for our working definition of quasi-integrability. 
	
	The other key result of the present paper is that it presents the first example of an analytical demonstration, and not only numerical, of the quasi-integrability of a truly non-integrable theory. This was discussed in section \ref{sec:analyticalmrlw} where we showed that the two-soliton sector of the mRLW theory was analytically  quasi-integrable. The proof of this result is valid for all 
	charges from their infinite set. Our numerical simulations have confirmed it for the case of the first non-trivial charge . 
	For next charges the results were also supportive; however, as the next charges involved expressions which contained terms with
	more derivatives, our numerical results had more numerical errors and so were less reliable. Clearly, better numerical techniques
	have to be developed to prove our claims more decisively ({\it i.e.} with the same accuracy as we had for $Q^{(-3)}$). This work
	lies beyond the present paper.
	
	We must emphasise, however, that the strongest support for our working definition of quasi-integrability comes from the results of our numerical simulations. In all simulations reported  in section \ref{sec:numerical} it has been shown that the charge $Q^{(-3)}$ is quasi-conserved, {\it i.e.} that it returns, after the scattering of two or three solitons, to the value it had before the scattering. This quasi-conservation has been tested to a very good precision. In addition, our simulations have shown what had been observed before in other theories; namely, that the charge only varies in time when the solitons come together and interact. The charge remains constant when the solitons are far apart. Another important and intriguing result of our simulations is related to the observation 
	that these simulations have shown that the interaction of radiation with the solitons also seems to respect quasi -integrability. In the cases, in which we did not have an analytical configuration to be used as a seed for these simulations, there was a reasonable amount of radiation sent out while the initial configuration gradually settled to a proper solution of the deformed theory. The radiation emitted by each soliton interacted non-linearly with the other solitons, and we observed that such a radiation-soliton interaction made the charge 
	$Q^{(-3)}$ to vary quite reasonably in time. However, when the radiation and the soliton have got separated, the charge returned to its original value.  So, we have observed that the radiation-soliton interaction also seems to respect quasi-integrability, but at this stage we do not have any clear idea why this property holds too. Our analytical parity arguments do not apply to such an interaction.
	
	The numerical methods used in this paper involved the specially 
	constructed numerical programs which were based on the original program of \cite{Eilbeck} and have been appropriately adapted to numerically approximate equation~(\ref{deformedqkdv}). More details are given in appendix A. The results of our numerical simulations supported very well our expectations. This was particularily true for the lowest charge but it was also true to higher charges. However, expressions for higher charges involve more derivatives of the fields and so our expressions are more prone to numerical errors.
	Thus, although their behaviour supported our claims, the reliability 
	of these results is not as sound as for the ones we have included in this paper. 
	
	{\bf Acknowledgments}: LAF and WJZ want to thank the Royal Society for
	its grant which supported their research. All three authors want
	to thank FAPESP and Durham University for their joint 'Collaboration grant' (SPRINT) that supported their reciprocal visits to S\~ao Carlos and Durham which have made the research described in this paper possible. LAF is partially supported by CNPq-Brazil.

	\appendix

	\section{Numerical methods} \label{Numerical_method}
	\label{app:numerical}
	
	\subsection{Numerical approximation of the deformations of the KdV equation}
		
		In order to perform numerical simulations of equation~(\ref{deformedqkdv}) we follow the techniques discussed by J. C. Eilbeck and G. R. McGuire~\cite{Eilbeck}. The equation necessitates the introduction
of implicit methods.  Hence first of all we introduce a new field $p(x,t)$ by
		\begin{equation}
			p = q_t \,, \label{A.1}
		\end{equation}
		so that equation~(\ref{deformedqkdv}) can be written as
		\begin{equation}
			p_{t} + p_{x} - 4p_{x}^2 - 2\varepsilon_2q_{xx}p_{t} + p_{xxx} - \varepsilon_1\(p_{xxt}+p_{xxx}\) = 0 \,. \label{A.2}
		\end{equation}
		We see that the main difference in approximating this equation compared to the numerical scheme used in~\cite{first_paper} is an extra term proportional to $p_{xxx}$.
		
		In order to adjust our scheme to this extra term, we first discretize in both~$x$ and~$t$ by taking a finite set of points $x_0, x_1, \ldots, x_N$ and~$t_0, t_1, \ldots, t_K$, where~$h$ and~$\tau$ denote the step size in space and time. Furthermore, we denote the grid points as~$(i h , m \tau) \equiv (i,m)$, where~$i = 0,1,2,\ldots, N$ and~$m = 0,1,2,\ldots, K$, and we employ the notation~$p_i{}^m \equiv p(ih,m \tau)$ and~$q_i{}^m \equiv q(ih,m \tau)$. Finally, let~$v_i{}^m$ denote our approximation to~$p_i{}^m$ and let~$w_i{}^m$ denote our approximation to~$q_i{}^m$. 
		
		Next we introduce the following central finite difference operators by their actions on $v_i{}^m$ :
		\begin{align}
			\delta_x^2 v_i{}^m & = (v^m_{i+1} - 2 v_i{}^m + v^m_{i-1})/h^2 \,, \\
			H_x v_i{}^m & = (v^m_{i+1}-v^m_{i-1}) / 2h \,, \\
			H_t v_i{}^m & = (v^{m+1}_i-v^{m-1}_i) / 2 \tau \,,
		\end{align}
		and similarly on~$w_i{}^m$. Applying these operators to equation~(\ref{A.2}) in a straightforward manner we get
		\begin{equation}
			H_t v_i{}^m + H_x v_i{}^m - 4 \left( H_x v_i{}^m \right)^2 - 2 \varepsilon_2 \delta_x^2 w_i{}^m H_t v_i{}^m + \delta_x^2 H_x v_i{}^m - \varepsilon_1(\delta_x^2 H_t v_i{}^m + \delta_x^2 H_x v_i{}^m) = 0 \,,
		\end{equation}
		which can be rewritten as
		\begin{equation}
			\begin{aligned}
				\begin{split}
					\frac{v_i^{m+1}}{2 \tau} & - \varepsilon_1 \frac{v_{i+1}^{m+1}-2v^{m+1}_i + v_{i-1}^{m+1}}{2h^2 \tau} - \varepsilon_2 \frac{w_{i+1}^m v_i^{m+1} - 2 w_i{}^m v_i^{m+1} +  w^m_{i-1}v_i^{m+1}}{h^2 \tau} \\& = \frac{v_i^{m-1}}{2 \tau} - \frac{v_{i+1}^m- v_{i-1}^m}{2h} - \varepsilon_1 \frac{v_{i+1}^{m-1}-2v_i^{m-1}+v_{i-1}^{m-1}}{2h^2 \tau}  \\& \hspace{4.5 mm} - \varepsilon_2 \frac{w_{i+1}^m v_i{}^{m-1} - 2w_i{}^m v_i^{m-1} + w_{i-1}^m v_i^{m-1}}{h^2 \tau}  \\& \hspace{4.5 mm} + \frac{(v_{i+1}^m)^2 - 2 v_{i+1}^m v_{i-1}^m + (v_{i-1}^m)^2}{h^2}  \\& \hspace{4.5 mm} - (1-\varepsilon_1) \frac{v_{i+2}^m-2v_{i+1}^m+2v_{i-1}^m-v_{i-2}^m}{2h^3} \,. \label{A.2.1}
				\end{split}
			\end{aligned}
		\end{equation}
		Next we introduce the following matrices and vectors
		\begin{equation}
		A \equiv \begin{pmatrix} 1 & 0 & 0 & 0 & \cdots &  & 0 & 0 \\
		0 & 1 & 0 & 0 & \cdots &  & 0 & 0 \\
		0 & \frac{1}{2h^2 \tau} & a_2{}^m & \frac{1}{2h^2 \tau} & 0 & & \vdots & \vdots \\
		0 & 0 & \frac{1}{2h^2 \tau} & a_3{}^m & \frac{1}{2h^2 \tau} & \\
		\vdots & & & \ddots & \ddots & \ddots & 0 & 0 \\
		&  &  &  & \frac{1}{2h^2 \tau} & a_{N-2}^m & \frac{1}{2h^2 \tau} & 0 \\
		0 & 0 & \cdots & & 0 & 0 & 1 & 0 \\  
		0 & 0 & \cdots & & 0 & 0 & 0 & 1  
		\end{pmatrix} \,,
		\end{equation}
		\begin{equation}
		B \equiv \begin{pmatrix} v_0^{m+1} \\ v_1^{m+1} \\ v_2^{m+1} \\ \vdots \\ v_{N-1}^{m+1} \\ v_N^{m+1} \end{pmatrix} \,,\quad  C \equiv \begin{pmatrix} c_0{}^{m} \\ c_1{}^{m} \\ c_2{}^{m} \\ \vdots \\ c_{N-1}^{m} \\ c_N^{m} \end{pmatrix} \,,
		\end{equation}
		where		
		\begin{equation}
			a_i{}^m \equiv \frac{1}{2\tau} - \varepsilon_2 \frac{w_{i+1}^m - 2 w_i{}^m +  w_{i-1}^m}{h^2 \tau} + \varepsilon_1 \frac{v^{m+1}_i}{h^2 \tau} \,,
		\end{equation}		
		\begin{equation}
			\begin{aligned}
				\begin{split}
					c_i{}^m \equiv & \frac{v_i^{m-1}}{2 \tau} - \frac{v_{i+1}^m- v_{i-1}^m}{2h} - \varepsilon_1 \frac{v_{i+1}^{m-1}-2v_i^{m-1}+v_{i-1}^{m-1}}{2h^2 \tau}  \\& \hspace{4.5 mm} - \varepsilon_2 \frac{w_{i+1}^m v_i{}^{m-1} - 2w_i{}^m v_i^{m-1} + w_{i-1}^m v_i^{m-1}}{h^2 \tau}  \\& \hspace{4.5 mm} + \frac{(v_{i+1}^m)^2 - 2 v_{i+1}^m v_{i-1}^m + (v_{i-1}^m)^2}{h^2}  \\& \hspace{4.5 mm} - (1-\varepsilon_1) \frac{v_{i+2}^m-2v_{i+1}^m+2v_{i-1}^m-v_{i-2}^m}{2h^3} \,,\quad i = 2,3,\ldots,N-2 \,,				
				\end{split}
			\end{aligned}
		\end{equation}
		with the following boundary conditions
		\begin{equation}
		c_0{}^{m} \equiv v_0^{m+1} \,,\quad c_1^{m} \equiv v_1^{m+1}\,,
		\end{equation}
		\begin{equation}
		c_{N-1}^{m} \equiv v_{N-1}^{m+1} \,,\quad c_N^{m} \equiv v_N^{m+1}\,,
		\end{equation}
		Note that we need $2$ elements per boundary, which is a consequence of the term proportional to $p_{xxx}$. Having introduced this notation, our problem reduces to solving the following matrix equation
		\begin{equation}
		AB = C \,.
		\end{equation}
		where we need to solve for vector~$B$. In our algorithm we have used the well-known~$LU$~decomposition method for this problem, see for instance the book~\cite{Schay} for more information. 
		
		Once we have solved for~$B$, we have the values of $v_i^m$ at the next time level. We then solve equation~(\ref{A.1}) to determine all the values of~$w_i^m$ at the next time level
		\begin{equation}
		v_i{}^m = \frac{w^{m+1}_i - w_i^{m-1}}{2 \tau} \implies w_i^{m+1} = 2 \tau v_i{}^m + w_i^{m-1} \,. \label{A.3}
		\end{equation}
		Repeating this procedure for many time levels allows us to determine the numerical time evolution of a system.
		
		It is not too difficult to verify that this scheme is both second-order accurate in~$\tau$ and in~$h$. Furthermore, this scheme is an extension of the scheme discussed in~\cite{first_paper}, and so that gives us an indication that we can trust the results of our simulations. 
		
		\subsection{Summary of parameters used to produce figures} \label{summary_of_variables}
		
		Tables~\ref{variables1},~\ref{variables2} and~\ref{variables3} shows all the parameters used to produce the figures shown in this paper.
		\begin{table}[H]
			\center
			\caption{This table summarizes the variables used to produce figures~\ref{plot0_2to0_7} to~\ref{plot3_8to3_9}.}
			\begin{tabular}{ l?{0.5mm} c | c | c | c }
				\toprule[1.5pt]
				& Figures~\ref{plot0_2to0_7} and~\ref{I_1_alpha_8_e1_1_e2_1_e3_1_1_analytical_and_numerical} & Figures~\ref{plot1_2to1_7} and~\ref{plot1_8to1_9}  & Figures~\ref{plot2_2to0_7} and~\ref{plot2_8to2_9} & Figures~\ref{plot3_2to3_7} and~\ref{plot3_8to3_9} \\
				\midrule
				$x_0$ & $-50$  & $-200$ & $-50$ & $-200$ \\
				$x_N$ &  $250$ & $200$ & $250$ & $200$ \\
				$h$ & $0.1$  & $0.1$  &  $0.1$ & $0.1$  \\
				$t_0$ & $0$  & $0$  & $0$ &  $0$ \\
				$t_K$ & $40$  & $70$ & $40$ & $70$ \\
				$\tau$ & $0.001$  & $0.001$  & $0.001$  & $0.001$  \\
				$\omega_1$ & $5.00$  & $0.80$ & $5.00$ & $0.80$ \\
				$\delta_1$ & $0.00$  & $27.71$ & $0.00$ & $27.71$  \\
				$\omega_2$ & $3.00$  & $3.07$ & $3.00$ &  $3.07$ \\
				$\delta_2$ & $-40.00$  & $106.28$ & $-40.00$ &  $106.28$ \\
				$\omega_3$ & N/A  & $4.28$ & N/A & $4.28$  \\ 
				$\delta_3$ & N/A  & $148.34$ & N/A &  $148.34$
				\label{variables1}
			\end{tabular}
		\end{table}		
		
		\begin{table}[H]
			\center
			\caption{This table summarizes the variables used to produce figures~\ref{plot11_2to11_7} to~\ref{plot5_11to5_13}.}
			\begin{tabular}{ l?{0.5mm} c | c | c }
				\toprule[1.5pt]
				& Figures~\ref{plot11_2to11_7},~\ref{plot11_8to11_10} and~\ref{plot11_11to11_13} & Figures~\ref{plot12_2to12_7},~\ref{plot12_8to12_10} and~\ref{plot12_11to12_13}  & Figures~\ref{plot4_2to4_7},~\ref{plot4_8to4_9},~\ref{plot5_8to5_10},~\ref{plot5_2to5_7} and~\ref{plot5_11to5_13}  \\
				\midrule
				$x_0$ & $-50$  & $-200$ & $-50$  \\
				$x_N$ & $300$  & $200$ & $300$  \\
				$h$ & $0.1$  & $0.1$  &  $0.1$  \\
				$t_0$ & $0$  & $0$  &  $0$  \\
				$t_K$ & $40$  & $70$ & $40$ \\
				$\tau$ &  $0.001$ & $0.001$  & $0.001$  \\
				$\omega_1$ & $5.00$  & $0.80$ & $5.00$ \\
				$\delta_1$ & $0.00$  & $27.71$ &  $0.00$ \\
				$\omega_2$ & $3.00$  & $3.07$ & $3.00$ \\
				$\delta_2$ & $-40.00$  & $106.28$ &  $-40.00$ \\
				$\omega_3$ & N/A  & $4.28$ &  N/A  \\ 
				$\delta_3$ & N/A  & $148.34$ &   N/A  
				\label{variables2}
			\end{tabular}
		\end{table}	
		
		\begin{table}[H]
			\center
			\caption{This table summarizes the variables used to produce figures~\ref{plot7_2to7_7} to~\ref{plot8_11to8_13}.}
			\begin{tabular}{ l?{0.5mm} c | c }
				\toprule[1.5pt]
				& Figures~\ref{plot7_2to7_7},~\ref{plot7_8to7_10} and~\ref{plot7_11to7_13} & Figures~\ref{plot8_2to8_7},~\ref{plot8_8to8_10} and~\ref{plot8_11to8_13}   \\
				\midrule
				$x_0$ &  $-50$ & $-200$    \\
				$x_N$ & $300$  & $300$    \\
				$h$ &  $0.1$ &  $0.1$     \\
				$t_0$ & $0$  &  $0$     \\
				$t_K$ & $40$  &  $70$  \\
				$\tau$ & $0.001$  & $0.001$     \\
				$\omega_1$ & $5.00$  &  $0.80$  \\
				$\delta_1$ &  $0.00$ &  $27.71$   \\
				$\omega_2$ & $3.00$ &  $3.07$  \\
				$\delta_2$ & $-40.00$ &  $106.28$   \\
				$\omega_3$ & N/A  &  $4.28$    \\ 
				$\delta_3$ & N/A  &    $148.34$  
				\label{variables3}
			\end{tabular}
		\end{table}

\section{The $\mathfrak{sl}(2)$ loop algebra}
\label{app:loopalgebra}

We use the $\mathfrak{sl}(2)$ finite algebra with commutation relations
\begin{equation}
		[T_3, T_{\pm}] = \pm T_{\pm}  \qquad\qquad \text{and} \qquad\qquad   [T_+, T_-] = 2\,T_3 \,,
		\end{equation}
		which is satisfied by the basis 
				\begin{equation}
		T_i = \frac{1}{2} \sigma_i \qquad\qquad \text{and} \qquad\qquad  T_{\pm} = T_1 \pm i T_2 \,.
		\end{equation}
		where $\sigma_i$, are the Pauli matrices. 
		To proceed further, we take the following basis for the corresponding loop algebra
		\begin{equation}
		b_{2m+1} = \lambda^m \left( T_+ + \lambda T_- \right) \,,
		\end{equation}
		and
		\begin{equation}
		F_{2m+1} = \lambda^m \left(T_+ - \lambda T_- \right) \quad \text{and} \quad F_{2m} = 2\lambda^m T_3 \,,
		\end{equation}
		where $m \in \mathbb{Z}$, and $\lambda$ is the so-called spectral parameter. The algebra is then given by
		\begin{align}
		[b_{2m+1}, b_{2n+1}] & = 0 \,, \label{5.3}
		\\ [F_{2m+1}, F_{2n+1}] & = 0 \,, \label{5.4}
		\\ [F_{2m}, F_{2n}] & = 0 \,, \label{5.5}
		\\ [b_{2m + 1}, F_{2n + 1}]  & = -2F_{2(m+n+1)} \,, \label{5.6}
		\\ [b_{2m + 1}, F_{2n}] & = -2F_{2(m+n)+1} \,, \label{5.7}
		\\ [F_{2m + 1}, F_{2n}] & = -2b_{2(m+n)+1} \,. \label{5.8}
		\end{align}
We introduce the grading operator as 
\begin{equation}
			d= T_3+ 2\,\lambda \frac{d}{d\lambda} \, \label{gradingop}
		\end{equation}
		such that
		\begin{equation}
			[d,\,b_{2m+1}]=(2m+1)\,b_{2m+1} \qquad \qquad \text{and} \qquad\qquad  [d,\,F_{m}]=m\,F_{m} \,.
		\end{equation}		
The important ingredient in such procedure is that the generator $b_1$ in $A_x$ given in (\ref{5.8.1}), is a semi-simple element of the algebra in the sense that its adjoint action on the loop algebra, splits it into kernel and image without an intersection, i.e. 
 \be
 {\cal G}= {\rm Ker} + {\rm Im}\; ;\qquad\qquad 
 \sbr{b_1}{{\rm Ker}}=0\; ;\qquad\qquad
 {\rm Im}=\sbr{b_1}{{\cal G}}\; ;\qquad\qquad
 {\rm Ker} \cap{\rm Im} = \emptyset
 \ee
 and we have that 
 \be
 {\rm Ker}=\{ b_{2n+1}\; |\; n\in \IZ\} \; ;\qquad\qquad {\rm Im}=\{ F_{n} \; |\; n\in \IZ\}
 \ee
 
 \section{The parameters of the gauge transformation  (\ref{5.9.1})}
 \label{app:parameters}
 
 We give here the explicit expressions for the parameters of the group element performing the gauge transformation (\ref{5.9.1}):
 \br
\zeta_1&=&0
\nonumber\\
\zeta_2&=&
-\frac{1}{24} \alpha u
\nonumber\\
\zeta_3&=&
\frac{1}{48} \alpha u^{(1,0)}
\nonumber\\
&=&\partial_x\left[\frac{1}{48} \alpha u\right]
\nonumber\\
\zeta_4&=&
\frac{1}{288} \left[-\alpha^2 u^2-3 \alpha u^{(2,0)}\right]
\lab{zetaresults}\\
\zeta_5&=&
\frac{1}{192} \left[\alpha^2 u u^{(1,0)}+\alpha u^{(3,0)}\right]
\nonumber\\
&=&\partial_x\left[\frac{1}{192} \left[\frac{\alpha^2}{2} u^2 +\alpha u_{xx}\right]\right]
\nonumber\\
\zeta_6&=&
\frac{1}{10368}\left[-4 \alpha^3 u^3-36 \alpha^2 u^{(2,0)} u-27 \alpha^2
   \(u^{(1,0)}\)^2-27 \alpha u^{(4,0)}\right]
\nonumber\\
\zeta_7&=&
\frac{1}{20736}\left[22 \alpha^3 u^2 u^{(1,0)}+90 \alpha^2 u^{(1,0)}
   u^{(2,0)}+45 \alpha^2 u u^{(3,0)}+27 \alpha
   u^{(5,0)}\right]
   \nonumber\\
&=&\partial_x\left[\frac{1}{20736}\left[\frac{22}{3} \alpha^3 u^3 +\frac{45}{2} \alpha^2 u_x^2+45 \alpha^2 u u_{xx}+27 \alpha
   u_{xxxx}\right]\right]   
\nonumber\\
\zeta_8&=&
\frac{1}{41472}\left[-2 \alpha^4 u^4-36 \alpha^3 u^{(2,0)} u^2-53 \alpha^3
   \(u^{(1,0)}\)^2 u-54 \alpha^2 u^{(4,0)} u-90 \alpha^2
   \(u^{(2,0)}\)^2
   \right. \nonumber\\
   &-& \left. 135 \alpha^2 u^{(1,0)} u^{(3,0)}-27 \alpha
   u^{(6,0)}\right]
   \nonumber
\er
where we have used the notation
\be
\partial_x^m\,\partial_t^n\, u\equiv u^{(m,n)}
\ee
The explicit expressions for the first few components of the time component $a_t$ of transformed connection given in (\ref{abelat}) are given by  
\br
a_t^{(-1)}&=& 
-\frac{\alpha^2}{72} \, u^2-\frac{G}{2}
\nonumber\\
a_t^{(-3)}&=&
\frac{\alpha^2}{288} \, \(u^{(1,0)}\)^2+\frac{ \alpha^2}{288}\, u^2-\frac{\alpha}{24} \, G\, u
\nonumber\\
a_t^{(-5)}&=&
\frac{\alpha^4 u^4}{13824}+\frac{\alpha^3 u^{(2,0)}
   u^2}{1728}+\frac{\alpha^3 u^3}{1728}-\frac{1}{192} \alpha^2 G
   u^2+\frac{\alpha^2 u^{(1,1)} u}{1152}+\frac{1}{576} \alpha^2
   u^{(2,0)} u
   \nonumber\\
   &-&\frac{\alpha^2 \(u^{(1,0)}\)^2}{1152}+\frac{\alpha^2
   \(u^{(2,0)}\)^2}{1152}-\frac{\alpha^2 u^{(0,1)}
   u^{(1,0)}}{1152}-\frac{1}{96} \alpha G u^{(2,0)}
   \nonumber\\
  a_t^{(-7)}&=& 
   \frac{\alpha^5 u^5}{62208}+\frac{5 \alpha^4 u^{(2,0)}
   u^3}{20736}-\frac{5 \alpha^4 \(u^{(1,0)}\)^2 u^2}{82944}+\frac{5
   \alpha^4 u^4}{55296}-\frac{5 \alpha^3 G
   u^3}{6912}+\frac{\alpha^3 u^{(1,1)} u^2}{3456}+\frac{5 \alpha^3
   u^{(2,0)} u^2}{6912}
   \nonumber\\
  & +&\frac{\alpha^3 u^{(4,0)}
   u^2}{6912}+\frac{\alpha^3 \(u^{(2,0)}\)^2 u}{1728}-\frac{\alpha^3
   u^{(0,1)} u^{(1,0)} u}{6912}-\frac{\alpha^3 u^{(1,0)}
   u^{(3,0)} u}{2304}+\frac{\alpha^3 \(u^{(1,0)}\)^2
   u^{(2,0)}}{2304}
   \nonumber\\
   &-&\frac{5 \alpha^2 G u^{(2,0)}
   u}{1152}-\frac{5 \alpha^2 G \(u^{(1,0)}\)^2}{2304}+\frac{\alpha^2
   u^{(3,1)} u}{4608}+\frac{\alpha^2 u^{(4,0)}
   u}{2304}+\frac{\alpha^2 \(u^{(2,0)}\)^2}{4608}-\frac{\alpha^2
   \(u^{(3,0)}\)^2}{4608}
   \nonumber\\
   &+&\frac{\alpha^2 u^{(1,1)}
   u^{(2,0)}}{4608}-\frac{\alpha^2 u^{(1,0)}
   u^{(2,1)}}{4608}-\frac{\alpha^2 u^{(0,1)}
   u^{(3,0)}}{4608}-\frac{\alpha^2 u^{(1,0)}
   u^{(3,0)}}{2304}+\frac{\alpha^2 u^{(2,0)}
   u^{(4,0)}}{2304}-\frac{\alpha G u^{(4,0)}}{384} 
   \nonumber
\er
and
\br
c_t^{(2)}&=& 0\nonumber\\
c_t^{(1)}&=& 0\nonumber\\
c_t^{(0)}&=& 0\nonumber\\
c_t^{(-1)}&=&
-\frac{\alpha^2}{36} \, u^2-\frac{\alpha}{12} \, u^{(2,0)}-\frac{ \alpha}{12}\, u+\frac{G}{2}
\nonumber\\
&=& \frac{X}{2}
\nonumber\\
c_t^{(-2)}&=&\frac{\alpha^2}{24}  \, u\, u^{(1,0)}+\frac{\alpha}{24} \, u^{(0,1)}
   +\frac{\alpha}{24}\,  u^{(1,0)}+\frac{\alpha}{24} \,u^{(3,0)}
   \nonumber\\
   &=&-\frac{1}{4}\, X_x
   \nonumber\\
  c_t^{(-3)}&=& 
   -\frac{1}{432} \alpha^3\, u^3-\frac{1}{36} \alpha^2 \, u^{(2,0)}\,u
   -\frac{1}{48} \alpha^2\, \(u^{(1,0)}\)^2-\frac{1}{144} \,\alpha^2\,u^2
   +\frac{1}{24} \alpha\, G\, u 
   \nonumber\\
  & -&\frac{1}{48}\, \alpha\, u^{(1,1)}
   -\frac{1}{48} \alpha\, u^{(2,0)}
   -\frac{1}{48} \alpha\,   u^{(4,0)}
   \nonumber\\
   &=& \frac{\alpha}{24}\,u\,X+\frac{1}{8}\,X_{xx}
\er


\begin{thebibliography}{1}
		
	\bibitem{afgzcharges}
	H.~Aratyn, L.~A.~Ferreira, J.~F.~Gomes and A.~H.~Zimerman,
	``The Conserved charges and integrability of the conformal affine Toda
	models,''
	Mod.\ Phys.\ Lett.\  A {\bf 9}, 2783 (1994)
	[arXiv:hep-th/9308086].
	
	\bibitem{Benjamin} T. B. Benjamin, J. L. Bona and J. J. Mahoney, \textit{Model equations for long waves in nonlinear dispersive systems}, Phil. Trans. R. Soc. A \textbf{272}, 47-78 (1972)
	
	\bibitem{first_paper} F. ter Braak and W.J. Zakrzewski, \textit{Aspects of modified regularized long-wave equation},  ... (2017)
	
	\bibitem{Bryan} A. Bryan and A. Stuart, \textit{Solitons and the regularized long wave equation: A nonexistence theorem}, Chaos, Solitons \& Fractals, \textbf{7} 11, 1881-1886 (1996)
	
	\bibitem{dickey} L.A. Dickey, Soliton Equations and Hamiltonian Systems, {\em Advanced Series in Mathematical Physics} Vol. 12, (World Scientific Publishing Co. Pte. Ltd. 1991).
		
	\bibitem{drinfeld} 
	V.~G.~Drinfeld and V.~V.~Sokolov,
	``Lie algebras and equations of Korteweg-de Vries type''; 
	J.\ Sov.\ Math.\  {\bf 30}, 1975 (1984);
	``Equations of Korteweg-De Vries type and simple Lie algebras''; Soviet. Mat. Dokl. {\bf 23}, 457 (1981). 
	
	\bibitem{Dunajski} M. Dunajski, \textit{Integrable Systems}, damtp.cam.ac.uk. (2017), [online] Available at: \newline http://www.damtp.cam.ac.uk/user/md327/ISlecture\_notes\_2012.pdf [Accessed 16 May 2017]
	
	\bibitem{Eilbeck} J. C. Eilbeck and G. R. McGuire, \textit{Numerical Study of the Regularized Longe-Wave Equation I: Numerical Methods}, J. Comp. Phys., \textbf{19} 1, 43-57 (1975)
	
	\bibitem{eilbeck2} J. C. Eilbeck and G. R. McGuire, \textit{Numerical Study Of The Regularized Long-Wave Equation II: Interaction Of Solitary Waves}, J. Comp. Phys., \textbf{23} 1, 63-73 (1977)
	
	\bibitem{us}
	L.~A.~Ferreira and W.~J.~Zakrzewski,
  ``The concept of quasi-integrability: a concrete example,''
  Journal of High Energy Physics, JHEP {\bf 1105}, 130 (2011); 
  [arXiv:1011.2176 [hep-th]].
  
  \bibitem{Gibbon} J. D. Gibbon, J. C. Eilbeck and R. K. Dodd, \textit{A modified regularized long-wave equation with an exact two-soliton solution}, J. Phys. \textbf{A} 9, L127-L130 (1976)
  

	\bibitem{kdv} D.J. Korteweg and G. de Vries, On the change of form of long waves advancing in a rectangular canal, and on a new type of long stationary waves, {\em Phil. Mag.}, {\bf 39}, 422-443, (1895).
	
	\bibitem{Kutluay} S. Kutluay, and A. Esen, \textit{A finite difference solution of the regularized long-wave equation}, Mathematical Problems in Engineering, 1 (2006)

\bibitem{oliveabelian}
  D.~I.~Olive and N.~Turok,
  ``Local Conserved Densities And Zero Curvature Conditions For Toda Lattice
  Field Theories'';
  Nucl.\ Phys.\  B {\bf 257}, 277 (1985);
  ``The Toda Lattice Field Theory Hierarchies And Zero Curvature Conditions In Kac-moody Algebras,''
  Nucl.\ Phys.\ B {\bf 265}, 469 (1986).

		\bibitem{Peregrine} D. H. Peregrine, \textit{Calculations of the development of an undular bore}, J. Fluid Mech. \textbf{25}, 321-330 (1966
		
		\bibitem{Schay} G. Schay, \textit{A Concise Introduction to Linear Algebra}, Birkh\"{a}user Boston, (2012)
		
		


	\end{thebibliography}
\end{document}